%% file: thesis.tex
\documentclass[fontsize=11pt, letterpaper, titlepage, abstracton]{scrreprt}

\usepackage[inner=1.in,outer=1.in,top=0.75in,bottom=1.25in]{geometry}	
\usepackage{setspace}
\usepackage[pdftex]{hyperref}
	\hypersetup{
		colorlinks=true,
		linkcolor= Maroon,
		citecolor=Blue,
		urlcolor=Cyan
	}

\usepackage{amsmath,amssymb}
\usepackage{graphicx}
\usepackage[usenames,dvipsnames]{color}
	\definecolor{light-gray}{gray}{0.95}

\usepackage{multicol,multirow}								
\usepackage{bigstrut} 										
\usepackage{rotating}										
\usepackage{threeparttable}									
\usepackage{wasysym}										
\usepackage{units}											
\usepackage{fancyhdr}										
\renewcommand{\vec}[1]{\mathbf{#1}}
\newcommand{\zcr}[0]{\ensuremath{z_{\text{cr}}}}
\newcommand{\tmax}[0]{\ensuremath{t_{\text{max}}}}
\newcommand{\dd}[0]{\ensuremath{{\rm d}}}						

\let\originalincludegraphics\includegraphics
\def\includegraphics[#1]#2{\originalincludegraphics[type=pdf,ext=.pdf,read=.pdf,#1]{#2}}

\allowdisplaybreaks

\setcapindent{0pt}								

\begin{document}

\include{00frontmatter/frontmatter}

\clearpage
\phantomsection
\addcontentsline{toc}{chapter}{Contents}
\begin{singlespace}
\tableofcontents
\end{singlespace}

\include{01intro/intro}
\include{02additional/additional}
\include{03secondaries/secondaries}
\include{04haze/haze}
\include{05epilogue/epilogue}

\include{06backmatter/backmatter}

\end{document}

%% file: 00frontmatter/frontmatter.tex
\begin{singlespace}

\begin{titlepage}

\centering

\vspace*{3cm}

\textsf{\textbf{\Huge Cosmic Ray Backgrounds for \\[0.3\baselineskip]
	Dark Matter Indirect Detection } }

\vspace{2cm}

\textsf{{\huge Philipp Mertsch}}\\[0.2 \baselineskip]

\vspace{3.5cm}

\includegraphics[height=4 cm]{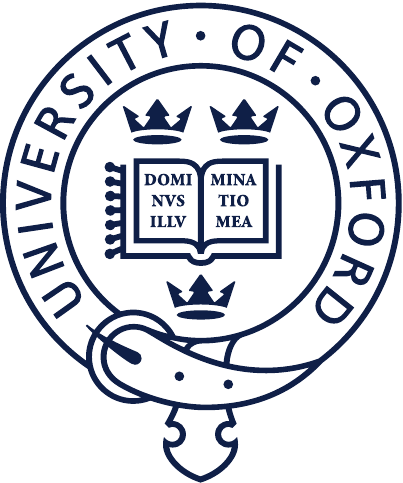}

\vspace{3cm}

\textsf{{\large Balliol College\\[0.2 \baselineskip]
University of Oxford}}

\vspace{1cm}

\textsf{{\large
A thesis submitted for the degree of\\[0.2 \baselineskip]
\textsl{Doctor of Philosophy}\\[0.2 \baselineskip]
Trinity 2010
}}
\vspace{1cm}

\end{titlepage}

\end{singlespace}

\clearpage
\phantomsection
\addcontentsline{toc}{chapter}{Abstract}
\chapter*{Abstract}

\thispagestyle{fancy}
\renewcommand{\headrulewidth}{0pt} 

\addtolength{\headheight}{3.6cm}
\addtolength{\headsep}{-3.6cm}
\rhead{ 
\begin{doublespace}
  \textsf{\textbf{\large Cosmic Ray Backgrounds for Dark Matter Indirect Detection.}}\\
  \textsf{Philipp Mertsch, Balliol College.}\\
  \textsf{A thesis submitted for the degree of \textsl{Doctor of
        Philosophy}. Trinity 2010.}
\end{doublespace}
}

\noindent The identification of the relic particles which presumably
constitute cold dark matter is a key challenge for astroparticle
physics. Indirect methods for their detection using high energy
astrophysical probes such as cosmic rays have been much discussed. In
particular, recent `excesses' in cosmic ray electron and positron
fluxes, as well as in microwave sky maps, have been claimed to be due
to the annihilation or decay of dark matter. In this thesis, we argue
however that these signals are plagued by irreducible astrophysical
backgrounds and show how plausible conventional physics can mimic the
alleged dark matter signals.

In chapter~\ref{chp:intro}, we review evidence of, and possible
particle candidates for, cold dark matter, as well as our current
understanding of galactic cosmic rays and the state-of-the-art in
indirect detection. All other chapters contain original work, mainly
based on the author's journal
publications~\cite{Mertsch:2009ph,Ahlers:2009ae,Mertsch:2010ga}. 
In particular, in chapter~\ref{chp:additional}, we consider the
possibility that the rise in the positron fraction observed by the
PAMELA satellite is due to the production through (hadronic) cosmic
ray spallation and subsequent acceleration of positrons, in the same
sources as the primary cosmic rays. We present a new (unpublished)
analytical estimate of the range of possible fluctuations in the high
energy electron flux due to the discreteness of plausible cosmic ray
sources such as supernova remnants. Fitting our result for the total
electron-positron flux measured by the Fermi satellite allows us to
fix the only free parameter of the model and make an independent
prediction for the positron fraction. Our explanation relies on a
large number of supernova remnants nearby which are accelerating
hadronic cosmic rays. Turning the argument around, we find
encouraging prospects for the observation of neutrinos from such
sources in km$^3$-scale detectors such as IceCube.

Chapter~\ref{chp:secondaries} presents a test of this model by
considering similar effects expected for nuclear secondary-to-primary
ratios such as B/C. A rise predicted above $\mathcal{O}(100) \, \text{GeV/n}$
would be an unique confirmation of our explanation for a rising
positron fraction and {\em rule out} the dark matter explanation.

In chapter~\ref{chp:haze}, we review the assumptions made in the
extraction of the `WMAP haze' which has also been claimed to be due to
electrons and positrons from dark matter annihilation in the Galactic
centre region. We argue that the energy-dependence of their diffusion
means that the extraction of the haze through fitting to templates of
low frequency diffuse galactic radio emission is unreliable. The
systematic effects introduced by this can, under specific
circumstances, reproduce the residual, suggesting that the `haze' may
be just an artefact of the template subtraction.

We present a summary and thoughts about further work in the epilogue.

\clearpage
\phantomsection
\addcontentsline{toc}{chapter}{Acknowledgements}
\chapter*{Acknowledgements}

First and foremost, I want to express my deeply felt gratitude towards
my DPhil advisor, Prof Subir Sarkar. Over the last three years, Subir
has granted me the freedom I wanted and given me the guidance I
needed, providing the best possible environment for the completion of
this thesis. I consider myself lucky having had such a committed,
diligent and supportive supervisor. His vast knowledge and the
enlightening discussions with him have left a lasting imprint on me.

As an Early Stage Researcher in the EU Research and Training Network
``UniverseNet'' I have benefited from generous financial support
(supplemented by an STFC Postgraduate Studentship). Being part of this
community and meeting young researchers from all over the world has
been an enriching experience which I am grateful for.

Furthermore, I thank Markus Ahlers for his fruitful collaboration and
the inspiring discussions we had during his time at Oxford and
afterwards. There are many more people at Oxford who have contributed
in one way or the other, from the helpful admin staff, my office mates
and the other ``first years'' to all of my friends who had to put up
with my going into hiding from time to time.

Finally, I want to thank my family, who I owe my education and my
critical mind to, for their continuous support. It is Anneliese who
has carried most of the burden over the last three years and whose
witty advice and loving support has been the most precious gift.

%% file: 01intro/intro.tex
\chapter{Introduction} 
\label{chp:intro}

\section{Dark Matter}

One of the most astounding results of modern astrophysics and
cosmology is the finding that about $20 \, \%$ of the energy density
of the Universe and about $80 \, \%$ of the matter in the Universe is
in the form of non-baryonic dark matter (DM). It turns out that DM
plays an important role in structures on a wide range of scales
ranging from faint satellite galaxies to the largest known structures
of the Universe. DM, however, also presents a veritable problem for
particle physics. Since the discovery of the discrepancy between the
amount of luminous matter and total matter, many different pieces of
evidence have been gathered, the most important of which we review in
Sec.~\ref{sec:EvidenceDM}. In Sec.~\ref{sec:DMcandidates} we discuss
some relic particle candidates for DM, emphasising their connection to
new particle physics beyond the Standard Model.

\subsection{Evidence for dark matter}
\label{sec:EvidenceDM}

The probably most commonly known piece of evidence for DM (evidence at
least for matter with a much larger mass-to-light ratio $M/L$ than
usual) comes from galaxy rotation curves, that is the variation of
rotational velocity of stars with distance $r$ from the galactic
centre. Most of the luminous mass of spiral galaxies is in fact
contained in the central bulge, and assuming Newtonian gravity and
dynamics, the rotational velocity of stars should increase up to a
certain radius $r_\text{lum}$ that contains most of the luminous
matter, and decrease beyond that like $1/\sqrt{r}$. What Vera Rubin
however discovered~\cite{Rubin:1970} in 1970 for the nearby spiral
galaxy M31, Andromeda, was a \emph{flattening} in the rotation curve
beyond the radius $r_\text{lum}$, implying that the Galaxy contains
mass with a distribution different from the distribution of luminous
matter. Flat or even increasing rotation curves have been also found
for many other spiral galaxies~\cite{Sofue:2001}, including the Milky
Way (for a recent analysis, see~\cite{Xue:2008se}).

Already almost 40 years earlier, the amount of luminous matter in the
Coma cluster had been determined~\cite{Zwicky:1933,Zwicky:1937} by
Fritz Zwicky, applying the virial theorem to measurements of the
velocity of individual galaxies. He discovered that the amount of
matter to explain the large velocity dispersion of up to $1000 \,
\text{km} \, \text{s}^{-1}$ would require $M/L$ up to hundred times
the solar value.

Another possibility of inferring the mass of a distant system, e.g. a
galaxy cluster, is by gravitational lensing (for a review,
see~\cite{Bartelmann:1999yn}). This makes use of the bending of light
from a distant bright source, like a quasar, by the gravitational
potential in between the source and the observer, as predicted by
General Relativity. If the distant source, the observer and the mass
in between are aligned, the picture of the source becomes strongly
distorted to multiple images or arcs which allows the determination of
the mass of the object in between. Such constellations that lead to
this so-called \emph{strong} lensing are however rare and the effect
on the shape of individual sources is mostly too small to determine
the gravitational potential. It is however possible to analyse the
systematic alignment of a set of background galaxies around the
interjacent mass. This technique, called \emph{weak} gravitational
lensing, is able to resolve the (DM dominated) mass distribution
purely by statistical means. The strength of the lensing signal
compared to the intrinsic, so-called ``shape noise'' depends on the
nature of the interjacent mass: Lensing by galaxy clusters usually
gives the strongest signals and can be used in conjunction with
observations of baryonic matter to determine the mass-to-light ratio
of these systems. Lensing by foreground galaxies has much smaller
effects (due to the smaller mass of galaxies compared to clusters) and
usually requires combining, called ``stacking'', the signals from many
galaxy lenses. Finally, it is also possible to determine the alignment
of background galaxy shapes along the large scale structure of the
Universe, called ``cosmic shear'' and resolve the three dimensional
(again DM dominated) mass distribution using also redshift
information. For instance, this allows the determination of the matter
power spectrum and to constrain important cosmological parameters. The
general advantage of gravitational lensing is that the mass of an
object can be estimated without any assumptions about its dynamical
state or even without any detectable baryonic matter in the first
place. In the cases where such information is available, the
inferences on the mass, from weak or strong lensing and from kinematic
measurements, usually agree very well.

Of course, the above observational results do not require that the
non-luminous form of missing matter is necessarily non-baryonic. In
fact, a certain fraction of the non-luminous matter may well be in the
form of heavy, compact objects, like black holes, neutron stars, white
dwarfs or brown dwarfs, collectively referred to as MACHOs (MAssive
Compact Halo Objects), and diffuse, hot, interstellar and
intergalactic gas. However, the total estimated mass (including DM) of
astronomical objects (galaxies, cluster, large scale structure) is
much higher than the total amount of \emph{baryonic} matter which can
be determined from cosmological measurements as follows. The abundance
of light elements as predicted by big bang nucleosynthesis
(BBN)~\cite{Alpher:1948,PDG2010} is a function of the baryon abundance
$\Omega_b$ where $\Omega_i$ denotes energy densities in terms of the
critical density $\rho_c = 3 H^2/(8 \pi G)$ with the Hubble parameter
$H$ and Newton's gravitational constant $G$. Observations of the
(primordial) Deuterium abundance constrain the baryon abundance to
$\Omega_b h^2 \approx 0.02$, where $h = H / (100 \, \text{km} \,
\text{s}^{-1} \, \text{Mpc}^{-1}) \approx 0.7$. Baryons therefore only
account for about $20 \, \%$ of the matter in the
Universe. Furthermore, the power spectrum of cosmic microwave
background (CMB) anisotropies is also sensitive to the amount of
baryonic and the total amount of matter through the acoustic
peaks. More precisely, the recent release~\cite{Larson:2010gs} of the
WMAP 7-year data reports $\Omega_\text{cdm} h^2 = 0.1109 \pm 0.0056$
and $\Omega_b h^2 =0.02258 ^{+0.00057}_{-0.00056}$ for $H = h \, 100
\, \text{km} \, \text{s}^{-1} \, \text{Mpc}^{-1} = 71.0 \pm 2.5 \,
\text{km} \, \text{s}^{-1} \, \text{Mpc}^{-1}$.

According to the concordance model, the same quantum fluctuations that
lead to the anisotropies in the CMB are also the seeds for the
formation of structure on all scales in the Universe today which
provides more evidence for DM. On the one hand, the amount of density
perturbations at the time of recombination is rather small (the CMB is
isotropic to one part in $10^5$). On the other hand, structure
formation through gravitational collapse could not start from baryonic
matter before the time of recombination because of electromagnetic
interactions. It turns out that the time that baryonic matter alone
therefore could have had to form structure is much too
short. Non-baryonic DM, however, would have started collapsing before
recombination and it turns out it can explain structure formation
rather accurately as shown by $N$-body simulations
~\cite{Springel:2005nw,Kuhlen:2008qj,Diemand:2005vz}. Furthermore,
structure formation requires that DM is non-relativistic at and after
the time of recombination, that is cold dark matter (CDM).

Other ideas, like MOdification of Newtonian Dynamics
(MOND)~\cite{Milgrom:1983ca} (for a review, see~\cite{Sanders:2002pf})
and particular theoretical representations, for example, Tensor Vector
Scalar (TeVeS) theory~\cite{Bekenstein:2004ne} can reproduce some of
the above mentioned results without the need for new, non-baryonic
DM. For example, MOND reproduces galactic rotation curves better than
DM, in particular needing less free parameters. Furthermore, it gives
an explanation of the ``Tully-Fisher relation'', that is the observed
correlation between luminosity and rotation velocity, but fails on the
scales of galaxy clusters~\cite{Sanders:2002pf}.

In particular, the `smoking gun' signature for DM comes from the
so-called `bullet cluster'~\cite{Clowe:2006eq}. Weak lensing and x-ray
expose the distribution of dark matter and hot gas (which makes up the
majority of baryonic matter), respectively, in two galaxy clusters
that have collided $\sim 100 \, \text{Myr}$ ago.  Not only are the
distributions clearly distinct, but they also prove that while the hot
gas has interacted (electromagnetically) during the collision, the DM
has just passed through without interacting. Such a behaviour is
basically impossible to imitate by modifying gravity.

Despite all the successes of the CDM paradigm in explaining the above
mentioned astrophysical as well as cosmological data, there are a
large number of questions/problems that CDM cannot answer (so far):
Why are the DM haloes predicted by (pure DM) $N$-body simulations much
``cuspier'' than those actually observed? Will including baryonic
matter in $N$-body simulations flatten the profile in the inner
kiloparsecs? Why do we observe less satellites (DM substructure) than
predicted by $N$-body simulations? Can galaxy dynamics, e.g. SN
feedback or (local) reionisation~\cite{Benson:2010de} reconcile this
discrepancy? Why are there less massive galaxies today than expected?
Although the quantitative discrepancy between the predictions and
observations in some of these cases might be rather large, the
expectation is that the uncertainty due to astrophysics is also still
quite large and more realistic modelling will resolve many, if not all
contradictions.

\subsection{Particle candidates}
\label{sec:DMcandidates}

What all the different pieces of evidence presented in the last
section have in common, is that they are based on the gravitational
interaction of DM only. However, from a particle physicist's viewpoint
it is desirable to describe DM by a fundamental particle, in a similar
way that the standard model (SM) of particle physics describes the
fundamental matter particles and force carriers. The properties that
characterise a particle in a relativistic quantum field theory are its
mass, spin, quantum numbers under (gauge) transformations and also its
coupling to other (SM) particles.

Even before considering any particular model, the astrophysical
evidence of DM gives us some constraints on its particle physics
nature. First of all, the DM particle needs to be stable on
cosmological timescales, that is its lifetime must be larger than the
age of the Universe. Secondly, zero results for searches for exotic,
heavy nuclei on Earth constrain DM to be at most weakly interacting
since DM would form such states if it was interacting via strong or
electromagnetic forces. Furthermore, DM must be cold to be able to
explain structure formation. This rules out the only known neutral and
purely weakly interacting particle, the neutrino, as at least one mass
eigenstate is relativistic today.  Hence, none of the SM particles can
accommodate for CDM and this clearly hints at physics beyond the
SM. Interestingly, many theories of new physics that address the
short-comings of the SM predict new particles including some therefore
well-motivated dark matter candidates.

\subsubsection{WIMPs} 

The SM has been very successful in providing a fundamental theory of
matter and its interactions that has been tested to a great accuracy
in colliders and non-accelerator experiments. However, although the
Higgs mechanism explains the breaking of electroweak symmetry, one of
the short-comings considered more serious is the so-called hierarchy
problem: How is the weak scale stabilised with respect to radiative
corrections that would normally boost it to the Planck scale? One of
the most elegant solutions to this problem is supersymmetry (SUSY)
which entangles the usual Poincar\'e algebra with a new set of
generators transforming fermionic into bosonic degrees of freedom and
{\it vice versa}. In its simplest implementation, this effectively
amounts to roughly doubling the particle content of the SM by
mirroring each SM particle in a supersymmetric partner (sparticle)
with a spin different by $\nicefrac{1}{2}$. These sparticles cancel
the quadratic divergencies arising in radiative corrections of the
Higgs mass parameter and hence stabilise the weak scale. To keep the
necessary cancellations natural, the new particles must have masses
$m_\text{DM}$ close to the weak scale. Among these new particles there
are a number of DM candidates.

If such new particles at the weak scale were also weakly interacting,
that is they are weakly interacting massive particles (WIMPs), then
their relic density today as predicted from production by freeze-out,
i.e. thermal
decoupling~\cite{Zeldovich:1965,Chiu:1966kg,Steigman:1979kw,Lee:1977ua,Dicus:1977nn,Vysotsky:1977pe},
is in excellent agreement with the constraints from astrophysics and
cosmology (in particular $\Omega_\text{cdm}$ from the WMAP experiment,
see above). More precisely, at early times, the DM particles are in
thermal equilibrium. As the Universe cools, their density is being
suppressed by the Boltzmann factor $\propto {\rm e}^{-m_\text{DM}/T}$
and for late times and hence low temperatures $T$, the abundance would
normally vanish. However, apart from cooling, the Universe also
expands, such that the rate of production/annihilation becomes smaller
than the Hubble rate and the WIMPs drop out of thermal
equilibrium. Quantitatively, the WIMP density $n$ is governed by the
Boltzmann (continuity) equation (first given
in~\cite{Zeldovich:1965}),
\begin{equation}
\frac{{\rm d} n}{{\rm d} t} = -3 H n - \langle \sigma_\text{ann} v
\rangle \left( n^2 - n_\text{eq}^2 \right)
\end{equation}
where $H$ is the Hubble rate, $\langle \sigma_\text{ann} v \rangle$
the thermal average of the WIMP annihilation cross section and
velocity and $n_\text{eq}$ the WIMP equilibrium density. In general,
this needs to be solved numerically but a simple analytical estimate
gives
\begin{equation}
\label{eqn:OmegaWIMP}
\Omega_\text{WIMP} \sim \frac{x_f T_0^3}{\rho_c M_\text{Pl}} \langle
\sigma_\text{ann} v \rangle^{-1} \, .
\end{equation}
with $x_f = m_\text{WIMP} / T_f \approx 20$, $T_f$ ($T_0$) the
temperature at freeze-out (today) and $M_\text{Pl}$ the Planck
mass. The WIMP mass does not enter $\Omega_\text{WIMP}$ directly but
in most theories it is the only mass scale that determines the
annihilation cross section. In particular, for weak interactions and a
mass at the weak scale, one finds an s-wave annihilation cross
section,
\begin{equation}
\label{eqn:sigmaAv}
\sigma_A v \approx \frac{g^4}{16 \pi^2 m_\text{WIMP}^2} \approx 3
\times 10^{-26} \, \text{cm}^3 \, \text{s}^{-1} \, ,
\end{equation}
that reproduces the relic density $\Omega_\text{WIMP} =
\mathcal{O}(1)$. This fact has been called the `WIMP miracle'.

One might wonder why such a new, weak-scale particle should not decay,
seeing that all SM particles above a GeV are unstable. Usually, one
postulates a discrete symmetry that forbids interactions leading to
decay of these particles. In the case of the minimal supersymmetric
standard model (MSSM) this symmetry is called $R$-parity. Originally,
it was introduced \cite{Farrar:1978xj} to satisfy bounds on the proton
lifetime. However, to protect the proton it is enough to forbid one of
the interactions necessary for its decay which can be achieved by,
e.g. baryon or lepton parity. Therefore, to make the lightest
supersymmetric particle (LSP) stable, $R$-parity has to be introduced
by hand. It was also argued \cite{Campbell:1990fa} that $R$-parity
violating interactions must be suppressed to prevent the washout of
the cosmic baryon/lepton asymmetry but this argument does not hold
when lepton-mass effects are included \cite{Dreiner:1992vm}.

Examples for WIMPs arise naturally in weak scale SUSY theories as
motivated above. If SUSY breaking is gravitationally mediated, the
lightest supersymmetric particle (LSP) is usually the
neutralino~\cite{Goldberg:1983nd,Ellis:1983ew}, a mixture of the
supersymmetric partners of the hypercharge gauge boson, the neutral
component of the W boson and the neutral higgs partners.

Theories of universal extra dimension (UED)~\cite{Appelquist:2000nn},
for example, do not try to address the hierarchy problem, but also
lead to new particles at the weak scale and thus WIMP DM
candidates~\cite{Cheng:2002ej,Servant:2002aq}. The general idea is
that all SM particles propagate in a higher dimensional space. In the
simplest version, a single additional dimension is compactified on an
$S_1/Z_2$ orbifold of radius $R \sim 10^{-8} \, \text{m}$ or
smaller. In 4 dimensions this leads to an infinite spectrum for each
SM particle, so-called Kaluza-Klein (KK)~\cite{Klein:1926tv}
particles, equally separated in mass by $R^{-1}$. KK particles posses
a discrete symmetry, $K$-symmetry, which makes the lightest KK
particle, the LKP, stable. In most models, the LKP is the first KK
state of the hypercharge gauge boson, $B^1$.

\subsubsection{superWIMPs} 

On the one hand, the big advantage of the WIMP ``miracle'' is that it
predicts the right DM abundance for stable, weak particles. On the
other hand, constraining ourselves to stable and weakly interacting
particles turns out to be too rigid a presumption. In fact, a number
of the new particles suggested by beyond standard model (BSM) theories
turn out to be either unstable or interact much more weakly.

However, if every WIMP produced by thermal decoupling would decay to a
``superWIMP'', that is a particle which is also at the weak scale but
is only ``super weakly'' interacting, the virtue of the right relic
density can be saved~\cite{Feng:2003xh,Feng:2003uy},
\begin{equation}
\Omega_\text{superWIMP} = \frac{m_\text{superWIMP}}{m_\text{WIMP}}
\Omega_\text{WIMP} \, .
\end{equation}
Another production mechanism is
reheating~\cite{Krauss:1983ik,Nanopoulos:1983up,Ellis:1984eq,Ellis:1984er}. It
can be shown that today's density is proportional to the reheating
temperature $T_\text{R}$ and one finds $T_\text{R} \simeq 10^{10} \,
\text{GeV}$ for a $100 \, \text{GeV}$ gravitino superWIMP.

Weak scale gravitinos are in fact a typical example of
superWIMPS~\cite{Ellis:1984er,Feng:2003xh,Feng:2003uy,Buchmuller:2004rq,Ellis:2003dn,Wang:2004ib,Roszkowski:2004jd},
realised in SUSY theories with gravity mediated SUSY breaking. As
there is no reason to believe that gravitinos are systematically
lighter or heavier than the other superpartners in these scenarios,
the gravitino is the LSP in about half the parameter space. The role
of the WIMP is then played by the next-to-lightest SUSY particle
(NLSP), for example the stau, which can decay to the gravitino with
lifetimes naturally of the order of hours to months.

Another possible example are axinos, the superpartner of axions (see
below), and both could be possibly contributing in a multi-component
DM scenario~\cite{Baer:2009vr}.

\subsubsection{Hidden dark matter}

The fact that DM is not interacting through electromagnetic or strong
forces has led us to the conclusion that it could at most be weakly
interacting. The alternative, i.e. that it is only gravitationally
interacting is usually disfavoured because of lack of predictivity,
missing connections to new physics and the loss of an automatic
prediction for the relic density. Generally speaking, constraints on
hidden sectors can only be obtained from their gravitational
interactions, e.g. from constraints on the expansion rate at
BBN~\cite{Kolb:1985bf}. There are however counter-examples resolving
some or all of these problems and in the following we mention one
particular class of examples that provide a DM candidate with the
correct relic density from thermal freeze-out, too.

In fact, going back to Eq.~\ref{eqn:sigmaAv}, one realises that every
extension of the SM that predicts similar ratios of $g^4/m^2$ will
reproduce the observed relic density, no matter whether $g$ is a SM
coupling. Let's consider for example a setup of gauge mediated SUSY
breaking which one could imagine is also mediated to a hidden sector
whose particles are not charged under any SM gauge group; string
theories, for example, predict many such hidden sectors. The masses of
the hidden sector superpartners, $m_h$, are however set in the same,
generation-independent way as the SM superpartner slepton masses, $m$,
by the factor $F/M_m$ from the SUSY breaking sector,
\begin{equation}
m \sim \frac{g^2}{16 \pi^2} \frac{F}{M_m} \quad \text{and similarly}
\quad m_h \sim \frac{g_h^2}{16 \pi^2} \frac{F}{M_m} \, ,
\end{equation}
with $g_h$ the hidden sector gauge coupling. Therefore, the ratio
$m_h/g_h^2$ that determines the relic density from freeze-out, see
Eqs.~\ref{eqn:OmegaWIMP} and~\ref{eqn:sigmaAv}, is universal and we
expect that the hidden sector contains (some) DM candidates with just
the right relic abundance. This relation is called the ``WIMPless
miracle''~\cite{Feng:2008ya}.

\subsubsection{Asymmetric dark matter}

It is questionable whether the freeze-out paradigm for DM production
is necessarily the right one. Applied to baryon production, for
example, it fails spectacularly -- not only, because the predicted
baryon abundance, $n_B / n_\gamma \approx 10^{-19}$, is about ten
orders of magnitude smaller than the observed one, but also because it
cannot explain the observed matter-antimatter asymmetry. Instead, the
baryon abundance can be explained from an asymmetry of baryons over
anti-baryons generated, for example, by non-perturbative sphaleron
processes. If the annihilation cross section of baryons and
antibaryons is large enough that all antibaryons will have annihilated
away, we are left with the right density of baryons only.

The fact that the abundance of DM and baryons is only different by a
factor 5, a relation which in a freeze-out scenario of DM would be
merely coincidental, may hint at a common physical
origin~\cite{Nussinov:1985xr,Kaplan:1991ah}. If one wants to explain
the DM abundance today from an asymmetry generated in a similar way,
DM must have a quantum number $B'$, similar to the baryon number
$B$. At temperatures above the electroweak phase transition, $T >
T_*$, electroweak anomalous processes are in thermal equilibrium and
equilibrate the lepton, baryon and DM asymmetries. Depending on wether
the DM mass $m_\text{DM}$ is smaller or larger than $T_*$, the $B'$
asymmetry gets frozen at $Y_{\Delta B}$ or Boltzmann suppressed as
$Y_{\Delta B'} / Y_{\Delta B} \sim {\rm e}^{- m_\text{DM}/T_*}$, and
the DM abundance, $\Omega_{B'}$, can be related to the baryon
abundance, $\Omega_B$,
\begin{align}
\frac{\Omega_{B'}}{\Omega_B} = c \left\{ \begin{array}{ll}
  \frac{m_\text{DM}}{m_B} & \text{for} \; m_\text{DM} \ll T_* \, ,
  \\ 12 \frac{m_\text{DM}}{m_B} \left( \frac{m_\text{DM}}{2 \pi T_*}
  \right)^{3/2} {\rm e}^{-m_\text{DM}/T_*} & \text{for} \, m_\text{DM}
  \gg T_* \, ,
\end{array} \right.
\end{align}
with $c$ an order one parameter and $m_B$ the nucleon mass. The
observed DM abundance \mbox{$\Omega_{B'} = 5 \, \Omega_B$} can
therefore be recovered by $m_\text{DM} \sim 5 \, \text{GeV}$ for
$m_\text{DM} \ll T_*$~\cite{Kaplan:1991ah} or $m_\text{DM} \sim 2 \,
\text{TeV}$ for $m_\text{DM} \gg
T_*$~\cite{Nussinov:1985xr}. Technicolour, a strongly interacting
theory suggested for electro-weak symmetry breaking, for example,
rather naturally predicts \emph{techni}-baryons with TeV masses.

\subsubsection{Axions}

The axion was originally postulated to solve the so-called `strong CP
problem'~\cite{Peccei:1977hh,Wilczek:1977pj,Weinberg:1977ma}. The QCD
Lagrangian contains a four-divergence proportional to the angle
$\bar{\theta}$ that would lead to P and CP violation through
non-perturbative effects. The non-observation of, e.g. a neutron
dipole moment, however, constrains $|\bar{\theta}|$ to $<
10^{-10}$. The most elegant way to guarantee a small value is the
Peccei-Quinn mechanism that promotes $\bar{\theta}$ to a dynamical
field with a classical potential that is minimised at $\bar{\theta} =
0$. This is achieved by adding an additional, chiral symmetry
$U(1)_\text{PQ}$, the Peccei-Quinn symmetry, which is spontaneously
broken at the scale $f_a$. The pseudo Nambu-Goldstone boson is the
axion $a$; ``pseudo'', because the global symmetry is not exact at the
quantum level and therefore the axion is not massless but has a mass
of order $\Lambda_\text{QCD}/f_a$,
\begin{equation}
m_a \simeq 0.6 \, \text{eV} \frac{10^7 \, \text{GeV}}{f_a} \, .
\end{equation}
Also, this anomaly leads to a potential for the axion, fixing it to
around $\langle a \rangle = - \bar{\theta} f_a / \text{const.}$ which
cancels the above four-divergence, thus solving the strong
CP-problem. Expanding the axion field around its minimum $\langle a
\rangle$ one obtains the axion Lagrangian which is still to be
complemented by the axion couplings to, e.g. photons and fermions,
\begin{equation}
\mathcal{L}_{a \gamma \gamma} = - g_{\gamma} \frac{\alpha}{\pi}
\frac{a}{f_a} \vec{E} \cdot \vec{B} \quad \text{and} \quad
\mathcal{L}_{a \bar{f} f} = i g_f \frac{m_f}{v} a \bar{f} \gamma_5 f
\, ,
\end{equation}
where $\alpha$ is the fine-structure constant, $\vec{E}$ and $\vec{B}$
the (colour) electric and magnetic field, respectively, and
$g_{\gamma}$, $g_f$ are model-dependent coefficients of order one.

There is a variety of constraints on the axion mass. Collider searches
for rare decays \mbox{$\pi^+ \rightarrow a(e^+ e^-) e^+ \nu_e$} rule
out very short-lived (lifetime $< 10^{-11} \, \text{s}$) axions with
masses above 1 GeV. For long-lived axions (lifetime $> 10^{-11} \,
\text{s}$) production ($p + N \rightarrow a + X$ or $e + N \rightarrow
a + X$) as well as interaction ($a + N \rightarrow X$) cross sections
are constraint from beam dumps and rule out axions heavier than 50
keV~\cite{Eichler:1986nj}. Astrophysical
constraints~\cite{Raffelt:1990yz} like bounds on the lifetime of red
giants limit the cooling due to axions and give $200 \, \text{keV}
\gtrsim m_a \gtrsim 0.5 \, \text{eV}$ ($\gtrsim 10^{-2} \, \text{eV}$
even, if the coupling to electrons is large). The duration of the
neutrino burst observed from SN 1987a finally also limits the fraction
of axion cooling and excludes $2 \, \text{eV} \gtrsim m_a \gtrsim 3
\times 10^{-3} \, \text{eV}$. Axions lighter than $10^{-6} \,
\text{eV}$ are ruled out because they would over-close the Universe
($\Omega_a \propto f_a \propto m_a^{-1}$). The axion is thus extremely
light and weakly interacting, making it a potential DM
candidate~\cite{Preskill:1982cy,Abbott:1982af,Dine:1982ah}.

Depending on whether the reheating temperature of inflation is smaller
or larger than the temperature $T_\text{PQ}$ of the spontaneous
breaking of the PQ symmetry, the axion field is homogenised over
enormous distances or carries strings and domain walls as topological
defects. Axions therefore get produced by vacuum-realignment
only~\cite{Bae:2008ue} or also by string and domain wall decay, and
the critical density today is
\begin{equation}
\Omega_a \approx 0.15 \left( \frac{f_a}{10^{12} \, \text{GeV}}
\right)^{7/6} \left( \frac{0.7}{h} \right)^2 \alpha_1^2 \quad \Bigg(
\Omega_a \approx 0.7 \left( \frac{f_a}{10^{12} \, \text{GeV}}
\right)^{7/6} \left( \frac{0.7}{h} \right)^2 \Bigg) \, ,
\end{equation}
with $\alpha_1$ the initial misalignment angle.

\section{Cosmic Rays} 
\label{sec:SMofGCRs}

\subsection{Galactic and extra-galactic cosmic rays} 
\label{sce:GalAndXgalCRs}

The Earth's atmosphere is constantly bombarded by a flux of elementary
particles: cosmic rays (CRs). The spectrum of these particles measured
on Earth covers 12 orders of magnitude in energy -- corresponding to
40 octaves in frequency! On the other hand, the steeply falling power
law spectrum between $E^{-2.7}$ and $E^{-3}$ implies quickly declining
flux rates of $1 \, \text{particle} \, \text{m}^{-2} \, \text{s}^{-1}
\, \text{sr}^{-1}$ above $100 \, \text{GeV}$, $1 \, \text{particle} \,
\text{m}^{-2} \, \text{yr}^{-1} \, \text{sr}^{-1}$ above $10^{16} \,
\text{eV}$ and $1 \, \text{particle} \, \text{km}^{-2} \,
\text{yr}^{-1} \, \text{sr}^{-1}$ above $10^{19} \, \text{eV}$.

\subsubsection*{Spectrum}

\begin{figure}
\begin{center}
\includegraphics[width=\textwidth]{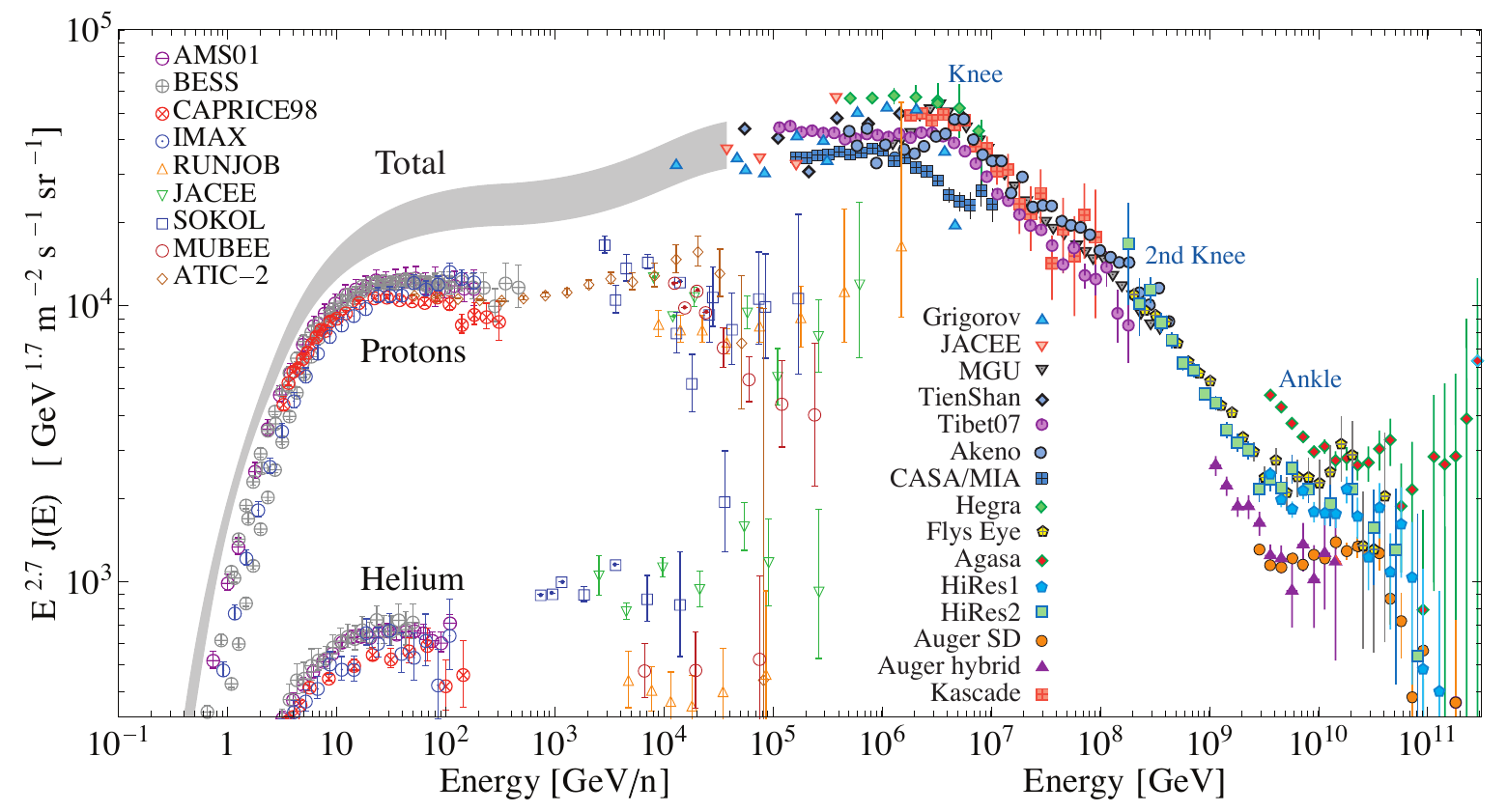}
\end{center}
\caption{Flux of galactic and extra-galactic cosmic rays, scaled by
  $E^{2.7}$ (from~\cite{PDG2010} and with additional
  data~\cite{GCR_data}). The energy for the helium flux is in
  $\text{GeV} / \text{n}$.}
\label{fig:CosmicRaySpectrum}
\end{figure}

The cosmic rays with energies between $\sim 1 \, \text{GeV}$ and (at
least) $\sim 3 \times 10^{15} \, \text{eV}$ are considered to be of
galactic origin and are hence called Galactic Cosmic Rays (CGRs). They
exhibit a rather featureless power law with spectral index $\alpha
\approx -2.75$. The low energy cut-off is due to solar modulation,
that is cosmic rays of energy lower than a few hundred MeV lose all
their energy by running up against the electric potential generated by
the solar wind~\cite{Gleeson:1968zz}. At $\sim 3 \times 10^{15} \,
\text{eV}$, a softening to $\alpha \approx -3.1$ is observed, a
feature called the ``knee''. At around $5 \times 10^{17} \, \text{eV}$
the spectrum further softens to $\alpha \approx -3.3$ (``second
knee''). As the Larmor radius at these energies starts exceeding the
spatial dimensions of the Galaxy of kiloparsecs, cosmic rays at least
beyond this energy must be of extra-galactic origin. At $3 \times
10^{18} \, \text{eV}$, the spectrum hardens again to $-2.7$
(``ankle''), before it gets cut off at $\sim 5 \times 10^{19} \,
\text{GeV}$~\cite{Abraham:2008ru}, probably due to the so-called GZK
cut-off, the suppression of the flux by photo-pion production on the
CMB~\cite{Greisen:1966jv,Zatsepin:1966jv}. The flux of cosmic rays
measured on Earth, scaled by $E^{2.7}$ to amplify the features, is
shown in Fig.~\ref{fig:CosmicRaySpectrum}.

Below a GeV cosmic rays must be of local, i.e. solar system origin. In
fact, the solar wind is a prime laboratory for testing models of
cosmic ray acceleration (e.g. by shock waves) observed by
interplanetary probes, like the Voyager spacecrafts. As these
low-energy particles are however not important for dark matter
indirect searches and cosmic ray backgrounds we do not consider solar
cosmic rays any further.

\subsubsection*{Composition}

Most chemical elements observed in GCRs occur in ratios that are
consistent with their relative abundance in the solar system, hinting
at a global validity of these abundances elsewhere in the Galaxy,
potentially even the local Universe. Some elements however, in
particular those which are rare in the solar system, are overabundant
in cosmic rays. The interpretation is that although the source
abundances are similar to solar system values, these so-called
`secondaries' get produced from the more abundant `primaries' by
spallation on the interstellar medium. This explanation gives in turn
estimates of the average matter or `grammage' of a few tens $\text{g}
\, \text{cm}^{-2}$ that primaries must traverse to produce the
observed number of secondaries. The energy dependence of the ratios of
secondaries-to-primaries is an important test of the propagation model
(see Sec.~\ref{sec:ObservationalResults}). It is worth noting that the
electronic component has a softer spectrum (see
Sec.~\ref{sec:Total_e-e+_Flux}) and is smaller than the nuclear one by
$\sim 10^{-2}$ at $10 \, \text{GeV}$.

The change of composition at spectral features is used to infer the
nature of the sources of ultra high energy cosmic rays
(UHECRs). Besides the observation of an increasingly heavy composition
around the knee (see above), particular interest has been generated by
recent contradictory observations by the HiRes~\cite{Abbasi:2009nf}
and Auger~\cite{Abraham:2010yv} experiments at the ankle. However, it
seems likely that this discrepancy could be resolved once hadronic
showers at the highest energies are understood better.

\subsubsection*{Anisotropies}

Another important information is of course encoded in the arrival
directions of cosmic rays. Up to $\sim 10^{19} \, \text{eV}$ these
directions get randomised by scattering on magnetic turbulence in the
galactic interstellar (but probably also in the intergalactic)
medium. The Auger collaboration has, for example, published evidence
of a correlation of their highest energy events with a catalogue of
nearby AGNs~\cite{Cronin:2007zz,Abraham:2007si}. The correlation has
however decreased ever since, as have claimed correlations of other
observations and source candidates. The question of the origin of
UHECRs therefore remains open.

\subsection{The standard model of galactic cosmic rays} 
\label{sec:SMGCRs}

Since their discovery by Victor Hess in 1912, a standard picture for
GCRs has emerged that can explain a large number of
observations. Usually, one assumes that the problem of GCRs factorises
in two parts that can be treated separately: acceleration in confined
sources and transport through the interstellar medium. This assumption
seems to be supported by the observed decrease of secondary-to-primary
ratios (see Sec.~\ref{sec:NuclearSecondaries}). In fact, if cosmic
rays were mostly accelerated in the interstellar medium by some
stochastic process (first order Fermi acceleration by interstellar
shock waves or second order Fermi acceleration by interstellar
turbulence) then these ratios would be logarithmically
rising~\cite{Eichler:1980hw,Cowsik:1980ApJ}.

The acceleration of both the hadronic (proton and nuclei) and the
electronic component of primary GCRs is assumed to take place in the
shocked shells of supernova remnants (SNRs) by first order Fermi
acceleration. Cosmic rays get accelerated as long as they are confined
to the shock region by up-stream turbulence and only some high-energy
particles might escape if no Alfv\'en waves of low enough wave number
are present for their
scattering~\cite{Reynolds:1998,Caprioli:2008sr}. In any case, once the
SNR has entered its radiative phase all high-energy particles are
released and diffuse through the interstellar medium by scattering on
magnetic turbulence in the form of Alfv\'en waves. It is expected that
convection by a possibly CR generated wind and reacceleration due to
stochastic, second-type Fermi acceleration also plays a role.

In the following sections, we describe the basic building blocks in
more detail. We start with a brief review of our understanding of
supernova remnants, in particular their dynamics and the properties of
the so-called Sedov-Taylor phase in which the bulk of particle
acceleration happens. We then derive the transport equation and apply
it to diffusive shock acceleration, claimed to be responsible for the
acceleration of cosmic ray electrons and nuclei up to TeV energies. We
also specialise the transport equation to cosmic ray propagation in
the Galaxy, outlining the different physical processes at work and
briefly review two simple analytical solutions which we extend on in
later chapters. We conclude this section by reviewing the main
predictions for fluxes and secondary-to-primary ratios and comparing
them to observations.

\subsection{Supernova remnants as sources}
\label{sec:SNRs}

The assumption that the population of GCRs is powered by SNRs is based
on three indications. The first one is the presence of non-thermal
populations of electrons as observed in radio and x-rays, for example,
for SN 1006~\cite{Koyama:1995rr}. This is usually explained as
synchrotron radiation of relativistic electrons on the ambient
magnetic fields, amplified by compression and possibly by the
Bell-Lucek mechanism~\cite{Lucek:2000,Bell:2001}. Estimates of the
highest energies reach up to hundreds of TeV (for a modelling of SN
1006, see~\cite{Reynolds:1998}).

Furthermore, there is a well-established theory that can explain the
generation of such a non-thermal population of particles by first
order Fermi acceleration in shocked shells, see Sec.~\ref{sec:DSA}. As
SNRs exhibit strong shocks with very high Mach numbers, applying this
theory to the parameters derived from, e.g. kinetic observations, it
can be shown that efficient particle acceleration is possible.

Finally, even as early as 1953, it was pointed
out~\cite{Shklovskii:1953} that SNRs with their benchmark total bulk
kinetic energy of $10^{51} \, \text{erg}$ could easily provide the
right order of magnitude energy needed to power the galactic
population of nuclear cosmic rays. More precisely, with a SN rate of
about $0.03 \, \text{yr}^{-1}$, a volume of the extended cosmic ray
halo of $\pi (15 \, \text{kpc})^2 \times 3 \, \text{kpc} = 5.7 \times
10^{67} \, \text{cm}^3$ and an average residence time of cosmic rays
in the Galaxy (that is the time until escape form the galactic cosmic
ray halo) of $20 \, \text{Myr}$, $\mathcal{O}(10) \%$ of the kinetic
energy must be transferred into GCRs to maintain the (local) energy
density of $0.3 \, \text{GeV} \, \text{cm}^{-3}$ -- a reasonable
efficiency.

The spectroscopic classification of supernovae (SNe) into type I
(without H lines) and type II (with H lines) does not coincide with
the nature of the progenitor system. While only type Ia SNe originate
in the thermonuclear burning of carbon-oxygen in a white dwarf, all
other types (Ib, Ic, IIP, IIL and IIn) are core collapse (CC)
supernovae. Interestingly, the kinetic energy of the ejecta of the
subsequent supernova remnants (SNRs) are quite similar, typically of
the order of $10^{51} \, \text{erg}$. Once the mass of the
interstellar medium swept up by the shock front exceeds the mass of
the ejecta, that is from the Sedov-Taylor phase (see below) onwards,
there should not be any phenomenological distinction between the
remnants of thermonuclear and CC SNe.

Conservation of mass, momentum and energy across a planar, adiabatic
shock front leads to the Rankine-Hugoniot
relations~\cite{Landau:1959}. In particular, for the compression ratio
$r \equiv n_1/n_2$ of the shock front, one finds,
\begin{equation}
\frac{u_2}{u_1} = \frac{1}{r} = \frac{\gamma - 1}{\gamma + 1} +
\frac{2}{\gamma + 1} \frac{1}{\mathcal{M}^2} \, ,
\end{equation}
and for strong shocks with a ratio of the specific heats $\gamma =
5/3$ and a Mach number $\mathcal{M} \gg 1$, $r = 4$. Furthermore, this
analysis is based on the assumption that radiative processes and loss
of non-thermal particles can be neglected. Also, back reaction of the
energetically important fraction of accelerated cosmic rays will
change the shock structure, modifying $\gamma$ towards the
fully-relativistic value of $4/3$.

Typical explosion velocities $\sqrt{2 E_\text{SN}/M_\text{ej}}$ of the
ejecta in the first phase of the SNR are $10^4 \, \text{km} \,
\text{s}^{-1}$ for type Ia and $5000 \, \text{km} \, \text{s}^{-1}$
for CC SNe. This is much higher than the estimated sound speed in the
surrounding medium, and therefore a blast wave with $\mathcal{M}
\gtrsim 10^3$ forms, that is a shock front followed by self-similar
($v(r) \propto r$), quickly cooling ejecta. After only a few days,
when the shock front has decelerated sufficiently, the SNR enters the
so-called ``ejecta-driven'' phase and a reverse-shock forms, that
reheats the ejecta of the SNR. A contact discontinuity (with constant
pressure) develops between the heated ejecta and the shock front, like
a piston pushing into the ISM. This discontinuity is expected to be
unstable to Raleigh-Taylor instability and therefore produces strong
turbulence.

The reverse shock will reach the centre of the SNR and disappear after
hundreds of years which marks the transition to the Sedov-Taylor
phase~\cite{Sedov:1959}. The time $t_\text{ch}$ at which this happens
can be estimated by dimensional analysis, assuming that the only
scales available to the problem are the energy $E$, the mass of the
ejecta $M_\text{ej}$ and the ambient density $\rho_0$,
\begin{equation}
t_\text{ch} = E^{-1/2} M_\text{ej}^{5/6} \rho_0^{-1/3} \, .
\end{equation}
The transition usually occurs when the mass swept up by the shock
front is a few times the mass of the ejecta and therefore the value of
$M$ cannot enter into the subsequent evolution. Only using $E$ and
$\rho_0$, the dynamics of the Sedov-Taylor phase can be again be
determined dimensionally~\cite{1999ApJS..120..299T},
\begin{align}
R = & 1.15 \, \left( \frac{E}{\rho_0} \right)^{1/5} t^{2/5} = 0.31
\left( \frac{E}{10^{51} \, \text{erg}} \right)^{1/5} \left(
\frac{\mu}{1.4} \right)^{-1/5} \left( \frac{n_0}{\text{cm}^3}
\right)^{-1/5} \left( \frac{t}{\text{yr}} \right)^{2/5} \, \text{pc}
\, , \\ u = & \frac{2}{5} \frac{R}{t} = 1.2 \times 10^5 \left(
\frac{E}{10^{51} \, \text{erg}} \right)^{1/5} \left( \frac{\mu}{1.4}
\right)^{-1/5} \left( \frac{n_0}{\text{cm}^3} \right)^{-1/5} \left(
\frac{t}{\text{yr}} \right)^{-3/5} \, \text{km} \, \text{s}^{-1} \, ,
\end{align}
where $\mu$ is the mean mass per particle in units of the proton
mass. Extensive analytical and numerical simulations of the
ejecta-driven and Sedov-Taylor phases have been presented, for
example, in~\cite{1999ApJS..120..299T}.

The Sedov-Taylor phase eventually comes to an end when the shock front
has slowed down enough, so that radiative processes can become
dominant. If the ejecta are hot enough at this point, the piston can
still be powered by the pressure although the expansion is not
adiabatic any more. If the ejecta have cooled too much, the former
shock front just continues outwards, conserving
momentum. Hydrodynamical simulations~\cite{Blondin:1998} show that
under realistic assumptions this happens at
\begin{align}
t_\text{tr} =& 2.9 \times 10^4 \left( \frac{E}{10^{51} \, \text{erg}}
\right)^{4/17} \left( \frac{n_0}{\text{cm}^3} \right)^{-9/17} \,
\text{yr} \, , \\
\label{eqn:SNRtrans}
R_\text{tr} =& 91 \left( \frac{E}{10^{51} \, \text{erg}}
\right)^{5/17} \left( \frac{n_0}{\text{cm}^3} \right)^{-7/17} \,
\text{pc} \, , \\ M_\text{tr} =& 10^3 \left( \frac{E}{10^{51} \,
  \text{erg}} \right)^{15/17} \left( \frac{n_0}{\text{cm}^3}
\right)^{-4/17} \, M_{\odot} \, .
\end{align}

After a few times $t_\text{tr}$ the velocity is as low as $100 - 300
\, \text{km}^{-1}$ and the Mach number is of order $3 - 6$ such that
particle acceleration comes to a halt.

\subsection{The transport equation}
\label{sec:TransportEquation}

In the following we sketch the derivation of the transport equation
governing the dynamics of a test-particle under magnetohydrodynamical
turbulence in a moving plasma. We closely follow the treatment given
in~\cite{Blandford:1987} which we refer the reader to for a more
detailed calculation.

The fundamental relation is the Vlasov
equation~\cite{Clemmow:1969,Lifshitz:1981} for the Lorentz invariant
phase space density $f = f(t, \vec{x}, \vec{p})$ at time $t$,
\begin{equation}
\label{eqn:Vlasov}
\frac{\partial f}{\partial t} + \vec{v} \cdot \frac{\partial
  f}{\partial \vec{x}} + \frac{\partial}{\partial \vec{p}} \cdot
\left( \vec{F} f \right) = 0\, .
\end{equation}
We incorporate the processes of momentum, pitch angle and spatial
diffusion by converting the microscopic Lorentz force term, encoding
the electromagnetic interactions of the test-particle with the plasma,
to macroscopic effective collision operators, thereby transforming the
Vlasov into a Boltzmann equation.

\subsubsection*{Momentum diffusion}

Scattering on magnetohydrodynamical turbulence leads to diffusion,
both in momentum $p$ and in pitch angle $\theta$, $\mu \equiv \cos
\theta = \cos{(\vec{p} \cdot \vec{B}/(p B))}$ where $\vec{B}$ is the
large scale magnetic field. We can treat the diffusion in momentum
space in a Fokker-Planck framework (see,
e.g.~\cite{Clemmow:1969,Lacombe:1977}), that is considering the
scattering of charged particles on magnetic inhomogeneities or
Alfv\'en waves as a Markov process. The Fokker-Planck equation,
\begin{equation}
\label{eqn:FokkerPlanckMomentumDiffusion}
\frac{\partial f}{\partial t} + \left( \vec{v} \cdot \vec{\nabla}
\right) f = \frac{\partial}{\partial \vec{p}} \cdot \vec{D}_{pp} \cdot
\frac{\partial f}{\partial \vec{p}} \quad \text{with} \quad
\vec{D}_{pp} = \frac{1}{2} \left\langle \frac{\Delta \vec{p} \Delta
  \vec{p}}{\Delta t} \right\rangle \, ,
\end{equation}
describes diffusion in momentum space and leads to second order Fermi
acceleration as originally envisaged by
Fermi~\cite{Fermi:1949}. Anticipating isotropy of the phase space
density with respect to pitch angle (see below), the RHS of
Eq.~\ref{eqn:FokkerPlanckMomentumDiffusion} simplifies to
\begin{align}
\label{eqn:RHSofFokkerPlanckMomentumDiffusion}
\frac{1}{p^2} \frac{\partial}{\partial p} p^2 D_{pp} \frac{\partial
  f}{\partial p} \, .
\end{align}
The momentum diffusion coefficient, $D_{pp}$, is calculated by
considering that the momentum gained on bouncing off scattering
centres in the plasma moving with velocity $V$, $\Delta p = - (\vec{p}
\cdot \vec{V})/v$, occurs every $\Delta t = L/v$ where $L$ is the
collision mean free path,
\begin{equation}
\label{eqn:DiffCoeffPP}
D_{pp} = \frac{1}{2} \left\langle 2 \, \frac{(\vec{p} \cdot
  \vec{V})^2}{v^2} \right\rangle \frac{v}{L} = \frac{1}{3} \frac{p^2
  \langle V^2 \rangle}{v L} \, .
\end{equation}

\subsubsection*{Pitch angle scattering}

The next necessary ingredient is pitch angle diffusion. Particles of
momentum $\vec{p}$ with Larmor radius $r_\text{L} = p / (Z eB) $
interact resonantly with Alfv\'en waves of similar wave-length,
$r_\text{L} \sim 1/k$, changing their pitch angle
$\theta$. Interacting with waves with random phases, the pitch angle
performs a random walk and after $\sim (B / \delta B)^2$ interactions
it has changed by $\pi$ and the particle can be considered as
scattered, that is having lost all information about its initial
direction. Here, $\delta B$ denotes the turbulent component of the
magnetic field. The Fokker-Planck equation for pitch angle scattering
reads,
\begin{equation}
\label{eqn:FokkerPlanckPitchAngleDiffusion}
\left( \frac{\partial f}{\partial t} \right)_\text{c} = \frac{1}{2}
\frac{\partial}{\partial \mu} \left( \left( 1 - \mu^2 \right) \nu
\frac{\partial f}{\partial \mu} \right) \, ,
\end{equation}
with the diffusion coefficient $\nu$ in pitch angle (i.e. the
scattering frequency),
\begin{equation}
\nu = \left\langle \frac{\Delta \theta^2}{\Delta t} \right\rangle =
\frac{\pi}{4} \left( \frac{k \mathcal{E}_k}{B^2/8 \pi } \right) \Omega
\, ,
\end{equation}
and where $\mathcal{E}_k$ is the energy in the mode $k$ while $B^2/8
\pi $ is the energy of the ambient magnetic field and $\Omega = e B /
(\gamma m_p)$ the Larmor frequency with $m_p$ the proton mass.  Having
assumed spatial uniformity, $\partial f / \partial x = 0$, pitch angle
diffusion can be implemented into the Vlasov equation~\ref{eqn:Vlasov}
by identifying the RHS of
Eq.~\ref{eqn:FokkerPlanckPitchAngleDiffusion} as an effective
collision operator.

\subsubsection*{Spatial diffusion}

If the mean free path length is short, spatial transport needs to be
treated in the diffusion equation. To prepare for the discussion of
acceleration across a shock front, we also want to consider the
possible motion of the scattering centres. The idea is to transform
the Vlasov equation into the plasma frame and determine the isotropic
part of the phase space density, $f$. It turns out that after
averaging over pitch angle, $f$ satisfies the transport equation,
\begin{equation}
\label{eqn:Transport}
\frac{\partial f}{\partial t} + \left( \vec{u} \cdot \vec{\nabla}
\right) f - \vec{\nabla} \cdot \left( D_\parallel \vec{\nabla} f
\right) = \frac{1}{3} \left( \vec{\nabla} \cdot \vec{u} \right) p
\frac{\partial f}{\partial p} \, ,
\end{equation}
where the coefficient for diffusion parallel to the magnetic field is
\begin{equation}
\label{eqn:DiffCoeffParallel}
D_\parallel \simeq \frac{v^2}{3 \nu} \propto \left( \frac{k
  \mathcal{E}_k}{B^2/8 \pi } \right)^{-1} \, ,
\end{equation}
and the coefficient for diffusion perpendicular to the magnetic field,
\begin{equation}
\label{eqn:DiffCoeffPerpendicular}
D_\perp \simeq \frac{v^2 \nu}{3 \Omega^2} \propto \left( \frac{k
  \mathcal{E}_k}{B^2/8 \pi } \right) \, ,
\end{equation}
has been ignored.

The second and third term on the LHS of Eq.~\ref{eqn:Transport} are
the usual convection and spatial diffusion term and the RHS describes
the energy loss (gain) of particles in a diverging (converging)
flow. Adding Eq.~\ref{eqn:RHSofFokkerPlanckMomentumDiffusion} to the
RHS completes the transport equation.

\subsection{Diffusive shock acceleration}
\label{sec:DSA}

Fermi's original idea~\cite{Fermi:1949} for the acceleration of cosmic
rays was stochastic scattering on magnetic turbulence in the form of
``magnetic clouds''. Unfortunately, the acceleration rate is only
second order in the speed $u$ of the scattering centres (hence the
name second order Fermi acceleration): Although the average energy
gained in head-on collisions is $\propto u/c$, the average energy lost
in an overtaking collision is equal but of opposite sign. In fact,
only an asymmetry between the probability for either type of
interaction, also of order $u/c$, leads to an overall energy gain, but
only of second order, i.e. $(u/c)^2$. It turns out that with the low
speed of magnetic interstellar turbulence this mechanism would take
much too long. Even for the much faster turbulence in the shocked
shells of SNRs, second order Fermi acceleration is still too
inefficient.

The standard theory for cosmic ray acceleration is therefore a first
order Fermi process in the converging flow of upstream and downstream
plasma across a shock front: diffusive shock acceleration (DSA). For a
contemporary review, see~\cite{Malkov:2001}.

\subsubsection{Macroscopic picture}

We consider steady diffusive shock acceleration (DSA) at a
one-dimensional, parallel shock (that is the ambient magnetic field is
parallel to the shock {\em normal}) in the test-particle
approximation, that is ignoring the back-reaction of a potentially
energetically important population of hadronic cosmic rays on the
shock front. This was worked out almost simultaneously by three
different groups~\cite{Krymskii:1977,Axford:1977,Blandford:1978}. (For
the microscopic treatment of DSA~\cite{Bell:1978a}, also published at
the same time, see~\ref{sec:DSAmicro}.) Here, we follow the treatment
of~\cite{Drury:1983}.

\begin{figure}
\begin{center}
\includegraphics[width=0.3\textwidth]{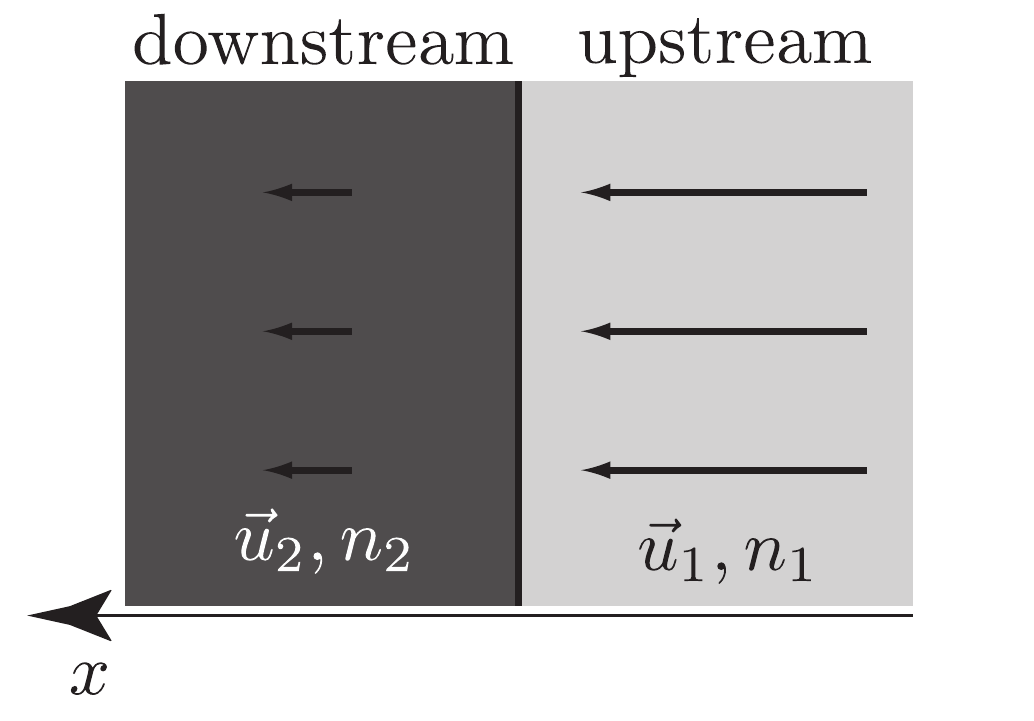}
\hspace{0.1\textwidth}
\includegraphics[width=0.3\textwidth]{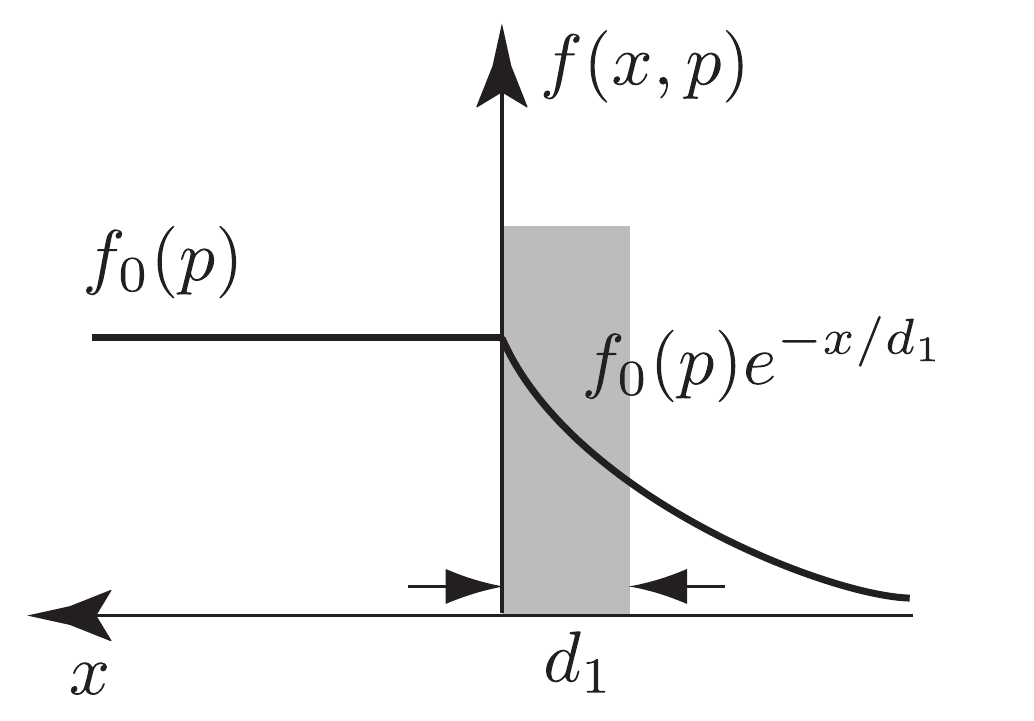}
\end{center}
\caption[]{{\bf Left:} DSA setup in the rest frame of the shock
  front. $u_1$ ($u_2$) and $n_1$ ($n_2$) denote upstream (downstream)
  plasma velocity and density, respectively.  {\bf Right:} Profile of
  the phase space density at a fixed momentum $p$. Particles within a
  momentum dependent distance $D/u$ of the shock front participate in
  the acceleration process.}
\label{fig:DSAsetup}
\end{figure}

The setup is as follows: The shock front is in its rest-frame in the
$y$-$z$-plane at $x=0$ and upstream plasma is flowing in from $x<0$
and downstream plasma flowing out to $x>0$ (see left panel of
Fig.~\ref{fig:DSAsetup}). The density (velocity) of the background
plasma are $n_1$ and $n_2$ ($u_1$ and $u_2$) in the upstream and
downstream region, respectively.

We now consider the transport equation~\ref{eqn:Transport} in this
one-dimensional setup, however, ignoring the momentum diffusion term,
\begin{equation}
\label{eqn:Transport4DSA}
\frac{\partial f}{\partial t} + u \frac{\partial f}{\partial x} =
\frac{\partial}{\partial x} \left( D \frac{\partial f}{\partial
  x}\right) + \frac{1}{3} \frac{\partial u}{\partial x} p
\frac{\partial f}{\partial p} \, .
\end{equation}
where we have set $D \equiv D_\parallel$. The general steady state
solution for $x \not = 0$ is,
\begin{equation}
\label{eqn:StdyState1}
f(x,p) = g_1(p) \exp{\left( \int_0^x \dd x' \frac{u}{D(x',p)} \right)}
+ g_2(p) \, .
\end{equation}

Imposing the usual boundary conditions, i.e. $f(x,p) \xrightarrow{x
  \rightarrow -\infty} f_1(p)$, where $f_1(p)$ is the phase space
density far upstream, that is the injection spectrum, and $|f(x,p)| <
\infty$ for $x \rightarrow \infty$, we find
\begin{equation}
\label{eqn:StdyState2}
f(x,p) = \left\{ \begin{array}{ll} f_1(p) + g_1(p) \exp{\left(
    \int_0^x \dd x' \, u/D \right) } & \text{for} \, x<0 \, ,
  \\ g_2(p) = f_2(p) & \text{for} \, x>0 \, .
\end{array} \right.
\end{equation}
The exponential term reflects the fact that only upstream diffusion
can counterbalance the convection by the background plasma in the
upstream half-plane. The spatial dependence is illustrated for a
homogeneous diffusion coefficient in the right panel of
Fig.~\ref{fig:DSAsetup}. Furthermore, the scale height of the
exponential, $d_1(p) = D_1(p) / u_1$, is the distance that particles
can on average diffuse away from the shock before being lost from the
acceleration process. As the diffusion coefficient is growing with
momentum, lower energy particles are confined more closely to the
shock than high energy particles.

To relate $f_1(p)$, $g_1(p)$ and $f_2(p)$, we need to find matching
conditions at the shock. These can be found most easily by multiplying
Eq.~\ref{eqn:Transport4DSA} by a test function and integrating over an
infinitesimal interval around $x=0$. For the test-function $\int_0^x
\dd x' / D(x')$, we find,
\begin{equation}
\label{eqn:Matching1}
\left[ f \right]_{- \varepsilon}^\varepsilon =
\mathcal{O}(\varepsilon) \, ,
\end{equation}
that is $f$ is continuous across the shock. For the test-function $1$
on the other hand,
\begin{equation}
\label{eqn:Matching2}
\left[ D \frac{\partial f}{\partial x} \right]_{-
  \varepsilon}^\varepsilon + \frac{1}{3} \left[ u \right]_{-
  \varepsilon}^\varepsilon \, p \frac{\partial f}{\partial p} =
\mathcal{O}(\varepsilon) \, ,
\end{equation}
which for $\varepsilon \rightarrow 0$ is the matching condition for
$\partial f/ \partial x$.

Substituting Eq.~\ref{eqn:StdyState2} into Eqs.~\ref{eqn:Matching1}
and~\ref{eqn:Matching2}, and eliminating $g_1$, we find
\begin{equation}
(r-1) p \frac{\partial f_2}{\partial p} = 3 r (f_1 - f_2) \, ,
\end{equation}
which is integrated to give
\begin{equation}
\label{eqn:MacResult}
f_2(p) = a p^{- a} \int_0^p \dd p' p'^{a - 1} f_1(p') + b p^{-a} \, ,
\end{equation}
where $a \equiv 3 r/(r-1)$ is 4 for a strong shock, $r=4$ (see
Sec.~\ref{sec:SNRs}). Ignoring the second term, which can be
interpreted as acceleration of particles from the thermal background,
it turns out that, as long as $f_1(p)$ is softer than $p^{- a}$, the
resulting spectrum $f_2(p)$ is a power law with spectral index
$a$. The coordinate space density of accelerated particles is then
$n(p) \dd p = 4 \pi p^2 f_2(p) \propto E^{-2}$ which is the well-known
result of DSA.

\subsubsection{Microscopic picture}
\label{sec:DSAmicro}

Although the macroscopic picture nicely proves that DSA leads to a
power law spectrum $\propto E^{-2}$, there is no physical intuition
involved that could explain {\em how} the power law form comes
about. Therefore, we re-derive the result of the last section,
following a more physical microscopic approach~\cite{Bell:1978a} (see
also~\cite{Peacock:1981,Michel:1981}).

As mentioned above, in second order Fermi acceleration, the
test-particles can gain or lose energy in approaching or following
scatterings, respectively. In DSA, a first order Fermi process,
however, the test-particles only encounter approaching scatterings:
Seen from either plasma frame (downstream or upstream), the other side
is always approaching. After crossing the shock, the test-particle
quickly isotropises in the new frame experiencing only approaching
scatterings which lead to energy gains. The energy gain rate is
therefore proportional to the energy gained in every scattering,
$u/c$.

To calculate the spectrum we need in fact both, the momentum gain
$\Delta p$ of a test-particle in a cycle across the shock-front,
i.e. coming from the upstream plasma, crossing to the downstream side
and crossing back to the upstream side; and the probability that a
particle gets advected downstream so that it is lost from the
acceleration process.

Consider a particle with momentum $p$, velocity $v$ and pitch angle
$\mu$ in the upstream frame. In the shock frame its momentum is $p(1 +
\mu u_1/v)$. Crossing the shock front, its momentum does not get
changed, but in the downstream frame it is measured as $p(1 + \mu (u_1
- u_2)/v)$. To calculate its average momentum change on crossing the
shock front {\em once}, we average over the pitch angle, including the
weighting factor $2 \mu$,
\begin{equation}
\langle \Delta p \rangle = p \int_0^1 \dd \mu \, \mu (u_1 - u_2)/v \,
2 \mu = \frac{2}{3} p \frac{u_1 - u_2}{v} \, .
\end{equation}
Crossing back, $u_1$ and $u_2$ get swapped but the integration also
runs from $0$ to $-1$, so the results is the same.

The probability that a particle gets advected downstream can be
calculated from the flux of particles to downstream infinity, $n_2
u_2$, divided by the flux of particles crossing the shock front from
upstream to downstream, $\int_0^1 \dd \mu \, \mu v n_2/2 = n_2
v/4$. Hence, the probability that a particle gets advected to
downstream infinity is $4 u_2/v$.

We now consider the cumulative spectrum $N_i(x,p) = \int_p^{\infty}
\dd p' \, 4 \pi p'^2 f_i(x,p)$ and assume that at upstream infinity
all particles have the same momentum $p_0$, that is $N_1(-\infty,p) =
N_0 \theta(p-p_0)$. Since all the particles injected will eventually
end up in the downstream region and can only have gained energy,
$N_2(\infty,p) = r N_0$ for $p < p_0$.

The momentum $p_n$ and velocity $v_n$ of a particle that has performed
$n$ cycles and has therefore crossed the shock $2n$ times, are of
course random variables but for large $n$, they are expected to peak
sharply around a mean (deterministic) value. Then,
\begin{equation}
\label{eqn:DSA_pn}
p_n \sim \prod_{i = 1}^n \left( 1 + \frac{4}{3} \left( u_1 - u_2
\right) / v _i \right) p_0 \quad \Rightarrow \quad \log{\left(
  \frac{p_n}{p_0} \right)} \sim \frac{4}{3} \left( u_1 - u_2 \right)
\sum_{i = 1}^n \frac{1}{v_i} \, .
\end{equation}

The probability $P_n$, that the particle has not been lost from the
acceleration after $n$ cycles is,
\begin{align}
\label{eqn:DSA_Pn}
P_n \sim \prod_{i = 1}^n \left( 1 - \frac{4 u_2}{v_i} \right) \quad
\Rightarrow& \quad \log{\left( P_n \right)} \sim -4 u_2 \sum_{i = 1}^n
\frac{1}{v_i} \, .
\end{align}
Combining Eqs.~\ref{eqn:DSA_pn} and~\ref{eqn:DSA_Pn} gives
\begin{align}
\log{\left( P_n \right)} \sim -3 \frac{u_2}{u_1 - u_2} \log{\left(
  \frac{p_n}{p_0} \right)} \Rightarrow \quad P_n = \left(
\frac{p_n}{p_0} \right)^{-3u_2/(u_1 - u_2)}
\end{align}

The cumulative spectrum of particles downstream is then just the total
density of particles $N_2 = r N_0$ times the probability $P_n$,
\begin{equation}
N_2(p_n) = P_n N_2 = r \left( \frac{p_n}{p_0} \right)^{-3 u_2/(u_1 -
  u_2)} N_0 \quad \text{for} \quad p_n > p_0 \,
\end{equation}
and
\begin{equation}
f_2(p) = -\frac{1}{4 \pi p^2} \frac{\partial N_2}{\partial p} =
\frac{N_0}{4 \pi} \frac{3 u_1}{u_1 - u_2}\left( \frac{p}{p_0}
\right)^{-3 u_1/(u_1 - u_2)}
\end{equation}

Although this result only reproduces the result of the macroscopic
approach, Eq.~\ref{eqn:MacResult}, it becomes clear that as in every
Fermi process, the power law spectrum is the result of the form of the
energy gained on every crossing which is proportional to the energy of
the incoming particle, and the probability to be lost from the
acceleration process. In the case of acceleration at a shock front,
both these number are fixed by the kinematics of the shock, as
expressed by the compression ratio (and the Mach number).

\subsection{Galactic propagation}
\label{sec:Transport}

The transport equation can, of course, not only be applied to the
acceleration of cosmic rays but also to cosmic ray transport in the
Galaxy. Expressing the original Eq.~\ref{eqn:Transport} in
differential particle density $n(t,\vec{x},E)$ instead of phase space
density $f(t\, \vec{x}, p)$, $n(t,\vec{x},E) \dd E = 4 \pi p^2 f(t\,
\vec{x}, p) \dd p$, and adding the injection of cosmic rays by SNRs as
well as their production and losses due to fragmentation and decay, we
find,
\begin{align}
\label{eqn:GCRtransport}
& \frac{\partial n_i}{\partial t} - \vec{\nabla} \cdot \left( D_{xx}
\cdot \vec{\nabla} n_i - \vec{u} \, n_i \right) -
\frac{\partial}{\partial p} p^2 D_{pp} \frac{\partial}{\partial p}
\frac{1}{p^2} n_i - \frac{\partial}{\partial p} \left( \frac{\dd
  p}{\dd t} n_i - \frac{p}{3} \left( \vec{\nabla} \cdot \vec{u}
\right) n_i \right) \nonumber \\ = & q + \sum_{i < j} \left( c \,
\beta \, n_\text{gas} \, \sigma_{j \rightarrow i} + \gamma \, \tau_{j
  \rightarrow i}^{-1} \right) \, n_j - \left( c \, \beta \,
n_\text{gas} \, \sigma_i +\gamma \, \tau_i^{-1} \right) \, n_i \, ,
\end{align}
where $\beta$ is the speed in units of the speed of light $c$ and
$\gamma = (1 + \beta^2)^{-\nicefrac{1}{2}}$. In the following, we
discuss the different terms and their physical meaning in some more
detail.

\begin{itemize}
\item {\bf diffusion:} $-\vec{\nabla} \cdot D_{xx} \cdot \vec{\nabla}
  n_i$ \\ The main transport mode of cosmic rays is resonant pitch
  angle scattering on magnetic irregularities. Diffusion is strongly
  anisotropic ($\delta B \ll B$) locally, but the particle density
  gets isotropised by the fluctuations of the magnetic field at larger
  scales of $\mathcal{O}(100) \, \text{pc}$. The parallel diffusion
  coefficient $D_{xx} \approx (\delta B_\text{res} / B)^{-2} v r_g /
  3$, with $r_g$ the Larmor radius and $\delta B_\text{res}$ the field
  strength at the resonant wave number $k~=~1/r_g$, is inversely
  proportional to the energy density in the turbulent component (see
  also Eq.~\ref{eqn:DiffCoeffParallel}), $w(k) \dd k \sim k^{-2 +
    \delta} \dd k$, which can be of Kolmogorov ($\delta =
  \nicefrac{1}{3}$) or Kraichnan ($\delta =\nicefrac{1}{2}$)
  type. Expressing the diffusion coefficient in energy $E$, one finds
  $D_{xx} \propto \beta E^ \delta$. Furthermore, with $\delta
  B_\text{res} \approx B \simeq \text{few} \, \mu\text{G}$ at the
  principal scale of $\sim (100 \, \text{pc})^{-1}$, $D_{xx0} \equiv
  D_{xx} (1 \, \text{GeV} ) \simeq 10^{28} \, \text{cm}^{2} \,
  \text{s}^{-1}$. The effective values for $\delta$ and $D_{xx0}$ are
  usually determined from local observations of secondary-to-primary
  ratios (see Sec.~\ref{sec:NuclearSecondaries}).
\item {\bf convection:} $\vec{\nabla} \cdot \vec{u} \, n_i$ \\ There
  is evidence of a (CR driven~\cite{Breitschwerdt:1999sv}) wind in
  other galaxies and one might wonder whether SNRs also power a
  similar wind in the Milky
  Way~\cite{Jokipii:1976jk,Owens:1977ut,Owens:1977uu,Jones:1978ap,Jones:1979,Bloemen:1993}. Direct
  observational evidence, however, only comes form x-rays close to the
  Galactic centre. A common way to achieve symmetry with respect to
  the galactic plane, $\vec{u}(-z) = \vec{u}(z)$, is to consider a
  velocity linearly increasing with distance from the plane, $\vec{u}
  \propto \vec{e}_z \, (\dd u / \dd z) \, z$. As the convection rate
  is decreasing with energy, convection can only play a role at lower,
  $\mathcal{O}(1) \, \text{GeV}$, energies and convection flattens out
  the steepening of source spectra (due to energy-dependent
  diffusion), observable for example in secondary-to-primary
  ratios. These also allow to constrain the convection velocity to $u
  = \mathcal{O}(10) \, \text{km} \, \text{s}^{-1}$ (for one-zone
  models~\cite{Maurin:2002hw,Jones:2001}).
\item {\bf adiabatic energy losses/gains:} $\frac{\partial}{\partial
  p} \frac{p}{3} \left( \vec{\nabla} \cdot \vec{u} \right)
  n_i$\\ Diverging flows lead to adiabatic energy losses. This effect
  is for example important in propagation models with non-uniform
  galactic winds.
\item {\bf reacceleration:} $- \frac{1}{p^2} \frac{\partial}{\partial
  p} p^2 D_{pp} \frac{\partial}{\partial p} \frac{1}{p^2}
  n_i$\\ Substituting for the spatial diffusion coefficient,
  Eq.~\ref{eqn:DiffCoeffParallel}, in Eq.~\ref{eqn:DiffCoeffPP} gives
  the relation $D_{xx} D_{pp} = p^2 v_A^2 / 9$ where the Alfv\`en
  velocity $v_A$ is somewhere around $30 \, \text{km} \,
  \text{s}^{-1}$. (In fact, the energy dependence of the diffusion
  coefficients can be modified if one accounts for the energy lost
  from the ISM turbulence into CRs which leads to damping and
  consequently a steep rise in the spatial diffusion coefficient at
  lower energies~\cite{Ptuskin:2005ax}.) As already mentioned,
  distributed acceleration in the ISM cannot be the main source of
  acceleration because of the observed energy dependence of
  secondary-to-primary ratios. However, as the diffusion-loss time,
  $z_\text{max}^2/D_{xx}$, is decreasing with energy and the
  time-scale of distributed acceleration, $p^2/D_{pp} \propto
  D_{xx}/v_A^2$, is increasing, second order Fermi acceleration by
  diffusion in momentum space, so-called reacceleration, can play a
  role at lower energies. For example, reacceleration can show up in
  secondary-to-primary ratios and can potentially explain a bump in
  the boron-to-carbon ratio around $1 \,
  \text{GeV}$~\cite{Simon:1986,Seo:1994}.
\item {\bf continuous energy losses:} $- \frac{\partial}{\partial p}
  \frac{\dd p}{\dd t} n_i$\\ All CRs lose energy due to ionisation and
  Coulomb interactions which are however only important at energies
  below a few GeV. In addition, electrons and positrons interact with
  the ISM emitting bremsstrahlung (again, only important at $\sim
  \text{GeV}$ energies), but also synchrotron radiation on the
  galactic magnetic fields and inverse Compton scattering (ICS) on
  interstellar radiation fields (ISRFs). In the Thomson approximation,
  the energy loss rate is proportional to the energy
  squared~\cite{Crusius:1988},
\begin{align}
\frac{\dd p}{\dd t} = - \frac{4 \sigma_\text{T} c}{3} \left(
\rho_\text{ISRF} + \rho_\vec{B} \right) \left( \frac{E}{m c^2}
\right)^2 \, .
\end{align}
where $ \sigma_\text{T}$ is the Thomson cross section. The energy
densities for a $3 \, \mu \text{G}$ magnetic field is $0.22 \,
\text{eV} \, \text{cm}^{-3}$; for the CMB, IR and stellar radiation
they are $0.26 \, \text{eV} \, \text{cm}^{-3}$, $0.2 \, \text{eV} \,
\text{cm}^{-3}$ and $0.45 \, \text{eV} \, \text{cm}^{-3}$,
respectively~\cite{Kobayashi:2003kp}. (See~\cite{Porter:2008ve} for a
sophisticated 3D modelling of ISRFs.)  At higher centre-of-mass
energies, the Thomson approximation is not valid any more and at even
higher energies the cross section is in the Klein-Nishina regime and
the loss rate becomes
suppressed~\cite{Kobayashi:2003kp,Schlickeiser:2009qq} (see
also~\cite{Stawarz:2009ig}). For ICS, the critical energy depends on
the average energy of the radiation background considered and is about
$1.1 \times 10^6 \, \text{GeV}$ for the CMB, $7.6 \times 10^4 \,
\text{GeV}$ for IR and $8.7 \times 10^2 \, \text{GeV}$ for
starlight~\cite{Delahaye:2010ji}.
\item {\bf primary injection:} $q$\\ As discussed in
  Sec.~\ref{sec:SMGCRs}, acceleration of primary CRs is supposed to
  take place in SNRs and the spectrum is a power law in energy (or
  rigidity) with spectral index close to $-2$ (in the test-particle
  approximation).
\item {\bf nuclear spallation}: $\sum_{i < j} \left( c \, \beta \,
  n_\text{gas} \, \sigma_{j \rightarrow i} + \gamma \, \tau_{j
    \rightarrow i}^{-1} \right) \, n_j - \left( c \, \beta \,
  n_\text{gas} \, \sigma_i +\gamma \, \tau_i^{-1} \right) \, n_i$
  \\ Spallation or decay of primary (secondary) cosmic rays during
  their propagation lead to their depletion and to the injection of
  secondary (tertiary) CRs. Furthermore, catastrophic energy losses
  can in principle be accounted for by these terms. Note that
  spallation is dominantly on interstellar hydrogen and helium,
  i.e. $n_\text{gas} \, \sigma_{j \rightarrow i} = n_H \, \sigma^H_{j
    \rightarrow i} + n_{He} \, \sigma^{He}_{j \rightarrow i}$ and
  similar for $n_\text{gas} \, \sigma_i$. Parametrisations for meson
  production are discussed in Refs.~\cite{Kamae:2006bf,Kelner:2006tc}
  and an overview of nuclear spallation cross section models is given
  in Ref.~\cite{Mueller:2001} and references therein.
\end{itemize}

\begin{figure}[!tb]
\begin{center}
\includegraphics[scale=1]{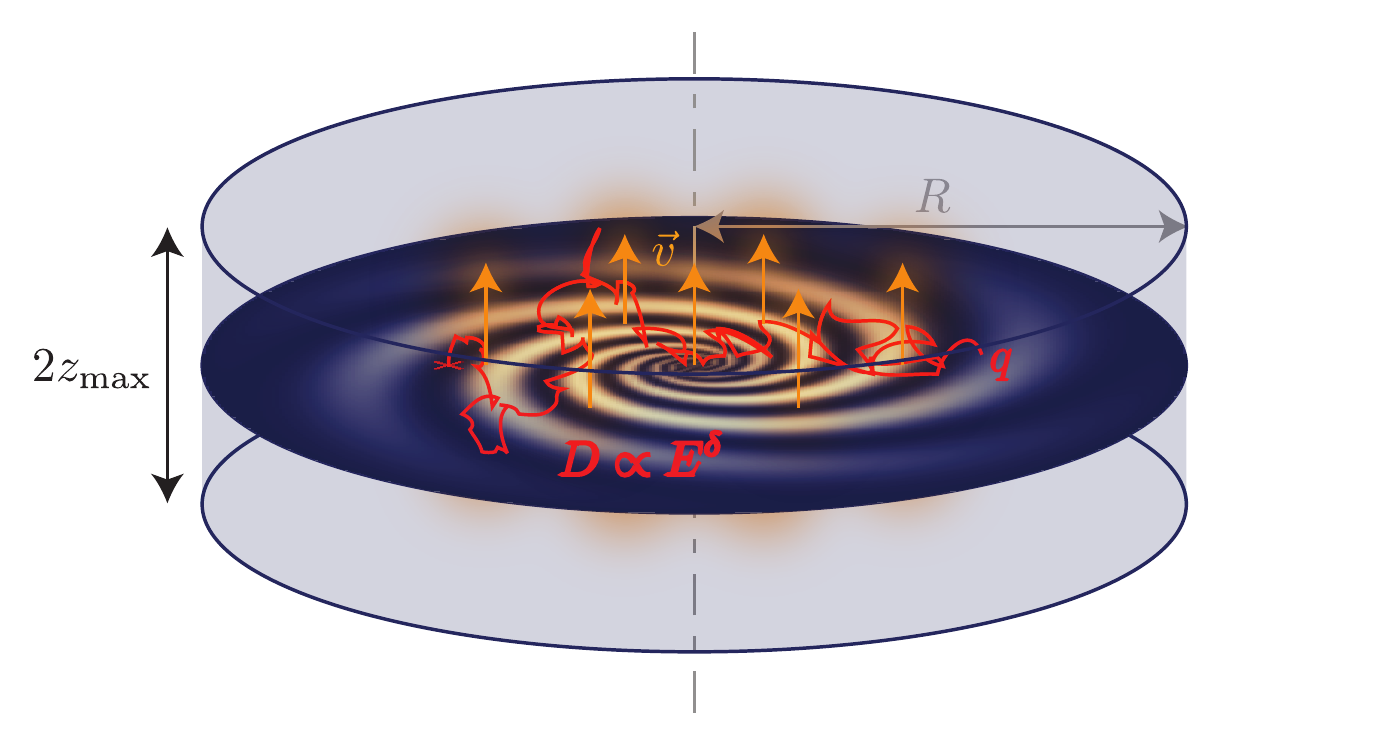}
\end{center}
\caption{Propagation setup. The thin disk of sources and interstellar
  gas is contained within a cylindrical cosmic ray halo of half-height
  $z_\text{max}$ and radius $R$. Within the halo cosmic rays diffuse
  in coordinate and momentum space, get convected, spallate and lose
  energy, depending on the details of the propagation model.}
\label{fig:PropagationGeometry}
\end{figure}

Usually, GCR propagation is considered to be limited to a cylindrical
volume of radius $R$ and half-height $z_\text{max}$ that the galactic
plane is contained in, see Fig.~\ref{fig:PropagationGeometry}. The
transport equation is thus required to satisfy the boundary
conditions,
\begin{align}
n(\vec{r}_\parallel, z, t) \equiv 0 \quad \text{for }
|\vec{r}_\parallel| = R \text{ and for } z = z_\text{max} \, ,
\end{align}
where $\vec{r}_\parallel = {\rm\vec{e}}_x x + {\rm\vec{e}}_y y $
. Alternatively, solutions without boundary conditions, but with a
quickly increasing diffusion coefficient at the halo boundaries have
been considered in~\cite{Shibata:2006} (see also~\cite{Evoli:2008dv}).

Solar modulation is usually modelled by a simple effective (electric)
potential $\Phi \sim \mathcal{O}(100) \, \text{MV}$
\cite{Gleeson:1968zz} which is assumed to be the effect of dynamical
effects in the solar wind which are still not fully understood. The
effect on charged CRs is that the flux at the top of the atmosphere
(TOA), $J_{\text{TOA}}(E)$, is suppressed with respect to the
interstellar flux $J_{\text{IS}}(E)$,
\begin{align}
J_{\text{TOA}}(E) = \frac{E^2 - m^2}{(E + |Z| e \Phi)^2 - m^2} \,
J_{\text{IS}}(E + |Z| e \Phi) \, ,
\end{align}
where $(Z \, e)$ is the charge and $m$ the mass of the CR particle.

\subsubsection{Leaky box model}
\label{sec:LeakyBoxModel}

One of the earlier attempts to solve the problem of cosmic ray
transport is the so-called Leaky Box model~\cite{Cowsik:1967}. In this
approach, all spacial dependencies are ignored and the Galaxy is
described as a system that particles can escape from with a certain --
in its simplest formulation, energy-independent -- probability. In the
steady state, this results in an exponential distribution of the
column depth or grammage that cosmic rays particles have
traversed. The average value of the order of a few $\text{g} \,
\text{cm}^{-2}$ must be a function of energy to explain the falling
secondary-to-primary spectra. The simple Boltzmann-type equation
describing this model,
\begin{align}
\frac{\dd n_i}{\dd t} = - \frac{n_i}{\tau_\text{esc}} -
\frac{n_i}{\tau_\text{cool}} + \sum_{i < j} \left( c \, \beta \,
n_\text{gas} \, \sigma_{j \rightarrow i} + \gamma \, \tau_{j
  \rightarrow i}^{-1} \right) \, n_j - \left( c \, \beta \,
n_\text{gas} \, \sigma_i +\gamma \, \tau_i^{-1} \right) \, n_i \, ,
\end{align}
can be derived from the full transport equation by integrating over
the cosmic ray halo and the contributions to the escape time
$\tau_\text{esc}$ from diffusion and convection are $\sim D_{xx} /
z_\text{max}^2$ and $\sim u / z_\text{max}$, respectively, and the
cooling time $\tau_\text{cool} \sim p / (\dd p / \dd t)$.

Applying the simple leaky box model to a network of stable nuclei, one
can compute the relative abundances of different isotopes assuming
ratios at source that are similar to solar system abundances. For
stable nuclei, it can be shown that the leaky box model reproduces the
results of a diffusion model in certain limiting
cases~\cite{Ptuskin:2009zz}. This can be understood considering that
nuclei have a rather long residence time in the halo and therefore
basically achieve the spatial averaging thereby justifying the leaky
box approach. A generalisation of the leaky box model is the so-called
weighted-slab technique~\cite{Protheroe:1981,Garcia:1987} that allows
for more flexible path-length distributions.

\subsubsection{Green's function approach}
\label{sec:GreensFunctionApproach}

A more sophisticated, but still analytic approach is to determine the
solution to the transport Eq.~\ref{eqn:GCRtransport} for a
$\delta$-like injection $\delta(t)\delta(\vec{r}) \delta(E - E_0)$,
that is, to find the Green's function $G(t, \vec{r}, E)$. The
differential spectral density for a continuous source distribution is
then given by its convolution with the Green's function. A
particularly nice and useful example has been worked out for the
propagation of electrons and positrons\footnote{In the following,
  ``electrons'' is meant to denote both, electrons and
  positrons.}~\cite{TheOriginOfCosmicRays1964} (see
also~\cite{Cowsik:1979,Aharonian:1995zz,Atoian:1995ux}), which we will
also employ in Chapter~\ref{chp:additional}.

As we are mainly interested in energies above $\sim 10 \, \text{GeV}$,
we can ignore convection ($\vec{u} \equiv 0$) and reacceleration
($D_{pp} \equiv 0$), i.e. the transport
equation~\ref{eqn:GCRtransport} for the differential spectral density
$n$ of electrons and positrons reads,
\begin{equation}
\label{eqn:TransportEquationFore+e-}
\frac{\partial n}{\partial t} - \vec{\nabla} \cdot \left( D \cdot
\vec{\nabla} \right) n - \frac{\partial}{\partial E} \left( b(E) n
\right) = q(\vec{r},t) \, ,
\end{equation}
where the source term $q$ now includes primary and secondary sources
and the diffusion coefficient $D \equiv D_{\parallel}$ is assumed to
be homogeneous in the following. If one could ignore the energy loss
term $b(E)$ the energy of individual electrons and positrons would not
change during their propagation, and the energy would simply be a
parameter of a diffusion problem of the heat equation type and the
corresponding Green's function is
\begin{equation}
\frac{1}{(4 \pi D t)^{3/2}} {\rm e}^{-\vec{r}^2/4 D t} \, .
\end{equation}

However, electrons and positron do lose energy and as the energy
losses are continuous, the energy of a particle is monotonously
decreasing from the energy at injection, $E_0 = E(t = 0)$, obeying,
\begin{equation}
\label{eqn:EnergyLossFunction}
\frac{\dd E}{\dd t} = b(E) = - b_0 E^2 \quad \Rightarrow \quad
\frac{1}{E_0} - \frac{1}{E} = - b_0 t \, .
\end{equation}
where we limit ourselves to the Thomson approximation. Given the time
$t$ since injection and the energy $E$ at observation, the energy at
injection $E_0$ is therefore unambiguously defined. As the diffusion
coefficient is energy-dependent, we need to average it over time or
equivalently over intermediate energies, $E'$. The diffusion length
squared $\ell^2 = 4 D t$ is a function of both, $E$ and $E_0$,
\begin{equation}
\ell^2(E, E_0) = \int_{0}^t \dd t' 4 D(E(t')) = 4 \int_{E_0}^E \dd E'
\frac{D(E')}{b(E')} \, .
\end{equation}
With Eq.~\ref{eqn:EnergyLossFunction} and $D(E) = D_0 E^\delta$, we
find,
\begin{align}
\label{eqn:ell2}
\ell^2(E, E_0) = 4 \int_{E_0}^E \dd E' \frac{D_0 E'^\delta}{b_0 E'^2}
= \frac{4 D_0}{b_0 (1 - \delta)} \left( E^{\delta - 1} - E_0^{\delta -
  1} \right) \, ,
\end{align}
or with $E_0 = E / (1- b_0 E t )$,
\begin{align}
\ell^2(E, t) = \frac{4 D_0}{b (1 - \delta)} \left[ E^{\delta - 1} -
  \left( \frac{E}{1-b_0 E t} \right)^{\delta - 1} \right] \, .
\end{align}

The contribution from an injection of electrons of energy $E_0$ at $t
= 0$, $\vec{r} = 0$ to the electrons of energy $E$ at $t$, $\vec{r}$
is therefore,
\begin{equation}
g(t,\vec{r},E) = \frac{{\rm e}^{-\vec{r}^2/\ell^2(E,t)}}{\left[ \pi
    \ell^2(E,t)\right]^{3/2}} \, \delta \left( E - \frac{E_0}{1 + b_0
  E_0 t} \right) \, ,
\end{equation}
and a burst-like injection of a source spectrum $Q(E_0)$ hence
contributes,
\begin{equation}
\label{eqn:GreensFunctionSpectrum}
G(t,\vec{r},E) = \int_0^{\infty} \dd E_0 \, Q(E_0) g(t,\vec{r},E) =
\frac{{\rm e}^{-\vec{r}^2/ \ell^2(E,t)} }{\left[ \pi
    \ell^2(E,t)\right]^{3/2}} Q \left( \frac{E}{1 - b_0 E t} \right)
\left( 1 - b_0 E t \right)^{-2} \, .
\end{equation}
To get the density for a steady source, this Green's function is
integrated over time,
\begin{equation}
G^{\text{stdy}}(\vec{r},E) = \int_0^{\infty} \dd t \, G(t,\vec{r},E) =
\frac{1}{b_0 E^2} \int_E^{\infty} \dd E_0 \frac{{\rm e}^{-\vec{r}^2 /
    \ell^2(E,E_0)} }{\left[ \pi \ell^2(E,E_0)\right]^{3/2}} Q(E_0) \,
.
\end{equation}
This result can also be obtained from the time-independent transport
equation, i.e. Eq.~\ref{eqn:TransportEquationFore+e-} with $\partial n
/ \partial t \equiv 0$, see~\cite{Baltz:1998xv}.

As electrons and positrons of tens and hundreds of GeV will lose most
of their energy before travelling more than a few kiloparsecs and
reaching the radial boundary of the cosmic ray halo, we ignore the
radial boundary condition $n(t, r = R, z) = 0$. The boundary condition
at $z = z_\text{max}$ can be implemented by the method of ``mirror
charges''~\cite{Baltz:1998xv},
\begin{singlespace}
\vspace{-\baselineskip}
\begin{align}
G_{\text{disk}}(t,\vec{r}, E) &= \sum_{n=-\infty}^{\infty}
G(t,\vec{r}_n, E) \quad \text{where } \vec{r}_n = \left(
\begin{array}{c} x\\ y\\ (-1)^n z + 2 z_\text{max} n \end{array} \right) \, .
\end{align}
\end{singlespace}
\noindent It is useful to factorise the spatial dependence into a
dependence on $(x,y)$ and $z$, $\vec{r} = \vec{r}_\parallel +
\vec{e}_z z$,
\begin{align}
G_{\text{disk}}(t,\vec{r}, E) &= \sum_{n=-\infty}^{\infty}
\frac{1}{(\pi \ell^2)^{3/2}} e^{-\vec{r}_n^2/\ell^2} Q(E_0) (1 - b_0 E
t)^{-2} \\ &= \frac{1}{\pi \ell^2} e^{-\vec{r}_{\parallel}^2/\ell^2}
Q(E_0) (1 - b_0 E t)^{-2} \frac{1}{\zcr} \chi \left( \frac{z}{\zcr} ,
\frac{\ell}{\zcr} \right)
\end{align}
where the sum has been expressed in terms of the elliptic theta
function, $\vartheta_3$,
\begin{align}
\chi( \hat{z} , \hat{\ell}) = \frac{1}{\pi} \left[ \vartheta_3 \left(
  \hat{z}, e^{-\hat{\ell}^2} \right) - \vartheta_3 \left( \hat{z} +
  \frac{\pi}{2}, e^{-\hat{\ell}^2} \right) \right] \, ,
\end{align}
and $\zcr = 4 z_\text{max} / \pi$. As most primary and secondary
sources as well as our position are basically in the thin galactic
disk, $z \approx 0$ and $\chi( \hat{z} , \hat{\ell} ) \rightarrow
\chi( \hat{\ell} ) \equiv \chi ( 0, \hat{\ell} )$.

The Green's function with boundary condition for a steady source is
\begin{align}
G_{\text{disk}}^{\text{stdy}} (\vec{r}, E) &= \frac{1}{b E^2}
\int_E^{\infty} \mathrm{d}E_0 \frac{1}{\pi \ell^2}
e^{-\vec{r}_{\parallel}^2/\ell^2} Q(E_0) \frac{1}{\zcr} \chi \left(
\frac{\ell}{\zcr} \right) \ .
\end{align}

\subsubsection{(Semi-)numerical codes}

Both these propagation models oversimplify the problem but their
assumptions are justified in certain limits: the leaky box model, for
example, is accurate when considering stable nuclei, and the Green's
function approach can be used for the propagation of electrons and
positrons.

A more realistic approach must take into account the morphology of the
galactic magnetic field (GMF) and the interstellar radiation fields
(ISRFs) as well as the resulting spatial dependence of the energy loss
rate and the production of secondary radiation like diffuse radio and
gamma-ray. It should further use the information on the gas density in
the ISM, which does not only enter through the energy loss rate but
also in the production of gamma-rays through $\pi^0$ decay. The
resulting transport equation cannot be solved analytically any more
and therefore numerical solutions are the only resort.

There are a number of numerical codes for GCR propagation in the
literature, some of them are publicly available. {\tt
  GALPROP}~\cite{Moskalenko:1997gh,galprop} numerically integrates the
transport equation~\ref{eqn:GCRtransport} on a spatial lattice by a
Crank-Nicholson scheme. It can be run in a time-dependent mode or
iteratively until a steady state, $\partial n_i / \partial t \approx
0$ is reached. {\tt GALPROP} uses input from 21-cm and CO (tracer of
$H_2$) maps for the ISM gas density and a sophisticated modelling of
the interstellar radiation fields. It therefore allows not only
calculations of nuclear and electron-positron fluxes but also more
realistic predictions for gamma-rays and synchrotron radiation. The
broad range of the predictions possible allows for important
cross-checks, however, at the moment the full code is still not fast
enough to allow for automated scans of the full multi-dimensional
parameter space of the diffusion model.

A similar, though slightly more general code is {\tt
  DRAGON}~\cite{Evoli:2008dv} which allows, for example, for position
dependent, anisotropic diffusion and separate injection spectra for
different species. As an example for a semi-analytical approach we
mention {\tt USINE}~\cite{Maurin:2001sj,Donato:2001ms,Donato:2001eq}
which employs analytical solutions found for a simplified setup with,
{\it e.g,} simplified gas maps. This allows for sophisticated studies
of the parameter space using MCMC techniques~\cite{Putze:2010zn}.

\subsection{Observational results on the nuclear component}
\label{sec:ObservationalResults}

In the following, we summarise the observational results in nuclear
GCRs which are, to a large extent, in agreement with the predictions
from the standard model of GCRs, as described above. We postpone the
discussion of the leptonic component as well as synchrotron and
gamma-rays, all of which have recently been claimed to show anomalies
possibly connected to DM, to Sec.~\ref{sec:DMIndirectDetection} where
DM indirect detection is discussed.

\subsubsection{Primary nuclei}

As mentioned before, it has been shown~\cite{Ptuskin:2009zz} that for
stable primary and secondary nuclei the diffusive-convective transport
equation~\ref{eqn:GCRtransport} is in some limits equivalent to a
leaky-box model. In particular, further neglecting secondary
production, energy losses and reacceleration, the ambient spectrum
$I_i$ of a primary species $i$ is~\cite{Ptuskin:2009zz},
\begin{align}
I_i \propto Q_i \frac{1}{1/ X_{\text{esc},i} + \sigma_i / m} \quad
\text{with} \quad \frac{\sigma_i}{m} = \frac{n_H \sigma_H + n_{He}
  \sigma_{He}}{n_H m_H + n_{He} \sigma_{He}} \, ,
\end{align}
where the column depth or grammage $X_{\text{esc},i} \propto 1 / D$ is
a power law in energy, $\propto E^{-\delta}$, above a few GeV/nucleon
and $1/ X_{\text{esc},i}$ dominates over the loss term $\sigma_i /
m$. With a source spectrum $Q \propto E^{-\gamma}$, $\gamma \approx
2$, the theoretical prediction for absolute primary fluxes is
therefore $I_i \propto E^\alpha$ with $\alpha = -\gamma -
\delta$. From the local observation of primary nuclei (protons, Carbon
etc.), we have $\alpha \approx -2.75$ which would imply $\delta
\approx 0.75$. However, from secondary-to-primary ratios (see below),
one can determine $\delta$ independent of the source spectrum and
finds $\delta \approx 0.6$ for plain diffusion, $\delta \approx 0.4$
when including reacceleration (see
Sec.~\ref{sec:NuclearSecondaries}). This implies a source spectrum of
$\gamma \approx 2.15$ and $\gamma \approx 2.35$, for plain diffusion
and diffusive reacceleration, respectively, which is only marginally
in agreement with the expectation from diffusive shock acceleration,
cf. Sec.~\ref{sec:DSA}. At energies below a few GeV, ionisation energy
losses need to be taken into account.

\begin{figure}[bth]
\begin{center}
\vspace{1cm} \includegraphics[width=0.45\textwidth]{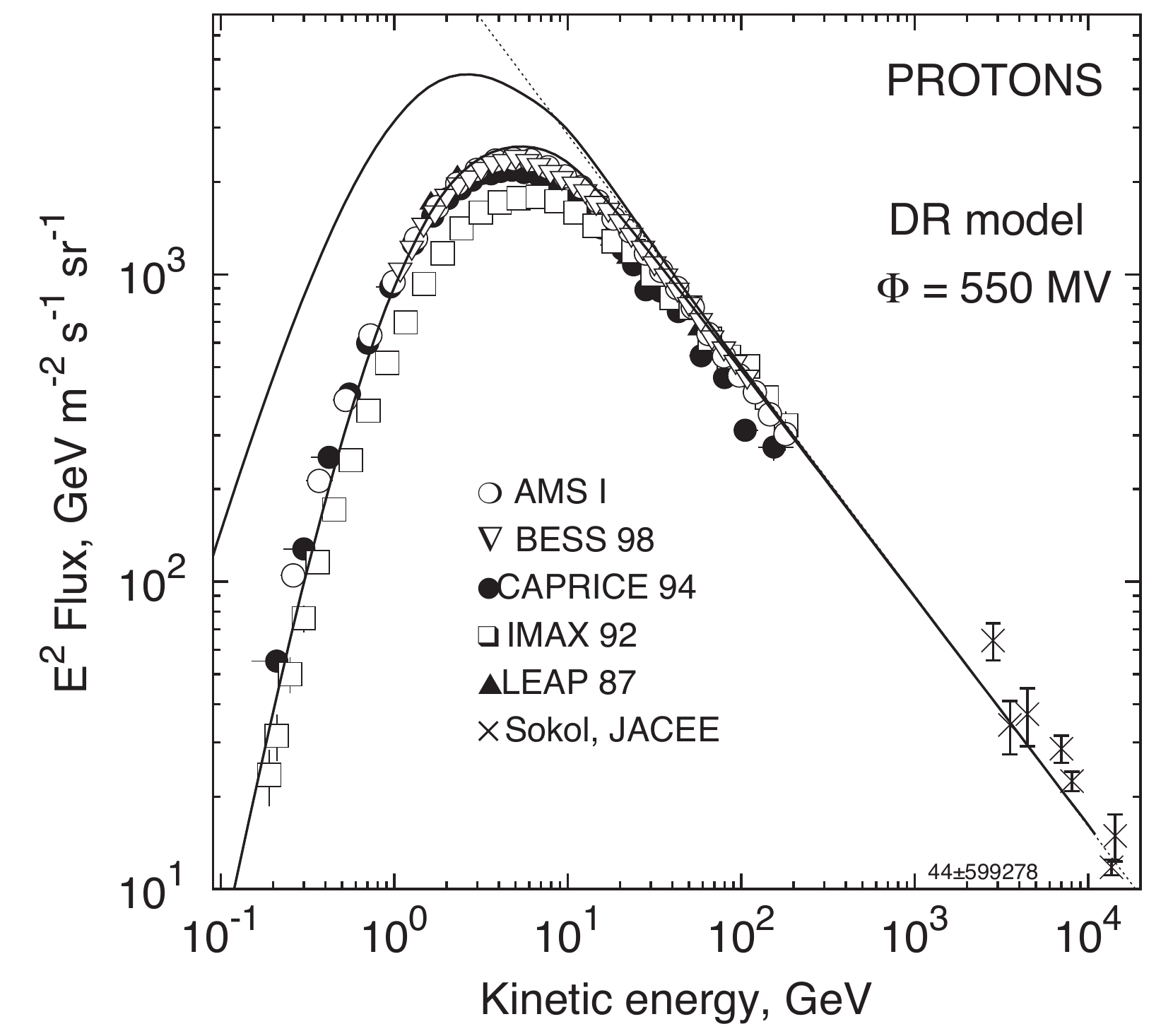}
\hspace{0.05\textwidth}
\includegraphics[width=0.45\textwidth]{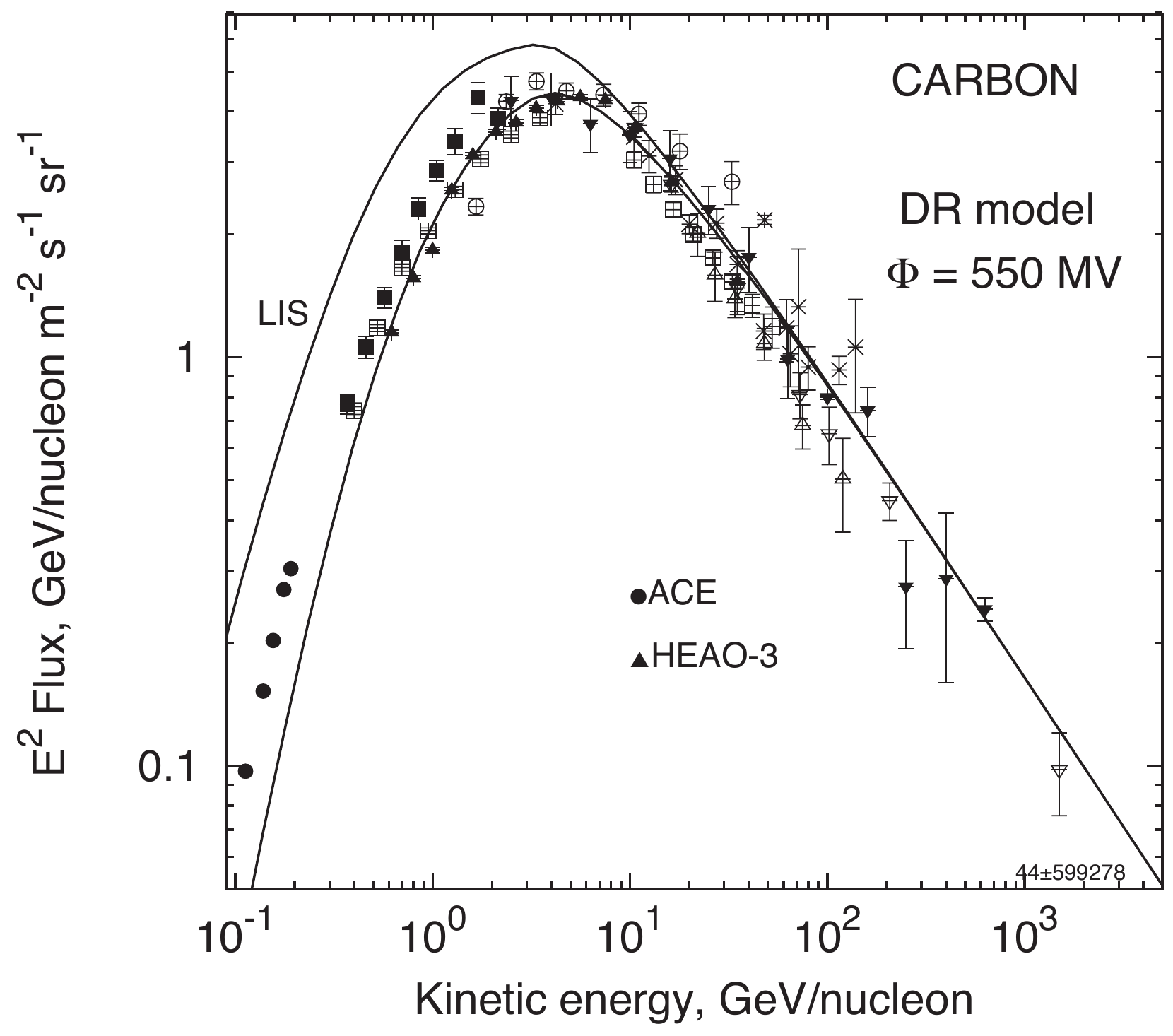}
\end{center}
\caption{ {\bf Left:} Absolute proton flux in a diffusive
  reacceleration (DR) model (from~\cite{Ptuskin:2005ax}). The upper
  line shows the local interstellar (LIS) flux, the lower line the
  solar modulated one with $\Phi = 550 \, \text{MV}$ and the thin
  dotted line the LIS spectrum best fitted to the data above 20
  GeV~\cite{M02}. Data: AMS~\cite{p_ams}, BESS 98~\cite{sanuki00},
  CAPRICE 94~\cite{Boez99}, IMAX 92~\cite{Menn00}, LEAP
  87~\cite{p_leap}, Sokol~\cite{sokol}, JACEE~\cite{jacee}.
{\bf Right:} Absolute carbon flux in a diffusive reacceleration model
(from~\cite{Ptuskin:2005ax}). Data from ACE~\cite{davis,davis01},
HEAO-3~\cite{Engelmann:1990zz}, for other references
see~\cite{StephensStreitmatter98} (symbols are changed).  }
\label{fig:AbsoluteProtonAndCarbonFluxes}
\end{figure}

In Fig.~\ref{fig:AbsoluteProtonAndCarbonFluxes} we show a number of
observations of the total proton and carbon flux, together with {\tt
  GALPROP} results~\cite{Ptuskin:2005ax}, for illustration.

\subsubsection{Stable secondary nuclei}
\label{sec:NuclearSecondaries}

Secondary-to-primary ratios are used as test of the propagation model
or to constrain its parameters, like the spectral index $\delta$ and
normalisation $D_{xx0}$ of the diffusion coefficient as well as the
height of the diffusion zone $z_\text{max}$. However, as $D_{xx}$ and
$z_\text{max}$ only enter into the fluxes of stable nuclei as the
ratio $D_{xx}/z_\text{max}$, there is a degeneracy between these
quantities.

The steady-state ambient spectrum of species $i$ of secondary cosmic
rays can also be understood in the framework of the leaky box model,
\begin{align}
I_j \propto \frac{\sum_{j<k} \left( \sigma_{k \rightarrow j} / m +
  \gamma \left( c \beta n_\text{gas} \tau_{k \rightarrow j}
  \right)^{-1} \right) I_k }{1/ X_{\text{esc},j} + \sigma_j / m} \, .
\end{align}
The numerator is dominated by primary species $I_k$ and above a few
GeV/nucleon, \mbox{$I_k(E) \propto E^{\gamma - \delta}$}. The
denominator behaves similarly as for primaries, $X_{\text{esc},j}
\propto E^{-\delta}$. Therefore, the secondary flux is softer than the
primary one by $E^{-\delta}$ and the secondary-to-primary ratio is
falling, $I_2/I_1 \propto E^{-\delta}$.

\begin{figure}[bth]
\begin{center}
\includegraphics[width=0.5\textwidth]{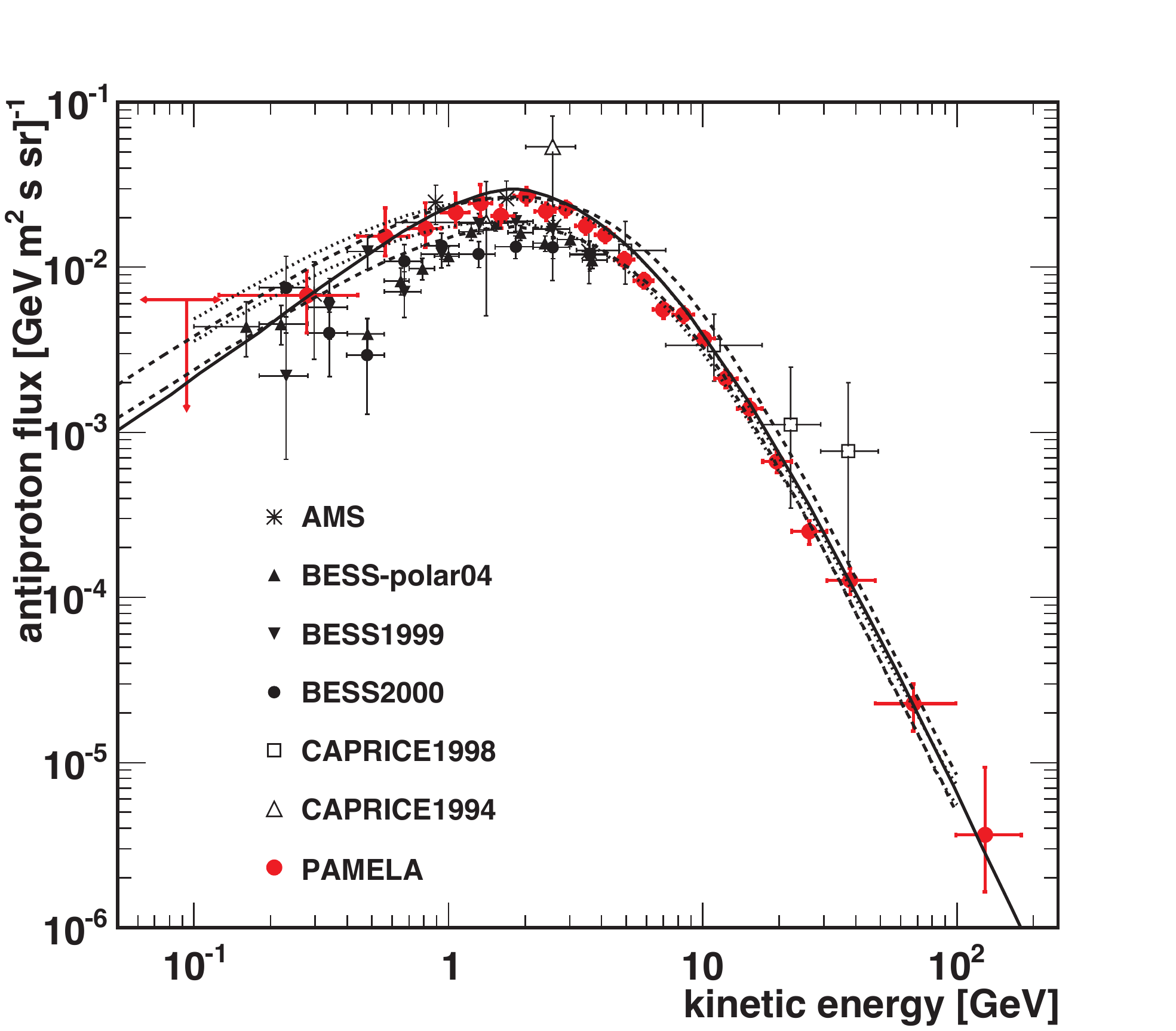}
\hspace{-0.02\textwidth}
\includegraphics[width=0.5\textwidth]{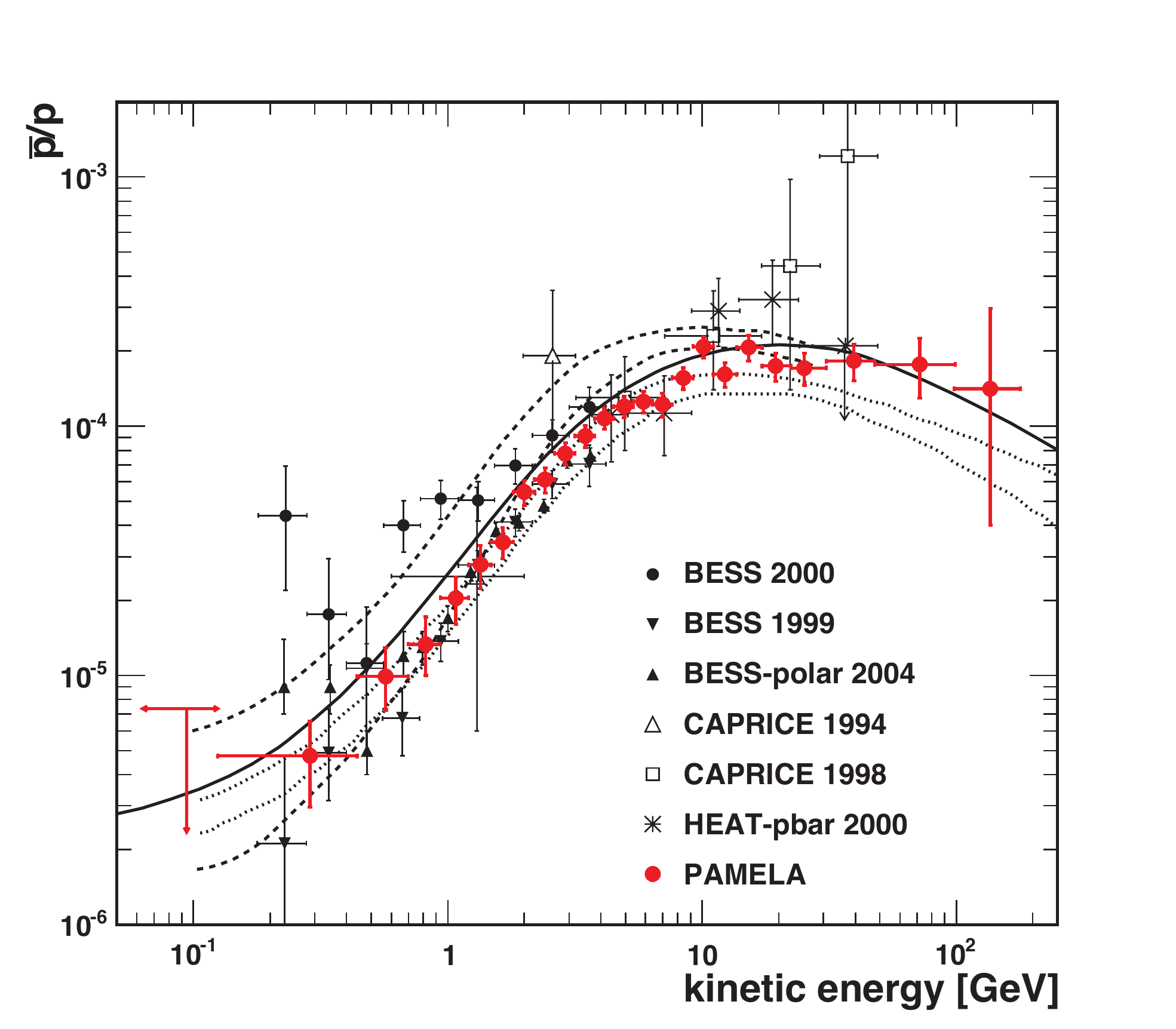}
\end{center}
\caption{{\bf Left:} Absolute antiproton flux from PAMELA and some
  other contemporary
  measurements~\cite{Boezio:1997,Boezio:2001,Asaoka:2001fv,Abe:2008sh,Aguilar:2002}
  (from~\cite{Adriani:2010rc}). The dotted and dashed lines show the
  upper and lower limits for a range of diffusion models, taking into
  account the uncertainties on diffusion model parameters and cross
  sections~\cite{Donato:2001ms} and the solid line shows the
  prediction from a different, plain diffusion
  model~\cite{Ptuskin:2005ax}. {\bf Right:} Antiproton-to-proton ratio
  as measured by PAMELA and some other contemporary
  experiments~\cite{Boezio:1997,Boezio:2001,Asaoka:2001fv,Abe:2008sh,Beach:2001}
  (from~\cite{Adriani:2010rc}). The dashed lines show the upper and
  lower limits for a leaky box model~\cite{Simon:1998}, the dotted
  lines for a diffusive reacceleration convection
  model~\cite{Donato:2008jk} and the solid line the prediction for a
  plain diffusion model~\cite{Ptuskin:2005ax}.  }
\label{fig:AntiprotonRatioAndFlux}
\end{figure}

\begin{figure}[tbh]
\begin{center}
\vspace{1cm} \includegraphics[width=0.45\textwidth]{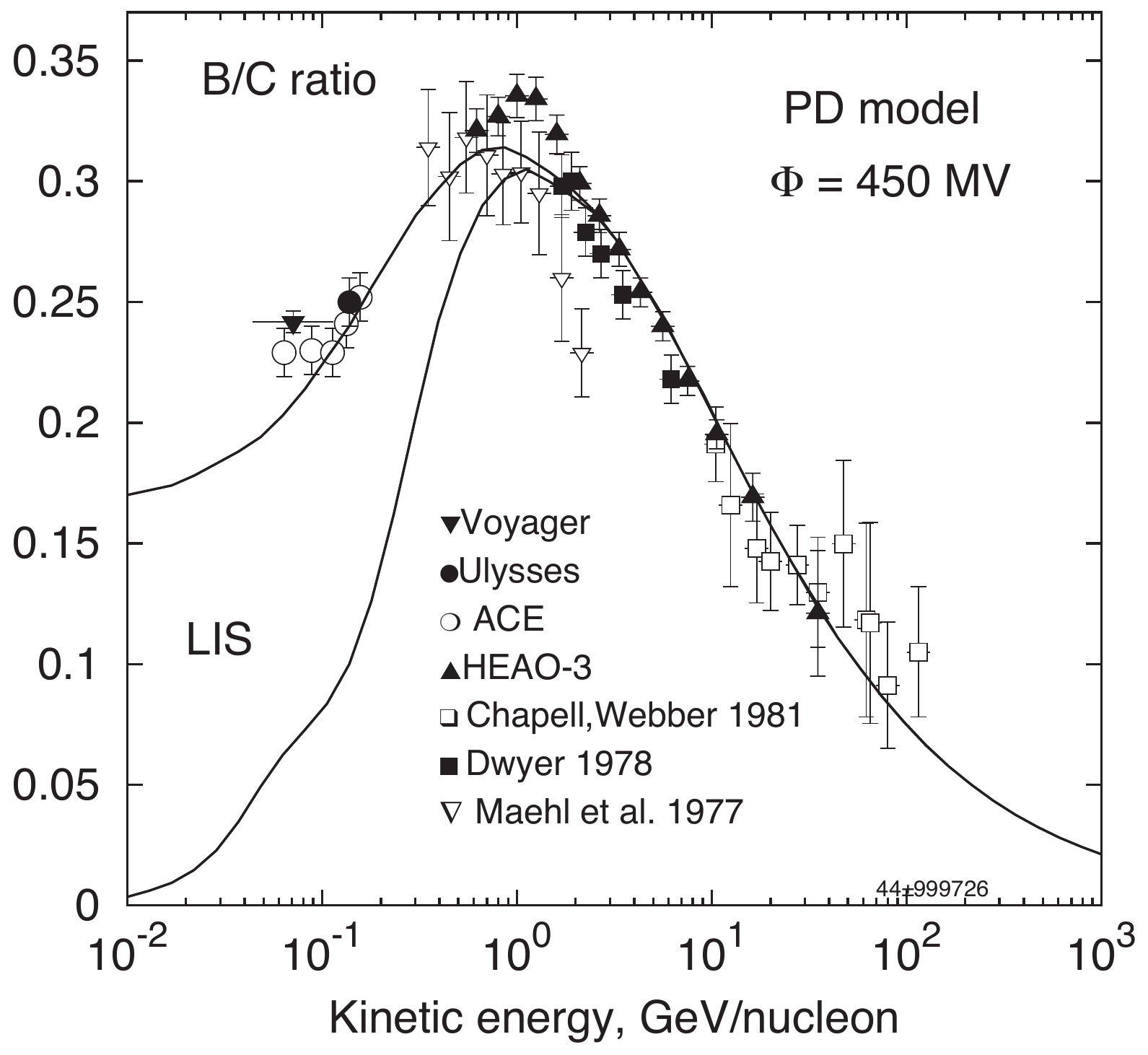}
\hspace{0.05\textwidth}
\includegraphics[width=0.45\textwidth]{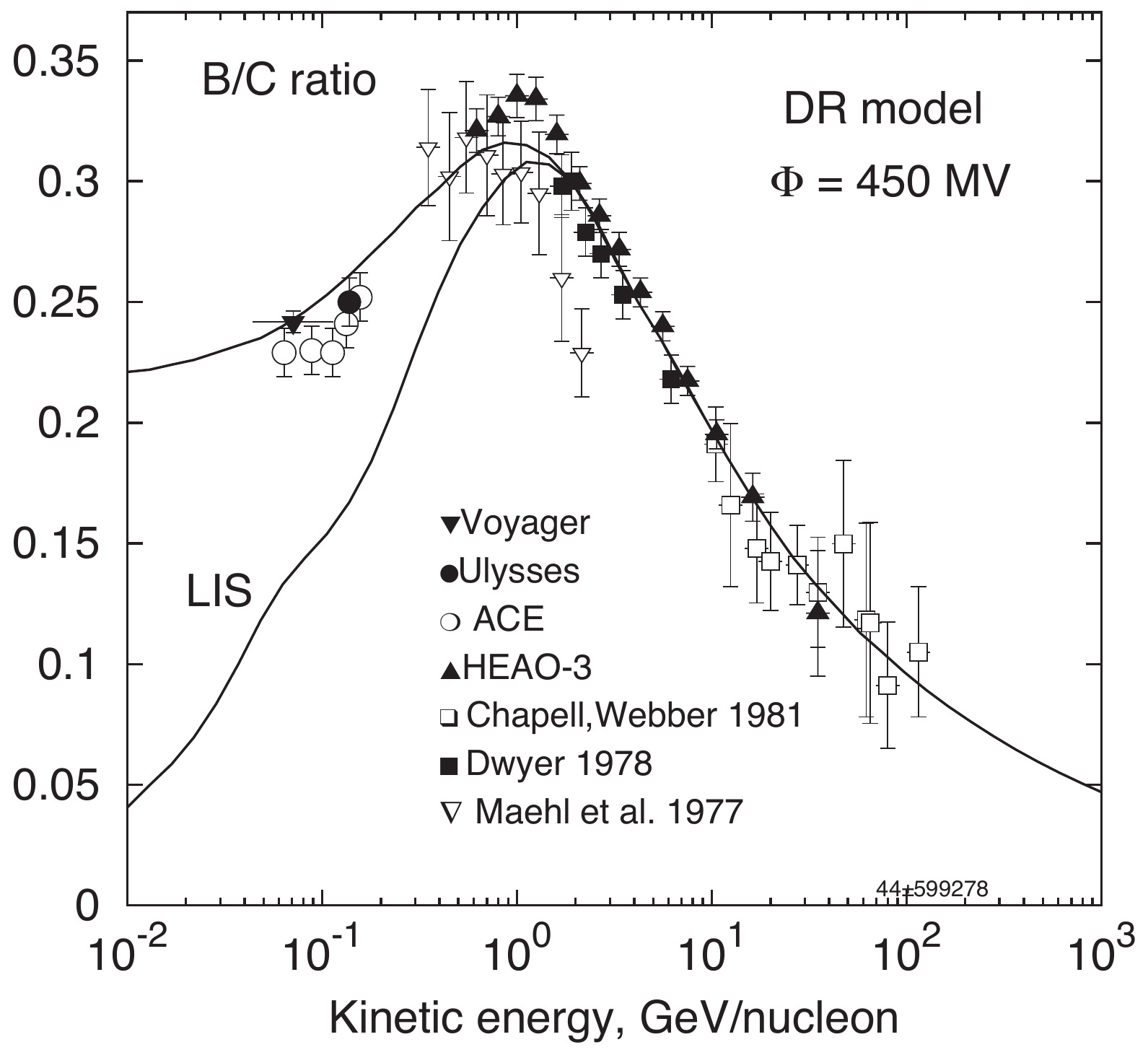}
\end{center}
\caption{Boron-to-carbon ratio as a function of kinetic energy per
  nucleon in a plain diffusion (PD) model (left panel) and a diffusive
  reacceleration (DR) model (right panel)
  (from~\cite{Ptuskin:2005ax}). The lower lines show the local
  interstellar (LIS) flux, the upper lines the solar modulated one
  with \mbox{$\Phi = 450 \, \text{MV}$}. The experimental data below
  200 MeV/nucleon are from ACE~\cite{davis}, Ulysses~\cite{ulysses_bc}
  and Voyager~\cite{voyager} and some high energy data are from
  HEAO-3~\cite{Engelmann:1990zz}. For other references
  see~\cite{StephensStreitmatter98}.}
\label{fig:B2C}
\end{figure}

Measurements of the antiproton-to-proton ratio and the absolute
antiproton flux are shown in Fig.~\ref{fig:AntiprotonRatioAndFlux}
together with some typical theoretical
predictions. Figure~\ref{fig:B2C} shows the B/C ratio, with
predictions from the same {\tt GALPROP} model as in
Fig.~\ref{fig:AbsoluteProtonAndCarbonFluxes} and for a diffusive
reacceleration model~\cite{Ptuskin:2005ax}. The range of spectral
indices allowed by the data ranges from $0.3$ for diffusive
reacceleration to $\sim 0.6$ for plain diffusion models.

\subsubsection{Unstable secondary nuclei}
\label{sec:UnstableNuclearSecondaries}

Stable nuclei propagate over kiloparsecs before escaping from the
cosmic ray halo, and hence the leaky box approximation is a good
approximation for, if not even equivalent to, a diffusion model. For
unstable nuclei the situation is however different. If their lifetime
is smaller than or of the order of the average residence time in the
Galaxy, their decays must be taken into account and escape is now
competing with nuclear decay. As the survival probability, however,
only depends on the time spent since production and not on the
distance (the amount of matter traversed), secondary-to-primary
ratios, like ${}^{10}\text{Be}/{}^9\text{Be}$, can in principle break
the degeneracy between diffusion coefficient $D_0$ and halo height
$z_\text{max}$. At the moment however, the quality of the data, in
particular the statistics, is not good enough to really allow
determining both parameters independently. Future cosmic ray
experiments will provide better data, that will hopefully allow for
the determination of either parameters, as both are in fact important
for other observables, for example diffuse radio or gamma-ray
backgrounds.

\section{Dark Matter Indirect Detection}
\label{sec:DMIndirectDetection}

If the DM particle is in fact a weakly interacting particle produced
by thermal freeze-out, see Sec.~\ref{sec:DMcandidates}, then DM
particles and antiparticles exist in equal amounts. By virtue of the
weak annihilation cross section, the rate of annihilation is
appreciable. A simple estimate of the annihilation rate per volume of
a solar system DM density $\rho_\odot$ gives,
\begin{align}
\frac{1}{2} \frac{\rho_\odot^2 \langle v \sigma
  \rangle}{m_\text{DM}^2} \simeq 1.4 \times 10^{-31} \,
\text{cm}^{-3}\, \text{s}^{-1} \left( \frac{\rho_\odot}{0.3 \,
  \text{GeV} \, \text{cm}^{-3}} \right)^2 \left(
\frac{m_\text{DM}}{100 \, \text{GeV}} \right)^{-2} \left(
\frac{\langle v \sigma \rangle}{3 \times 10^{-26} \, \text{cm}^3 \,
  \text{s}^{-1} } \right) \, .
\end{align}
This amounts to a total rate of $1.6 \times 10^{34} \, \text{s}^{-1}$
in a $1 \, \text{kpc}$ sphere around the Sun or $4.1 \times 10^{37} \,
\text{s}^{-1}$ for an NFW DM density profile, $\rho(r) = 2 \rho_\odot
(r/r_\odot)^{-1} (1 + (r / r_\odot))^{-2}$~\cite{Navarro:1995iw}.

Another possibility is DM decay. Although many DM models invoke a
discrete symmetry to make the DM particle stable what is needed for an
effectively stable DM candidate is actually only that its lifetime is
much longer than the age of the Universe. In order to produce
observable fluxes, the lifetime should however not be too large. A
similar estimate as above gives,
\begin{align}
\frac{\rho_\odot}{m_\text{DM}} \frac{1}{\tau_\text{DM}} \simeq 3
\times 10^{-30} \, \text{cm}^{-3}\, \text{s}^{-1} \left(
\frac{\rho_\odot}{0.3 \, \text{GeV} \, \text{cm}^{-3}} \right) \left(
\frac{m_\text{DM}}{100 \, \text{GeV}} \right)^{-1} \left(
\frac{\tau_\text{DM}}{10^{27} \, \text{s}} \right)^{-1} \, .
\end{align}
Lifetimes of $10^{26} \, \text{s}$ are naturally expected from,
e.g. dimension-6 operators in grand unifying theories (GUTs) which are
suppressed by two powers of the GUT scale
$M_\text{GUT}~=~2~\times~10^{16}~\,~\text{GeV}$,
\begin{equation}
\tau \sim \frac{M_\text{GUT}^4}{m_\text{DM}^5} \sim 10^{26} \,
\text{s} \, ,
\end{equation}
for $m_\text{DM} = 1 \, \text{TeV}$.

The particles produced this way, i.e. cosmic ray protons, antiprotons,
electrons, positrons, gamma-rays and neutrino, could in principle be
observed on or around the Earth. Furthermore, annihilation products
can produce secondary radiation like radio/microwaves from synchrotron
or gamma-rays from ICS in addition to those from astrophysical
CRs. The idea of indirect DM detection is then to extract the
(possible) signal of DM annihilation or decay from the astrophysical
backgrounds and use their spectral or spatial information to constrain
the particle physics model of DM.

Of course, a crucial question and the main purpose of this work is the
determination of the contribution from backgrounds, i.e. the fluxes
from purely astrophysical sources, to the new signals from the
annihilation or decay of DM. On first sight, the constraint on these
fluxes seem to be rather robust: According to the standard paradigm
(cf. Sec.~\ref{sec:SMGCRs}), the primary sources of GCRs are supernova
remnants (SNRs) which are expected to produce power law spectra (see
Sec.~\ref{sec:DSA}) and even after propagation, these fluxes should
have rather featureless spectra. DM annihilation or decay is however
expected to have very different injection spectra: if DM annihilated
in a $2 \rightarrow 2$ process or decayed to two particles, $1
\rightarrow 2$, the energy of the annihilation/decay products would be
$M_\text{DM}$ or $M_\text{DM}/2$, respectively. Hadronisation, for
example of $q \bar{q}$ pairs directly produced or from gauge bosons,
$W^+ \rightarrow q \bar{q}'$, $Z \rightarrow q \bar{q}$ etc., would
considerably broaden the energy spectrum. In any case, the injection
would be much more concentrated in energy and even after propagation,
the spectra should be different from the generic power law type of
astrophysical origin.

In the following sections, we will briefly review the ideas, prospects
and current status of DM indirect detection in different channels
assuming the {\em standard} astrophysical backgrounds as defined by
the standard picture of GCRs. We present the claims of DM signature
made in the charged lepton channels as well as the microwave and
gamma-ray sky maps.

\subsection{Antimatter} 
\label{sec:Antimatter}

Using antimatter produced by annihilation or decay of DM for indirect
detection harnesses the standard predictions for antimatter from
astrophysical sources: a generally much lower abundance than matter
and a particular energy dependence of the ratio of antimatter to
matter. Due to their secondary nature, the spectrum of antimatter
cosmic rays, like antiprotons or positrons, is softer than (primary)
matter spectra and the secondary-to-primary ratios are falling above a
few GeV, $\propto E^{-\delta}$, see Sec.~\ref{sec:NuclearSecondaries}.

In the following, we will briefly review the prospects for detection
in two possible secondary-to-primary ratios, the positron fraction and
the antiproton-to-proton ratio and quickly mention antideuterons.

\subsubsection{Positron fraction} 
\label{sec:PosFrac}

The positron fraction is defined as the ratio of the flux of positrons
$J_{e^+}$ to the total flux of electrons and positrons $(J_{e^+} +
J_{e^-})$,
\begin{equation}
\label{eqn:DefPosFrac}
\text{positron fraction} \equiv \frac{J_{e^+}}{J_{e^+} + J_{e^-}} \, .
\end{equation}
As mentioned above, in the standard picture of GCRs, positrons are
only produced as secondaries by spallation of primary protons and
nuclei which leads to lower abundances and softer spectra for the
positrons and hence a falling positron fraction above a few
GeV~\cite{Serpico:2008te}. The benchmark prediction
in~\cite{Moskalenko:1997gh} was one of the first applications of the
{\tt GALPROP} code.

\begin{figure}[!bt]
\begin{center}
\includegraphics[width=0.5\textwidth]{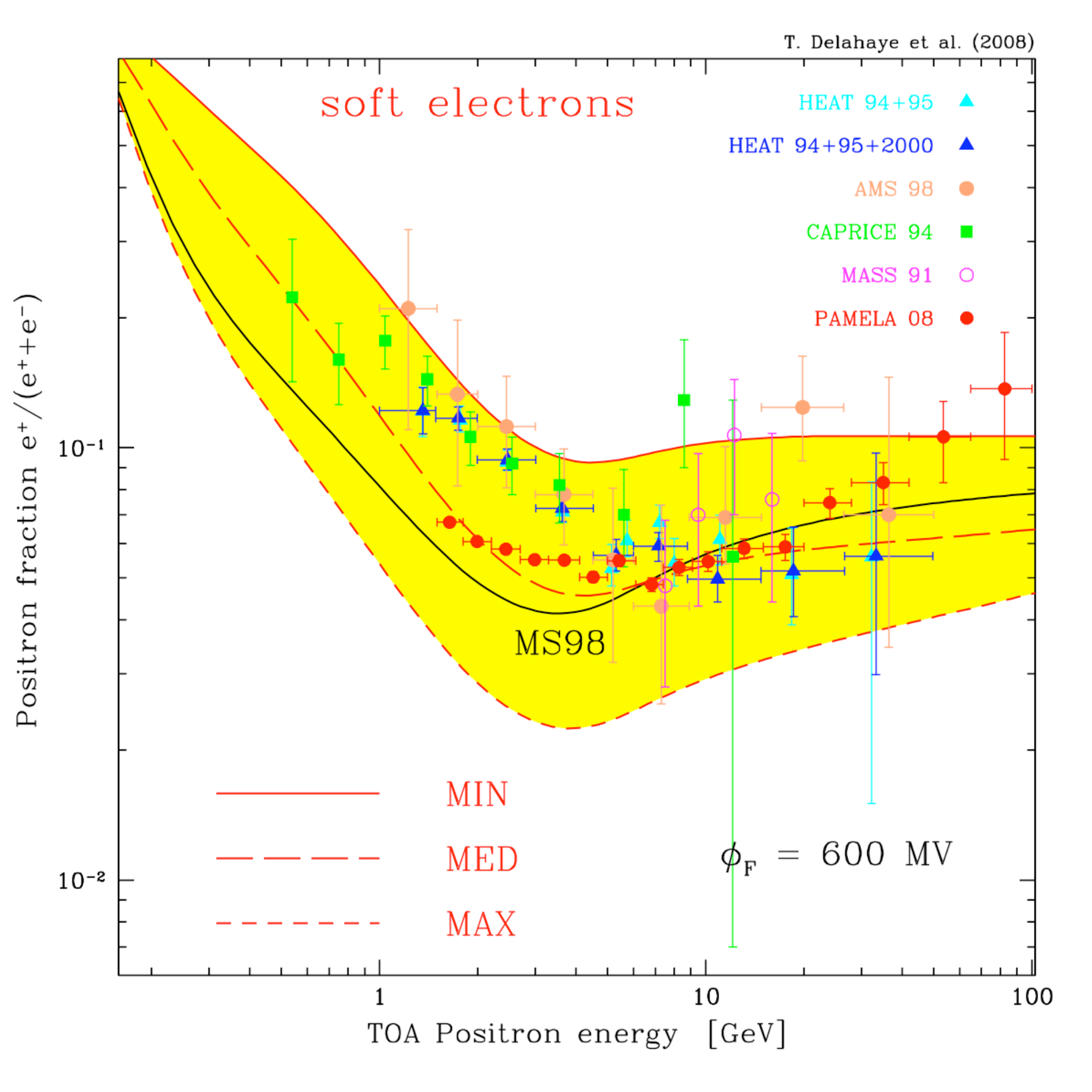}
\hspace{-0.02\textwidth}
\includegraphics[width=0.5\textwidth]{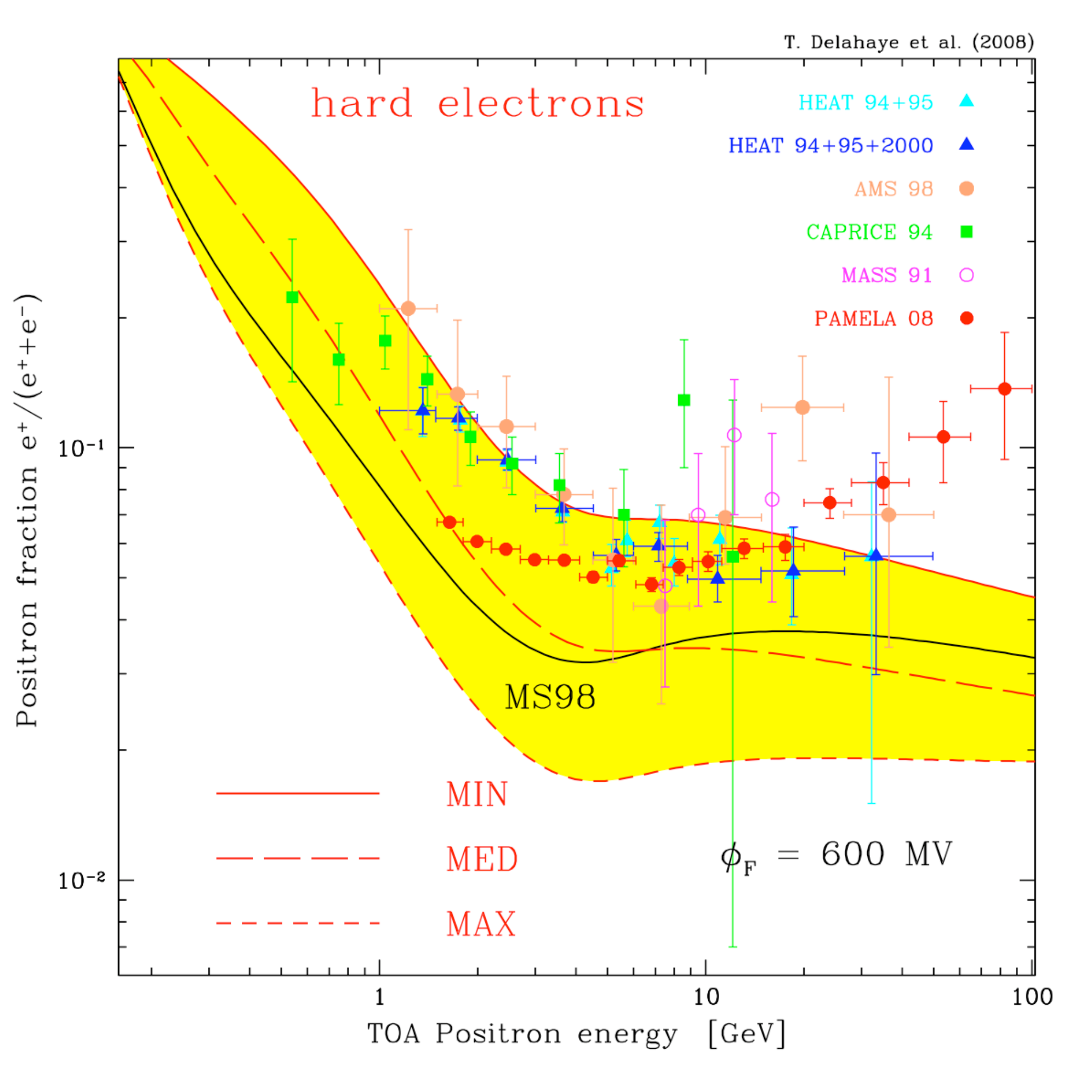}
\end{center}
\caption{The positron fraction as predicted in different cosmic ray
  propagation models for a rather soft (left panel) and harder (right
  panel) electron spectrum (from~\cite{Delahaye:2008ua}). The yellow
  band is spanned by the fluxes from a range of sets of propagation
  model parameters. Three particular models, MIN, MED and MAX (named
  according to the effect on secondary antiprotons) are shown by the
  red solid, long dashed and short dashed lines. The data are from
  CAPRICE~\cite{boe00}, HEAT~\cite{bar97},
  AMS~\cite{Aguilar:2007yf,alc00}, MASS~\cite{2002A&A...392..287G} and
  PAMELA~\cite{Adriani:2008zr}.}
\label{fig:PositronFraction}
\end{figure}

The various sources of uncertainties and their impact on both the
absolute positron flux and the positron fraction have been assessed
in~\cite{Delahaye:2008ua}. The authors calculate the possible range in
the absolute positron flux and the positron fraction from a broad
range of models which differ in the primary proton fluxes adopted, the
cross sections used and the primary electron spectrum assumed. The
range induced by the uncertainty in the propagation model covers a
factor of six between the lowest and the highest possible flux at $1
\, \text{GeV}$ but decreases to a factor 2.9 at $100 \,
\text{GeV}$. In particular, two different spectral indices for the
locally measured electron flux are used, $\alpha = - 3.35$ (hard) and
$\alpha = - 3.53$ (soft), and the positron fraction is always found to
be falling for the hard electron spectrum and only very slightly
rising for the soft electron spectrum,
cf. Fig.~\ref{fig:PositronFraction}. In fact, the (still preliminary)
absolute electron flux measured by PAMELA~\cite{Boezio:2010}, can be
fitted with a power law with spectral index $\alpha = - 3.226 \pm
0.020$ between $10$ and $60 \, \text{GeV}$ and is even harder
below. Therefore, the soft electron spectrum is ruled out and even
within the rather large uncertainties, the positron fraction from
secondary production is always falling.

This is however in disagreement with the recent
findings~\cite{Adriani:2008zr} from the PAMELA experiment which shows
a \emph{rising} positron fraction above $\sim 5 \, \text{GeV}$, see
Fig.~\ref{fig:PositronFraction}. This trend was already apparent in
the combined (1994, 1998, 2000) HEAT data~\cite{bea04}, although with
much larger uncertainties. Even more than 20 years ago, the rise in
the positron fraction was realised \cite{Boulares:1989} and tried to
explain with some exotic contribution, possibly from pulsars.

Of course, in particular in connection with claims of excesses in the
absolute electron plus positron flux by PPB-BETS~\cite{Torii:2008xu},
ATIC~\cite{Chang:2008zzr} and Fermi-LAT~\cite{Abdo:2009zk} (see also
Sec.~\ref{sec:Total_e-e+_Flux} below), this result has generated much
attention, mainly prematurely interpreted as evidence for WIMP
annihilation. In fact, within a year of the presentation and
publication of the PAMELA data, a multitude of models was put forward
(see, for example,~\cite{Profumo:2008ms} for a comprehensive, though
non-exhaustive list of references) that could explain the upturn in
terms of the onset of a new, harder primary positron component from DM
annihilation or decay.

The contribution from DM annihilation can be taken into account both
in (semi) analytic and in fully numerical computations by an
additional source term on the RHS of the transport
equation~\ref{eqn:GCRtransport},
\begin{equation}
\label{eqn:SourceTermFromDMannihilation}
q_{e^\pm}^\text{DM} (r,z,E_{e^\pm}) = \frac{1}{2} \langle
\sigma_\text{ann} v \rangle g(E_{e^\pm}) \left(
\frac{\rho_\chi(r,z)}{m_\text{DM}} \right)^2 \, .
\end{equation}
Here, $\langle \sigma_\text{ann} v \rangle$ denotes the averaged
annihilation cross section times velocity, $\rho_\chi(r,z)$ the DM
halo profile and $g(E_{e^\pm})$ is the differential production
spectrum. The latter encodes the particle physics (obviously dependent
on the DM model considered). Adding this flux to the one from
astrophysically produced secondary positrons, predictions for the
positron fraction can be made.

However, with a thermal annihilation cross section, \mbox{$\langle
  \sigma_\text{ann} \, v \rangle \approx 3 \times 10^{-26} \,
  \text{cm}^3 \, \text{s}^{-1}$}, and standard astrophysical
assumptions about the DM halo, e.g. NFW \cite{Navarro:1995iw} or
isothermal \cite{Bahcall:1980fb} profile, local DM density of $\sim
0.3 \, \text{GeV} \, \text{cm}^{-3}$, it turns out that the additional
fluxes fall short by between a factor of 10 to 1000, depending on the
DM mass (see, e.g.~\cite{Cirelli:2008pk}). It would be very difficult
to detect such small deviations from the astrophysical backgrounds, in
particular considering the uncertainties involved.

In fact under certain assumptions one can expect an additional ``boost
factor'' to appear in Eq.~\ref{eqn:SourceTermFromDMannihilation},
either from astrophysics or particle physics. Astrophysical boost
factors are induced by overdensities in the distribution of DM as for
example predicted by $N$-body simulations. This does not only lead to
a global amplification compared to a smooth DM density, $\int \dd V \,
\rho^2 > (\int \dd V \, \rho)^2$, but in particular to potentially
``bright'' nearby clumps of DM. However, the probability of such a DM
overdensity close enough to the Earth to explain the excess in the
positron fraction is very small~\cite{Brun:2009aj}. Particle physics
boost factors can come from a low-velocity enhancement of the
annihilation cross section, for example by a resonance just below $2
M_\text{DM}$~\cite{Ibe:2008ye} or so-called Sommerfeld
enhancement~\cite{Sommerfeld:1931,Hisano:2003ec,Hisano:2006nn}. It is
also possible to overcome the helicity suppression of the $s$-wave
annihilation into two fermions by emitting a photon. Alternatives
include giving up the WIMP paradigm by, e.g. assuming non-thermal DM
production (see~\cite{Kane:2009if} for the example of Wino-like
neutralino DM).

In terms of annihilation channels, the PAMELA positron fraction
\emph{alone} allow annihilation to charged leptons for a wide range of
DM masses, $m_\text{DM} \gtrsim 100 \,
\text{GeV}$~\cite{Cirelli:2008pk}. For light DM, $m_\text{DM} \sim 100
\, \text{GeV}$, annihilation into W bosons is also possible.

\subsubsection{Absolute antiproton flux and antiproton-to-proton ratio} 

For a generic WIMP model, there is {\it a priori} no reason why DM
should only annihilate or decay into electrons and positrons or other
charged leptons. In fact, heavy quarks possibly produced by DM
annihilation or decay will hadronise to all sorts of baryons,
ultimately also yielding protons and antiprotons. The
antiproton-to-proton ratio,
\begin{equation}
\label{eqn:DefAntiProtonRatio}
\text{antiproton-to-proton ratio ratio} \equiv \frac{J_{\bar{p}}}{J_p}
\, ,
\end{equation}
and the absolute antiproton flux can therefore be used as a cross
check for DM explanations for the positron excess.

First of all, one should note that predictions for the background of
antiprotons from the spallation of (mostly) GCR protons on the
interstellar H and He are able to reproduce the measurements by
BESS~\cite{Abe:2008sh}, CAPRICE~\cite{Boezio:2001} and
AMS-01~\cite{Aguilar:2002} on the absolute antiproton flux as well as
the most recent PAMELA data (see Sec.~\ref{sec:NuclearSecondaries} and
Fig.~\ref{fig:AntiprotonRatioAndFlux}). A detailed study of the
background investigating the uncertainties from the cross sections
used and the diffusion parameters adopted, has been presented
in~\cite{Donato:2001ms}.

The potential contribution from the annihilation of DM into
antiprotons can be calculated within the theoretical frameworks
presented in~\ref{sec:Transport} or using fully numerical codes like
{\tt GALPROP}, starting from a spatially varying injection term,
\begin{equation}
q_{\bar{p}}^\text{DM} (r,z,E_{\bar{p}}) = \frac{1}{2} \langle
\sigma_\text{ann} v \rangle g(E_{\bar{p}}) \left(
\frac{\rho_\chi(r,z)}{m_\text{DM}} \right)^2 \, .
\end{equation}
Here, $g(E_{\bar{p}})$ is the differential production spectrum,
assembled from the branching ratios $B_{h}$ into quarks or gluons $h$
in different channels F, as well as the fragmentation and
hadronisation functions $\dd N_{\bar{p}}^h / \dd E_{\bar{p}}$,
\begin{equation}
g(E_{\bar{p}}) = \sum_{\text{F},h} B_{h}^{(\text{F})} \frac{\dd
  N_{\bar{p}}^h}{\dd E_{\bar{p}}} \, ,
\end{equation}
where $E_{\bar{p}}$ is the antiproton's kinetic energy.

Considering again the uncertainties introduced by the GCR propagation
parameters but also by the DM halo model adopted, the semi-analytic
analysis of~\cite{Donato:2003xg} finds a large proportion of a scan
over a particular MSSM parameter space to be consistent with the
antiproton fluxes. We note that in this study no boost factors were
introduced.

However, antiproton measurements can give stringent
constraints~\cite{Cirelli:2008pk,Donato:2008jk} on a number of models
invoking boost factors to explain the anomalies in the lepton
channels, e.g. the positron excess. Lepton channels are still
available at all masses, but the antiproton data basically exclude
hadronic annihilation channels with $m_\text{DM} \lesssim \text{a few}
\, \text{TeV}$. Above $10 \, \text{TeV}$, both leptonic and hadronic
channels (excluding, perhaps, direct annihilation into quarks) give
both good fits to positron and antiproton data. For light and
intermediate masses, however, annihilation into gauge and Higgs bosons
must somehow be suppressed. For an example of a model building way out
of this, see~\cite{ArkaniHamed:2008qn}.

\subsubsection{Anti deuterons}

Following the idea of DM indirect detection in rare nuclei it was
suggested~\cite{Donato:1999gy} to look for anti deuterons $\bar{\rm
  D}$ from the annihilation of DM. Secondary antideuterons get
produced by GCR $p$, He and $\bar{p}$ impinging on interstellar H and
He, with $p \,$H and $p \,$He interactions dominating but
contributions from $\bar{p}$ becoming comparable below $1 \,
\text{GeV}/n$. Uncertainties from the cross section and again from the
cosmic ray model amount to an order of magnitude at $\sim 0.1 \,
\text{GeV}/n$, decreasing to about a factor four at $\sim 100 \,
\text{GeV}/n$~\cite{Donato:2008yx}, similar to the behaviour for the
antiproton flux~\cite{Donato:2001ms}.

Although currently only upper limits on the $\bar{\rm D}$ flux
exist~\cite{Fuke:2005it}, it turns out that with reasonable
assumptions for the propagation model the ratio of DM signal to
astrophysical secondary background is usually larger than $0.5$ below
$1 \, \text{GeV}/n$ for DM masses up to hundreds of GeV. Anti
deuterons are therefore one of the most promising detection channels
for light and intermediate WIMP masses, and it has been
shown~\cite{Donato:2008yx} that with the sensitivity of the
forthcoming GAPS long-duration balloon flight
experiment~\cite{Fuke:2008} a large fraction of a low-energy MSSM
parameter space is accessible.

\subsection{The total electron-positron flux}
\label{sec:Total_e-e+_Flux}

In contrast to stable nuclei, electrons and positrons lose their
energy quickly through synchrotron radiation and inverse Compton
scattering, see Sec.~\ref{sec:Transport}. They can therefore only
travel finite, energy-dependent distances before losing all their
energy. The flux on Earth at the highest, $\mathcal{O}(1) \,
\text{TeV}$ energies, is therefore dominated by the closest and
youngest sources. In particular, considering that due to the
discreteness of the sources in space and time there is/are necessarily
a (few) source(s) of minimum distance and age, one predicts a
propagation cut-off in energy. The older and further sources however
add up to a smooth spectrum below $\sim 100 \, \text{GeV}$. For a more
detailed and quantitative discussion see
Sec.~\ref{sec:DiscreteSources}.

\begin{figure}[th]
\begin{center}
\includegraphics[width=0.8\textwidth]{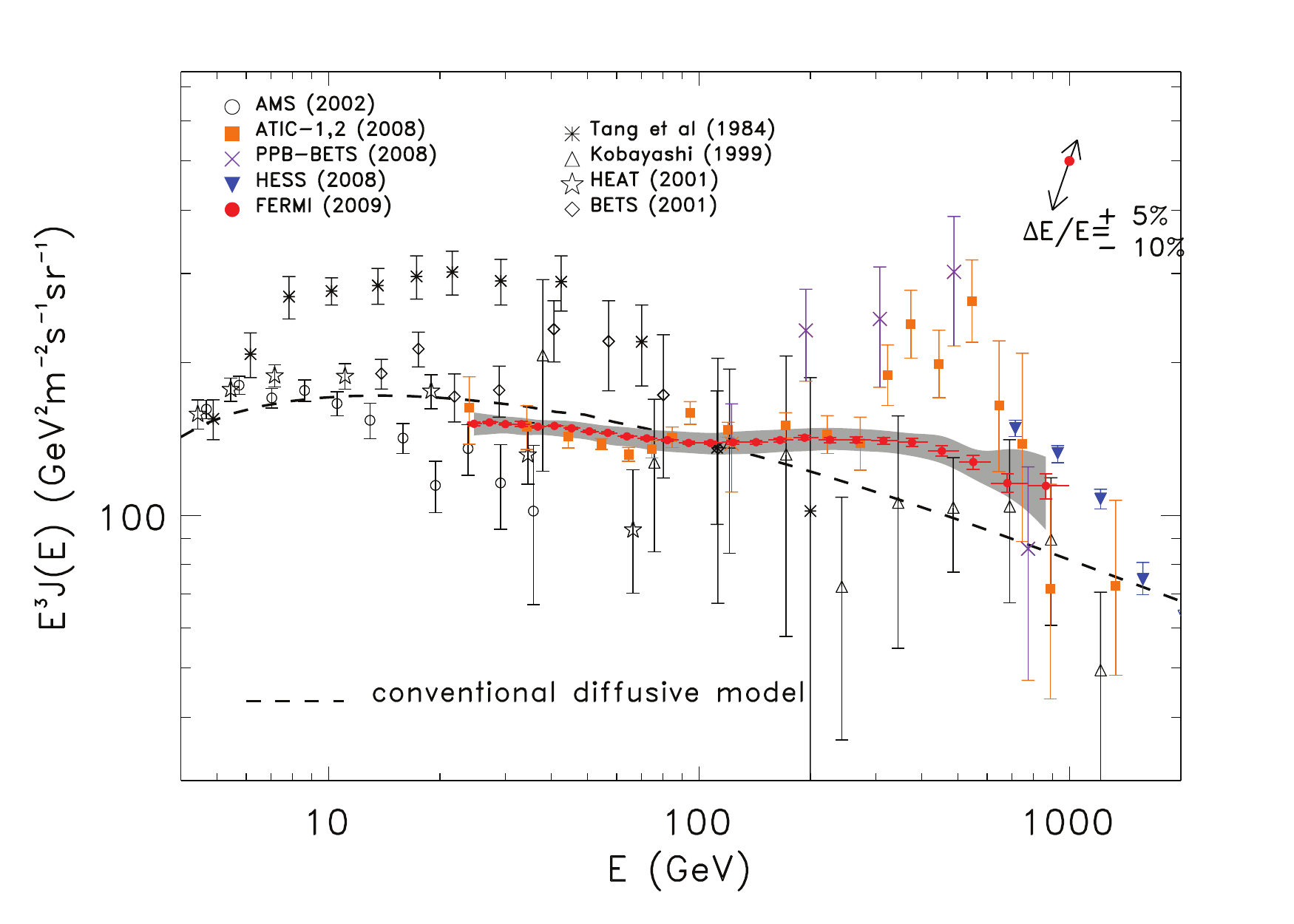}
\end{center}
\caption{Absolute electron plus positron flux $J(E)$, scaled by $E^3$
  (from~\cite{Abdo:2009zk}). The red data points with statistical
  error bars are from the Fermi-LAT measurement~\cite{Abdo:2009zk},
  the grey band denotes the systematic uncertainty and the double
  headed arrow the energy scale uncertainty. Other measurements
  \cite{Aguilar:2002,Chang:2008zzr,Torii:2008xu,Collaboration:2008aaa,Tang:1984ev,Kobayashi:1999,DuVernois:2001,Torii:2001}
  are referenced in the legend. The dashed line is from a {\tt
    GALPROP} conventional diffusive model~\cite{Strong:2004de}.}
\label{fig:AbsElPosFlux}
\end{figure}

The combined differential flux of GCR electrons and positrons,
$(J_{e^-} +J_{e^+})$, has been measured by a number of experiments
over the last decades, however, with considerable scatter, see
Fig.~\ref{fig:AbsElPosFlux}. Here, we focus on two more recent, but
somewhat contradictory results which both hint at some sort of excess
with respect to pre-Fermi, standard GCR propagation
models~\cite{Moskalenko:1997gh,Grasso:2009ma}.

On the one hand, a sharp feature, i.e. a hardening from an $E^{-3}$
spectrum at $\sim~100 \, \text{GeV}$ and a rather strong shoulder at
$\sim 800 \, \text{GeV}$, has been observed independently by
PPB-BETS~\cite{Torii:2008xu} and ATIC~\cite{Chang:2008zzr}. On the
other hand, the Fermi collaboration has performed an
analysis~\cite{Abdo:2009zk} with data from their Large Area Telescope
(LAT) which shows a smooth continuation of the $E^{-3}$ spectrum up to
hundreds of GeV where the spectrum starts to soften. This is in
agreement with a determination of the flux by
HESS~\cite{Collaboration:2008aaa,Aharonian:2009ah} which also shows a
softening to $E^{-4}$ around a TeV and is basically also inconsistent
with the PPB-BETS/ATIC findings. Both signals, i.e. either a sharp
feature or a broader excess, can in principle be interpreted as a
contribution from DM annihilation or decay into leptonic channels in
the galactic DM halo, cf. Sec.~\ref{sec:PosFrac}.

We note that an $E^{-3}$ total electron-positron flux {\it per se}
does not constitute an excess with respect to the standard picture of
GCRs. For example a steady, homogeneous $E^{-2.2}$ injection of
electrons together with a diffusion coefficient $\propto E^{-0.6}$
could reproduce such a spectrum. The pre-Fermi models were adjusted to
earlier measurements which indicated a softer spectrum. On the other
hand, the slight dip and bump in the Fermi-LAT data at $\sim 100 \,
\text{GeV}$ and $\sim 300 \, \text{GeV}$, respectively, as well as a
rather soft primary injection spectrum derived from, e.g. gamma-rays
from SNRs (see \ref{sec:PrimaryElectrons}) can however be interpreted
as indications for the presence of an additional component in the
total electron-positron flux.

Again presuming that the excesses seen in the total electron-positron
flux are DM induced, we can further constrain possible annihilation
channels and DM masses. Whereas the PPB-BETS/ATIC data favour
annihilation into $\mu^+ \mu^-$ with $m_\text{DM} \sim 1 \,
\text{Tev}$, the smoother Fermi-LAT spectrum is better fit by
annihilation to $\tau^+ \tau^-$ with $m_\text{DM} \sim 2 \,
\text{TeV}$~\cite{Meade:2009iu}. For a harder background the Fermi-LAT
data also allow the $\mu^+ \mu^-$
channel~\cite{Bergstrom:2009fa}. Heavier DM can however not reproduce
either excess and therefore, DM models leading to hadronic excesses
(even beyond the current reach of antiproton measurements) are ruled
out.

\subsection{Gamma-rays} 

The major appeal of DM indirect detection through gamma-rays is that
in contrast to charged particles, gamma-rays free-stream through the
Galaxy and allow to be traced back to their sources. This allows, for
example, to investigate directions in the Galaxy that are particularly
apt for DM detection, e.g. due to low astrophysical
backgrounds. Furthermore, gamma-rays reach the Earth without
attenuation or energy losses, apart from redshift losses. (This is
true at least up to energies of hundreds of TeV; above, the mean free
path for gamma-rays sharply drops due to $e^+ \, e^-$ pair production
on the CMB down to $10 \, \text{kpc}$ at a few PeV.)

The physical processes possibly contributing to gamma-rays from DM
annihilation or decay are prompt emission, i.e. internal
bremsstrahlung (final state radiation and virtual internal
bremsstrahlung) and neutral pion decay (from hadronic decay modes), or
ICS of electrons and positrons, all of which have continuous
spectra. The annihilation or decay to $\gamma \gamma$, $\gamma Z$ or
$\gamma H$ is, although loop-suppressed (DM is electrically neutral)
and with a branching fraction of $10^{-3}$ or smaller, important
because the resulting mono-energetic line emission is a very clean
signature.

Again, we distinguish between DM annihilation and decay because of the
different dependence on the DM density. As the annihilation rate
depends on the DM density square, the differential flux of gamma-rays
$\phi_\gamma^\text{ann}$ in a particular solid angle $\Delta \Omega$
around a direction $(\ell, b)$ in the sky contains the line of sight
(l.o.s.) integral of the density square, $\rho^2(r)$,
\begin{equation}
J_\gamma^\text{ann}(\ell, b) = \frac{1}{4 \pi} \frac{1}{2}
\frac{\langle \sigma_\text{ann} v \rangle}{m_\text{DM}^2} \sum_f
\frac{\dd N_\gamma^f}{\dd E_\gamma} B_f \int_{\Delta \Omega} \dd
\Omega \int_\text{l.o.s.} \dd s \, \rho^2(r(s,\ell,b)) \, ,
\label{eqn:GammaRaysFromAnnihilation}
\end{equation}
where $\langle \sigma_\text{ann} v \rangle$ is the thermal average
annihilation rate, $\dd N_\gamma^f / \dd E_\gamma$ is the differential
spectrum for final state $f$ and $B_f$ is the branching ratio for
gamma-rays.

The decay rate on the other hand is proportional to the DM density, so
the differential gamma-ray flux $\phi_\gamma^\text{dec}$ from DM decay
reads,
\begin{equation}
J_\gamma^\text{dec}(\ell, b) = \frac{1}{4 \pi}
\frac{\Gamma}{m_\text{DM}} \sum_f \frac{\dd N_\gamma^f}{\dd E_\gamma}
B_f \int_{\Delta \Omega} \dd \Omega \int_\text{l.o.s.} \dd s \,
\rho(r(s, \ell, b)) \, ,
\end{equation}
with $\Gamma$ the DM decay rate.

The recent data from the Large Area Telescope (LAT) on board the Fermi
satellite~\cite{Atwood:2009ez} allow for various searches for possible
DM signatures with a so far unprecedented accuracy. The LAT is a
pair-conversion gamma-ray detector consisting of $4 \times 4$ towers
of tungsten trackers on top of a electromagnetic calorimeters with a
thickness of 7 radiation lengths and is surrounded by anti-coincidence
detectors. It has a broad energy range of $20 \, \text{MeV}$ to $300
\, \text{GeV}$ and with its large field of view of $2.4 \, \text{sr}$
it covers the whole sky in two orbits, i.e. three hours. Its
energy-resolution is around $10 \, \%$ and its point spread function
decreases from a few degree at $20 \, \text{MeV}$ to less than
$0.1^\circ$ at hundreds of GeV.

In the following, we briefly review the different targets for DM
indirect detection in gamma-rays.

\subsubsection{Galactic centre}

The galactic centre is a prime target for DM searches since it
contains the highest DM density and is therefore expected to be the
brightest DM source in the sky. Of course, astrophysical backgrounds
are also very bright close to the galactic centre, and so the
prospects strongly depend on the form of the DM density profile in the
inner kiloparsecs. On the one hand, density profiles fitted to
``observations'' in $N$-body simulations hint at a ``cuspy'' NFW
profile~\cite{Navarro:1995iw,Navarro:1996gj} with a rather steep,
$r^{-1}$ decline in the central kpc. On the other hand, kinematical
observations show rather flat, iso-thermal
profiles~\cite{Bahcall:1980fb}. Again, the amplification due to the
$\rho^2(r)$ factor leads to much larger signals from DM annihilation
than from decay.

\begin{figure}[!tb]
\begin{center}
\includegraphics[width=0.45\textwidth]{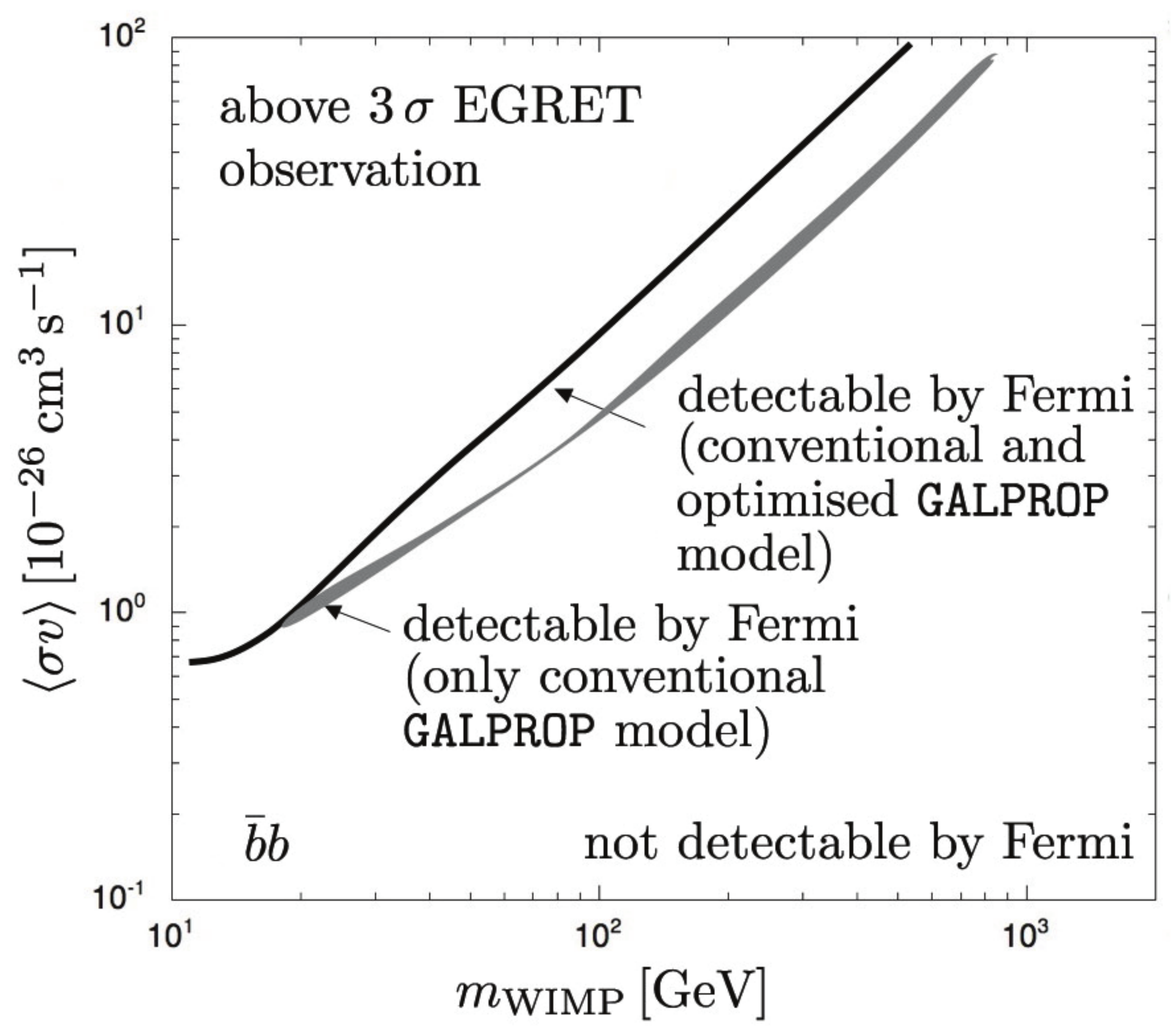}
\hspace{0.05\textwidth}
\includegraphics[width=0.45\textwidth]{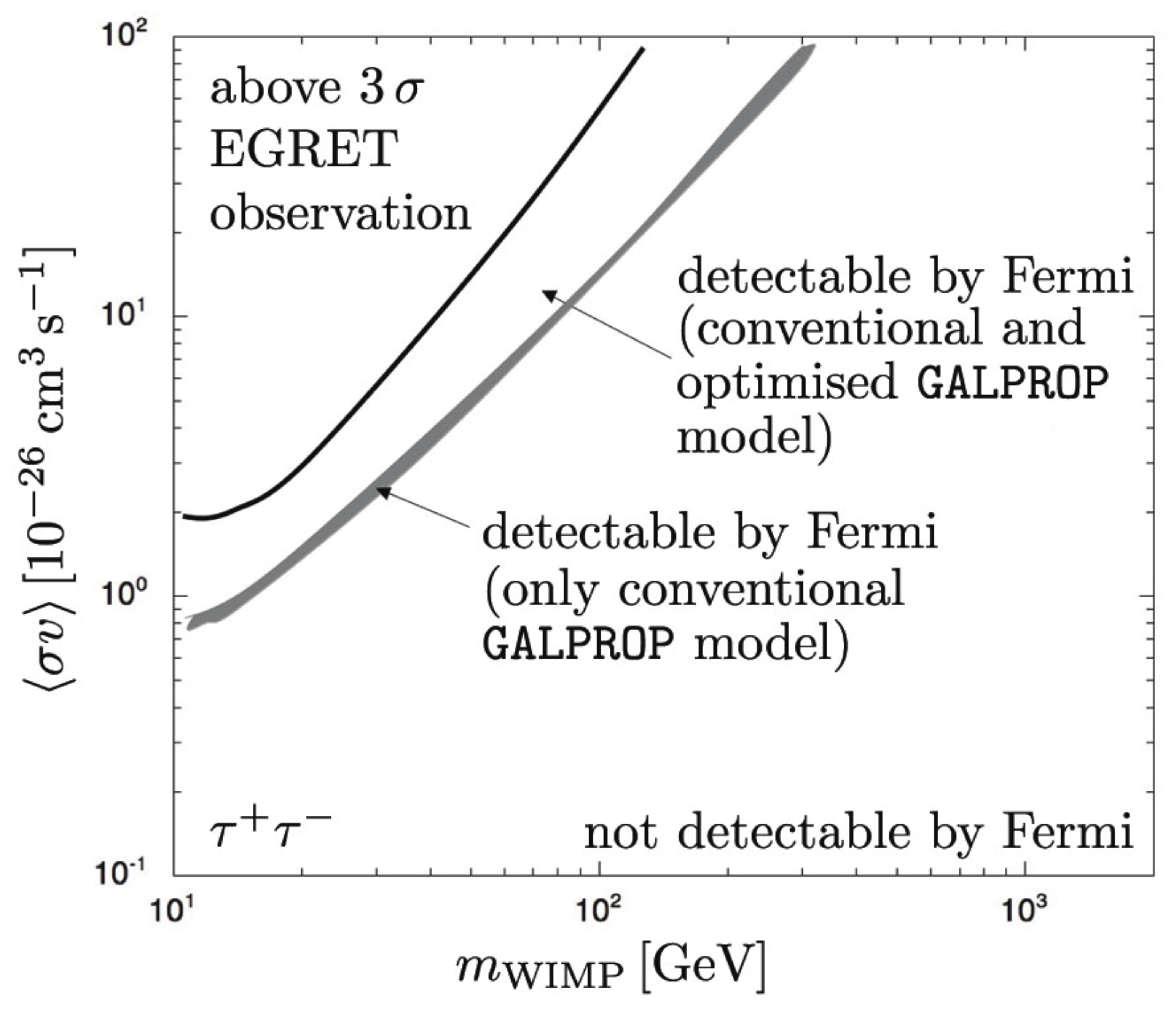}
\end{center}
\caption{Significance for detection in $b \bar{b}$ (left panel) and
  $\tau^+ \tau^-$ (right panel) in the $m_\text{DM} - \langle
  \sigma_\text{ann} v \rangle$ plane (from~\cite{Baltz:2008wd}). The
  region above the black line is excluded by EGRET, the region below
  the grey line is not detectable by Fermi-LAT and the region in
  between is detectable by Fermi-LAT if the
  `conventional'~\cite{Strong:1998fr} or
  `optimised'~\cite{Strong:2004de} background model is realised. The
  shaded region can only be detected under assumption of the
  `conventional' background~\cite{Strong:1998fr}.}
\label{fig:BaltzPreFlightStudy}
\end{figure}

A sensitivity study~\cite{Baltz:2008wd}, considering two different
{\tt GALPROP} diffuse backgrounds~\cite{Strong:1998fr,Strong:2004de},
but no point sources, has been performed pre-flight. For generic DM
annihilation into $W^+ W^-$, $b \bar{b}$, $t \bar{t}$ and $\tau^+
\tau^-$, an NFW density profile and accounting for detector response,
the sensitivity after 5 years of data is good enough to dig deeper
into the $m_\text{DM} - \langle \sigma_\text{ann} v \rangle$ plane
than possible with
EGRET~\cite{MayerHasselwander:1998hg,Dodelson:2007gd}. In
Fig.~\ref{fig:BaltzPreFlightStudy}, we reproduce the sensitivity plots
for gamma-rays for annihilation to $b \bar{b}$ and $\tau^+ \tau^-$
from~\cite{Baltz:2008wd}.

Shortly after the public release of the first 11 month data, an excess
above astrophysical backgrounds has been claimed. A bump-like feature
around $\sim 2 \, \text{GeV}$ above power-law background has been
identified~\cite{Goodenough:2009gk} within the inner $3^\circ$. The
excess has a steeper radial profile and is more spherically symmetric
than astrophysical backgrounds. It could be explained by $25-30 \,
\text{GeV}$ dark matter particle, annihilating to $b \bar{b}$ with
$\langle \sigma_\text{ann} v \rangle \sim 9 \times 10^{-26} \,
\text{cm}^3 \, \text{s}^{-1}$ in a cusped density profile.

\begin{figure}[!th]
\begin{center}
\includegraphics[width=1\textwidth]{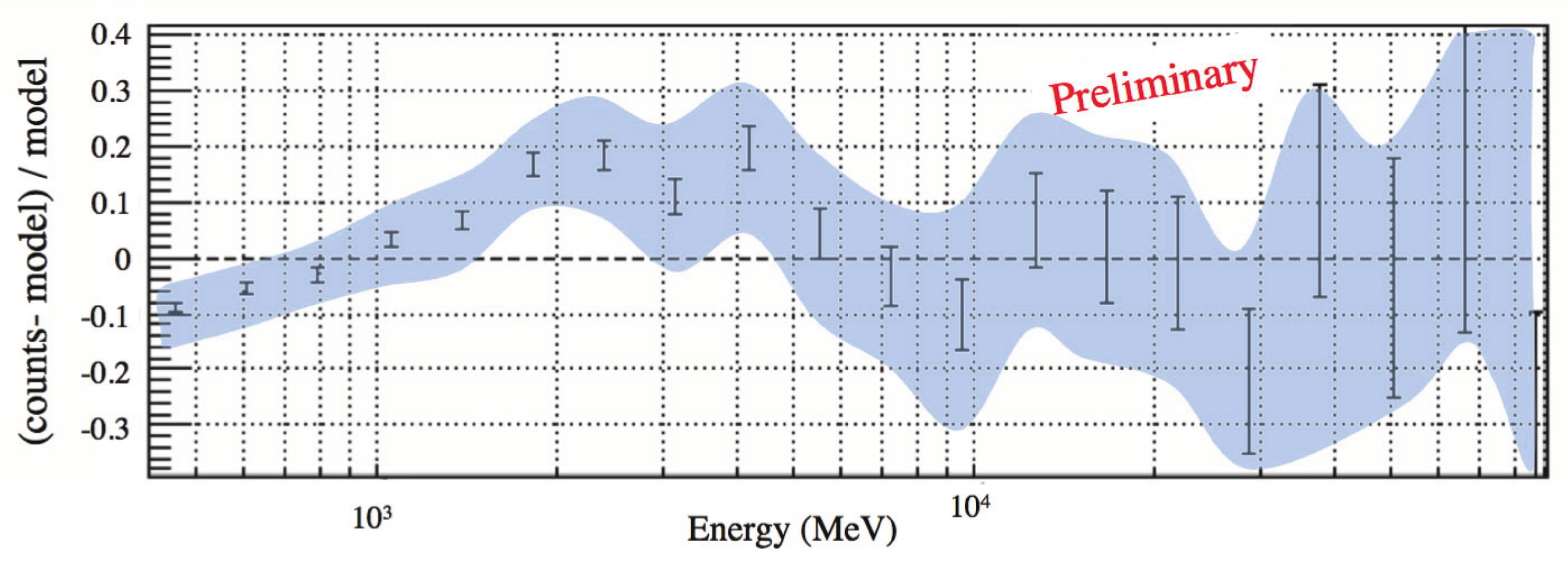}
\end{center}
\caption{Residual spectrum from the likelihood analysis
  of~\cite{Vitale:2009hr}, integrated over the $7^\circ \times
  7^\circ$ ROI (from~\cite{Vitale:2009hr}). The blue band shows the
  systematic error from the uncertainty on the effective area.}
\label{fig:Vitale09}
\end{figure}

The Fermi collaboration is of course performing their own studies of
the emission around the Galactic centre but has, so far, only
presented very preliminary results. A binned likelihood
analysis~\cite{Vitale:2009hr} with a simultaneous spatial and spectral
fit has been performed within a region of interest (ROI) of $7^\circ
\times 7^\circ$ around the galactic centre, using only high-quality
reconstructed events gathered during the first 11 month of
operation. Various contributions have been considered, in particular,
point sources have been modelled and subtracted, the diffuse galactic
emission is accounted for by a {\tt GALPROP} model and an isotropic,
extra-galactic background is allowed for. The residuals, see
Fig.~\ref{fig:Vitale09}, also show an unmodelled excess in the $2 - 5
\, \text{GeV}$ range. The Fermi collaboration however concludes, that
better modelling of the galactic diffuse emission and possibly
unresolved point sources is necessary before any excess can be
confirmed.

\subsubsection{Galactic halo}
\label{sec:GalacticHalo}

Although DM contribution to gamma-ray fluxes from the galactic halo
are comparatively lower than from the galactic centre, the
astrophysical backgrounds, in particular from (unresolved) sources,
are also less strong. Still, diffuse backgrounds remain a formidable
challenge for these studies. Furthermore, the spectrum from DM
annihilation or decay is the same anywhere in the halo and is expected
to have a particular shape with sharp cut-off at the DM mass (half the
DM mass for decay) which is difficult to explain
astrophysically. Usually, one excludes a region around the galactic
centre or the galactic plane, thereby also limiting the influence of
the uncertainty of the halo profile in the inner kiloparsecs.

The pre-flight study~\cite{Baltz:2008wd} has estimated the necessary
$\langle \sigma_\text{ann} v \rangle$ for a detection with $3 \,
\sigma$ sensitivity after only one year of data as a function of the
DM mass. The uncertainty in the background model turns out to be large
and the necessary $\langle \sigma_\text{ann} v \rangle$ increases from
\mbox{$7 \times 10^{-26} \, \text{cm}^{3} \, \text{s}^{-1}$} (for
$m_\text{DM} = 50 \, \text{GeV}$) to $6 \times 10^{-26} \,
\text{cm}^{3} \, \text{s}^{-1}$ (for $m_\text{DM} = 250 \,
\text{GeV}$) for the `conventional' background
model~\cite{Strong:1998fr}. It was concluded that Fermi-LAT could
probe a large region of MSSM or mSUGRA parameter spaces.

Constraining the analysis to intermediate galactic latitudes $10^\circ
\leq |b| \leq 20^\circ$, the Fermi collaboration has investigated
~\cite{Abdo:2009mr} the diffusive background, maximising the fraction
of background produced within a few kiloparsecs from the solar system
and thereby minimising the dependence on the uncertainty of ISM and CR
densities elsewhere in the Galaxy. The shape of the spectrum is
consistent with an {\it a priori} diffusive emission model, an update
of the `conventional' {\tt GALPROP} model~\cite{Strong:1998fr}. The
excess measured by EGRET~\cite{Hunter:1997we} is also not reproduced
and the deviation is therefore likely to be an instrumental effect of
the EGRET experiment~\cite{Stecker:2007xp}.

A conservative analysis that does not attempt to subtract or fit any
astrophysical backgrounds is presented
in~\cite{Cirelli:2009dv}. Model-independent two-body annihilation or
decay to leptonic and hadronic channels are considered and the
resulting gamma-ray flux from final state radiation and $\pi^0$-decay
as well as from ICS of $e^\pm$ are required not too exceed the
Fermi-LAT measurements by more than $3 \, \sigma$ in different,
selected regions of the sky. This gives already quite considerable
constraints in the $m_\text{DM}$ -- $\langle \sigma_\text{ann} v
\rangle$ plane. Interestingly, a large fraction of the parameter space
needed to explain the electron-positron excesses from DM, see
Secs.~\ref{sec:PosFrac} and~\ref{sec:Total_e-e+_Flux}, are ruled
out. In particular, for NFW or Einasto profiles all fits to PAMELA,
Fermi-LAT and HESS $e^\pm$-data from DM annihilation are excluded, and
even for a cored isothermal profile, only annihilations into muons
remains marginally consistent.

A different analysis~\cite{Dobler:2009xz} is based on template
subtraction of foregrounds that are assumed to be traced by spatial
templates. An excess, called the `Fermi haze', is found up to $|b|
\approx 40^\circ$ above the galactic centre which is satisfactorily
fit by a bivariate Gaussian. This is argued to be most likely due to
inverse-Compton scattering (ICS) by relativistic electrons, and that
the underlying electron distribution is compatible with the `WMAP
haze'~\cite{Dobler:2009xz}, see Sec.~\ref{sec:Radio&Microwaves}. While
such a signature in ICS is naturally expected if there is indeed an
additional population of electrons with a hard spectrum, it was
pointed out~\cite{Linden:2010ea} that the template maps applied in
Ref.~\cite{Dobler:2009xz} are in fact inappropriate and underestimate
both the $\pi^0$ decay and ICS contributions to the gamma-ray
emission, in particular in the galactic centre region. The `Fermi
haze' may therefore be an artefact due to incorrect foreground
removal. Furthermore, with $1.6 \, \text{years}$ of data from
Fermi-LAT, a recent analysis~\cite{Su:2010qj} finds this excess to be
distributed above the galactic centre in an hourglass-shaped
morphology. Both because of this elongation and the peculiar angular
profile, the authors conclude that a DM explanation of this signal
seems to be disfavoured.

The Fermi collaboration has so far not confirmed this excess but is
performing analyses with more sophisticated foreground models and is
investigating possible correlations with (local) structures seen in
radio maps, like the North polar spur, see~\cite{Casandjian:2009wq}.

\subsubsection{Milky Way satellites}
\label{sec:Satellites}

The CDM paradigm and in particular $N$-body
simulations~\cite{Springel:2005nw,Kuhlen:2008qj,Diemand:2005vz}
predict a large number of bound substructures. The masses of these
so-called Milky Way satellites go down to $(10^{-4} \mathellipsis
10^{-12}) \, M_{\odot}$, depending on the free-streaming length of the
particular DM model considered. However, the minimum of the satellite
masses that will be observable with Fermi is rather $10^6 \,
M_{\odot}$.

A simple estimate of the significance expected from a satellite with
truncated NFW profile and typical WIMP annihilation cross section sets
the number of Milky Way satellites detectable by Fermi to $5 \,
\sigma$ within 5 years to $\sim 12$~\cite{Baltz:2008wd}. Particular
care must be taken not too misidentify statistical fluctuations in the
gamma-ray flux as DM annihilation or decay from substructure but it
was shown in~\cite{Baltz:2008wd} that a log-likelihood analysis can
distinguish between both cases for a `$5 \, \sigma$'
satellite. Furthermore, the DM substructure needs to be distinguished
from astrophysical sources.

It is worth stressing that in the case of DM annihilation,
substructures are not only targets in themselves but also increase the
diffuse flux by the $\rho^2$ dependence (see
Eq.~\ref{eqn:GammaRaysFromAnnihilation}). Estimates relying on the
frequency of such structures seen in $N$-body simulations however
limits the boost factor to $\mathcal{O}(10)$~\cite{Lavalle:2008}.

\subsubsection{Dwarf spheroidal galaxies}
\label{ref:dSphs}

Dwarf spheroidal galaxies (dSphs) are particularly faint companion
galaxies of the Milky Way or Andromeda. They may well be the most
abundant type of galaxies in the Universe but difficult to detect due
to their faintness. Before 2005, there were only 9 known
dSphs~\cite{Willman:2004kk,Zucker:2006bf,Irwin:2007jz,Walsh:2007tm,Belokurov:2009hw}
but the Sloan Digital Sky Survey (SDSS)~\cite{York:2000gk} has
increased their number by 11, also improving our understanding of this
type of galaxies. From stellar kinematics it can be inferred that some
have mass-to-light ratios of up to $\mathcal{O}(1000)$, that is many
times more than in conventional types of galaxies. Furthermore, dSphs
contain only little neutral or ionised gas which could otherwise
contribute to its gamma-ray emission. Therefore, dSphs are the most
extremely DM dominated environments known which makes them an
interesting target for DM searches.

A recent study~\cite{Abdo:2010ex} of 14 local group dSphs with data
from the first eleven month of Fermi-LAT operation does not find any
significant gamma-ray emission above $100 \, \text{MeV}$ and sets
upper limit on their gamma-ray fluxes, both for power-law injection
and spectra motivated by different WIMP models. Using stellar velocity
data to model the DM content of a subset of 8 of the above dSphs, the
Fermi team can constrain the cross section for annihilation into
gamma-rays by $b\bar{b}$, $\tau^+ \tau^-$ and a mixture of both, as
motivated from neutralino DM. More precisely, the limits are starting
to be competitive with cross sections from a scan over mSUGRA
parameter space and already exclude a fraction of the MSSM parameter
space. A model of UED with $B^{(1)}$ KK DM is not constrained but SUSY
models with Wino-like DM in the context of anomaly mediated SUSY
breaking (AMSB), see e.g.~\cite{Kane:2009if}, can be ruled out for
$m_\text{DM} \lesssim 300 \, \text{GeV}$. Interestingly, models
invoked to explain the cosmic ray lepton excesses (see
Secs.~\ref{sec:PosFrac} and~\ref{sec:Total_e-e+_Flux}), can be
constrained in so far as masses above $\sim 1 \, \text{TeV}$ can be
ruled out by considering the ICS emission produced.

\subsubsection{Line emission}
\label{sec:Line}

Although DM is found to be electrically neutral and annihilation or
decay to photons can therefore only occur at loop order, the preferred
smoking-gun signature of DM in indirect detection is the gamma-ray
line. The internal width of the line is for most annihilation
(excluding, perhaps annihilation via a Z boson resonance) and decay
(due to the necessarily long lifetime) processes rather small. The
width of the line, $\sim 10^{-3}$, is therefore mostly due to Doppler
broadening although this effect is also small because of the
non-relativistic WIMP velocities today. The lack of astrophysical
sources with lines in the GeV - TeV regime finally makes this
signature virtually background free. Theoretical frameworks include
SUSY WIMP and gravitino DM models. Unfortunately, the branching
fraction of the necessary processes is found to be $10^{-3}$ or
smaller.

A recent Fermi study~\cite{Abdo:2010nc} has found no significant
excess and has set limits on gamma-ray lines between $30$ and $200 \,
\text{GeV}$. By considering three different halo profiles,
NFW~\cite{Navarro:1996gj}, Einasto~\cite{Graham:2005xx,Navarro:2008kc}
and isothermal~\cite{Bahcall:1980fb}, upper limits on the annihilation
and decay cross section have been derived. Although for the
annihilation cross sections the bound is about an order of magnitude
weaker than the cross section for thermally produced WIMPs, some
non-thermal production models, e.g.~\cite{Kane:2009if}, can again be
ruled out. The upper limits on the life-times are also constraining
for some models with monoenergetic lines from decay of gravitinos.

\subsubsection{Other targets}

Other targets include so-called cosmological DM~\cite{Abdo:2010dk},
that is extra-galactic haloes and large-scale structure which
contribute to the isotropic gamma-ray flux, and DM signals from
clusters of galaxies~\cite{Ackermann:2010rg}.

\subsection{Radio and microwaves} 
\label{sec:Radio&Microwaves}

The measurement of anisotropies in the cosmic microwave background
(CMB) by COBE and WMAP has ushered in an exciting new era in
cosmology. The study of the cosmic signal requires careful subtraction
of galactic foreground emissions and this will become even more
crucial for studies of the `B-mode' polarisation signal by PLANCK and
the proposed CMBPol satellites~\cite{Dunkley:2008am}.  A by-product of
this foreground subtraction is the study of diffuse galactic microwave
emission in its own right. (In fact most of the diffuse foreground
emission is of galactic origin, a counter-example being unresolved
extra-galactic radio sources.) Usually, one distinguishes three
different physical components which have different underlying
processes and frequency behaviour:

\begin{enumerate}
{\bf \item Free-free emission} \\ Thermal electrons produce
bremsstrahlung on the gas in the ISM. The intensity produced by
free-free emission, $I_\nu \propto \nu^{-0.15}$, is proportional to
the line of sight integral of the ISM gas density squared which is why
it is believed to be traced by recombination H$\alpha$ line
maps~\cite{Finkbeiner:2003yt}. Dust can absorb some of this emission
and usually regions close to the galactic plane with dust optical
depth too high are excised.
{\bf \item Dust-correlated emission} \\ Dust grains vibrating in
equilibrium with the surrounding radiation fields are producing
emission in the microwave range. This `thermal dust' emission has been
mapped~\cite{Schlegel:1997yv} and evaluated at $94 \,
\text{GHz}$~\cite{Finkbeiner:1999aq}. In addition, larger dust grains
can have an electric dipole moment and be excited by collisions with
ions~\cite{Draine:1998gq}. While individual dust grains radiate with a
bump like spectrum, the superposition of many dust grains with
different frequencies can generate a soft power law spectrum. To a
first approximation, this `spinning dust' is expected to spatially
correlated with thermal dust and in fact an anomalous, soft component
was found in Ref.~\cite{Dobler:2007eg}. Although the presence of an
anomalous component is widely accepted, its interpretation as spinning
dust is still being debated, for example by the WMAP team itself who
do not conclusively confirm the spinning dust hypothesis.
{\bf \item Galactic synchrotron} \\ GCR electrons and positrons
produce synchrotron radiation on the galactic magnetic field
(GMF). For a power law spectrum of relativistic electrons, $n_{e^\pm}
\propto E^{-\alpha}$, the synchrotron emissivity is also a power law,
$I_\nu \propto \nu^{(1 - \alpha)/2}$. For an electron spectral index
$\alpha \approx 3$, \mbox{$I_\nu \propto \nu^{-1}$} which is close to
the observed $\nu^{-1.2}$ \cite{Gold:2008kp}. Tracer maps include
radio maps or difference maps at microwave frequencies, used, for
example by the WMAP collaboration~\cite{Hinshaw:2006ia}.
\end{enumerate}

\begin{figure}[!thb]
\begin{center}
\includegraphics[width=0.75\textwidth]{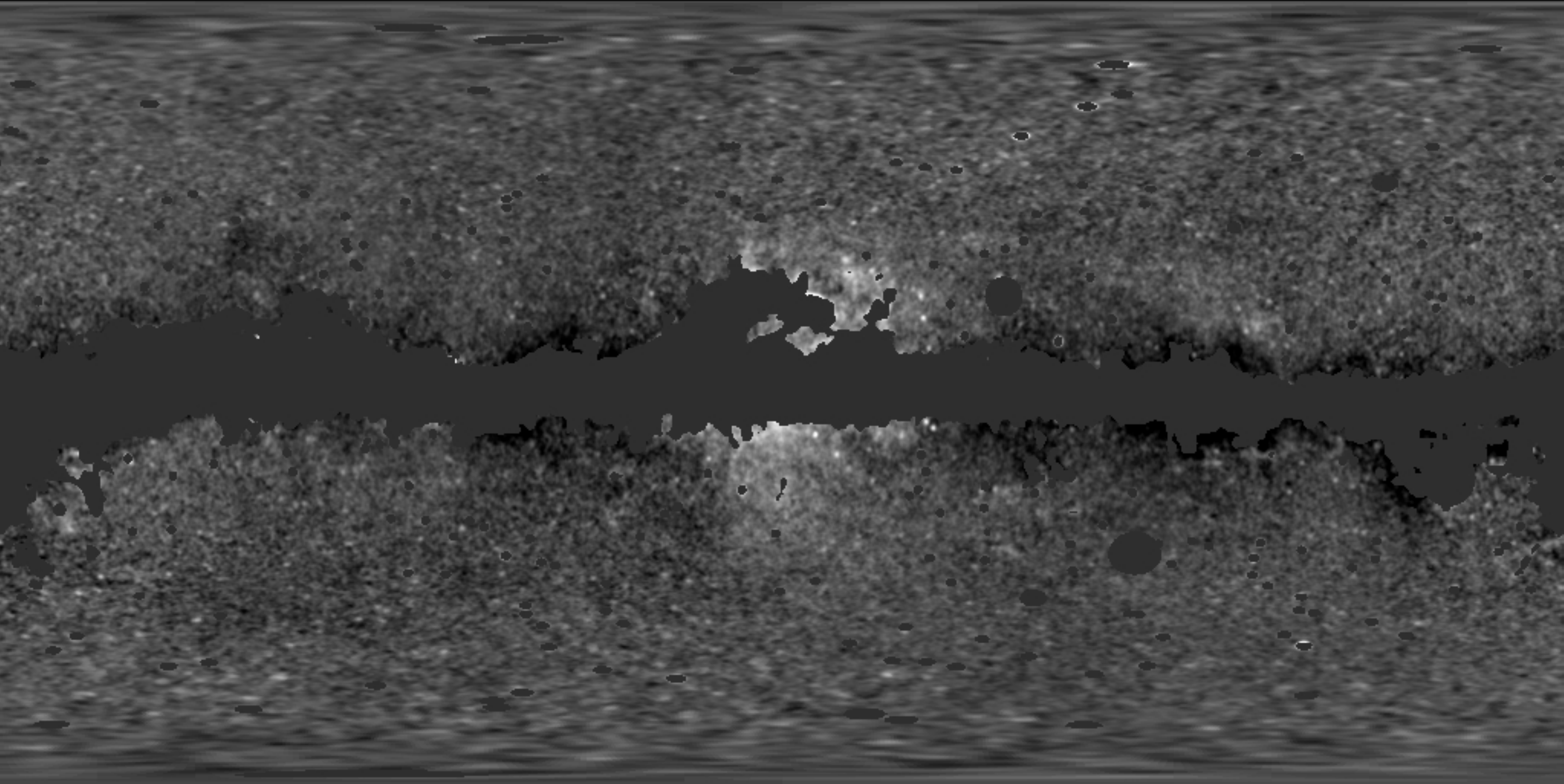}
\end{center}
\label{fig:WMAPhaze}
\caption{Sky map of the `WMAP haze' (from~\cite{Dobler:2007wv}).}
\end{figure}

For dark matter indirect searches more important is, however, the
claimed presence~\cite{Finkbeiner:2003im} of a new foreground that
does not correlate spatially with any of the above processes. Using
the template fitting technique (see~\ref{sec:TempSub}), a residual
remains after subtraction of the known foregrounds -- the so-called
`WAMP haze'. This residual has a roughly spherical morphology
localised around the centre of the Galaxy, see
Fig.~\ref{fig:WMAPhaze}, and a harder spectrum \cite{Dobler:2007wv}
than synchrotron radiation by relativistic GCR electrons from standard
astrophysical sources like SNRs. An independent analysis has confirmed
the existence of the haze~\cite{Bottino:2009uc}, but
others~\cite{Dickinson:2009yg,Cumberbatch:2009ji} do not find the
evidence to be significant, including the analysis~\cite{Gold:2010fm}
by the WMAP collaboration itself.

Initially, it was believed that the haze is free-free emission from
ionised gas too hot to be traced by recombination line maps but too
cold to be visible in X-rays~\cite{Finkbeiner:2003im}. However it was
suggested later that it is in fact synchrotron emission from a new
population of relativistic electrons and positrons, produced by dark
matter annihilation \cite{Finkbeiner:2004us}.  It is indeed possible
to explain the haze~\cite{Hooper:2007kb} by the synchrotron radiation
of an additional population of relativistic electrons and positrons,
produced by WIMP annihilation with a thermal cross section.  Other
authors however argue that the annihilation cross-section needs to be
significantly boosted over the usual value~\cite{Cumberbatch:2009ji}
or that the parameters of the diffusion model used are somewhat
non-standard~\cite{Linden:2010eu}. There have also been attempts to
fit both the morphology and spectrum of the haze by ascribing it to
electrons emitted by pulsars with a hard spectrum
\cite{Kaplinghat:2009ix, Harding:2009ye}; however the expected haze is
then less spherical since most pulsars are in the galactic disk. This
is also true of SNRs which have in fact recently been invoked
\cite{Blasi:2009hv, Ahlers:2009ae} as sources of positrons with a hard
spectrum to explain the rise in the cosmic ray positron fraction at
high energies measured by PAMELA~\cite{Adriani:2008zr}.

More conservative constraints on DM annihilation or decay can be
derived by demanding that the possible contribution from DM does not
exceed the observed fluxes in radio and microwaves (including
astrophysical backgrounds). For examples of such studies,
see~\cite{Borriello:2008gy,Zhang:2009pr}.

\subsection{Neutrinos} 

Dark matter indirect detection with neutrinos benefits from the same
advantages as indirect detection with gamma rays: Neutrinos propagate
through the Universe truly unimpeded and also point back to their
sources. Their small interactions however require detector volumes of
cubic kilometre size and provide event rates of a few per year, both
for conventional astrophysical application as well as for DM
searches. The weak interaction cross section are however also an
advantage as they could possibly make neutrinos the only messengers
that can escape very dense environments.

One such situation can lead to a particularly interesting idea for DM
indirect detection. The Sun is constantly moving through the DM
halo. Although the scattering cross section of WIMPs off the nucleons
in the Sun is small, occasionally, a WIMP can scatter elastically and
become gravitationally trapped inside the sun. Over the lifetime of
the Sun, a large number of WIMPs can become trapped and an equilibrium
between WIMP capture and annihilation will be reached. Neutrinos
detected from the Sun with energies of tens to hundreds of GeV can be
unambiguously assigned to these annihilation processes and used to
constrain WIMP annihilation cross sections and DM models.

Such neutrinos could be detected by the SUPER-Kamiokande experiment
which has the best sensitivity for WIMP masses of about a GeV, whereas
for larger masses the neutrinos could be detected in
IceCube~\cite{Ahrens:2003ix} (in particular in its low-energy
extension, ``DeepCore''), currently under construction under the South
pole. IceCube consists of a cubic kilometre of clean ice, about $2000
\, \text{km}$ under ground and instrumented with thousands of photo
multiplier tubes. A muon converted from a muon neutrino in the
surrounding material or the detector itself leaves a track of
\v{C}erenkov light in the detector which allows the reconstruction of
the neutrino direction and energy.

The beauty of detection of neutrinos from WIMP annihilation inside the
Sun is its relative robustness against astrophysical uncertainties;
the flux depends only the local DM density and velocity distribution
which also affect DM direct searches since in equilibrium the neutrino
production only depends on the capture rate. The rate has
contributions from both spin-independent (SI) and spin-dependent (SD)
cross sections and it turns out that for the SI cross section, upper
limits from direct detection experiments basically rule out any
prospects for event rates $\mathcal{O}(1) \, \text{yr}^{-1}$.

SD cross sections, on the other hand, can become much larger than SI
ones, e.g. for certain regions of the MSSM parameter space that are
experimentally not too well constrained and could lead to event rates
of up to $1000 \, \text{yr}^{-1}$. Large SD cross sections for
neutralino WIMPs are usually associated with large Z couplings and
hence a large Higgsino mixing. Therefore, the focus point region of
the $m_0$ -- $m_{\nicefrac{1}{2}}$ plane is particularly promising.

IceCube has already provided limits on the elastic scattering cross
section better by 2-3 orders of magnitude for the spin-dependent
case~\cite{Abbasi:2009uz}. For spin-independent cross sections,
however, the event rates allowed by direct detection limits are much
too low to be discovered by IceCube.

Of course, similar to gamma-ray searches, neutrinos can be searched
for from other targets, for example the galactic
centre~\cite{Hisano:2008ah,Meade:2009iu}. For preliminary result from
IceCube, see~\cite{Rott:2009hr}.

\section{Conclusion} 

Indirect detection of DM, that is the search for deviations in the
fluxes of cosmic ray nuclei and charged leptons as well as
radio/microwaves, gamma-rays and neutrinos, is a very promising
idea. Not only capitalises it on the large amount of dark matter
contained in the Milky Way halo and possibly also in extra-galactic
haloes, but it also uses a huge variety of astrophysical environments
or targets, ranging from substructures of the halo of a few solar
masses to large galaxy clusters.

Astrophysical backgrounds, however, play a crucial role for the
prospect of detecting DM in these channels. In particular, with most
DM explanations of the positron and electron-positron excesses
starting to be ruled out by gamma-ray constraints, for example, it is
clear that these signatures must be of astrophysical
origin. Therefore, a better understanding of the astrophysical
backgrounds is crucial both to resolve these problems in our
understanding of current excesses but also for future indirect
searches.

Even when DM indirect detections provide a (first) evidence of DM
annihilation or decay, provided it had some non-gravitational
interactions, it is unlikely that we could identify the interesting DM
parameters, mass, couplings etc., let alone the full DM model. In most
cases, the uncertainties inherent to the DM modelling itself as well
as uncertainties induced by the astrophysics of its detection are
considerable and all we can hope for are some rough indications. DM
indirect detections can therefore only unfold its true potential in
combination with other approaches to DM detection. Direct detection
(for a recent review, see, e.g.~\cite{Gaitskell:2004gd}) and
accelerator searches (e.g.~\cite{Bertone:2010}) are two other, equally
important approaches that, unfortunately, could not be discussed in
this introduction due to page limitations. Other possible constraints
can come from other effects that DM would have, for example, on
cosmological observables: DM with large annihilation rates or short
lifetimes could also affect big bang
nucleosynthesis~\cite{Jedamzik:2004er,Jedamzik:2004ip,Hisano:2009rc,Jedamzik:2009uy,Iocco:2008va}. One
of the most stringent constraints however comes from the effect of DM
annihilation or decay on the CMB and during the epoch of
reionisation~\cite{Galli:2009zc,Slatyer:2009yq,Huetsi:2009,Kanzaki:2009hf}.

For the phenomenology of indirect searches, what is needed at the
moment are a better understanding of the backgrounds that go beyond
the (often) oversimplified assumptions and hence predictions of the
standard picture of GCRs. In the rest of this work we present our
humble contribution to this ongoing challenge.

%% file: 02additional/additional.tex
\chapter{Additional Electrons/Positrons from Astrophysical Sources}
\label{chp:additional}

\section{Introduction}

The excess in the positron fraction recently measured by the PAMELA
collaboration clearly hints at some additional positron component. The
astrophysical background is expected to be falling at energies above
$\sim 1 \, \text{GeV}$ if positrons are indeed mainly produced as
secondaries by decay of charged pions from the spallation of cosmic
ray protons and nuclei on the interstellar medium (ISM). In addition,
recent measurements of the total electron-positron flux by the Fermi
collaboration have shown it to be harder than previously expected.

The evaluation of the astrophysical predictions has gained in
importance due to the possibility of these excesses being due to
exotic physics, in particular due to additional electrons and
positrons produced in annihilation or decay of dark matter (DM) in the
galactic halo. A large number of these models is starting to be ruled
out by constraints from other messenger, e.g. radio waves from
synchrotron radiation or gamma-rays from inverse Compton scattering
(ICS). It is therefore of utmost importance to investigate how these
excesses can be understood in terms of astrophysical sources.

One of the most minimal approaches is to consider whether the
additional electrons/positrons could possibly be produced in the
standard setup \emph{without} making any additional assumptions about
new sources, new interactions or the like. Such minimality is a basic
principle of scientific heuristics as most famously expressed in
`Occam's razor': {\it `Numquam ponenda est pluralitas sine
  necessitate'}~\cite{Ockham:1967}. (Plurality ought never be posited
without necessity.) or {\it `Frustra fit per plura quod potest fieri
  per pauciora'}~\cite{Ockham:1974}. (It is futile to do with more
things that which can be done with fewer.) Instead, the assumptions of
the current model of galactic cosmic rays (GCRs) are to be
reconsidered, possibly premature arguments about the importance of
different effects are to be revisited and the available parameter
space is to be reassessed.

In particular, we are going to question the assumption that positrons
get exclusively produced by spallation of CR protons and nuclei on the
ISM and, instead, we consider the production of secondary electrons
and positrons by protons and nuclei \emph{inside} the paradigm
sources: supernova remnants (SNRs). Not only can the total fluxes of
these secondaries reach levels comparable to the primaries but the
secondary spectra will also differ quite drastically from the primary
ones. Hereby, we follow a recent idea to explain the rise in the
positron fraction by such a harder injection of secondaries from
SNRs~\cite{Blasi:2009hv}, however improving on this model in several
respects.

The effect of secondaries produced inside the GCR sources had so far
always been neglected due to a crude column depth argument, recently
reiterated \cite{Katz:2009yd}: The average residence time of primary
nuclei in a SNR is necessarily shorter than the SNR lifetime,
$\mathcal{O}(10^4) \, \text{yr}$, and therefore happens to be much
smaller than the residence time in the ISM, $\sim \mathcal{O}(10) \,
\text{Myr}$. Even if the gas density inside the SNR is higher than in
the ISM, due to compression by the shock, the grammage experienced
inside the SNR, $0.2 \, (n_\text{gas,SNR}/ (10 \, \text{cm}^{-3}) ) \,
\text{g} \, \text{cm}^{-2}$, is still much smaller than the average
grammage of the ISM, $\text{a few} \, \text{g} \,
\text{cm}^{-2}$. Therefore, the production of secondaries inside SNRs
was believed to be negligible.

What this argument ignores is that the secondaries produced inside the
source get also accelerated and their spectrum becomes considerably
harder than that of the secondaries produced during propagation of the
primaries in the ISM. As we will see below, the accelerated secondary
positrons in sources have a power law index of $-\gamma + 1$ where
$-\gamma$ is the index of the primaries at source. After propagation
this gives a spectral index $-\gamma$ at energies above $10 \,
\text{GeV}$ where cooling is the dominant loss process. The secondary
positrons from propagation, however, have a spectral index of $-\gamma
- \delta - 1$. We remind ourselves that the value predicted for
$\gamma$ by diffusive shock acceleration (DSA) is $\gamma \approx 2$
and $\delta$ is determined from nuclear secondary to primary ratios as
$\delta \approx 0.6$. We now compare the fluxes $J_\text{src} =
J_\text{src}^0 (E/E_0)^{-\gamma}$ of secondaries from the source and
$J_\text{ISM} = J_\text{ISM}^0 (E/E_0)^{-\gamma - \delta - 1}$ of
secondaries from propagation. Although the ratio of the total number
of particles is indeed small,
\begin{align}
\frac{N_\text{src}}{N_\text{ISM}} = \frac{\int_{E_0}^{E_\text{max}}
  \dd E \, J_\text{src}(E)}{\int_{E_0}^{E_\text{max}} \dd E \,
  J_\text{ISM}(E)} \simeq \frac{J_\text{src}^0}{J_\text{ISM}^0}
\frac{\gamma + \delta}{\gamma - 1} \simeq 2.6 \times 10^{-3} \, ,
\end{align}
if we chose, for example $J_\text{src}^0 / J_\text{ISM}^0 = 2 \times
10^{-3}$ and $E_\text{max} \gg E_0$, the differential fluxes become
comparable at $(J_\text{ISM}^0 / J_\text{src}^0)^{1/(1+\delta)} E_0
\simeq 50 \, E_0$ already.

The appeal of the acceleration of secondaries, besides its minimality,
is that its effect is guaranteed, in the sense that secondaries will
definitely be produced in cosmic ray source and are subject to
subsequent acceleration. However, the normalisation of the flux of
these secondaries, for example, with respect to the secondaries from
propagation is the crucial question.

In this chapter, we will study the acceleration of secondaries in
sources in some detail. In Sec.~\ref{sec:AccnOfSecs} we will calculate
the spectra of the secondaries in the simple test particle
approximation of DSA and explain why the secondary spectrum is harder
than the primary one. We will then consider the effect of discreteness
of sources and suggest a realistic distributions of sources in the
spiral arms of the galactic disk and how to determine it from cosmic
ray spectra alone in
Sec.~\ref{sec:DiscretenessOfSources}. Sec.~\ref{sec:Spectra} contains
a summary of the different populations of electrons and positrons and
some comments on how to normalise them by fitting to data.
Previously, the flux of secondary $e^-$ and $e^+$ in the sources has
been normalised with respect to the primary electrons in an {\it ad
  hoc} fashion \cite{Blasi:2009hv}. Instead, we exploit the hadronic
origin of these secondaries and normalise using the gamma-ray fluxes
(assumed to be from $\pi^0$ decay) detected from known SNRs by Imaging
Air \v{C}erenkov Telescopes (IACTs) like HESS. We can thus fix the
only free model parameter by fitting the total $e^- + e^+$ flux to
Fermi-LAT and HESS data. The $e^+$ fraction is then \emph{predicted}
up to TeV energies and provides a good match to PAMELA data
(Sec.~\ref{sec:Additional_Results}). Having constrained the
distribution of the closest SNRs via the measured $e^-$ and $e^+$
spectra, we present in Sec.~\ref{sec:GammaRays&Neutrinos} an example
of a likely source distribution in order to illustrate that there are
good prospects for IceCube to detect neutrinos from nearby SNR.  In
Sec.~\ref{sec:TimeDependentPicture}, we counter some criticism on this
model, in particular the claim that the acceleration of mechanism does
not hold in a time-dependent treatment.  A consistent picture thus
emerges for all presently available data in the framework of the
standard DSA/SNR origin model of GCR.  Some alternative, astrophysical
explanations of the excesses in GCR leptons are being mentioned in
Sec.~\ref{sec:AlternativeAstrophysicalExplanations}.  We conclude and
comment on some open issues and grounds for concern in
Sec.~\ref{sec:DiscussionSummary}.

\section{Acceleration of Secondaries in the Sources}
\label{sec:AccnOfSecs}

Similar to the diffusive shock acceleration of primary cosmic rays, we
describe the acceleration of secondaries produced and accelerated in
the source in the simple test particle approximation, that is ignoring
the back reaction of the energetically important nuclear component of
the cosmic rays on the structure of the shock front.

The setup is similar to the one discussed in Sec.~\ref{sec:DSA}. We
consider the phase space density, $f_\pm$, of secondary $e^-$ and
$e^+$ produced by the primary GCR, both undergoing DSA, which is
described by the steady state transport equation,
\begin{equation}
  u \frac{\partial f_{\pm}}{\partial x} = \frac{\partial}{\partial x}
  \left( D \frac{\partial}{\partial x} f_\pm \right) + \frac{1}{3}
  \frac{\text{d} u}{\text{d} x} p \frac{\partial f_{\pm}}{\partial p}
  + q_{\pm} \,,
\label{eqn:TransportEqReminder}
\end{equation}
where the source term $q_\pm$ is determined by solving an analogous
equation for the primary GCR protons (see
Sec.~\ref{sec:DSA}). (Ideally we should solve the time-dependent
equation, however we do not know the time-dependence of the parameters
and can extract only their effective values from observations. This
ought to be a good approximation for calculating {\em ratios} of
secondaries to primaries from a large number of sources which are in
different stages of evolution.) We consider the usual setup in the
rest-frame of the shock front (at $x=0$) where $u_1$ ($u_2$) and $n_1$
($n_2$) denote the upstream (downstream) plasma velocity and density,
respectively (cf. left panel of Fig.~\ref{fig:DSAsetup}). The
compression ratio of the shock $r=u_1 / u_2 = n_2 /n_1$ determines the
spectral index, $a = 3r/(r - 1)$, of the GCR primaries in momentum
space.

For $x \neq 0$, Eq.~\ref{eqn:TransportEqReminder} reduces to an
ordinary differential equation in $x$ that is easily solved taking
into account the spatial dependence of the source term,
cf. Eq.~\ref{eqn:StdyState2},
\begin{equation}
q_{\pm}^0(x,p) = \left \{
\begin{array}{ll}
q_{\pm, 1}^0 (p) \text{e}^{x \, u_1 / D (p_\text{p})} & \text{for } x
< 0 ,\\ q_{\pm, 2}^0 (p) & \text{for } x > 0 ,\\
\end{array}
\right.
\label{eqn:injection}
\end{equation}
where the proton momentum $p_\text{p}$ should be distinguished from
the (smaller) momentum $p$ of the produced secondaries, the two being
related through the inelasticity of $e^\pm$ production: $\xi \simeq
1/20$. Assuming $D \propto p$ (Bohm diffusion) in the SNR, the
solution to the transport equation~\ref{eqn:TransportEqReminder}
across the shock, satisfying the boundary conditions,
\begin{equation}
\lim\limits_{x \to -\infty}f_\pm = 0 \, , \lim\limits_{x \to
  -\infty}\frac{\partial f_\pm}{\partial x} = 0\,\,
\text{and}\,\,\left| \lim\limits_{x \to \infty} f_\pm \right| < \infty
\,,
\end{equation}
can then be written:
\begin{equation}
f_{\pm} = \left \{
\begin{array}{ll}
f_\pm^0 \text{e}^{x/d_1} -\frac{q_{\pm, 1}^0}{u_1}d_1
\left(\frac{\text{e}^{x/d_1} - \text{e}^{\xi x/d_1}}{\xi -
  \xi^2}\right) & \text{for } x < 0, \\ f_\pm^0 + \frac{q_{\pm,
    2}^0}{u_2} x & \text{for } x > 0, \\
\end{array}
\right.
\label{eqn:fpm}
\end{equation}
where $d_1 \equiv D/u_1$ is the effective size of the region where
$e^-$ and $e^+$ participate in DSA (see right panel of
Fig.~\ref{fig:DSAsetup}).

Continuity at the shock front $x=0$ requires:
\begin{equation}
\left. D \frac{\partial f_\pm}{\partial x} \right|_{x=0^-} -\left.  D
\frac{\partial f_\pm}{\partial x}\right|_{x=0^+} = \frac{1}{3} (u_2 -
u_1) p \frac{\partial f_\pm^0}{\partial p} \,,
\end{equation}
yielding the differential equation,
\begin{equation}
  p \frac{\partial f_\pm^0}{\partial p} = -a f_\pm^0 + a
  \left(\frac{1}{\xi} + r^2\right) \frac{D q_1^0}{u_1^2} \, .
\end{equation}
This is readily integrated with boundary condition $f^0_\pm (0) = 0$
and gives
\begin{align}
  f_\pm^0 (p) = a \left(\frac{1}{\xi} + r^2 \right) \int_0^p
  \frac{\text{d}p'}{p'} \left(\frac{p'}{p}\right)^a \frac{D(p')
    q_{\pm, 1}(p')}{u_1^2}.
\label{eqn:fpm0}
\end{align}
Assuming Feynman scaling for the pp interaction, i.e. $p\,\text{d}
\sigma_\text{pp}/\text{d}p \propto \Sigma_\pm $ we can express the
momentum dependence of the source term as
\begin{align}
\label{eqn:defqpm}
  q_{\pm, 1}(p) &= \frac{c\,n_{\text{gas}, 1}}{4\pi p^2} \int_p^\infty
  \text{d}p' N_\text{CR} (p') \frac{\text{d}\sigma_{\text{pp} \to
      \text{e}^\pm + X}}{\text{d}p} \simeq
  \frac{c\,n_{\text{gas},1}}{4\pi p^2} N_\text{CR} (p)
  \frac{\Sigma_\pm}{\gamma - 2} \,.
\end{align}

We can easily interpret the solution Eq.~\ref{eqn:fpm} in terms of
power laws in momentum. The second term downstream, $(q_2^0/u_2) x$,
follows the spectrum of the primary GCRs ($\propto p^{-a}$) and
describes the production of secondary $e^-$ and $e^+$ that are then
advected away from the shock front. However, secondaries that are
produced within a distance $\sim D/u$ from the shock front are subject
to DSA (see Eq.~\ref{eqn:injection} and right panel of
Fig.~\ref{fig:DSAsetup}). The fraction of secondaries that enters the
acceleration process is thus given by the ratio of the relevant
volumes, i.e.~$(D/u_1)/(u_2 \, \tau_{\text{SNR}})$, and the number
density injected into the acceleration process is $ (1/\xi + r^2 ) D
q_{\pm, 1}/u_1^2$. This rises with energy because of the momentum
dependence of the diffusion coefficient ($D(p) \propto p$) so the
first term downstream in Eq.~\ref{eqn:fpm} gets harder: $f_\pm^0 (p)
\propto p^{-a + 1}$.

The total injection spectrum $R_\pm$ from one SNR is obtained by
integrating the steady state solution over the volume of the SNR:
\begin{equation}
R_\pm = 4\pi p^2 4\pi \int_0^{u_2 \tau_\text{SNR}} \text{d}x \, x^2
f_\pm (x, p).
\end{equation}
The resulting source spectrum, $R_{\pm}$, is thus the sum of two power
laws,
\begin{equation}
  R_\pm \simeq R_\pm^0 \, p^{-a + 2} \left[1 +
    \left(\frac{p}{p_\text{cross}}\right)\right] ,
\end{equation}
where the ``cross-over'' momentum, $p_\text{cross}$, satisfies
\begin{equation}
\label{eqn:pxdef}
D (p_\text{cross}) = \frac{3}{4} \frac{ru_1^2\tau_\text{SNR}}{a(1/\xi
  + r^2)}\,.
\end{equation}
As has been noted~\cite{Blasi:2009hv}, this mechanism is most
efficient for {\it old} SNRs where field amplification by the shock
wave is not very effective any more. We therefore introduce a fudge
factor $K_\text{B}$ that parametrises the effect of the smaller field
amplification on the otherwise Bohm-like diffusion coefficient in the
SNR,
\begin{equation}
D (E) = 3.3 \times 10^{22} K_\text{B} \, \bigg(\frac{B}{\mu{\rm
    G}}\bigg)^{-1} \bigg(\frac{E}{\text{GeV}}\bigg) \, \text{cm}^2
\text{s}^{-1}.
\label{eqn:D(p)}
\end{equation}

The number of particles entering the acceleration process can of
course not exceed the total number of secondaries produced inside the
SNR. This effectively caps the growth of the term $D(p') q_{\pm, 1}^0
(p')/u_1^2$ once $(D/u_1)/(u_2 \, \tau_{\text{SNR}})$ becomes larger
than unity, a relation that defines a characteristic momentum scale
$p_\text{break}$. We therefore substitute in Eq.~\ref{eqn:fpm0},
\begin{equation}
\frac{D(p) q_{\pm, 1}^0 (p)}{u_1^2} \rightarrow \left \{
\begin{array}{ll}
\frac{D(p) q_{\pm, 1}^0 (p)}{u_1^2} & \quad \text{for } p <
p_\text{break}, \\ \frac{D(p_\text{break}) q_{\pm, 1}^0 (p)}{u_1^2} &
\quad \text{for } p > p_\text{break}.
\end{array}
\right.
\end{equation}
The source spectrum $R_\pm$ thus returns to a $p^{-\gamma}$ dependence
around $p = p_\text{break}$. At even higher energies the secondary
spectrum cuts off at the same $E_\text{cut}$ as for primary electrons
(see Sec.~\ref{sec:PrimaryElectrons}).

\section{The Discreteness of Sources}
\label{sec:DiscretenessOfSources}

In the standard picture, see Sec.~\ref{sec:SMofGCRs}, the explanation
for the observed galactic cosmic rays (GCRs) is factorised into two
parts which can be treated separately: the acceleration in supernova
remnants (SNRs) and the diffusive-convective transport through the
Galaxy.

Although SNRs can become quite large (up to $\mathcal{O}(100) \,
\text{pc}$ towards the end of their lifetime, see Sec.~\ref{sec:SNRs}
and Eq.~\ref{eqn:SNRtrans}), they are under most circumstances still
much smaller than the scale on which the propagation of GCRs take
place, i.e. kiloparsecs, and their extent can usually be neglected for
all practical purposes.

We are therefore not dealing with a continuous distribution of sources
within the galactic disk but with a large number of discrete
sources. The usual assumption is that diffusion will wash out this
source distribution. This is however only true as long as the
diffusion length is long compared to the average distance of sources
(which is of the same order as the average distance of an observer to
the closest source). For protons and (stable) nuclei the diffusion
length is indeed quite long -- of the order of tens of kpc -- since
energy losses are negligible above a few GeV and therefore escape from
the cosmic ray halo is the dominant loss process. For charged leptons
on the other hand, in particular for electrons and positrons, the
diffusion length is strongly energy dependent. The dominating energy
losses are through synchrotron radiation and ICS and the
diffusion-loss length, $\lambda(E) \equiv \sqrt{\ell^2(E,2E)} \propto
E^{(\delta - 1)/2}$, (see Eq.~\ref{eqn:ell2}), is quickly
declining. Therefore we expect to start seeing contributions from
individual, nearby sources above the energy at which the diffusion
length drops below the average source distance. Depending on the
particular values of the diffusion parameters, one predicts ``bumps'',
that is Green's functions of individual sources, to appear above $(50
\mathellipsis 200) \, \text{GeV}$. The superposition of a few such
bumps leads to features which we will investigate in the following
section.

In addition to the discreteness in space, GCR sources like SNRs also
have a finite lifetime. A common assumption is that the bulk of cosmic
rays will only be released after the SNR has entered the radiative
stage and DSA comes to an end.\footnote{In more sophisticated
  modelling, one expects that a certain fraction of the particles
  accelerated will be able to escape the SNR during the Sedov-Taylor
  phase already~\cite{Reynolds:1998,Caprioli:2008sr}. For example, due
  to a lack of turbulence at low wave numbers ahead of the shock, high
  energy particles cannot be confined to the shock region and are
  instead injected into the interstellar medium.} This makes the
distribution of sources also finite in the time-domain and will lead
to even stronger features.

\subsection{Fluctuations in the electron flux}
\label{sec:DiscreteSources}

Apart from some inferences about (only the observed) near SNRs we do
not know much about the position of sources of GCRs. Even in the limit
of a continuous source distribution, which is justified when
considering old, far away sources and hence fluxes at lower energies,
our knowledge of the global source distribution is limited, see also
Chapter \ref{chp:haze}. We have however argued that at higher
energies, the contribution to the electron-positron flux from
particular sources depend strongly on their distances. Our ignorance
of the source positions therefore introduces an uncertainty into our
prediction for the electron-positron fluxes at Earth. Put differently,
the electron-positron flux stochastically fluctuates between different
realisations of a statistically smooth source distribution.

In the following we will use central limit theorems and the analytical
form of the Green's function for the diffusion-energy loss problem
described in sec.~\ref{sec:Transport} to infer the statistical
properties of the flux, i.e. its average, its quantiles and the energy
dependence of both. For numerical studies of this effect based on a
Monte Carlo approach,
see~\cite{Pohl:1998ug,Strong:2001qp,Swordy:2003ds}.

The flux from a source that injected a spectrum $Q(E) = Q_0
E^{-\gamma}$ of electrons or positrons a time $t$ ago at a distance
$L$ from the observer is given by the Green's function $G_\text{disk}$
(cf. Eq.~\ref{eqn:GreensFunctionSpectrum}) of the diffusion equation,
\begin{align}
\label{eqn:GreensDiscreteSrc}
J_i(E) = \frac{c}{4 \pi} G_\text{disk} = \frac{c}{4 \pi} \left( \pi
\ell^2 \right)^{-1} {\rm e}^{-L^2 / \ell^2} Q_0 E^{-\gamma} (1 - b_0 E
t)^{\gamma - 2} \frac{1}{\zcr} \chi \left( \frac{\ell}{\zcr} \right)
\, ,
\end{align}
where $c/(4 \pi)$ denotes the ``flux factor'' for relativistic
particles and $\ell^2(E, t)$, $\chi(\ell/\zcr)$ and $\zcr$ are defined
in Sec.~\ref{sec:GreensFunctionApproach}.

The flux $J$ of $N$ identical\footnote{Here and in the following we
  assume that all the sources considered have a common injection
  spectrum $Q(E)$. Of course, in reality the sources differ in total
  power output and spectrum which will introduce additional
  fluctuations into the spectrum observed at Earth. Here, we however
  constrain ourselves to investigate the effects of the discreteness
  of sources.} sources at distances $\{ L_i \}$ and times $\{ t_i \}$
is the sum of the individual fluxes,
\begin{align}
\label{eqn:SumOfFluxes}
J = \sum_{i=1}^N J_i(E) = \frac{c}{4 \pi} \sum_{i=1}^N \left( \pi
\ell^2 \right)^{-1} {\rm e}^{-L_i^2 / \ell^2} Q_0 E^{-\gamma} (1 - b_0
E t_i)^{\gamma - 2} \frac{1}{\zcr} \chi \left( \frac{\ell}{\zcr}
\right) \, .
\end{align}
To investigate the fluctuations introduced by the stochasticity of
source distances and times, we calculate the expectation value and
standard deviation of the total flux $J$ exploiting the central limit
theorem~\cite{Feller:1971}.

If the central limit theorem was applicable, at a fixed energy $E$ the
fluxes $J(E)$ for different realisations of the same source density
would follow a normal distribution with mean $\mu_J$ and standard
deviation $\sigma_J$,
\begin{align}
\mu_J =& \frac{c}{4 \pi} N \mu_Z = \frac{c}{4 \pi} N \langle Z \rangle
\, , \\ \sigma_J =& \frac{c}{4 \pi} \sqrt{N} \sigma_Z = \frac{c}{4
  \pi} \sqrt{N} \sqrt{\langle Z^2 \rangle - \langle Z \rangle^2} \, ,
\end{align}
were $\langle Z^m \rangle$ denotes the moments of the Green's function
$Z \equiv G_\text{disk} (E, L, t)$,
\begin{align}
\langle Z^m \rangle &= \int \dd z f_Z(z) z^m \, .
\end{align}
As a function of the random variables $L$ and $t$, $Z$ itself is a
random variable with the probability density $f_Z$. In the case under
consideration, $L$ and $t$ are assumed to be independent random
variables with probability densities $f_L$ and $f_t$, respectively,
and thus the joint probability density $f_{L,t}$ factorises,
$f_{L,t}(L,t) = f_L(L) f_t(t)$.

\begin{figure}[bt]
\begin{center}
\includegraphics[scale=0.8]{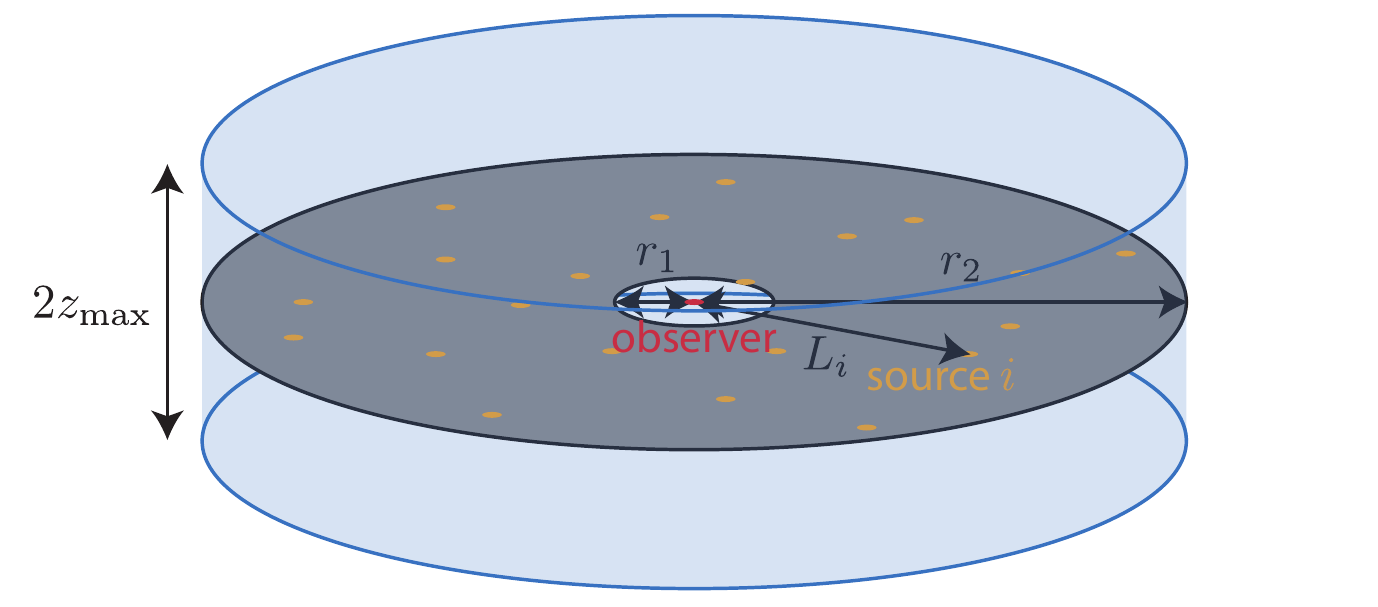}
\end{center}
\caption{Propagation setup. The sources are discrete, but
  homogeneously distributed around the observer in a ring of inner
  radius $r_1$ and outer radius $r_2$. The disk is contained in a
  diffusion volume of half-height $z_\text{max}$. We neglect the
  boundary condition in radial direction.}
\label{fig:HomSrcDistDiskWithHole}
\end{figure}

Let's assume that the sources are, although discrete, homogeneously
distributed in a ring around the observer with an inner and outer
radius $r_1$ and $r_2$, respectively (see
Fig.~\ref{fig:HomSrcDistDiskWithHole}). The distance $L$ to some
source is a random variable with the probability density,
\begin{equation}
f_L(L) = \left \{
\begin{array}{cl}
2 L / (r_2^2 - r_1^2) & \text{for } r_1 \leq L \leq r_2 \, , \\ 0 &
\text{otherwise} \, .
\end{array}
\right.
\end{equation}
%
%
Also, we assume that the source rate, e.g. the supernova rate, is
constant over the time scales considered, $\mathcal{O}(100) \,
\text{Myr}$ (electrons and positrons of GeV energies can diffuse over
$\mathcal{O}(100) \, \text{Myr}$ before losing their energy but are,
of course, subject to escape losses). Therefore, we take the sources
to be equally distributed up to a maximum time $\tmax = 1/ \left( b_0
E_{\text{min}} \right)$, set by the minimum energy $E_\text{min}$
which is to be considered,
\begin{equation}
f_t(t) = \left \{
\begin{array}{cl}
1/\tmax & \text{for } 0 \leq t \leq \tmax \, , \\ 0 & \text{otherwise}
\, .
\end{array}
\right.
\end{equation}

Rewriting the $m$-th moment of the Green's function as an integral
over $L$ and $t$ (see Appendix~\ref{app:Moments}), we have,
\begin{align}
\langle Z^m \rangle &= \int \dd z f_Z(z) z^m \\ &= \int_{r_1}^{r_2}
\dd L \int_0^{t_\text{max}} \dd t f_{L,t}(L,t) \, G^m(L,t) \\ &=
\int_{r_1}^{r_2} f_{L}(L) \dd L \int_0^{t_\text{max}} \dd t f_{t}(t)
\, G^m(L,t) \\ &= \frac{1}{t_\text{max}} \int_0^{t_\text{max}} \dd t
\int_{r_1}^{r_2} \dd L \frac{2 L}{r_2^2 - r_1^2} \, \left( \left( \pi
\ell^2 \right)^{-1} {\rm e}^{-L^2 / \ell^2} Q_0 E^{-\gamma} (1 - b_0 E
t)^{\gamma - 2} \frac{1}{\zcr} \chi \left( \frac{\ell}{\zcr} \right)
\right)^m \\ &= \frac{1}{t_\text{max}} \int_0^{t_\text{max}} \dd t
\int_{r_1^2}^{r_2^2} \frac{\dd L^2}{r_2^2 - r_1^2} \, {\rm e}^{- m L^2
  / \ell^2} \left( \left( \pi \ell^2 \right)^{-1} Q_0 E^{-\gamma} (1 -
b_0 E t)^{\gamma - 2} \frac{1}{\zcr} \chi \left( \frac{\ell}{\zcr}
\right) \right)^m \\ &= \frac{1}{t_\text{max}} \frac{1}{r_2^2 - r_1^2}
\int_0^{t_\text{max}} \dd t \left[ \frac{\ell^2}{m} {\rm e}^{- m L^2 /
    \ell^2} \right]_{r_2^2}^{r_1^2} \left( \left( \pi \ell^2
\right)^{-1} Q_0 E^{-\gamma} (1 - b_0 E t)^{\gamma - 2} \frac{1}{\zcr}
\chi \left( \frac{\ell}{\zcr} \right) \right)^m
\end{align}

We now substitute,
\begin{align}
\label{eqn:Subst1}
E_0 = \frac{E}{1 - b_0 E t} \quad \Rightarrow \quad \frac{\dd t}{(1 -
  b_0 E t)^2} = \frac{\dd E_0}{b_0 E^2} \, ,
\end{align}
to obtain an integral in energy $E_0$ at source,
\begin{align}
\label{eqn:Gm}
\langle Z^m \rangle &= \frac{1}{t_\text{max}} \frac{1}{r_2^2 - r_1^2}
\int_E^\infty \frac{\dd E_0}{b_0 E_0^2} \left[ \frac{\ell^2}{m} {\rm
    e}^{- m L^2 / \ell^2} \right]_{r_2^2}^{r_1^2} \left( \left( \pi
\ell^2 \right)^{-1} Q_0 E_0^{-\gamma} \left( \frac{E_0}{E} \right)^2
\frac{1}{\zcr} \chi \left( \frac{\ell}{\zcr} \right) \right)^m \, ,
\end{align}
with
\begin{align}
\ell^2 = \frac{4 D_0}{b_0 (1 - \delta)} \left( E^{\delta -1} -
E_0^{\delta - 1} \right) \, .
\end{align}

For $m=1$, this is,
\begin{align}
\langle Z \rangle &= \frac{1}{t_\text{max}} \frac{1}{r_2^2 - r_1^2}
\int_E^\infty \frac{\dd E_0}{b_0 E^2} \left[ \ell^2 {\rm e}^{- L^2 /
    \ell^2} \right]_{r_2^2}^{r_1^2} \left( \pi \ell^2 \right)^{-1} Q_0
E_0^{-\gamma} \frac{1}{\zcr} \chi \left( \frac{\ell}{\zcr} \right) \,
,
\end{align}
which can, for example, be evaluated numerically. The expectation
value $\mu_J = c/(4 \pi) N \langle Z \rangle$ for the flux is
identical to the flux from a continuous, homogeneous and steady source
distribution,
\begin{align}
\frac{N}{\pi (r_2^2 - r_1^2)} \frac{Q_0 E_0^{-\gamma}}{t_\text{max}}
\end{align}
in a ring with inner and outer radius $r_1$ and $r_2$,
respectively. The first term is the surface density of sources and the
second one the power spectrum per source (although we are, strictly
speaking, comparing to a continuous source distribution).

Expanding the term $\chi(\ell/\zcr)$ in Eq.~\ref{eqn:Gm} (associated
with the diffusion in the $z$ direction) for $\ell \ll \zcr$, which is
only justified for energies above $\sim 50 \, \text{GeV}$ for
$z_\text{max} \simeq 4 \, \text{kpc}$ and amounts to ignoring the
boundary (condition) in the $z$ direction,
\begin{align}
\frac{1}{\zcr} \chi \left( \frac{\ell}{\zcr} \right) \simeq
\frac{1}{\sqrt{\pi \ell^2}} \, ,
\end{align}
we can get analytical estimates for the moments $\langle Z^m
\rangle$. To this end we further substitute,
\begin{align}
\lambda^2 &= \frac{b_0 (1 - \delta)}{4 D_0} E^{1- \delta} \ell^2 =
\left[ 1- \left( \frac{E_0}{E} \right)^{\delta - 1} \right] \nonumber
\\ \Rightarrow E_0 &= E \left( 1 - \lambda^2 \right)^{\frac{1}{\delta
    - 1}} \quad \Rightarrow \quad \mathrm{d}E_0 = \frac{E}{1 - \delta}
\left( 1 - \lambda^2 \right)^{\frac{1}{\delta - 1} - 1} \mathrm{d}
\lambda^2 \, ,
\label{eqn:Subst2}
\end{align}
similarly define,
\begin{align}
\Lambda^2 = \frac{b_0 (1 - \delta)}{4 D_0} E^{1- \delta} L^2 \, ,
\quad \rho_i^2 = \frac{b_0 (1 - \delta)}{4 D_0} E^{1- \delta} r_i^2 \,
,
\end{align}
and find,
\begin{align}
\label{eqn:MomentIntegral}
\langle Z^m \rangle =& \frac{1}{t_\text{max}} \frac{1}{r_2^2 - r_1^2}
\frac{(4 D_0)^{1 - \frac{3}{2} m}}{(b_0 (1 - \delta))^{2 - \frac{3}{2}
    m}} \frac{Q_0^m}{m \pi^{\frac{3}{2} m}} E^{- 2 + \delta +
  \frac{3}{2} m (1 - \delta) - m \gamma} \nonumber \\ & \times
\int_0^{1} \dd \lambda^2 (1 - \lambda^2)^{\frac{m (\gamma - 2) +
    \delta}{1 - \delta}} (\lambda^2)^{1 - \frac{3}{2} m} \left[ {\rm
    e}^{- m \Lambda^2 / \lambda^2} \right]_{\rho_2^2}^{\rho_1^2} \, .
\end{align}

Eventually, we want to send $r_1$ to zero since the distance to the
nearest source is not physically limited. At the energies for which
the above expansion holds, we also have $\ell^2 \ll r_2^2$, and
therefore the exponential term in the integrand is $\approx 1$. For
$m=1$, the integral converges,
\begin{align}
\int_0^{1} \dd \lambda^2 (1 - \lambda^2)^{\frac{\gamma - 1}{1 -
    \delta} - 1} (\lambda^2)^{-1/2} = \frac{\sqrt{\pi} \, \Gamma
  \left( \frac{\gamma - 1}{1 - \delta} \right) }{ \Gamma \left(
  \frac{2 \gamma - \delta - 1}{2 (1 - \delta)} \right) } \, ,
\end{align}
where $\Gamma$ is the gamma function and the average flux is,
\begin{align}
\mu_J = \frac{c}{4 \pi} N \mu_Z = \frac{c}{4 \pi} \frac{1}{\sqrt{4 D_0
    b_0 (1 - \delta)}} \frac{N}{\pi r_2^2} \frac{Q_0}{t_\text{max}}
E^{- \gamma - 1 + (1 - \delta) / 2} \frac{\Gamma \left( \frac{\gamma -
    1}{1 - \delta} \right) }{ \Gamma \left( \frac{2 \gamma - \delta -
    1}{2 (1 - \delta)} \right) } \, .
\end{align}
With the parameters chosen as shown in Table
\ref{tbl:FluctuationsParameters} this gives,
\begin{align}
E^3 \mu_J \simeq 150 \, \text{GeV}^{-1} \, \text{cm}^{-2} \,
\text{s}^{-1} \, \text{sr}^{-1} \, ,
\end{align}
close to the featureless $E^{-3}$ spectrum measured with
Fermi-LAT~\cite{Abdo:2009zk}.

\begin{table}[!b]
\centering
\begin{threeparttable}[b]
\caption{Summary of parameters used in the Monte Carlo simulation.}
\label{tbl:FluctuationsParameters}
\begin{tabular}{c c c}
\hline\hline \multicolumn{3}{l}{Diffusion Model}\\ \hline

$D_0$ & $10^{28}\,\text{cm}^2\,\text{s}^{-1}$ &
\multirow{3}{*}{$\Bigg\}$ \begin{minipage}[c]{0.5\columnwidth}
    \flushleft from GCR nuclear secondary-to-primary
    ratios\end{minipage}} \\ $\delta$ & $0.6$ & \\ $z_\text{max}$ &
$3$ $\text{kpc}$ \\ $b_0$ & $10^{-16}\,\text{GeV}^{-1} \,
\text{s}^{-1}$ & CMB, IBL and $\vec{B}$ energy densities \\ \hline
\multicolumn{3}{l}{Source Distribution}\\ \hline $t_\text{max}$ & $3
\times 10^8\,\text{yr}$ & from $E_{\text{min}} \simeq 1 \, \text{GeV}$
\\ $N$ &$1 \times 10^7$ & from supernova rate and $t_\text{max}$
\\ \hline \multicolumn{3}{l}{Source Model}\\ \hline $Q_0$ & $8.4
\times 10^{49}\,\text{GeV}^{-1}$
&\multirow{2}{*}{$\Big\}$ \begin{minipage}[c]{0.5\columnwidth}
    \flushleft fit to absolute $e^+ + e^-$ flux \end{minipage}}
\\ $\gamma$ & $2.2$ & \\ $r_2$ & $15 \, \text{kpc}$ & \\ \hline \hline
\end{tabular}
\end{threeparttable} 
\end{table}

For $m=2$, the integral in Eq.~\ref{eqn:MomentIntegral} however
diverges; the second moment, $\langle Z^2 \rangle$, and hence the
variance $\sigma^2_Z$ is infinite. This could be cured by reinstating
$\exp{ (- 2 \rho_1^2 / \lambda^2)}$ with a finite $\rho_1$ as a
cut-off in the integral for small $\lambda^2$, however, for the price
of an additional and, even worse, physically unmotivated
parameter. For $r_1 = 0$, the standard deviation of the flux therefore
does not exist and cannot be used as an indicator of the amplitude of
the stochastic fluctuations. Furthermore, the usual central limit
theorem cannot be applied any more.

The reason that the expectation value is finite but not the variance
is that the probability density $f_Z(z)$ has a broad power-law
tail. For such cases, a generalised central limit theorem
\cite{Gnedenko:1954} is applicable: The centred and normalised sum
$X_N$ of $N$ independent and identically distributed (iid) random
variables $Z_i$ converges against a stable distribution
$\mathcal{S}(\alpha, \beta, 1, 0, 1)$~\cite{nolan:2010} if the
probability density $f_Z(z)$ behaves like $\left| z \right|^{-\alpha -
  1}$ for $z \rightarrow \infty$. In general, the distribution
function for $\mathcal{S}$ is not known analytically but can be
calculated as the inverse Fourier transform of its characteristic
function. To determine the parameters $\alpha$ and $\beta$, we need to
find the asymptotic behaviour of the probability density $f_Z(z)$ for
large $z$.

The distribution function $F_Z(z)$ satisfies,
\begin{align}
F_Z(z) = \iint_{\mathcal{D}_Z} \, \dd t \, \dd L \, f_t(t) f_L(L)
\end{align}
with the region $\mathcal{D}_Z$ given by the condition $Z < z$. With
$Z = G(t,L,E)$ this can be transformed into a condition on $L$:
\begin{align}
Z < z \quad \Leftrightarrow \quad L^2 > L_\text{min}^2 \equiv -\ell^2
\log z + \ell^2 \log \left( (\pi \ell^2)^{-3/2} Q_0 E^{-\gamma} (1 -
b_0 E t)^{\gamma - 2} \right) \, ,
\end{align}
and hence
\begin{align}
F_Z(z) &= \frac{1}{t_\text{max}} \frac{1}{r_2^2} \int_0^{t_\text{max}}
\, \dd t \int_{\max [0; L_\text{min}]}^{r_2} \, \dd L \, 2 L \\ &=
\frac{1}{t_\text{max}} \frac{1}{r_2^2} \left( \int_0^{t_*} \, \dd t
\int_{L_\text{min}^2}^{r_2^2} \, \dd L^2 + \int_{t_*}^{t_\text{max}}
\, \dd t \int_0^{r_2^2} \, \dd L^2 \right) \, ,
\end{align}
where $L_{\text{min}} \geq 0$ for $0 \leq t \leq t_*$ and
$L_{\text{min}} < 0$ for $t > t_*$. The probability density can then
be obtained by differentiating,
\begin{align}
f_Z(z) = \frac{\dd F_Z(z)}{\dd z} &= \frac{1}{t_\text{max}}
\frac{1}{r_2^2} \int_0^{t_*} \, \dd t \left( - \frac{\partial
  L_\text{min}^2}{\partial z} \right) \, .
\end{align}
With the substitutions, Eqs.~\ref{eqn:Subst1} and \ref{eqn:Subst2},
this reads,
\begin{align}
\label{eqn:fZ}
f_Z(z) &= \frac{1}{t_\text{max}} \frac{1}{r_2^2} \frac{4 D_0}{\left(
  b_0 (1 - \delta) \right)^2} E^{\delta - 2} \int_0^{\lambda_*^2} \dd
\lambda^2 \left( 1 - \lambda^2 \right)^{-\delta / (\delta - 1)}
\lambda^2 \frac{1}{z}
\end{align}
with $\lambda_*^2$ defined by $\Lambda_{\text{min}}^2 |_{\lambda^2 =
  \lambda_*^2} = 0$ where
\begin{align}
\Lambda^2_{\text{min}} = - \lambda^2 \left( \log z - \log \left[
  \left( \frac{4 \pi D_0}{b_0 (1 - \delta)} \right)^{- \frac{3}{2} }
  E^{\frac{3}{2} (1 - \delta) -\gamma} Q_0 \right] + \frac{3}{2} \log
\lambda^2 + \frac{\gamma - 2}{\delta - 1} \log \left( 1 - \lambda^2
\right) \right) \, .
\end{align}
As we are looking for the asymptotic behaviour of $F_Z(z)$ for large
$z$ which we anticipate corresponds to $\lambda_*^2 \ll 1$, we can
neglect the $( 1 - \lambda^2)$ term in $\Lambda^2_{\text{min}}$ and
solve $\Lambda^2_{\text{min}}$ for $\lambda_*^2$,
\begin{align}
\lambda_*^2 \simeq \frac{b_0 (1 - \delta)}{4 \pi D_0} E^{ (1 - \delta)
  - \frac{2}{3} \gamma} Q_0^{\nicefrac{2}{3}} z^{-\nicefrac{2}{3}} \,
.
\end{align}
With the same approximation, the integral in Eq.~\ref{eqn:fZ} gives
$(\lambda_*^2)^2/2$ and
\begin{align}
f_Z(z) \simeq \frac{1}{t_\text{max}} \frac{1}{r_2^2} \frac{1}{8 \pi^2
  D_0} E^{- \delta - \frac{4}{3} \gamma} Q_0^{\nicefrac{4}{3}}
z^{-\nicefrac{7}{3}} \, .
\end{align}
The asymptotic behaviour of the distribution function $F_Z(z)$ is
therefore
\begin{align}
1 - F_Z(z) \sim c^+ z^{-\nicefrac{4}{3}} \quad \text{with} \quad c^+ =
\frac{3}{4} \frac{1}{t_\text{max}} \frac{1}{r_2^2} \frac{1}{8 \pi^2
  D_0} E^{- \delta - \frac{4}{3} \gamma} Q_0^{\nicefrac{4}{3}} \, ,
\end{align}
and $F_Z = 0$ for $z < 0$.

The generalised central limit theorem \cite{Uchaikin:1999} then states
that the centred and normalised sum $X_N$,
\begin{align}
X_N = \frac{1}{b_N} \left( \sum_i^N X_i - a_N \right) \, ,
\end{align}
weakly converges to the stable distribution $\mathcal{S}(\alpha,
\beta, 1, 0, 1)$ with $\alpha = 4/3$ and $\beta = 1$. The
normalisation constants are
\begin{align}
a_N = N \mu_Z \quad \text{and} \quad b_N = \left( \frac{\pi c^+}{2
  \Gamma(\frac{4}{3}) \sin \left( \frac{\pi}{2} \frac{4}{3} \right) }
\right)^{\nicefrac{3}{4}} N^{\nicefrac{3}{4}} \, .
\end{align}

\begin{figure}[t]
\begin{center}
\includegraphics[width=0.7\textwidth]{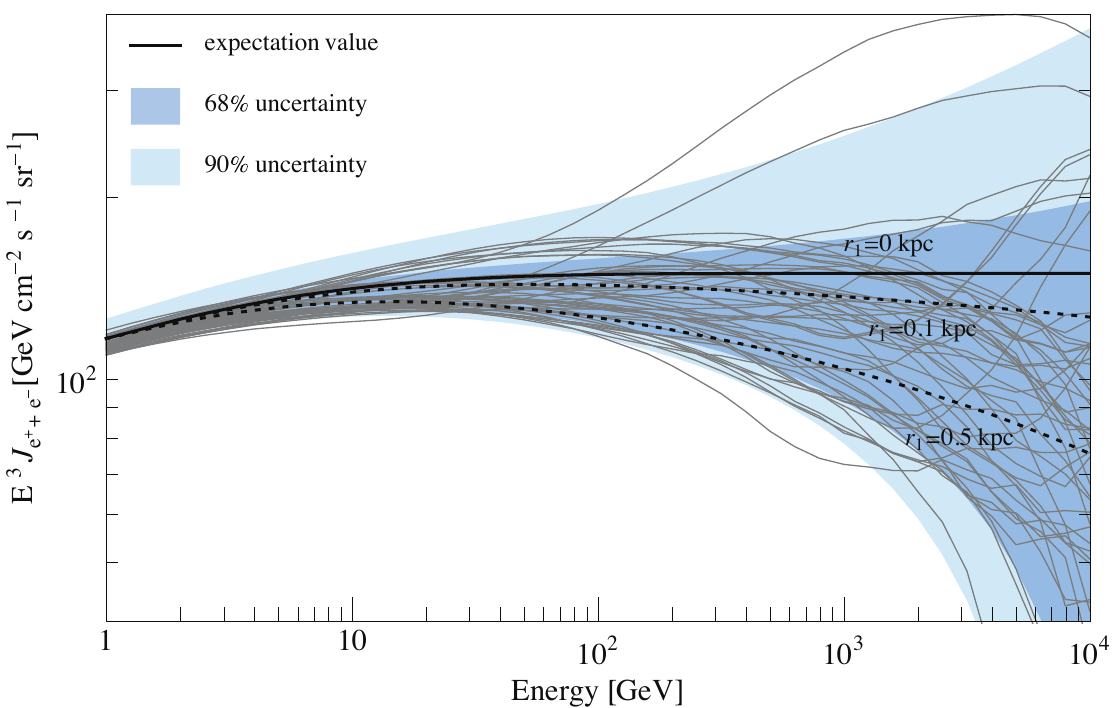}
\end{center}
\caption{Fluxes of Cosmic Ray electrons from ensembles of sources
  uniformly distributed in a disk around the observer. The solid line
  denotes the expectation value for the sum of fluxes from $N$
  discrete, transient sources. The dashed lines show the expectation
  values if the sources are limited to a ring with inner radius $r_1$
  (normalised to the expectation value for $r_1 = 0$ at $1 \,
  \text{GeV}$. The coloured bands quantify the fluctuations and
  contain the fluxes $68 \, \%$ and $90 \, \%$ of the time,
  respectively. The fluxes from 50 realisations of an ensemble of $N$
  individual sources are shown by the grey lines.}
\label{fig:nuFluctuationsSpec}
\end{figure}

The expectation value $\mu_J$ for the flux on Earth is therefore still
$c/(4 \pi) N \mu_Z$ but instead of the standard deviation, we use
quantiles of the stable distribution to quantify the level of
fluctuations. For $\mathcal{S}(4/3, 1, 1, 0, 1)$ the $5 \, \%$, $16 \,
\%$, $84 \, \%$ and $95 \, \%$ quantiles are approximately $-3.3$,
$-2.6$, $1.0$ and $4.7$, respectively. The energy dependence of the
quantiles of $J$ is contained in $b_N \propto
{(c^+)}^{\nicefrac{3}{4}} \propto E^{-\gamma - 3 \delta / 4}$, which
is always harder than $\mu_J \propto E^{-\gamma - (1+\delta)/2}$, so
the fluctuations are growing with energy. For the particular values
from Table~\ref{tbl:FluctuationsParameters}, $b_N \propto
E^{-2.65}$. Fig. \ref{fig:nuFluctuationsSpec} shows the uncertainty
bands that $68 \, \%$ and $90 \, \%$ of the total fluxes should be
contained in, that is $\big[ \mu_J + c/(4 \pi) b_N x_{16 \%}, \mu_J +
  c/(4 \pi) b_N x_{84 \%} \big]$ and $\big[ \mu_J + c/(4 \pi) b_N x_{5
    \%}, \mu_J + c/(4 \pi) b_N x_{95 \%} \big]$, together with the
expectation value for the flux $\mu_J$.

One notices that a spectral softening occurs in the spectrum around
$10 \, \text{GeV}$. Below $\sim 50 \, \text{GeV}$, the expansion for
$\chi(\ell/\zcr)$ made above is not valid any more and the spectrum is
not dominated by energy losses but by losses due to the escape from
the halo. We also show the the total fluxes from a Monte Carlo
calculation of 50 realisations of a uniform source distribution with
the same model parameters, see
Table~\ref{tbl:FluctuationsParameters}. Although the bands do indeed
contain the right fraction of fluxes, the majority appears to be not
only lower but also more curved downwards than the expectation value
for $r_1 = 0$.

The deficit in the majority of realisations is in fact a consequence
of the stable distribution which is highly asymmetric with a long
tail. On the one hand, it is because of the asymmetry that a majority
of the fluxes from different realisations of the source density is
below the expectation value. The long tail of the probability density
on the other hand makes the average from a large number of
realisations finally and slowly converge to the expectation
value. This convergence is slowing down with increasing energy as
because of the decreasing diffusion-loss length less sources
effectively contribute at higher energies.

A more physical way of explaining the deficit at higher energies is
that every realisation of a finite number of sources necessarily
contains one closest source at $L_\text{cl} = \min_i{ [ L_i ] }$ which
contributes the most at the highest energies. Once the diffusion-loss
length $\lambda(E)$ becomes however shorter than the distance
$L_\text{cl}$ even this contribution gets cut-off by the exponential
term in the Green's function, Eq.~\ref{eqn:GreensDiscreteSrc},
$\exp{(-L_\text{cl}^2/\ell^2)}$. The curving of the fluxes, called
propagation cut-off, is therefore nothing but the shoulder of (a few)
Green's function(s) from the closest source(s). We note that such a
cut-off cannot simply be modelled by reinstating the minimum radius
$r_1$ in Eq.~\ref{eqn:MomentIntegral} as shown by the dashed lines in
Fig.\ref{fig:nuFluctuationsSpec} for $r_1 = 0.1$ and $0.5 \,
\text{kpc}$.

\subsection{A realistic source distribution}
\label{sec:SourceDistribution}

Of course, the sources of cosmic rays are not evenly distributed in a
disk around the solar system. Above, this was only assumed to simplify
the calculation of the effect of the discreteness of sources. In fact,
SNRs are expected to spatially correlate with star formation activity
and to be mainly based in the thin galactic disk, tracing the spiral
structure. As discussed above, we have seen that effects of their
actual distribution will play a role at energies above $\sim 100 \,
\text{GeV}$.

\begin{figure}[tb]\centering
\begin{minipage}[t]{0.5\columnwidth}\centering
\includegraphics[width=0.95\columnwidth]{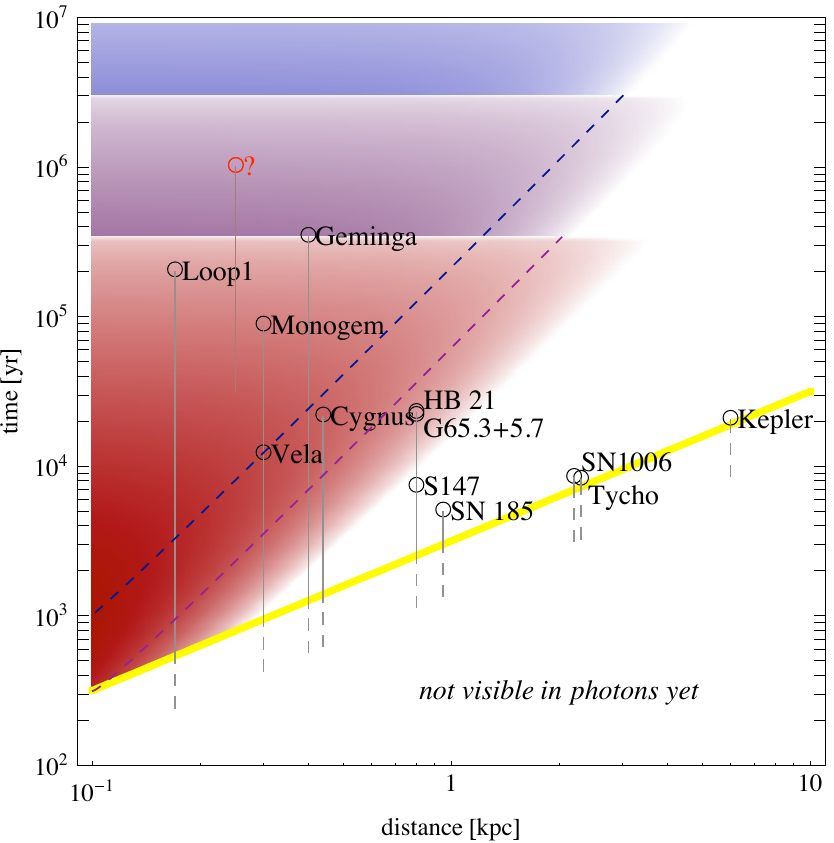}
\end{minipage}\hfill
\begin{minipage}[t]{0.5\columnwidth}\centering
\includegraphics[width=0.95\columnwidth]{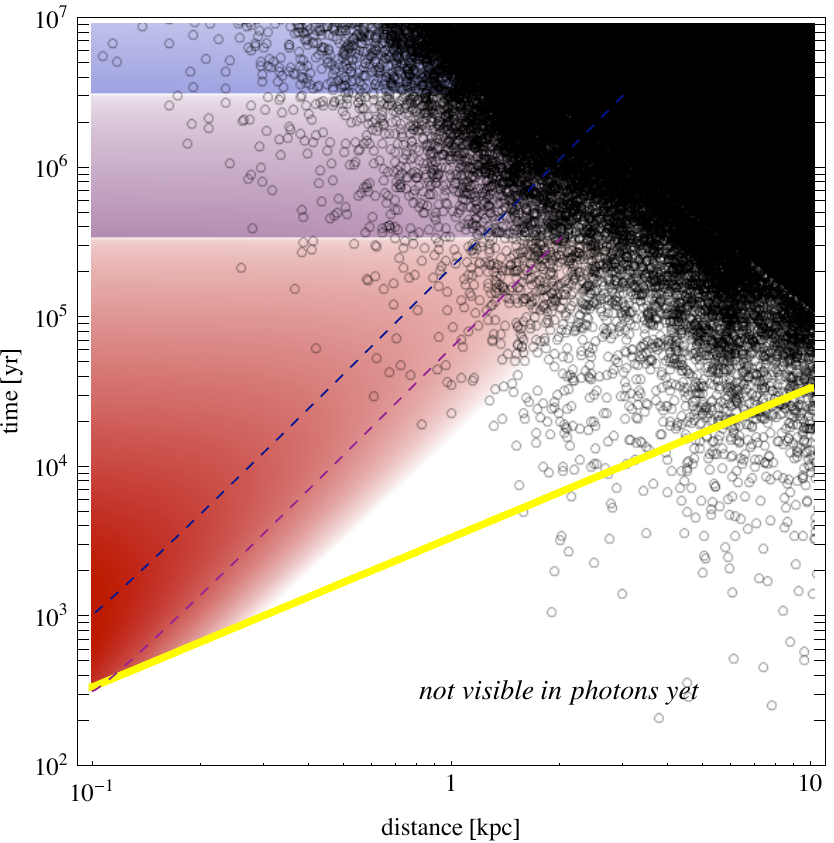}
\end{minipage}
\caption[]{Distance-time diagram for nearby SNRs (after
  Ref.~\cite{Swordy:2003ds}). {\bf Left:} The open circles mark
  supernova events and the world-lines of the discovered remnants are
  indicated. The thick yellow line is our past light-cone; all events
  lying on it, e.g. the SNR world-lines touching it, can be observed
  presently. The blue, purple and red shadings (top to bottom) show
  the relative contribution of sources to the diffuse $e^-$ and $e^+$
  flux observed at Earth at 10, 100 and 1000 GeV, respectively. The
  open red circle is an example of a hypothetical supernova whose
  remnant is too old to be visible any longer but which might still be
  contributing to the diffuse $e^-$ and $e^+$ flux. {\bf Right:} A
  distance-time diagram for hypothetical nearby SNRs. The open black
  circles are an example of a possible realisation of the supernova
  density (the world lines have been suppressed) as simulated by our
  Monte Carlo calculation (see Section~\ref{sec:SourceDistribution}).}
\label{fig:distance-time-diagram}
\end{figure}

Some authors~\cite{Kobayashi:2003kp,Shaviv:2009bu} have assumed a
continuous distribution of sources for distances beyond a few hundred
parsecs, supplemented by a set of SNRs, known from x-ray or radio
surveys, for smaller distances. This approach is however biased by the
choice of young, nearby sources which have been detected in radio
and/or X-rays. Older sources may not be visible in photons any longer
but still be contributing to the GCR electron flux. This relation is
illustrated in Fig.~\ref{fig:distance-time-diagram}. We note that the
effect of this incomplete assumed source distribution is a dip in the
electron flux seen in both
analyses~\cite{Kobayashi:2003kp,Shaviv:2009bu}, although at different
energies because of the different diffusion model parameters chosen.

Determining the complete distribution of sources in our vicinity
(i.e. up to a few kpc) from observations seems challenging. However it
turns out that we do not need to know the exact distribution in order
to make a prediction for the $e^+$ flux and fraction but require only
a limited amount of information, all of which is encoded already in
the total $e^- + e^+$ flux. By including the recent measurements by
Fermi-LAT \cite{Abdo:2009zk} and HESS
\cite{Collaboration:2008aaa,Aharonian:2009ah} of the total $e^- + e^+$
flux in the energy region of interest, we have sufficient information
at hand to make a prediction for the positron fraction under the
assumption that the additional positrons originate in the same
sources.

We perform a Monte Carlo calculation by considering a large number of
random distributions of sources drawn from a probability density
function that reflects our astronomical knowledge about the
distribution of SNRs in the Galaxy. The better the flux of $e^-$ and
$e^+$ from such a realisation of the source density reproduces the
measured fluxes, the closer is the underlying distribution of sources
likely to be to the actual one. We do not consider any scatter in the
parameters of the SNRs but assume a prototypical set of source
parameters that we determine from a compilation of gamma-ray SNRs, see
Sec.~\ref{sec:PrimaryElectrons}. Of course all SNRs are not the same,
however variations of the source parameters would only introduce
additional fluctuations into the fluxes without altering their
average. We can choose the best ``fit'' to the data and thus determine
the $e^+$ flux.

\begin{figure}[tb]
\begin{tabular*}{\columnwidth}{@{} p{0.49\columnwidth} @{} p{0.02\columnwidth} @{} p{0.49\columnwidth} @{}}
\includegraphics[width=0.45\columnwidth]{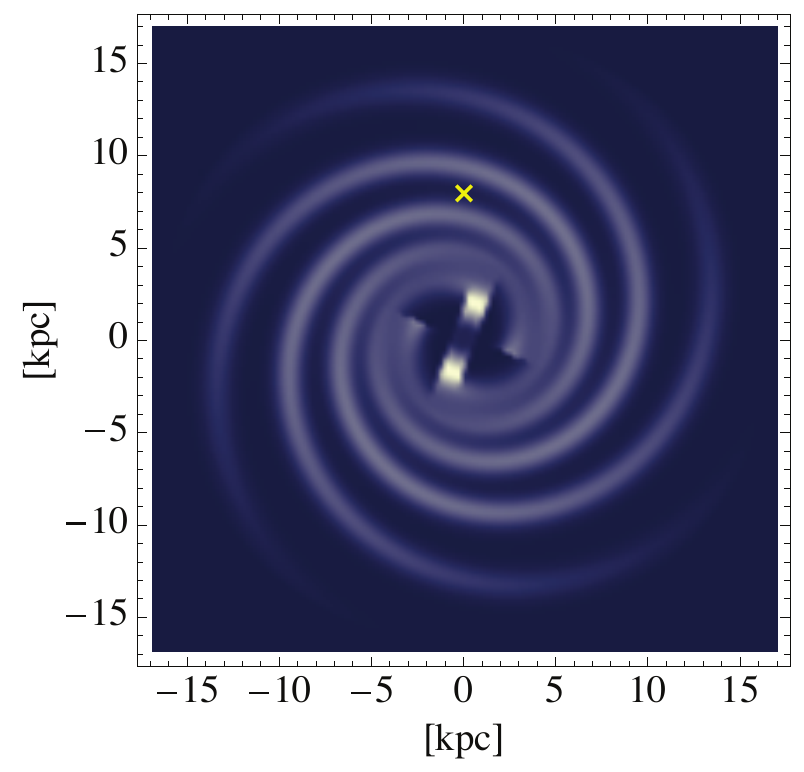} & &
\includegraphics[width=0.432\columnwidth]{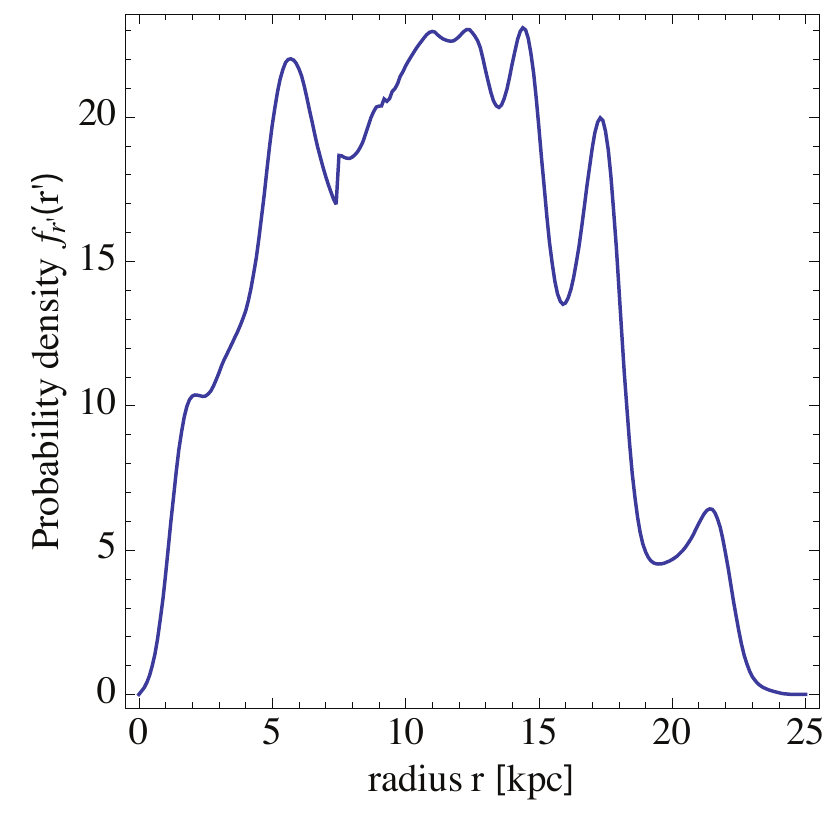} \\
\begin{minipage}[t]{\linewidth}
\caption{The assumed distribution of SNRs in the Galaxy; the cross
  denotes the position of the Sun in between two spiral arms.}
\label{fig:spiral_structure}
\end{minipage}
& &
\begin{minipage}[t]{\linewidth}
\caption{The probability density for the distance of a SNR from the
  Sun.}
\label{fig:SNR_density}
\end{minipage}
\end{tabular*}
\end{figure}

The smoothed radial distribution of SNRs in the Galaxy is well
modelled by~\cite{Case:1998qg}
\begin{equation}
f (r) = A \sin{\left(\frac{\pi r}{r_0} + \theta_0\right)}
\text{e}^{-\beta r},
\label{eqn:SNR_density}
\end{equation}
where $A = 1.96\,\text{kpc}^{-2}$, $r_0 = 17.2\,\text{kpc}$, $\theta_0
= 0.08$ and $\beta = 0.13$. To obtain a realistic probability density
for the distance between the Earth and a SNR we have to also take into
account the spiral structure of the Galaxy. We adopt a logarithmic
spiral with four arms of pitch angle $12.6^\circ$ and a central bar of
6 kpc length inclined by $30^\circ$ with respect to the direction Sun
- galactic centre~\cite{Vallee:2005}. The density of SNRs is modelled
by a Gaussian with 500 pc dispersion for each arm
\cite{Pohl:1998ug}. The resulting distribution $g (r, \phi)$ (see
Fig.~\ref{fig:SNR_density}) has been normalized with respect to
azimuth in such a way that the above radial distribution
Eq.~\ref{eqn:SNR_density} is recovered. To obtain the probability
density for the distances we transform to the coordinates $(r',
\phi')$ centred on the Sun. As the $e^-$ and $e^+$ fluxes are assumed
to be isotropic, we can average over the polar angle $\phi'$, such
that the probability density $f_{r'}$ depends only on the distance
$r'$ to the source,
\begin{equation}
 f_{r'} (r') = \frac{1}{2 \pi} \int_0^{2\pi} \text{d}\phi' r' g (r
 (r', \phi'), \phi (r', \phi')).
\label{eqn:dF}
\end{equation}
This function is shown in Fig.~\ref{fig:SNR_density}.

We assume that the sources are uniformly distributed in time,
i.e. their probability density $f_t (t)$ is
\begin{equation}
 f_t (t) = \left\{\begin{array}{ll} 1/t_\text{max} & \text{for } 0
 \leq t \leq t_\text{max}, \\ 0 & \text{otherwise}, \end{array}\right.
\end{equation}
with $t_\text{max}$ denoting the earliest time considered, which is
again related to the minimum energy for which our calculation is valid
through
\begin{equation}
t_\text{max} = \left(b E_{\text{min}}\right)^{-1}.
\end{equation}
The total number $N$ of sources that are needed in the Monte Carlo
simulation to reproduce the (observed) number $\mathcal{N}~\simeq~300$
of SNRs active in the galaxy at any given time depends on the average
lifetime of a SNR, $\tau_\text{SNR}$, which is suggested to be $\sim
10{^4} \, \text{yr}$ \cite{Reynolds:2008}, hence
\begin{align}
  N = 3 \times 10^6 \left(\frac{\mathcal{N}}{300}\right)
  \left(\frac{t_\text{max}}{10^8\,\text{yr}}\right)
  \left(\frac{\tau_\text{SNR}}{10^4\,\text{yr}}\right)^{-1}.
\end{align}

We take a Green's function approach to calculate the contribution from
these $N$ discrete, burst-like sources to the total electron-positron
flux in the solar system today for a halo of extent $\pm z_\text{max}$
in $z$ direction, again neglecting the boundaries in the radial
direction.  We remind ourselves that the Green's function for the flux
of electrons from a source at $\vec{r}$ that went off a time $t$ ago
with a spectrum $Q(E)$, is
\begin{align}
\label{Greensfunction}
G_\text{disk}(E, \vec{r}, t) \nonumber =& \sum_{n=-\infty}^\infty
(\pi\ell^2)^{-\nicefrac{3}{2}} \text{e}^{-\vec{r}_n^2/\ell^2} Q
\left(\frac{E}{1-b_0 E t}\right) (1 - b_0 E t)^{-2} \nonumber \\ =&
(\pi\ell^2)^{-1} \text{e}^{-\vec{r}_{\parallel}^2/\ell^2} Q
\left(\frac{E}{1-b_0 E t}\right) (1 - b_0 E t)^{-2}
\frac{1}{\ensuremath{\zcr}} \chi \left( \frac{z}{\zcr},
\frac{\ell}{\zcr} \right) \, ,
\end{align}
where
\begin{align}
\label{functionchi}
\chi (\hat{z}, \hat{\ell}) &\equiv \frac{1}{\sqrt{\pi} \hat{\ell}}
\sum_{n=-\infty}^\infty \text{e}^{-\hat{z}_n^2/\hat{\ell}^2},
\end{align}
and the diffusion length $\ell$ is defined by
\begin{align}
\label{difflength}\nonumber
\ell^2 &= 4 \int_E^{E/(1 - b(E)t)} \text{d}E' \, \frac{D(E')}{b(E')} =
\frac{4 D_0}{b_0 (1 - \delta)} \left[E^{\delta - 1} -
  \left(\frac{E}{1-b_0 E t} \right)^{\delta - 1}\right],
\end{align}
with $\ensuremath{\zcr} \equiv 4 z_\text{max}/\pi$.  If we neglect the
spatial extent of the disk and set $z = 0$, the function $\chi(
\hat{\ell}) \equiv \chi(0,\hat{\ell})$ is approximately:
\begin{equation}
\chi (\hat{l}) \simeq \left\{\begin{array}{ll} \frac{4}{\pi}
e^{-\hat{l}^2} & \text{for } \hat{l} \gg \frac{\pi}{4},
\\ \frac{1}{\sqrt{\pi} \hat{l}} & \text{for } \hat{l} \ll
\frac{\pi}{4}.
\end{array} \right.
\end{equation}
In practice both limits can be connected at $\hat{l} \simeq 0.66$ such
that the approximated $\chi(\hat{\ell})$ has a relative error of at
most 0.5\%.  We motivate the choice of the parameters of our diffusion
model from an analysis of nuclear secondary-to-primary
ratios~\cite{Strong:2007nh}: $D_0 = 10^{28} \,\text{cm}^2
\text{s}^{-1}$, $\delta = 0.6$, $L = 3$~kpc, and from the galactic
magnetic field and interstellar radiation fields
\cite{Kobayashi:2003kp}: $b_0 = 10^{-16} \, \text{GeV}^{-1} \,
\text{s}^{-1}$.

\section{Fitting the Total Electron-Positron spectrum}
\label{sec:Spectra}

\begin{figure}[tb]
\centering \includegraphics[width=1.0
  \columnwidth]{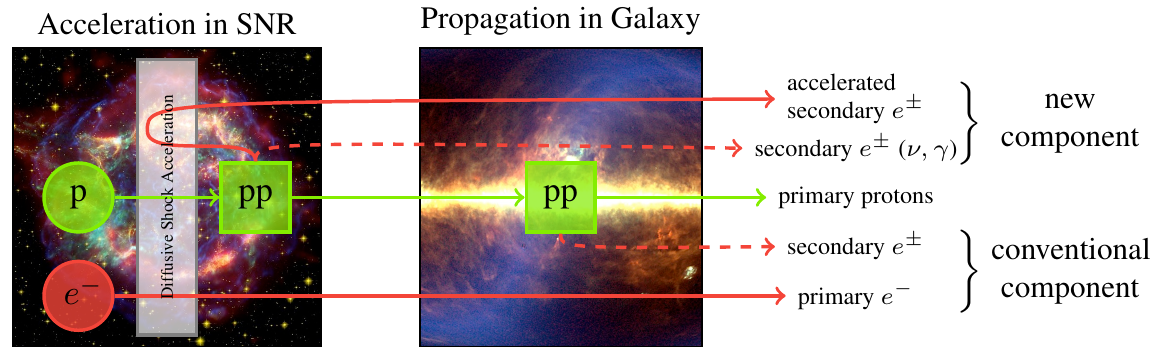}
\caption[]{Schematic description of contributions to the galactic
  cosmic rays observed at Earth in the present framework.}
\label{fig:SNR_mechanism}
\end{figure}

A schematic description of the present framework is shown in
Fig.~\ref{fig:SNR_mechanism}. Cosmic rays are shock accelerated in
SNRs and then diffuse through the Galaxy to Earth undergoing
collisions with interstellar matter {\it en route} and creating
secondary $e^+$. As discussed, the ratio of the secondary $e^+$ to the
primary $e^-$ from the sources should {\it decrease} with energy, in
contrast to the behaviour seen by PAMELA. We follow
Ref.~\cite{Blasi:2009hv} in explaining this by invoking a new
component of $e^+$ which is produced through cosmic ray interactions
in the SNRs, and then shock {\it accelerated}, thus yielding a harder
spectrum than that of their primaries. We discuss these components in
turn below and calculate their relative contributions by normalising
to the gamma-ray flux from the SNRs, which provides an independent
measure of the hadronic interactions therein.

\subsection{Primary electrons}
\label{sec:PrimaryElectrons}

The radio and X-ray emission observed from SNRs is interpreted as
synchrotron radiation of electrons accelerated up to energies of
${\cal O}(100)$~TeV~\cite{Reynolds:2008}. The spectrum of this
radiation then determines the spectrum of the underlying relativistic
electrons. Moreover, the theory of diffusive shock acceleration
\cite{Blandford:1987pw,Malkov:2001} predicts similar spectra for the
accelerated protons and nuclei as for the electrons. If the gamma-ray
emission observed by HESS from a number of identified SNRs is assumed
to be of hadronic origin, we can use the measured spectra to constrain
both the relativistic proton and electron population.

\begin{sidewaystable}[p]
  \begin{center}
\begin{threeparttable}[b]
    \caption{Summary of spectral parameters for SNRs detected in
      gamma-rays from a power-law fit to the spectrum, $J_\gamma =
      J_\gamma^0 (E/\text{TeV})^{-\gamma}$, with an exponential
      cut-off at $E_\text{max}$ in the case of HESS~J1713.7-397. The
      errors shown are statistical only --- the systematic error is
      conservatively estimated to be 20\% on the flux $J_\gamma$ and
      $\pm 0.2$ on the spectral index $\gamma$. Also shown is the
      estimated distance $L$ and the injection rate $Q_\gamma^0$
      derived from Eq.~\ref{eqn:GammaLumi}.}
    \label{tbl:GammaRays}
    \begin{tabular}{ l l r@{$\:\pm\:$} l r@{$\:\pm\:$}  l c  c  c   r }
      \hline\hline Source & Other name(s) &
      \multicolumn{2}{c}{$\gamma$} & \multicolumn{2}{c}{$J_\gamma^0
        \div 10^{-12}$} & $E_\text{max}$ & $L$ & $Q_{\gamma}^0 \div
      10^{33}$ & Ref.  \\ & &\multicolumn{2}{c}{ }
      &\multicolumn{2}{c}{\hspace{-0.2cm} \footnotesize [$({\rm
            cm}^{2}\,{\rm s}\,{\rm TeV})^{-1}$]} & {\footnotesize
        [TeV]} & {\footnotesize [kpc]} & {\footnotesize [$({\rm
            s}\,{\rm TeV})^{-1}$]} & \\ \hline HESS~J0852$-$463 & RX
      J0852.0-4622 (Vela Junior) & 2.1 & 0.1 & 21 & 2 & $> 10$ & 0.2 &
      0.10 &~\cite{Aharonian:2005sz} \\ HESS~J1442$-$624 &RCW 86, SN
      185 (?)  & 2.54 & 0.12 & 3.72 & 0.50 & $\gtrsim$ 20 & 1 & 0.46
      &~\cite{Aharonian:2008nw}\\ HESS~J1713$-$381 & CTB 37B,
      G348.7+0.3 & 2.65 & 0.19 & 0.65 & 0.11 & $\gtrsim$ 15 & 7 &
      3.812 &~\cite{Aharonian:2008ka}\\ HESS~J1713$-$397 & RX
      J1713.7-3946, G347.3-0.5 & 2.04 & 0.04 & 21.3 & 0.5 & 17.9 $\pm$
      3.3 & 1 & 2.55 &~\cite{Aharonian:2005qm, Aharonian:2006ws}
      \\ HESS~J1714$-$385 & CTB 37A & 2.30 & 0.13 & 0.87 & 0.1 &
      $\gtrsim$ 12 & 11.3 & 13.3 &~\cite{Aharonian:2008km}
      \\ HESS~J1731$-$347 & G 353.6-07 & 2.26 & 0.10 & 6.1 & 0.8 &
      $\gtrsim$ 80 & 3.2 & 7.48 &~\cite{Aharonian:2007zj, Tian:2008tr}
      \\ HESS~J1801$-$233\tnote{1} & W 28, GRO J1801-2320 & 2.66 &
      0.27 & 0.75 & 0.11 & $\gtrsim$ 4 & 2 & 0.359
      &~\cite{Aharonian:2008fp} \\ HESS~J1804$-$216\tnote{2} & W 30,
      G8.7-0.1 & 2.72 & 0.06 & \multicolumn{2}{c}{5.74} & $\gtrsim$ 10
      & 6 & 24.73 &~\cite{Aharonian:2005kn} \\ HESS~J1834$-$087 &W 41,
      G23.3-0.3 & 2.45 & 0.16 & \multicolumn{2}{c}{2.63} & $\gtrsim$ 3
      & 5 & 7.87 &~\cite{Aharonian:2005kn} \\ MAGIC J0616+225 & IC 443
      & 3.1 & 0.3 & \multicolumn{2}{c}{0.58} & $\gtrsim 1 $ & 1.5 &
      0.156 &~\cite{Albert:2007tr} \\ Cassiopeia A & & 2.4 & 0.2 & 1.0
      & 0.1 & $\gtrsim 40$ & 3.4 & 1.38
      &~\cite{Albert:2007wz}\tnote{3} \\ J0632$+$057 & Monoceros &
      2.53 & 0.26 & 0.91 & 0.17 & {$\cdots$} & 1.6 & 0.279
      &~\cite{Fiasson:2007bk} \\ \hline \multicolumn{2}{l }{Mean}&
      \multicolumn{2}{c}{$\sim 2.5$} & \multicolumn{2}{c}{} & $\gtrsim
      20$ & & $\sim 5.2$ & \\ \multicolumn{2}{l }{Mean, excluding
        sources with $\gamma > 2.8$}& \multicolumn{2}{c}{$\sim 2.4$} &
      \multicolumn{2}{c}{} & $\gtrsim 20$ & & $\sim 5.7$ &
      \\ \multicolumn{2}{l }{Mean, excluding sources with $\gamma >
        2.6$}& \multicolumn{2}{c}{$\sim 2.3$} & \multicolumn{2}{c}{} &
      $\gtrsim 20$ & & $\sim 4.2$ & \\ \hline\hline
    \end{tabular}
\begin{tablenotes}
    \item[1] We assume that W 28 powers only the emission from
      J1801$-$233 (and not the nearby J1800$-$240 A, B and C).
    \item[2] W30 is taken to be the origin of the VHE (very high
      energy) emission~\cite{Fatuzzo:2006ua}.
    \item[3] Cas A was first detected by
      HEGRA~\cite{Aharonian:2001mz}.
\end{tablenotes}
\end{threeparttable}
  \end{center}
\end{sidewaystable}

Table~\ref{tbl:GammaRays} shows a compilation of gamma-ray sources
observed by HESS that have been identified as SNRs. We have included
all identified shell-type SNRs and strong SNR candidates in the HESS
source catalogue~\cite{HESS:SourceCatalogue} (as of September 2009),
and also added the SNRs IC 443, Cassiopeia A and Monoceros. Actually
it is not clear that the acceleration of secondaries does occur in all
the SNRs considered, especially when the gamma-ray emission is
associated with a neighbouring molecular cloud rather than coming from
the vicinity of the shock wave. In fact the gamma-rays could equally
well be due to inverse-Compton scattering by the relativistic
electrons responsible for the observed synchrotron radio and X-ray
emission. Therefore, we have considered three possibilities ---
including all sources implies a mean power-law spectral index for the
protons of $\langle \gamma \rangle = 2.5$, while excluding steep
spectrum sources with $\gamma > 2.8$ gives $\langle \gamma \rangle =
2.3$ and excluding sources with $\gamma > 2.6$ yields $\langle \gamma
\rangle = 2.4$. In the following we adopt the central value, $\gamma
=2.4$, for the electron population too, unless stated otherwise. This
requires a compression factor of $r \leq 3.3$ in contrast to the value
of $r = 4$ expected for a strong shock, so there is clearly some
tension between the DSA theory and observations. This can possibly be
resolved if we consider only a subset of the SNRs in Table 1 to be
hadronic accelerators, or if the gamma-ray spectrum is steepened,
e.g. by the onset of an exponential cut-off in the electron
spectrum. Our model assumptions are intimately connected to the
production of neutrinos, the detection of which will therefore provide
an independent test, see Sec.~\ref{sec:GammaRays&Neutrinos}. In this
work we adopt a cut-off of $E_\text{cut} \simeq 20$~TeV which is
consistent with DSA theory~\cite{Reynolds:2008}. The source spectrum
of primary electrons is then:
\begin{equation}
  R_{e^-} = R_{e^-}^0 \left(\frac{E}{\text{GeV}}\right)^{-\gamma}
  \text{e}^{-E/E_\text{cut}}.
\end{equation}
The normalisation $R_{e^-}^0$ is determined by fitting the electron
flux at Earth resulting from our Monte Carlo computation to the
preliminary measurement by PAMELA at 10~GeV~\cite{Mocchiutti:2009};
the secondary fluxes can be neglected for this normalisation. We find
$R_{e^-}^0 = 1.8 \times 10^{50} \,\text{GeV}^{-1}$ for $\gamma = 2.4$
which corresponds to a total injection energy of
\begin{align}
\int_{1 \,\text{GeV}}^{20 \,\text{TeV}} \text{d}E \,E \,R_{e^-}(E)
\simeq 7 \times 10^{47} \,\text{erg}.
\end{align}
This compares well to the value of $9.2 \times 10^{47}$ erg said to be
required to power the GCR electrons~\cite{Reynolds:2008}.

\begin{figure}\centering
  \includegraphics[width=0.65\columnwidth]{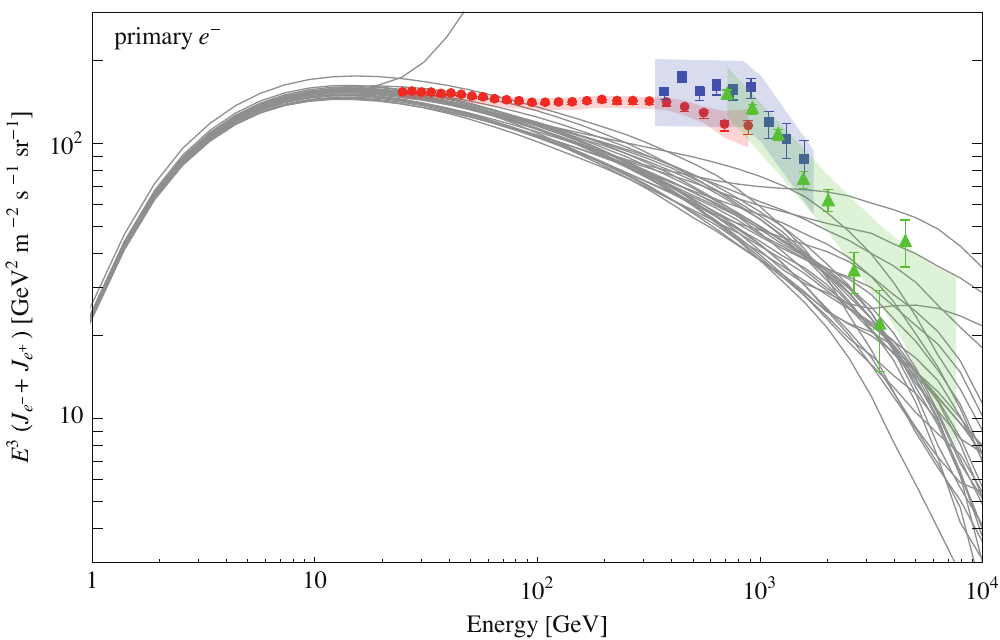}\\[0.3cm]
  \includegraphics[width=0.65\columnwidth]{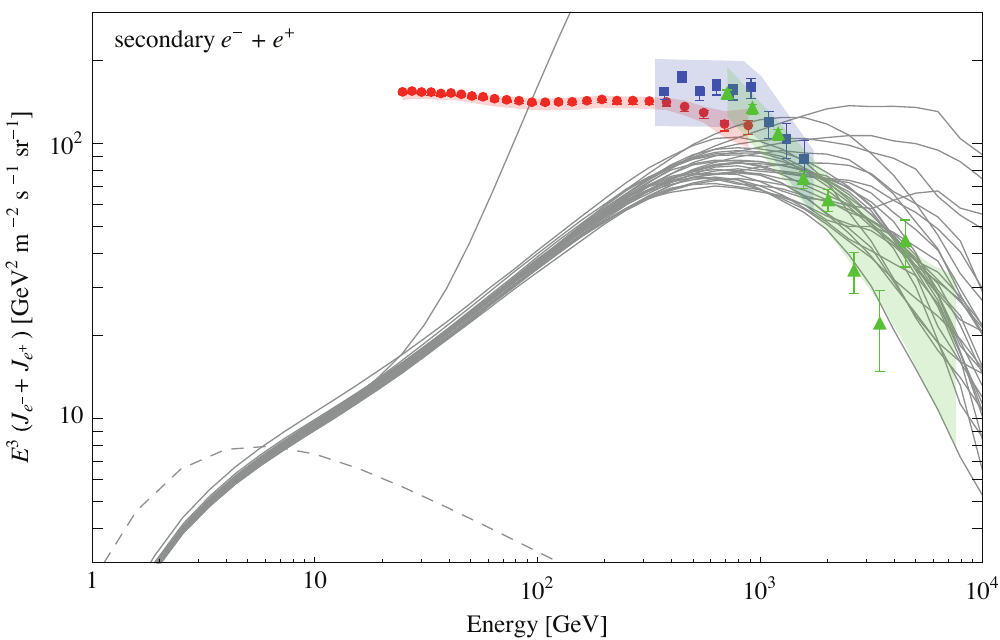}\\[0.3cm]
  \includegraphics[width=0.65\columnwidth]{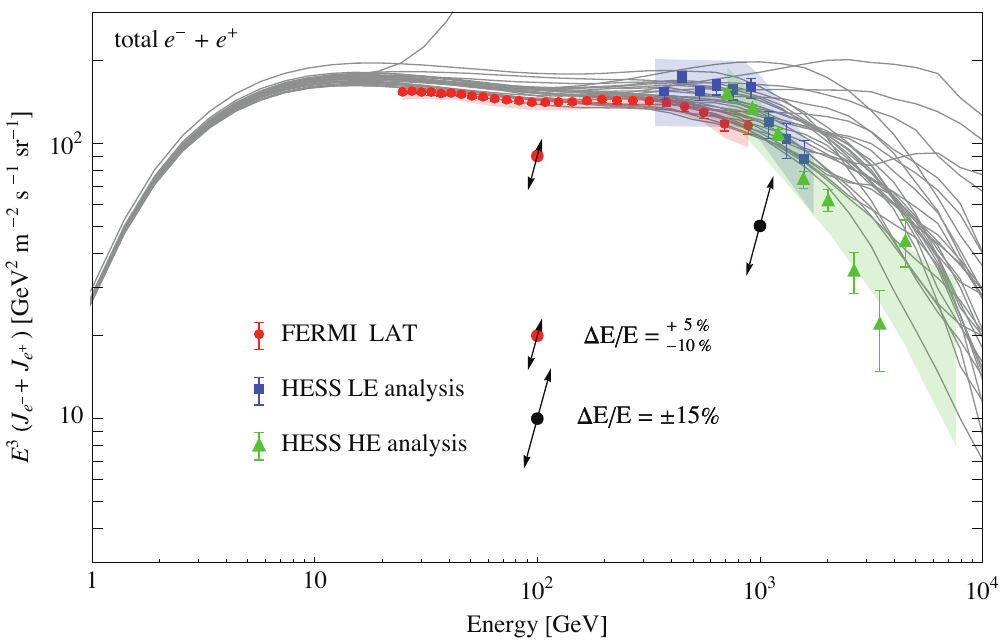}\\[0.3cm]
\caption{Predicted spectra of electrons and positrons with data from
  Fermi-LAT~\cite{Abdo:2009zk} (red circles) and HESS
  ~\cite{Collaboration:2008aaa, Aharonian:2009ah} (blue squares \&
  green triangles). The diagonal arrows show the energy scale
  uncertainty.  {\bf Top:} Primary electrons after propagation to
  Earth. {\bf Middle:} Secondary electrons and positrons from cosmic
  ray interactions, created during propagation (dashed line) and
  created during acceleration in SNRs (full lines).  {\bf Bottom:} The
  sum of primary and secondary electrons and positrons.}
  \label{fig:E3J}
\end{figure}

Solar modulation which is important below $\sim 10$~GeV, has been
accounted for using the force field approach~\cite{Gleeson:1968zz},
with a charge-independent potential of $\phi = 600 \, \text{MV}$.
However, our simple model ignores convection and (re)acceleration in
the interstellar medium which become important below $\sim 5
\,\text{GeV}$, hence the electron flux cannot be predicted at lower
energies. The primary $e^-$ fluxes as measured on Earth for 30
different source configurations are shown in the top panel of
Fig.~\ref{fig:E3J}. With an injection power-law index $\gamma \simeq
2.4\pm0.1$ as required for consistency with the gamma-ray data, there
clearly is a deficit at high energies compared to the $e^+ + e^-$ flux
measured by Fermi-LAT and HESS.

\subsection{Secondary electrons and positrons from propagation}

Positrons in GCR are generally assumed to be of purely secondary
origin, arising through the decay of pions and kaons produced in the
interactions of GCR protons (and nuclei) with the interstellar medium
(ISM)~\cite{Moskalenko:1997gh}. The neutral pions decay into
gamma-rays which then contribute to, if not dominantly constitute, the
galactic gamma-ray background. The charged pions on the other hand
decay into neutrinos and muons, the latter subsequently decaying into
electrons and positrons. Assuming that spatial and temporal variations
in the GCR proton flux $J_\text{p}$ and the ISM gas density
$n_\text{ISM}$ are small, the source density of these secondary
background $e^-$ and $e^+$ is also homogeneous, both in space and in
time:
\begin{equation}
 q^\text{ISM}_\pm = n_\text{ISM}\, c \int_{E_\text{thr}}^\infty
 \text{d}E' \frac{4\pi}{\beta c} J_\text{p} (E')
 \frac{\text{d}\sigma_{\text{pp} \rightarrow \text{e}^\pm +
     X}}{\text{d}E},
\end{equation}
where $\text{d}\sigma_{\text{pp} \rightarrow \text{e}^{\pm} +
  X}/\text{d} E$ is the partial differential cross-section for $e^\pm$
production and $\beta \simeq 1$ is the velocity of the GCR. We can
then integrate the Green's function for a single source over space and
time to calculate
\begin{equation}
 J_{\pm}(E) \simeq \frac{c}{4\pi} \frac{1}{|b(E)|} \int_E^{\infty}
 \text{d}E' q^\text{ISM}_\pm (E') \frac{2h}{\ensuremath{\zcr}} \chi
 \left(0, \frac{\ell}{\ensuremath{\zcr}}\right),
\end{equation}
where $h \sim 0.1$~kpc is the height of the galactic disk.

We calculate the flux of secondary background $e^-$ and $e^+$ from the
Solar-demodulated flux of GCR protons as derived from the BESS data
\cite{Shikaze:2006je} and model the cross-sections according to
Ref.~\cite{Kamae:2006bf}. The contribution from kaon decay is
subdominant and is therefore neglected. The presence of He both in
GCRs and in the ISM is taken into account by multiplying the proton
contribution by a factor of 1.2. Our results are in good agreement
with Ref.~\cite{Delahaye:2008ua}, taking into account the different
diffusion model parameters and keeping in mind that convection and
reacceleration have been neglected here. These fluxes are shown
(dashed line) in the middle panel of Fig.~\ref{fig:E3J} and are
clearly a subdominant component which cannot account for the deficit
at high energies.

Moreover, the positron flux is {\it falling} faster than the primary
$e^-$ flux at all energies whereas the PAMELA
data~\cite{Adriani:2008zr} clearly show a {\it rise} above a few
GeV. One way this can be resolved is if there is a dip in the electron
spectrum between $\sim 10$ and 100 GeV. It has been suggested that
Klein-Nishina corrections to the Thomson cross section for inverse
Compton scattering~\cite{Stawarz:2009ig} or inhomogeneities in the
distribution of sources~\cite{Shaviv:2009bu} can produce such a
dip. However the former would require a rather enhanced interstellar
background light (IBL) field~\cite{Stawarz:2009ig}, while the latter
calculation~\cite{Shaviv:2009bu} assumes an incomplete source
distribution (see Sec.~\ref{sec:SourceDistribution}) and moreover
adopts diffusion model parameters quite different from those derived
from the measured nuclear secondary-to-primary ratios
\cite{Strong:2007nh} and the measured galactic magnetic field and IBL
\cite{Kobayashi:2003kp}.

The other, perhaps more straightforward possibility is to consider an
additional component of GCR positrons with a {\it harder} source
spectrum that results in a harder propagated spectrum and therefore
leads to an increase in the positron fraction.

\subsection{Secondary electrons and positrons from the sources}
\label{sec:SecondarySource}

Following Refs.~\cite{Blasi:2009hv, Blasi:2009bd}, the parameters are
those typical of an old SNR: $u_1 = 0.5 \times 10^8 \,\text{cm}
\,\text{s}^{-1}$, $n_{\text{gas}, 1} = 2 \,\text{cm}^{-3}$, $B = 1
\,\mu\text{G}$. Choosing $r = 3.1$ to recover $\gamma = 2.4$ the
characteristic momenta $p_\text{cross}$ and $p_\text{break}$ (see
Sec.~\ref{sec:AccnOfSecs}) turn out to be,
\begin{align}\label{eqn:px}
p_\text{cross} &= 427 \, K_\text{B}^{-1}
\left(\frac{\tau_{\text{SNR}}}{10^4 \, \text{yr}} \right) \,
\text{GeV},\\
\label{eqn:pc}
p_\text{break} &= 7.7 \, K_\text{B}^{-1} \left(
\frac{\tau_{\text{SNR}}}{10^4 \, \text{yr}} \right) \, \text{TeV}.
\end{align}

What is still missing is the normalization of the injection spectrum,
$R_{+}^0$, in the sources which is proportional to the normalisation
of the GCR protons, $N_{\text{GCR}}$, through
Eq.~\ref{eqn:injection}. Usually a factor $K_{\text{ep}} \simeq
10^{-4} - 10^{-2}$ is introduced to normalize the electron component
with respect to the protons; this depends on how particles are
injected from the thermal background into the acceleration process and
is not reliably calculable from first principles. We can get around
this by assuming that the gamma-rays detected from known SNRs by HESS
are of hadronic origin, as is expected in this framework. Thus we can
use the total luminosity of individual sources in gamma-rays,
\begin{equation}
Q_{\gamma} = 4 \pi L^2 J_{\gamma} \, ,
\label{eqn:GammaLumi}
\end{equation}
to determine the normalization of the proton component and therefore
also the secondary injection rate $q_\pm^0$ if we know their distance
$L$.

The compilation of $\gamma$ ray data on SNRs from HESS, see
Table~\ref{tbl:GammaRays}, suggests an average value $Q_\gamma^0
\simeq 5.7 \times 10^{33} \,\text{s}^{-1} \,\text{TeV}^{-1}$.  We find
then for the total spectrum
\begin{align}
R_+^0 = \tau_\text{SNR} Q_+^0 \simeq \tau_\text{SNR}
\frac{\Sigma_{+}}{\Sigma_\gamma} Q_\gamma^0 \,,
\end{align}
where $\Sigma_{+}$ (and analogously $\Sigma_\gamma$) is defined by
Eq.~\ref{eqn:defqpm}, or explicitly
\begin{align}
R_+^0 = 7.4 \times 10^{48} \bigg(\frac{\tau_\text{SNR}}{10^4
  \text{yr}}\bigg) \bigg(\frac{Q_\gamma^0}{5.7 \times 10^{33}
  \text{s}^{-1}\text{TeV}^{-1}}\bigg) \text{GeV}^{-1} .
\end{align}
In the Monte Carlo code we have explicitly input the experimentally
measured pp cross-section which gives a similar normalisation as the
estimate presented above assuming Feynman scaling. Also, we have taken
into account the cut-off of the underlying primary protons. Their
maximum energy is determined from the average maximum gamma-ray energy
$E_{\text{max}} \simeq 20 \, \text{TeV}$ (see
Table~\ref{tbl:GammaRays}) through the inelasticity of the $\text{pp}
\rightarrow \gamma + X$ process as $\sim 20\,\text{TeV}/0.15 \approx
100 \, \text{TeV}$~\cite{Aharonian:2006ws}. The normalisation for
secondary electrons is computed similarly.

The middle panel of Fig.~\ref{fig:E3J} shows an example of the flux of
secondary source $e^-$ and $e^+$ for $30$ realisations of the SNR
density in our Galaxy. Clearly this component can potentially match
the high energy Fermi-LAT and HESS data.

We note that in our model, the contribution from secondary electrons
and positrons to the total flux is about twice as large as in
Ref.~\cite{Blasi:2009hv} where the primary injection spectrum was
assumed to be $\propto E^{-2}$, motivated by DSA theory. However this
is not consistent with gamma-ray observations of SNRs as seen from
Table~\ref{tbl:GammaRays}.

\begin{figure}[p]
\includegraphics[width=1\textwidth]{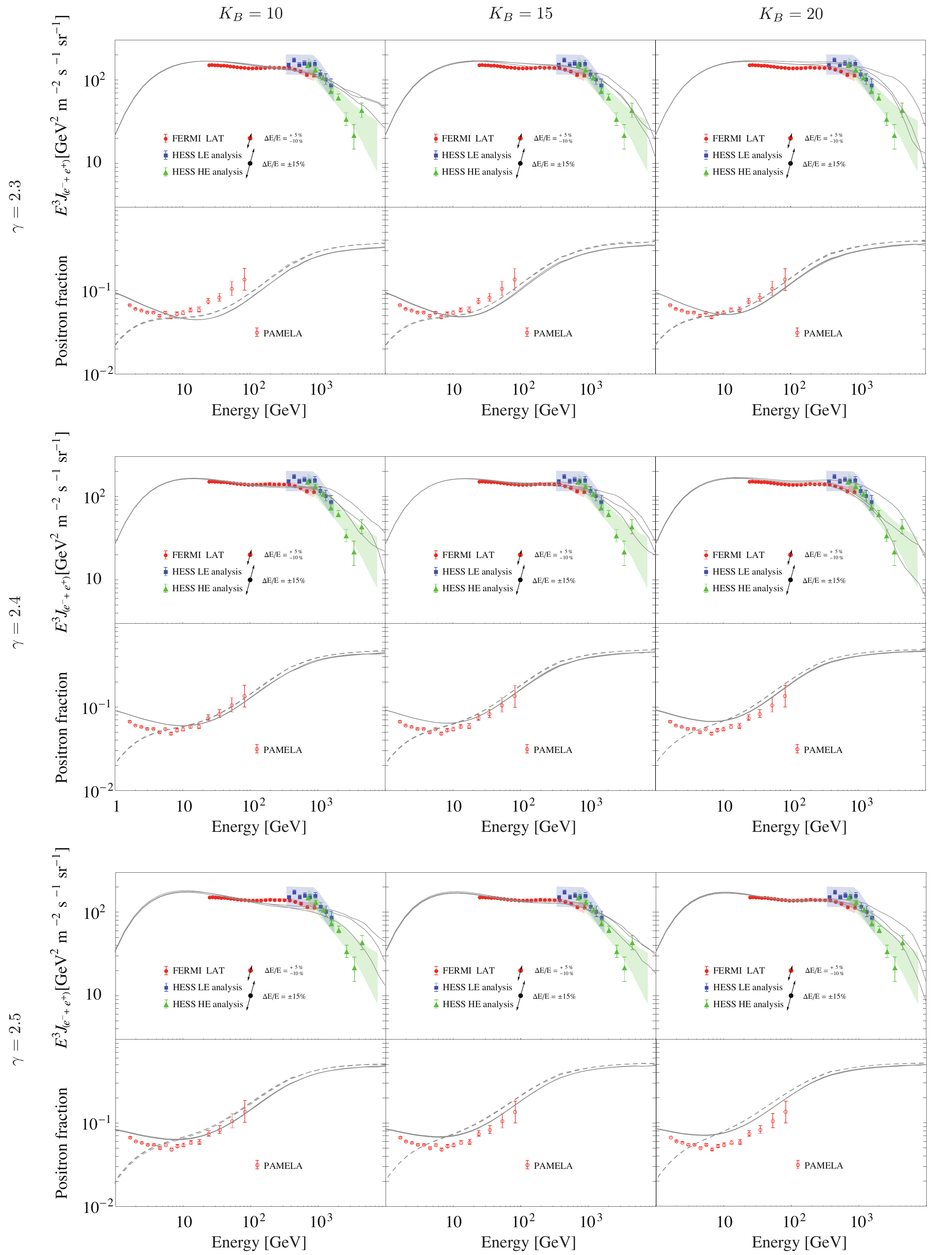}
\caption{The three best fits (out of 30 source realisations) to the
  total spectrum of electrons and positrons measured by
  Fermi-LAT~\cite{Abdo:2009zk} (red circles) and
  HESS~\cite{Collaboration:2008aaa, Aharonian:2009ah} (blue squares \&
  green triangles), and the corresponding prediction for the positron
  fraction for different values of $\gamma$ and $K_\text{B}$, for both
  charge-sign independent (full line) and charge-sign dependent
  (dashed line) Solar modulation (see text for details). The PAMELA
  data~\cite{Adriani:2008zr} is shown for comparison (open red
  circles).}
  \label{fig:bestfit}
\end{figure}

\section{Results}
\label{sec:Additional_Results}

\begin{table}[t]
\centering
\begin{threeparttable}[b]
\caption{Summary of parameters used in the Monte Carlo simulation, for
  an injection spectral index $\gamma \simeq 2.4$.}
\label{tbl:Parameters}
\begin{tabular}{c c c}
\hline\hline \multicolumn{3}{l}{Diffusion Model} \bigstrut \\ \hline

$D_0$ & $10^{28}\,\text{cm}^2\,\text{s}^{-1}$ &
\multirow{3}{*}{$\Bigg\}$ \begin{minipage}[c]{0.5\columnwidth}
    \flushleft from GCR nuclear secondary-to-primary
    ratios\end{minipage}} \bigstrut[t] \\ $\delta$ & $0.6$ &
\\ $z_\text{max}$ & $3$ $\text{kpc}$ \\ $b$ &
$10^{-16}\,\text{GeV}^{-1} \, \text{s}^{-1}$ & CMB, IBL and $\vec{B}$
energy densities \bigstrut[b] \\ \hline \multicolumn{3}{l}{Source
  Distribution} \bigstrut \\ \hline $t_\text{max}$ & $1 \times
10^8\,\text{yr}$ & from $E_{\text{min}} \simeq 3.3 \, \text{GeV}$
\bigstrut[t] \\ $\tau_{\text{SNR}}$ & $10^4\,\text{yr}$ & from
observations \\ $N$ &$3 \times 10^6$ & from number of observed SNRs
\bigstrut[b] \\ \hline \multicolumn{3}{l}{Source Model} \bigstrut
\\ \hline $R_{e^-}^0$ & $1.8 \times 10^{50}\,\text{GeV}^{-1}$ &fit to
$e^{-}$ flux at $10 \, \text{GeV}$ \bigstrut[t] \\ $\gamma$ & $2.4$
&average gamma-ray spectral index\\ $E_\text{max}$ & $20\,\text{TeV}$
& typical gamma-ray maximum energy \\ $E_{\text{cut}}$ &
$20\,\text{TeV}$ & DSA theory \\ $R_{+}^0$ & $7.4 \times
10^{48}\,\text{GeV}^{-1}$ & cf.~Sec.~\ref{sec:SecondarySource}
\\ $K_\text{B}$ & $15$ & only free parameter (for fixed
$\gamma$)\bigstrut[b] \\ \hline \hline
\end{tabular}
\end{threeparttable}
\end{table}

The parameters used in the Monte Carlo simulation are given in Table
\ref{tbl:Parameters}. For an assumed injection spectral index
$\gamma$, the only free parameter is $p_\text{cross}$
(cf.~Eq.~\ref{eqn:pxdef}) or, equivalently, the factor $K_\text{B}$
(cf.~Eq.~\ref{eqn:D(p)}) which is determined by fitting the total flux
of electrons and positrons to the Fermi-LAT and HESS data (see
Fig.~\ref{fig:E3J}).

We have calculated the $\chi^2$ with respect to the combined Fermi-LAT
and HESS data for each realisation $m$ of source distances and times,
$\{L_i, t_i \}_m$, over all energy bins. The three best ``fits'' are
shown in Fig.~\ref{fig:bestfit} for different values of $K_\text{B}$
and for $\gamma =$ 2.3, 2.4 and 2.5 (see
Table~\ref{tbl:Parameters}). The corresponding predictions for the
$e^+$ fraction are shown in the bottom panels.
Adopting $\gamma = 2.4$, we find good agreement for $K_\text{B} \simeq
15$, which corresponds to a cross-over of the accelerated and the
non-accelerated secondary components from the sources at
$p_\text{cross} \simeq 28$~GeV and a spectral break at $p_\text{break}
\simeq 510$~GeV (cf.~Eqs.~\ref{eqn:px} and \ref{eqn:pc}).
The predictions for the $e^+$ fraction agrees reasonably well with the
data down to 6~GeV; we would not expect agreement at lower energies
since we have neglected convection and reacceleration during
interstellar propagation. In fact the PAMELA measurements of the $e^+$
fraction are systematically lower than previous measurements,
e.g. AMS-01~\cite{Aguilar:2002} or HEAT~\cite{bea04}, and it has been
noted that this discrepancy can be resolved by considering charge-sign
{\em dependent} Solar modulation with $\phi_+ = 438 \, \text{MV}$ for
$e^+$ and $\phi_- = 2 \, \text{MV}$ for $e^-$~\cite{Beischer:2009zz}
(rather than $\phi_+ = \phi_- = 600$~MV). This however seems to be at
odds with preliminary PAMELA data on the absolute electron flux
\cite{Mocchiutti:2009} which \textit{does} show substantial Solar
modulation. Accordingly in Fig.~\ref{fig:bestfit} we have shown the
predicted $e^+$ fraction for both cases; note that this does not
affect our predictions for energies above 10 GeV.

\section{Gamma-Rays and Neutrinos}
\label{sec:GammaRays&Neutrinos}

Our fits to both the PAMELA (absolute $e^-$) and the Fermi-LAT (total
$e^- + e^-$) spectra, allow to predict the PAMELA positron fraction by
including secondary $e^+$ accelerated in SNRs and thus provides a
consistent picture of current data on cosmic ray $e^-$ and $e^+$
between a few GeV and tens of TeV. Turning the argument around, since
a large fraction of the $e^-$ and $e^+$ observed in GCR above hundreds
of GeV are required to be secondaries in this model, there \emph{must}
be a large number of hadronic cosmic ray accelerators in our Galaxy,
some of which should be quite nearby.

\begin{figure}[bth]
\centering
\includegraphics[width=0.6\columnwidth]{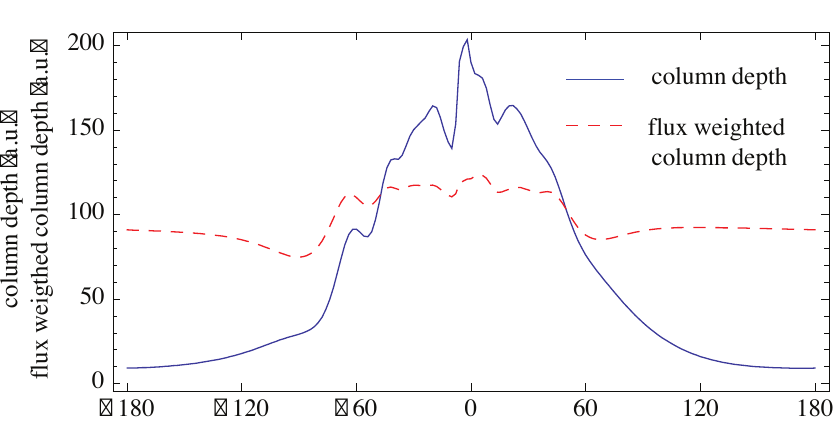}\\[0.4cm]
\hspace{0.5cm}\includegraphics[width=0.6\columnwidth]{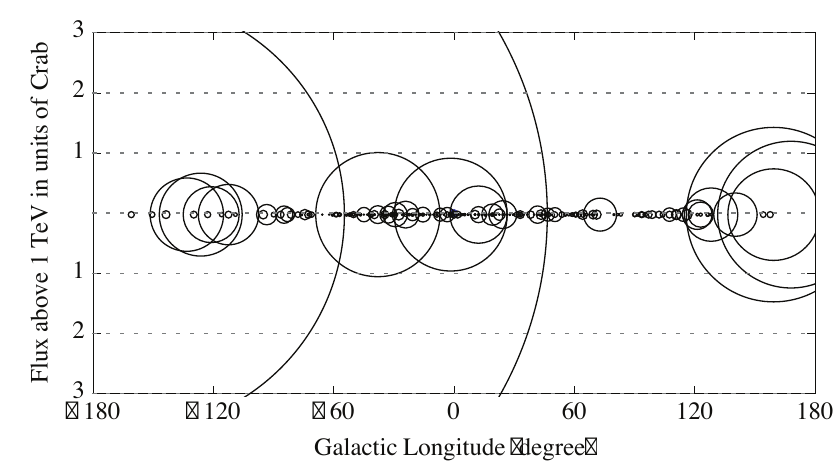}
\caption{{\bf Top:} The column depth and flux weighted column depth of
  the SNR density in the galactic plane.  {\bf Bottom:} Example of a
  distribution of SNRs in gamma-rays/neutrinos from the Monte Carlo
  simulation. The position of a circle denotes the galactic longitude
  of the source and the radius is proportional to the brightness in
  units of the Crab nebula. One source whose circle exceeds the
  vertical scale is $\sim 500$~pc from Earth and has a total
  integrated flux above 1~TeV of $\sim 6$ times the Crab Nebula.}
  \label{fig:SNRcolumndepth}
\end{figure}

An independent test of the model is provided by the usual `messengers'
of such hadronic acceleration environments, namely gamma-rays and
neutrinos. Taking the known distribution of SNRs in the Galaxy (see
Sec.~\ref{sec:SourceDistribution}) we have calculated the column depth
in SNRs in the galactic disk as seen from Earth,
\begin{equation}
X (\phi') = \int_0^\infty \text{d}r' \, r' g (r (r', \phi'), \phi(r',
\phi')),
\label{eqn:ColumnDepth}
\end{equation}
and show this in the top panel of Fig.~\ref{fig:SNRcolumndepth}. As
expected, the column depth is largest towards the galactic
centre. However, the quantity that is more important for observations
is the brightness of sources. We have therefore weighted the integrand
in Eq.~\ref{eqn:ColumnDepth} by $1/r'^2$ and this flux weighted column
depth is also shown in the top panel of
Fig.~\ref{fig:SNRcolumndepth}. We note that although the maximum
brightness is still expected around the galactic Centre, the
brightness in other directions is smaller by only $\sim 30\%$ because
the sources in the closest spiral arms are then dominant (if they are
actually there, of course).

This is illustrated in the bottom panel of
Fig.~\ref{fig:SNRcolumndepth} by an example distribution of SNRs from
the Monte Carlo simulation, denoted by circles. The position of the
circle denotes the galactic longitude and the radius is proportional
to the brightness in units of the Crab Nebula, i.e. an integrated flux
of $(1.98 \pm 0.08) \times 10^{-11} \,\text{cm}^{-2} \,\text{s}^{-1}$
above $1 \,\text{TeV}$~\cite{Aharonian:2004wa}. For a source of
luminosity of $Q_{\gamma}^0 = 5.7 \times 10^{33} \, \text{TeV}^{-1}
\,\text{s}^{-1}$ (see Table~\ref{tbl:GammaRays}) at distance $L$, the
integrated flux above 1~TeV is,
\begin{align}
F_\gamma (> 1\text{TeV}) = \frac{1}{4\pi L^2} \int_{1 \text{TeV}}
\text{d}E\, Q_\gamma \simeq 8.5 \times 10^{-12} \left(\frac{L}{2
  \,\text{kpc}}\right)^{-2} \,\text{cm}^{-2} \,\text{s}^{-1} \,,
\end{align}
i.e.~about 40 \% of the Crab Nebula flux at $L = 2$ kpc. It is seen
that although most of the sources are clustered towards the galactic
centre, there are several bright sources at large longitudes as
well. We find typically $\sim 3$ sources brighter than the Crab (or
$\sim 7$ brighter than 50 \% Crab).

The adopted distribution of SNRs (Sec.~\ref{sec:SourceDistribution})
and the average luminosity per source determined from a compilation of
known sources (Table~\ref{tbl:GammaRays}) thus leads to the prediction
of several nearby SNRs with fluxes of the order of the Crab
Nebula. Note, however, that close sources could be rather extended and
thus have escaped detection by HESS in one of its surveys of the Milky
Way~\cite{Aharonian:2005kn, Chaves:2009zza, Chaves:2009bq}. For
example, a diameter of $\sim50$~pc which is a typical value for a very
old SNR, corresponds to $1.5^\circ$ at 2~kpc.

Extended gamma-ray luminous SNRs can however be detected by MILAGRO
\cite{Abdo:2007ad} with its larger field of view. A survey in galactic
longitude $l \in [30^\circ, 220^\circ]$ and latitude $b \in
[-10^\circ, 10^\circ]$ has revealed 6 new sources at a median energy
of 20~TeV, several of which are spatially extended. The flux from a
SNR of the above luminosity at $L = 2 \,\text{kpc}$ is
$Q_\gamma^0/(4\pi d^2) \simeq 1.2 \times 10^{-11} \,\text{TeV}^{-1}
\,\text{cm}^{-2} \,\text{s}^{-1}$ at 1 TeV. Scaled with a spectral
index of 2.4 to 20~TeV, this gives $Q_\gamma^0/(4\pi d^2) \, 20^{-2.4}
\simeq 9.0 \times 10^{-15} \,\text{TeV}^{-1} \,\text{cm}^{-2}
\,\text{s}^{-1}$ which is in the range of the unidentified MILAGRO
sources~\cite{Abdo:2007ad}. We note that the MILAGRO source MGRO
J1908$+$06 was recently confirmed by HESS
\cite{deOnaWilhelmi:2009zza}, though with a smaller angular extent of
$\sim 0.7^\circ$. However, correlating unidentified MILAGRO sources
with the FERMI Bright Source List~\cite{Abdo:2009ku, Abdo:2009mg}
seems to favour associations with pulsars, although several new
unidentified extended sources have also been found.

Hadronic sources of cosmic rays should also be visible by their
neutrino emission. On general grounds, the neutrino luminosity (from
$\pi^\pm$ decay) can be directly related to the gamma-ray luminosity
(from $\pi^0$ decay) and should be of the same order of magnitude
since p-p interactions produce $\pi^+$, $\pi^0$ and $\pi^-$ in roughly
equal numbers. Each of the three neutrinos produced in the decay
chains $\pi^+\to\mu^+\nu_\mu\to e^+\nu_e\bar\nu_\mu\nu_\mu$ and
$\pi^-\to\mu^-\bar\nu_\mu\to e^-\bar\nu_e\nu_\mu\bar\nu_\mu$ carries
about half of the energy of each photon produced in the decay
$\pi^0\to\gamma\gamma$. Hence, the ratio of neutrinos to photons
produced on average is $\sim 3:1$ and the total neutrino luminosity is
\begin{align}\nonumber
  Q_{\text{all }\nu} (E_\nu) &\simeq 6 \, Q_\gamma (2E_\nu) \simeq 6
  \times 2^{-\gamma} Q_\gamma^0
  \left(\frac{E_\nu}{\text{TeV}}\right)^{-\gamma}.
\end{align}
Presently the largest cosmic neutrino detector is the IceCube
observatory~\cite{Ahrens:2003ix} under construction at the South
Pole. IceCube observes high energy neutrinos via their interactions
with nucleons in the vicinity of the detector and subsequent
\v{C}erenkov light emission of energetic charged particles in the
transparent glacial ice. The most important signal for neutrino
astronomy is the \v{C}erenkov radiation by muons produced via charged
current interactions of muon neutrinos. Since the muon inherits the
large boost of the initial neutrino the point source resolution is
$\sim 1^\circ$. The large background signal of atmospheric muons is
efficiently reduced for upward-going muons, i.e.~neutrino sources
which are somewhat below the horizon. Hence, IceCube is mainly
sensitive to neutrino point sources in the northern sky, which
excludes SNRs in the direction of the galactic centre.

Neutrino emission associated with galactic TeV gamma-ray sources has
been investigated by many
authors~\cite{Costantini:2004ap,Vissani:2006tf,Villante:2008qg,Anchordoqui:2006pb,Kistler:2006hp,Beacom:2007yu,Halzen:2007ah,Halzen:2008zj,GonzalezGarcia:2009jc}
including also the HESS sources used in our analysis. In particular,
Ref.~\cite{Kistler:2006hp} investigates the prospects of neutrino
detection for the SNRs HESS J0852.0--463, J1713--381, J1804--216,
J1834--087 (see Table~\ref{tbl:GammaRays}) in the proposed KM3NeT
detector in the Mediterranean which will see the galactic centre
region. The muon neutrino rate is expected to be a few events per year
for such sources.

Due to flavour oscillations of neutrinos with large mixing angles, the
initial flavour composition $Q_{\nu_e}:Q_{\nu_\mu}:Q_{\nu_\tau} \simeq
1:2:0$ from pion decay is expected to become $\sim 1:1:1$ at
Earth. The TeV muon neutrino point flux from a hadronic gamma-ray
source located at a distance $L$ and with a power-law index $\gamma
\simeq 2.4$ is thus $F_{\nu_\mu}(>1~\text{TeV})\simeq 2^{1 - \gamma}
F_{\gamma}(>1~\text{TeV})$, hence
\begin{align}
F_{\nu_\mu}(>1~\text{TeV}) \simeq 3.2 \times 10^{-12}
\left(\frac{L}{2~\text{kpc}} \right)^{-2} \, \text{cm}^{-2} \,
\text{s}^{-1}.
\end{align}
This should be compared to the results of searches for neutrino point
sources in the northern sky, in particular the close-by SNR Cassiopeia
A (see Table~\ref{tbl:GammaRays}), using data taken with AMANDA-II
(the predecessor of IceCube) during 2000--2006~\cite{Abbasi:2008ih}
and, more recently, with the first 22 strings of IceCube during
2007--08~\cite{Abbasi:2009iv}. The average 90\% C.L.~upper limit on
the integrated $\nu_\mu$ flux in the energy range 3~TeV to 3~PeV
is~\cite{Abbasi:2009iv}
\begin{equation}
F_{\nu_\mu} \leq 4.7\times10^{-12}~{\rm cm}^{-2}\,{\rm s}^{-1},
\end{equation} i.e. still above $\sim7\times10^{-13}~{\rm cm}^{-2}\,{\rm s}^{-1}$ 
expected from a SNR at 2~kpc, assuming $\gamma=2.4$.

The full 80 string configuration of IceCube thus has excellent
prospects to identify these SNRs. A point source in the northern sky
with an $E^{-2}$ muon neutrino flux,
\begin{equation}
\label{eqn:icecubesen}
F_{\nu_\mu} \simeq 7.2 \times 10^{-12} \text{cm}^{-2}\,\text{s}^{-1},
\end{equation}
in the TeV-PeV range can be detected with a $5\sigma$ significance
after three years of observation. This does depend somewhat on the
spectral index and energy cut-off, since the signal (after ``level 2
cuts'') peaks at an energy of $\sim10$~TeV~\cite{Ahrens:2003ix}. As
mentioned previously, our analysis predicts on average $\sim 3$ nearby
gamma-ray sources stronger than Crab with corresponding muon neutrino
fluxes larger than $\sim 7 \times
10^{-12}\,\text{cm}^{-2}\,\text{s}^{-1}$. Note that although the
galactic centre is not in the field of view of IceCube, SNRs following
the spiral arm structure of the Galaxy are expected to be detected
also in the galactic anti-centre direction, as seen in the example
distribution shown in the bottom panel of
Fig.~\ref{fig:SNRcolumndepth}.

\section{Comments on the Time-Dependent Picture}
\label{sec:TimeDependentPicture}

It was recently questioned~\cite{Kachelriess:2010gt} whether the
acceleration of secondaries also holds true in the time-dependent
picture of DSA. Intuitively, one would expect that the
energy-dependence of the diffusion coefficient still leads to an
increase of the size of the diffusion zone and therefore to a harder
spectrum for charged secondaries. The authors
of~\cite{Kachelriess:2010gt} however argue that both numerical
simulations and a simple analytical argument show that the secondary
fractions do not rise. In the following, we critically examine both
these claims.

\subsection{An analytical argument}

The authors of~\cite{Kachelriess:2010gt} present a simple
`gedankenexperiment' which they claim shows that the acceleration of
secondary mechanism does not lead to a rise in any
secondary-to-primary ratio. The argument compares the flux of
secondaries produced and accelerated in the SNR to the flux of
primaries in the case where no secondaries get produced (which would
be the case, for example for $n_\text{gas} = 0$). For the particular
case of the anti-proton fraction, assuming that in an inelastic pp
collision, the anti-proton produced takes all the energy of the
incoming proton, the interactions would in fact only convert part of
the proton flux into an anti-proton flux. As both particles are
affected in the same way by DSA, the total anti-proton spectrum would
be proportional to the proton spectrum in a secondary-less case.

This comparison is however not appropriate. What needs to be compared
with the proton spectrum $f_p^*$ in a secondary-less case is not the
antiproton flux $f_{\bar{p}}$ but the \emph{sum} of the proton and
antiproton flux $(f_p + f_{\bar{p}})$ for the case with
secondaries. If the assumptions about the inelasticity being $100 \,
\%$ were in fact correct, the sum of antiproton and proton flux $(f_p
+ f_{\bar{p}})$ should indeed be comparable to the proton flux
$f_p^*$. This, however, does not tell us anything about the individual
spectra, $f_p$ and $f_{\bar{p}}$. In fact, the hardness in the
antiproton spectrum, $f_{\bar{p}} \propto p f_p^*$ would be balanced
by a softer proton spectrum $f_p = (1 - \kappa \, (p/p_\text{cr}))
f_p^*$ where $p_\text{cr}$ is the momentum at which $f_p =
f_{\bar{p}}$ and $\kappa$ is a normalisation constant. Therefore, the
antiproton-to-proton ratio, $f_{\bar{p}} / f_p$ would become even
harder and in this sense our above argument is conservative in that it
ignores the $-p f_p^*$ reduction in the proton flux. However, for the
case of antiprotons, the latter correction is negligible as $f_p \gg
f_{\bar{p}}$, i.e. $p_\text{cr} \gg p_\text{max}$.

\subsection{The numerical model}

\begin{figure}[bt]
\begin{tabular*}{\columnwidth}{@{} p{0.49 \columnwidth} @{} p{0.02 \columnwidth} @{} p{0.49 \columnwidth} @{}}
\includegraphics[width=0.49\columnwidth]{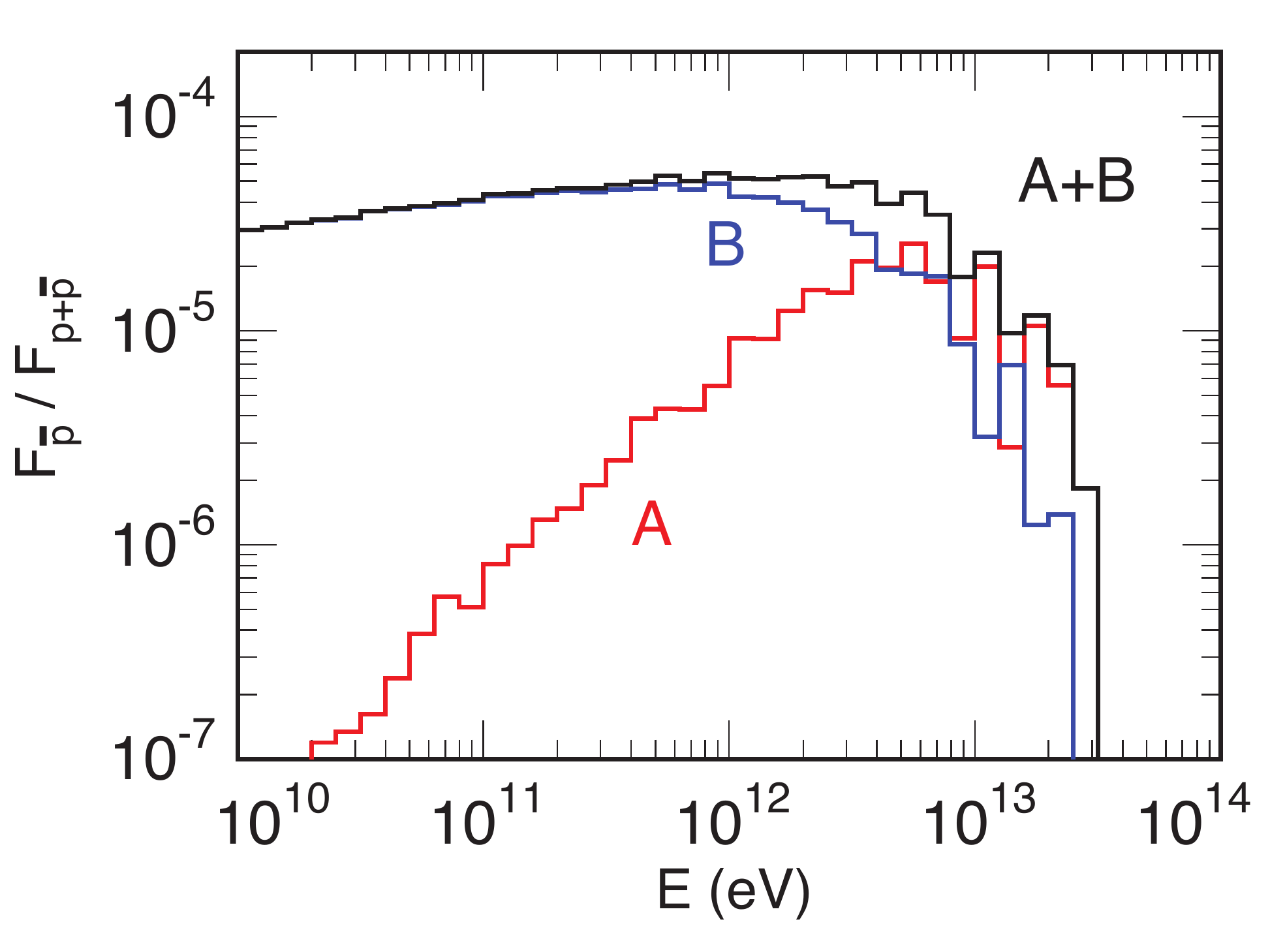} & &
\includegraphics[width=0.49 \columnwidth]{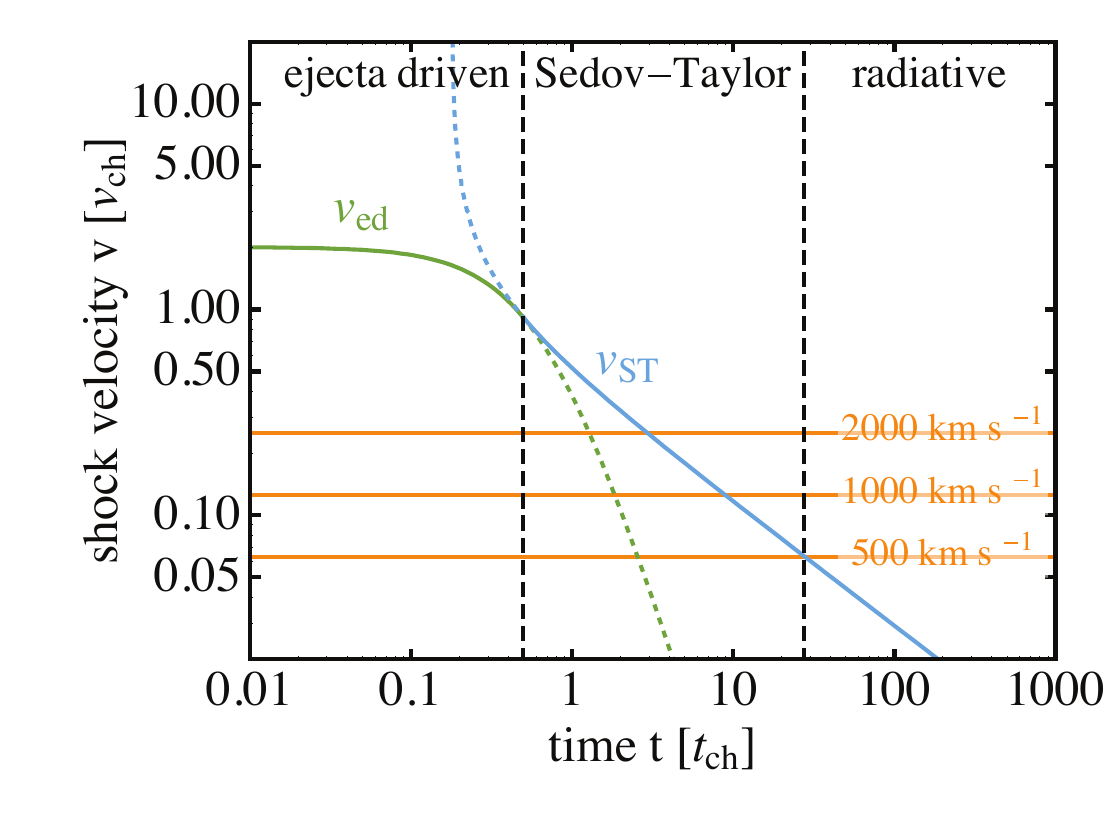} \\
\begin{minipage}[t]{0.49 \textwidth}
\caption{Antiproton-to-proton ratio produced in a time-dependent Monte
  Carlo calculation of diffusive shock acceleration in the test
  particle approximation (from~\cite{Kachelriess:2010gt}). The red (A)
  and blue (B) lines denote the ratio from the secondaries which have
  been produced upstream and downstream, respectively, and the black
  line (A+B) is the sum of both.}
\label{fig:Kachelriess}
\end{minipage}
& &
\begin{minipage}[t]{0.49 \textwidth}
\caption{Shock velocity of the SNR as a function of time, both in
  units of values characteristic for a particular SNR model, see
  \cite{1999ApJS..120..299T}. The blue solid and green solid lines
  show the analytical model of~\cite{1999ApJS..120..299T} for the
  ejecta-driven (ed) and the Sedov-Taylor (ST) phase,
  respectively. The effective value for the shock velocity adopted in
  our calculation is $500 \, \text{km} \, \text{s}^{-1}$.}
\label{fig:TimeDependentVsConstantVelocity}
\end{minipage}
\end{tabular*}
\end{figure}

The authors of~\cite{Kachelriess:2010gt} further present the results
of a Monte Carlo calculation of time-dependent DSA of electrons and
protons including the production and subsequent acceleration of
secondary antiprotons and positrons. Although the secondary-to-primary
ratios of the total spectra $N(p) \propto \int \dd V p^2 f(p)$ clearly
show the presence of a harder component compared to the primary one,
the former one seems to be subdominant and not influence the total
spectra too much. This is shown in their Fig.~2 which we here
reproduce in Fig.~\ref{fig:Kachelriess}.

However, we believe that this discrepancy is merely due to a different
set of parameters adopted. As shown in Eq.~\ref{eqn:fpm0}, the
normalisation of the harder component depends on the diffusion
coefficient $D$, the velocity of the shock front squared, $u_1^2$ and
the gas density $n_\text{gas}$ through $q_\pm$. Not only is the value
of the diffusion coefficient in~\cite{Kachelriess:2010gt},
$D_\text{Bohm}$, smaller than the one adopted in our calculation, $K_B
D_\text{Bohm}$ with $K_B \sim 20$, but also is the velocity larger:
The dynamical SNR model adopted~\cite{1999ApJS..120..299T} predicts a
time-dependent velocity, both for the free expansion and in the
Sedov-Taylor phase of a SNR. Our time-independent analysis, however,
needs to adopt an \emph{effective} velocity. As we show in
Fig.~\ref{fig:TimeDependentVsConstantVelocity}, the time-dependent
velocity is higher than our effective value most of the time. As the
velocity enters as the inverse square into the normalisation of the
harder component, a small difference in velocity can change the
relative importance rather drastically.

We therefore conclude that the discrepancy between our analytical
result and the outcome of the Monte Carlo calculation performed
in~\cite{Kachelriess:2010gt} is most likely due to different
parameters for the adopted SNR model. We note that it has not been
shown that the time-dependence introduces any additional physical
effects to the time-independent picture discussed above.

The true value of the parameters is, however, a separate issue. One
possibility, in fact the one pursued in~\cite{Kachelriess:2010gt}, is
to adopt an (analytic) model of the SNR dynamics that is believed to
reflect the physical behaviour of actual SNRs and feed it into a
numerical calculation of the secondary spectra. However, even the
parameters of the analytical model are affected by
uncertainties. Therefore, one needs to consider different SNR models
to cover all possibilities of effects.

Another possibility is to adopt the simplest analytical picture
possible, like the simple time-independent test particle approximation
above, and to treat the parameters, e.g. diffusion coefficient,
velocities, maximum energy, as effective ones. As explained above,
these effective values only enter the normalisation in a certain
combination. Choosing a value that allows to reproduce measured data,
i.e. basically fitting these parameters to the data, allows to make
independent, testable predictions for other observables. We have
followed this approach already for the charged lepton channels in
Sec.~\ref{sec:Spectra}, and we will further use this approach for
nuclear secondary-to-primary ratios in Chapter~\ref{chp:secondaries}.

\section{Alternative Astrophysical Explanations}
\label{sec:AlternativeAstrophysicalExplanations}

We have already addressed the possibility of DM annihilation or decay
accounting for the excesses in the positron fraction and total
electron-positron flux. Furthermore, we have discussed some of the
constraints, for example, from antiprotons, radio waves and
gamma-rays, and we have argued that a large fraction of the parameter
space necessary to explain the leptonic anomalies are already ruled
out.

To round off our discussion of the acceleration of secondaries in old
SNRs, we present some alternative astrophysical explanations that have
been claimed to explain the excesses.

\subsection{Pulsars}

The idea that pulsars might be responsible for the lepton
excess~\cite{Aharonian:1995zz,Atoian:1995ux} has been around since the
first indications of a rise in the positron
fraction~\cite{Mueller:1990}. Pulsars are highly-magnetised and
quickly spinning neutron stars that transfer (part of) their
rotational energy into a magnetised wind. Electrons are accelerated to
very high energies by the electric fields present and
synchrotron-radiate gamma-rays in the extreme ambient magnetic
fields. These gamma-rays are energetic enough to pair-produce high
energy electrons and positrons on the ambient magnetic fields or the
thermal X-rays form the pulsar itself. The high energy gamma-ray
emission is usually interpreted as curvature radiation or ICS from
high-energy electrons and positrons. There is disagreement about where
exactly the gamma-ray emission is produced: close to the surface of
the neutron star (polar cap models) or further away in the light
cylinder along the last open field lines (outer/slot gap models).

The total output in electrons and positrons is usually connected to
the spin-down power of the pulsar and with an efficiency of a few per
cent, most of the pulsars in the ATNF
catalogue~\cite{Manchester:2004bp} are predicted to provide outputs of
$10^{46 \mathellipsis 50} \, \text{erg}$. In fact, probably only
mature pulsars, i.e. with ages $\gtrsim 10^5 \, \text{yr}$, can inject
electrons and positrons into the ISM as for younger ones all charged
particles are trapped inside the pulsar wind nebula (PWN). The
spectral index $\alpha$ of the electrons and positrons can be
estimated from radio emission interpreted as synchrotron ($\gamma = 1
\mathellipsis 1.6$) or from the gamma-rays ($\gamma = 1.4
\mathellipsis 2.2$). The range usually considered is $\gamma = 1.4
\mathellipsis 2.2$.

As already mentioned, the idea was first developed in
Ref.~\cite{Aharonian:1995zz}, stressing the importance of young and
nearby pulsars for the electron-positron flux and the positron
fraction at high energies and suggesting Geminga as one of these
possible sources. The positron excess can be reproduced for a few
mature, nearby pulsars, e.g. Geminga and Monogem, complemented by a
distribution of pulsars further away~\cite{Hooper:2008kg}. This is in
supported by~\cite{Profumo:2008ms} where it is concluded that the
scenario with a single pulsar is disfavoured. Other authors focus on
the one-source model~\cite{Yuksel:2008rf} although this might require
too high an efficiency for the conversion of spin-down power into
electrons and positrons. It was also shown that the sum of all pulsars
covering a broad range of parameters can reproduce the Fermi-LAT and
PAMELA data~\cite{Grasso:2009ma}.

The way to pinpoint to a (few) pulsar(s) as the origin of the lepton
excesses suggested is to look for anisotropies in the arrival
directions of electrons and positrons, e.g. with Fermi-LAT. A simple
estimate of the anisotropy shows that most favoured scenarios could be
detected within a few years time to $2$ or $3 \, \sigma$
significance~\cite{Hooper:2008kg,Profumo:2008ms}. Nearby pulsars as
the source of the positrons are of course quite consistent with the
absence of antiprotons.

\subsection{GRBs}

Any astrophysical positron source with an injection distinctively
harder than the primary electrons from SNRs can in principle explain
the PAMELA positron excess. It was recently
suggested~\cite{Ioka:2008cv} that a galactic gamma-ray burst (GRB)
could be such a source. The GeV electrons and positrons would be
pair-produced by scattering of TeV gamma-rays on the afterglow eV
background. The energy contained in TeV gamma-rays is about $5 \, \%$
of the total GRB energy, typically $\sim 10^{50} \, \text{erg}$ and
the spectral index of the electron-positron pairs is $1.8 \pm
0.7$~\cite{Ioka:2008cv}.

In fact, an injection of $\mathcal{O}(10^{50}) \, \text{erg}$ a few
times $10^5 \, \text{yr}$ ago and with a spectral index between $1.6$
and $2.2$ nicely fits the data. A harder spectral index can also
reproduce a ATIC/PPB-BETS-like feature in the total electron-positron
flux while a softer index is in agreement with the Fermi-LAT data. If
in fact a GRB or another singular event of similar characteristics
(pulsar, hard SNR, microquasar) was responsible for the lepton
excesses one would again expect to see a rather large anisotropy in
electron and positron fluxes which would be detectable by Fermi-LAT or
AMS-02~\cite{ams02} in a few years' time~\cite{Ioka:2008cv}.

\subsection{Very old supernova remnants}

Another idea~\cite{Fujita:2009wk} invokes dense gas clouds around very
old SNRs as sites of a production of harder positrons. The authors
consider several simultaneous SN explosions $10^{5\mathellipsis6} \,
\text{yr}$ ago inside a dense gas cloud that need to have taken place
only hundreds of parsecs from the solar system but could explain the
local bubble or other similar structures like Loop
I~\cite{Berghoefer:2002jp,Breitschwerdt:2006pq}. Once the Sedov-Taylor
phase is coming to an end, the compression ratio $r$ (see
Sec.~\ref{sec:SNRs}) is not given by the Rankine-Hugoniot relation any
more but by the ratio of shock velocity and upstream Alfv\`en
velocity, $r \sim \sqrt{2} v_s/v_A$, and is much larger than $4$. The
spectral index $\gamma$ of the protons accelerated by DSA is therefore
smaller than $2$ and the interaction of the protons with the gas in
the surrounding cloud produces positrons of a similarly hard spectrum.

For gas clouds of tens of parsecs radius and $\sim 100 \,
\text{cm}^{-3}$ density, both the positron fraction and, depending on
the exact value of the proton spectral index $\gamma$ considered, even
the Fermi-LAT or ATIC/PPB-BETS total electron-positron spectrum can be
fitted. However, the maximum proton energy assumed is $100 \,
\text{TeV}$ -- a somewhat unrealistic situation, given that such high
energy protons can only be accelerated in the Sedov-Taylor phase when
the magnetic field amplification is strong and even the highest energy
particles are confined to the shock region.

The model predicts a similar rise in the antiproton-to-proton ratio,
due to the contributions from antiprotons produced in a similar
manner. Testing this model by the boron-to-carbon ratio might not be
conclusive, as the metallicity can vary throughout the Galaxy.

\subsection{Inhomogeneous source distribution}

We have already alluded to the suggestion that the inhomogeneous
distribution of SNRs in the galactic disk is responsible for the rise
in the positron fraction~\cite{Shaviv:2009bu}. The idea is that the
SNR density in the spiral arms is amplified by a factor of $4$ with
respect to the disk, and the propagation cut-off from the nearest
spiral arm at $\sim 1 \, \text{kpc}$ leads to a softening of the
primary electron spectrum. The secondary production is spatially more
homogeneous because it follows the cosmic ray proton density and is
therefore not affected by the details of the source distribution. The
softening of the electron spectrum therefore leads to a rise in the
positron fraction.

While we agree with the call for a realistic source distribution, we
argued above in detail that its effect is only to be seen at hundreds
of GeV. In particular, we expect the propagation cut-off from sources
at distances of $\sim 1 \, \text{kpc}$ to occur at $\mathcal{O}(1) \,
\text{TeV}$, i.e. more than a magnitude \emph{above} the energies at
which the positron fraction starts rising. This disagreement can be
traced back to propagation parameters used in~\cite{Shaviv:2009bu}
that are rather unrealistic. In particular, the diffusion coefficient
adopted, $D(E) = 6 \times 10^{27} \, \text{cm}^2 \, \text{s}^{-1}$, is
too low and the energy loss rate at $1 \, \text{GeV}$, $b_0 = 1.8
\times 10^{-16} \, \text{GeV}^{-1} \, \text{s}^{-1}$ too high, which
strongly affects the diffusion-loss length $\ell^2$, see
Eq.~\ref{eqn:ell2}. In addition, the predicted soft electron spectrum
seems to be in disagreement with preliminary PAMELA data on the
absolute electron flux~\cite{Boezio:2010}.

\section{Conclusion}
\label{sec:DiscussionSummary}

Supernova remnants have long been suspected to be the sources of
galactic cosmic rays. We have discussed a recent
proposal~\cite{Blasi:2009hv} that proton-proton interactions in the
shocks of SNRs followed by the diffusive shock acceleration of the
secondary positrons produced can flatten the spectrum of the
secondaries relative to that of the primaries. These hard spectra may
be the origin of the recently observed cosmic ray ``excesses'' ---
both the $e^+$ fraction observed by PAMELA~\cite{Adriani:2008zr} and
the $e^{-} + e^{+}$ flux measured by Fermi-LAT~\cite{Abdo:2009zk} and
HESS~\cite{Collaboration:2008aaa,Aharonian:2009ah}.

We have investigated how gamma-ray emission of SNRs -- assumed to be
of the same hadronic origin as the positrons -- together with cosmic
ray data, constrain the acceleration of positrons. We have accounted
for the spatial and temporal discreteness of SNRs via a Monte Carlo
exercise, drawing samples from a realistic galactic distribution with
the observed SN rate. For the diffusion parameters we have adopted
standard values derived from cosmic ray nuclear-to-primary ratios, as
well as the energy densities of galactic radiation and magnetic
fields.

We have compiled a list of all gamma-ray emitting SNRs observed by
HESS and determined the mean value of the flux, which fixes the
hadronic interaction rate in the SNR. Low energy data from PAMELA on
the absolute $e^-$ flux were used to normalize the primary flux of
$e^-$. The contribution from accelerated $e^+$ was then found by
fitting the $e^- + e^+$ flux to Fermi-LAT and HESS data, adjusting the
(only) free parameter $K_\text{B}$ which determines the diffusion rate
near SNR shocks.

The spectra of $e^+$ and $e^-$ thus derived agree well with the $e^+$
fraction observed by PAMELA in the range $5-100$~GeV. The apparent
discrepancy at lower energies can be attributed to the uncertainty in
solar modulation (charge-sign dependent or independent). Furthermore,
convection and diffusive reacceleration of primary electrons that
become important at these energies were neglected in our analysis. The
flux of $e^+$ and $e^-$ becomes dominated by the accelerated secondary
component at high energies; the corresponding $e^+$ fraction levels
out at $\sim0.4$, reflecting the relative multiplicity of $e^+$ and
$e^-$ produced by p-p interactions.

To be consistent with our overall framework the gamma-rays observed
from SNRs have been assumed to be of hadronic origin. The known
spatial distribution of SNRs then implies (on average) several nearby
sources with a gamma-ray flux comparable to the Crab. We have
speculated that some unidentified MILAGRO sources~\cite{Abdo:2007ad}
might correspond to such old SNRs. Moreover, the same hadronic
processes in SNRs will inevitably produce high energy neutrinos which
can be detected in cubic-km telescopes such as
IceCube~\cite{Ahrens:2003ix}. The neutrino luminosity can be directly
related to the gamma-rays and is not connected to the hypothetical
acceleration of $e^+$ and $e^-$ in the sources as in our present
model. Nevertheless, similarly to the previous argument, we expect on
average a few nearby sources, some of which may also lie within the
field of view of IceCube and can thus be detected with high
statistical significance after three years of data taking.

While our calculational framework is based on first-order Fermi
acceleration by SNR shock waves, we have noted that in detail the
observations do not fit the theoretical expectations, e.g.~the shock
compression ratio inferred from the observed gamma-ray spectrum ($\sim
E^{-2.4}$) is 3.1 rather than 4 as is expected for a strong
shock~\cite{Blandford:1987pw}. Going beyond the test particle
approximation, the generic expectation in such a process is for
particle spectra which are much flatter than those observed ($\sim
E^{-1.4}$ and slightly {\it concave}), when the back reaction of the
cosmic rays on the shock is taken into account~\cite{Malkov:2001}. By
contrast, the observed radio spectrum of Cassiopeia A is slightly
\emph{convex} and this, as well as the morphology and time evolution
of radio emission from such young SNRs, can be well explained in terms
of \emph{second-order} Fermi acceleration by plasma turbulence behind
the shock wave~\cite{CowsikSarkar:1984}. Moreover the observed spatial
correlation between the gamma-ray emission and the hard X-ray emission
from some SNRs argues for a leptonic rather than hadronic origin and
further observations are necessary to resolve this issue
\cite{Pohl:2008pq}. It has been argued that cosmic ray protons and
nuclei may well have different sources (e.g. ``superbubbles'' formed
by multiple supernovae) than the cosmic ray electrons
\cite{Butt:2009zz}.  The additional predictions made in this work
concerning the visibility of hadronic accelerators in gamma-rays and
neutrinos, tied to the expectations for the fluxes of the accelerated
{\it secondary} positrons in cosmic rays, will hopefully enable
further consistency tests of the SNR origin hypothesis for galactic
cosmic rays.

%% file: 03secondaries/secondaries.tex
\chapter{Acceleration of Secondary Nuclei}
\label{chp:secondaries}

\section{Introduction}

As we discussed in Chapter~\ref{chp:additional}, there is a wealth of
suggestions on how to explain the apparent excesses in the positron
fraction and in the total electron-positron flux as measured by PAMELA
and Fermi-LAT, respectively. In particular, we have devoted our
analysis to the investigation of the acceleration of secondaries
mechanism which can explain the excesses by considering the production
and subsequent acceleration of secondary electrons and positrons in
the cosmic ray sources, i.e. supernova remnants (SNRs). This process
is guaranteed in the sense that secondary particles will necessarily
be generated by the spallation of primary cosmic rays on ambient
matter in the sources. As discussed above, the crucial point is the
normalisation of this harder component which depends on many
parameters that are not directly accessible to observations, for
example, the ambient gas density. One parameter in particular, the
diffusion coefficient in the cosmic ray source, plays an important
role in determining the relative size of the effect of acceleration of
secondaries.

One possible way to resolve this problem is to determine the
normalisation from other observables. If it is in fact acceleration of
secondaries that leads to a rise in the positron fraction then both
the ambient density and the diffusion coefficient are large enough
such that other secondaries are produced in equally important
abundances and also get accelerated. Therefore, we should see similar
features of a rise in other secondary-to-primary ratios. We will use
observations of such observables as the antiproton-to-proton ratio,
titanium-to-iron (Ti/Fe) and boron-to-carbon (B/C) to determine the
normalisation of the harder electron-positron component and thereby
test the model of the acceleration of secondaries.

In particular, nuclear secondary-to-primary ratios like B/C not only
allow us to test this particular model but also to discriminate it
against other explanations of the rise in the positron fraction
$e^+/e^-$, e.g. dark matter annihilation or decay (see
Sec.~\ref{sec:Antimatter}) and pulsars (see
Sec.~\ref{sec:AlternativeAstrophysicalExplanations}). For instance, if
a feature similar to the rise in the positron fraction was also
observed in the antiproton-to-proton ratio $\bar{p}/p$, this would
rule out\footnote{There is a caveat: If in fact neither a dark matter
  model nor the acceleration of secondaries could explain the full
  positron excess when fitted to the potential $\bar{p}/p$ feature
  there might still be room for a contribution from pulsars.} the
pulsar explanation of the positron excess as pulsars are not expected
to produce any antiprotons. Conversely, and as such a rise is not
observed up to the maximum energies of current observations, $\sim 130
\, \text{GeV}$~\cite{Adriani:2010rc}, this sets some constraints on
models which in principle predict such a rise in $\bar{p}/p$. There
have, however, been efforts to build dark matter (DM) models that are
leptophilic, that is DM predominantly annihilating or decaying into
leptons, see Sec.~\ref{sec:Antimatter}. Similarly, a rise in nuclear
secondary-to-primary ratios, like B/C would rule out the DM
explanation of the positron excess as DM annihilation/decay should not
produce nuclei. Again, the non-observation of such a rise up to
energies higher than currently accessible, would in turn rule out the
acceleration of secondaries model\footnote{The acceleration of
  secondaries model could be saved by arguing that GCR protons and
  heavier nuclei are produced by different source populations and that
  the acceleration of secondaries only plays a role in the sources of
  protons.}. These relations are summarised in
Table~\ref{tbl:TruthTable}.

\begin{table} [tb]
\centering
\begin{threeparttable}[b] 
\caption{Truth table representing the logical possibilities to explain
  rises in different secondary-to-primary ratios.}
\label{tbl:TruthTable}
\begin{tabular}{c c c p{0.5cm} c c c}
\hline\hline \multicolumn{3}{c}{observed rise in} &&
\multicolumn{3}{c}{possible explanation} \\ $e^+/e^-$ & $\bar{p}/p$ &
nuclei && DM & pulsars & accn. of secs. \\ \hline t & t & t && f & f &
t \\ t & t & f && t & f & f \\ t & f & t && f & f & f \\ t & f & f &&
t & t & f \\ \hline\hline
\end{tabular}
\end{threeparttable} 
\end{table} 

\section{Antiproton-to-Proton Ratio}
\label{sec:AntiprotonRatio}

The calculation for the antiproton-to-proton ratio~\cite{Blasi:2009bd}
is conceptually very similar to that for the positron fraction. Of
course, both the production cross section and the inelasticity are
different for antiprotons than for positrons. Furthermore, as both
protons and antiprotons are stable and their energy losses are
negligible above a few GeV, their fluxes can be calculated in the
simple leaky box model with a grammage parametrised similarly as in
Sec.~\ref{sec:ObservationalResults},
\begin{equation}
X_{\text{esc}}(E) = X_0 \left(\frac{E}{10\,{\rm GeV}}\right)^{-\delta}
\, ,
\end{equation}
with $X_0 = 6 \, \text{g} \, \text{cm}^{-2}$ and $\delta = 0.6$. Above
$10 \, \text{GeV}$, solar modulation is negligible, too.

\begin{figure}[tb]
\begin{center}
\includegraphics[width=0.7\textwidth]{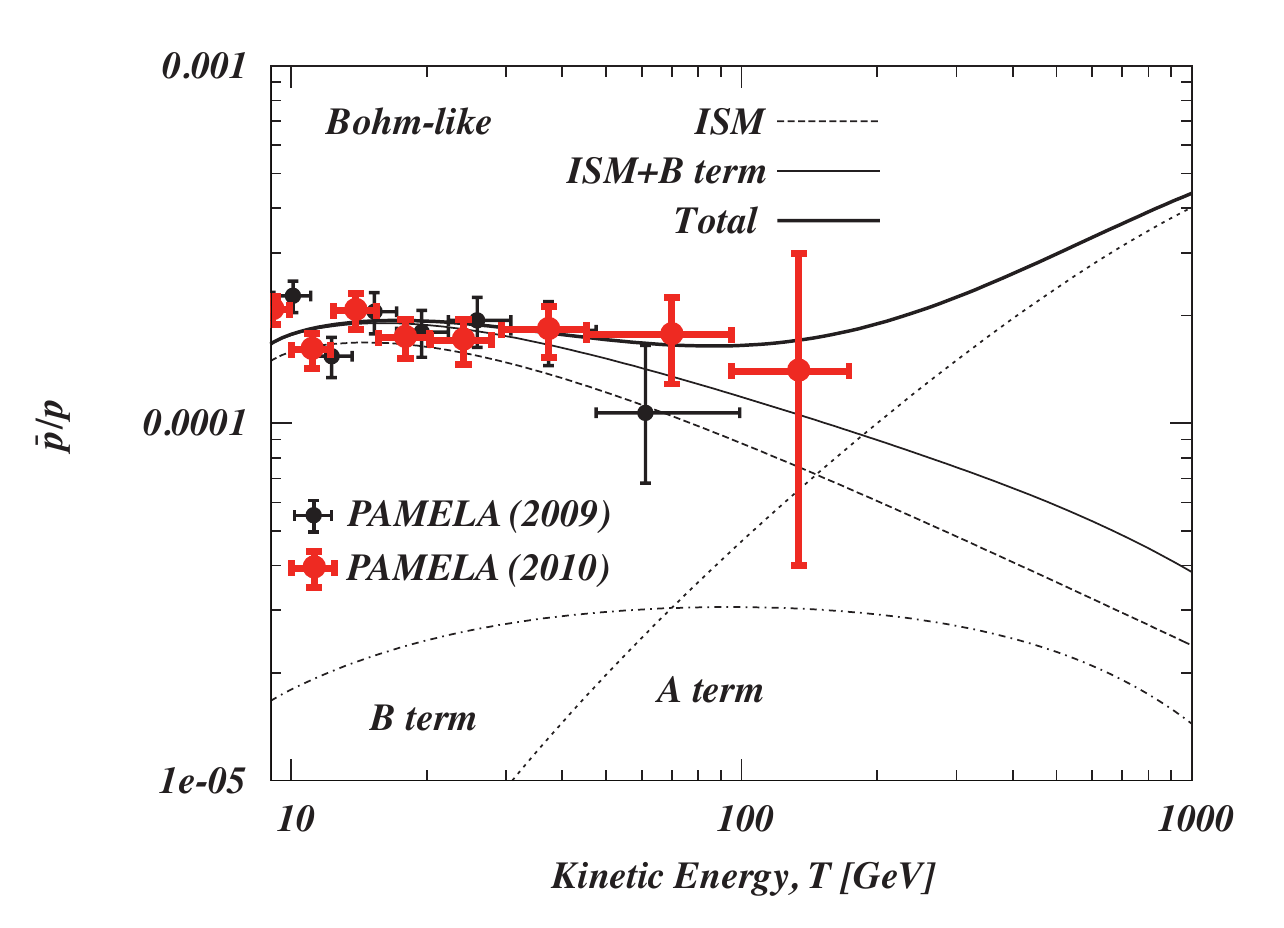}
\end{center}
\vspace{-0.5cm}
\caption[]{The antiproton-to-proton ratio (adapted
  from~\cite{Blasi:2009bd}). The $\mathcal{A}$ term (dotted line) and
  $\mathcal{B}$ term (dot-dashed line) are the secondaries produced in
  the diffusion zone and in the downstream region, respectively. The
  data points are from PAMELA
  measurements~\cite{Adriani:2008zq,Adriani:2010rc}. The antiprotons
  produced by spallation of GCR nuclei on the ISM are shown by the
  dashed line. The sums of $\mathcal{B}$ term and ISM contribution do
  not show a qualitatively different behaviour than the ISM
  contribution alone. The $\mathcal{A}$ term finally leads to the
  upturn in the total antiproton-to-proton ratio.}
\label{fig:pbar2p}
\end{figure}

The antiproton-to-proton ratio now has two contributions: the usual
one from the antiprotons produced in spallation of GCR nuclei on the
ISM,
\begin{equation}
\frac{J_{\bar{p},ISM}(E)}{J_{p}(E)} \simeq
\frac{2\,\varepsilon\,X_{\text{esc}}(E)}{m_p\,{E}^{2-a-\delta}}
\int_E^\infty \dd E_0 {E_0}^{2-a-\delta}
\sigma_{p\bar{p}}(E_0,E) \label{ISMsec} \, ,
\end{equation}
and one from the antiprotons produced inside the SNRs,
\begin{equation}
\frac{J_{\bar{p},SNRs}(E)}{J_{p}(E)} \simeq
2\,n_{1}\,c\,[\mathcal{A}(E)+\mathcal{B}(E)]\label{master2} \, ,
\end{equation}
where
\begin{align} 
\mathcal{A}(E)=a\left(\frac{1}{\xi}+r^2\right)
\times\int_m^{E}\dd\omega \,{\omega}^{a-3}
\frac{D_{1}(\omega)}{u_1^2}\int_{\omega}^{E_{\rm max}} \dd
E_0\,{E_0}^{2-a}\,\sigma_{p\bar{p}}(E_0,\omega) \label{SNRpbarA}
\end{align}
is from antiprotons produced in the diffusion zone and
\begin{equation} 
\mathcal{B}(E)=\frac{\tau_{SN}\,r}{2\,E^{2-a}}\int_{E}^{E_{\rm max}}
\dd E_0\,{E_0}^{2-a}\,\sigma_{p\bar{p}}(E_0,E)\,\label{SNRpbarB}.
\end{equation}
is from the antiprotons produced downstream. Here, $r = 3.8$ is the
compression ratio of the shock chosen such that the proton's source
spectral index in phase space $a = 3r/(r-1) \simeq 4.1$. The
inelasticity of the antiproton production $\xi \simeq 0.17$, the SNR
lifetime \mbox{$\tau_\text{SNR} = 10^4 \, \text{yr}$}, the upstream
velocity $u_1 = 0.5 \times 10^8 \, \text{cm} \, \text{s}^{-1}$ and the
fudge factor $K_B = 20$ in the diffusion coefficient $D = 3.3 \times
10^{22} K_B (E/ \text{GeV})^\delta \, \text{cm}^2 \,
\text{s}^{-1}$. The maximum energy is fixed to $E_\text{max} = 10 \,
\text{TeV}$ and for the production cross section $\sigma_{p \bar{p}}$
a parametrisation \cite{Kamae:2006bf} is used. The factor $\varepsilon
\simeq 1.26$ takes into account the contribution from CR nuclei and
nuclei in the ISM on the production of antiprotons. We reproduce the
antiproton-to-proton ratio calculated in Ref.~\cite{Blasi:2009bd} in
Fig.~\ref{fig:pbar2p}.

\section{Nuclear Secondary-to-Primary Ratios}

\subsection{Timescales of the problem}

The transport equation for any nuclear species $i$ reads
\begin{equation}
  u \frac{\partial f_i}{\partial x} = D_i \frac{\partial^2
    f_i}{\partial x^2} + \frac{1}{3} \frac{\mathrm{d}u}{\mathrm{d}x} p
  \frac{\partial f_i}{\partial p} - \Gamma_i f_i + q_i ,
\label{eqn:TransportReminder}
\end{equation}
where $f_i$ is the phase space density and the different terms from
left to right describe convection, spatial diffusion, adiabatic energy
losses as well as losses and injection of particles from spallation or
decay. We consider the acceleration of {\em all} species in the usual
setup: in the frame of the shock front the plasma upstream ($x<0$) and
downstream ($x>0$) is moving with velocity $u_{-}$ and $u_{+}$
respectively. We solve Eq.~\ref{eqn:TransportReminder} analytically
for relativistic energies $\varepsilon_k$ greater than a few
$\text{GeV/nucleon}$ such that $p \approx E$, $\beta \approx 1$ and
$N_i \mathrm{d}E \approx 4 \pi p^2 f_i \mathrm{d}p$. At these energies
ionization losses can be neglected and the spallation cross sections
become energy independent.

There are three relevant timescales in the problem:
\begin{enumerate}
\item Acceleration time $\tau_{\text{acc}}$ (cf .\cite{Malkov:2001}):
\begin{align}
  \tau_\text{acc} = \frac{3}{u_- - u_+} \int_0^p \left(
  \frac{D_i^+}{u_+} + \frac{D_i^-}{u_-} \right) \frac{\text{d}p'}{p'}
  \simeq 8.8 \, E_{\text{GeV}} Z^{-1} B_{\mu \text{G}} \, \text{yr}
\end{align}
for Bohm diffusion and the parameter values mentioned later.
\item Spallation and decay time $\tau_i$.
\begin{equation}
 \tau^\text{spall}_i \equiv 1/\Gamma^\text{spall}_i \sim 1.2 \times
 10^7 \, \left( \frac{n_{\text{gas}}}{\text{cm}^{-3}} \right)^{-1} \,
 \text{yr} .
\end{equation}
where an average $\sigma_i$ of ${\cal O}(100)$~mb has been
assumed. The rest lifetime $\tau^\text{dec}_i$ of the isotopes
considered ranges between $4 \times 10^{-2} \, \text{yr}$ and $10^{17}
\, \text{yr}$.
\item Age of the SNR under consideration (see
  Sec.~\ref{sec:AccnOfSecs})
\begin{equation}
\tau_\text{SNR} = x_\text{max}/u_+ \sim 2 \times 10^4 \, \text{yr} .
\end{equation}
\end{enumerate}

There are two essential requirements for a SNR to efficiently
accelerate nuclei by the DSA mechanism:
\begin{trivlist}
\item (a) $\tau_{\text{acc}} \ll \tau^{\text{spall}}_{i}$, which is
  equivalent to
\begin{equation}
  20 \frac{\Gamma_i^{-} D_i}{u_{-}^2} \ll 1 \quad \Rightarrow \quad
  \epsilon_{\text{k}} \ll 6.4 \times 10^5 \frac{Z_i}{A_i} \, B_{\mu
    \text{G}} \, \text{GeV} \, .
\label{eqn:cond1}
\end{equation}
\item (b) $\tau_{\text{SNR}} \ll \tau_i$ which implies,
\begin{equation}
  \frac{x_{\text{max}}}{u_{+}} \ll \frac{1}{\Gamma_i} \quad
  \Rightarrow \quad x \frac{\Gamma_i}{u_{+}} \ll 1\, .
\label{eqn:cond2}
\end{equation}
\end{trivlist}

The isotopes for which condition (b) is not satisfied at the lowest
energy considered viz.  ${}^{56}\text{Ni}$, ${}^{57}\text{Co}$,
${}^{55}\text{Fe}$, ${}^{54}\text{Mn}$, ${}^{51}\text{Cr}$,
${}^{49}\text{V}$, ${}^{44}\text{Ti}$ and ${}^{7}\text{Be}$ do {\em
  not} contribute significantly, so their decays in the source region
are neglected.

\subsection{Nuclear spectra at source}

We find that the general solution to Eq.~\ref{eqn:TransportReminder}
for $x \neq 0$ is
\begin{eqnarray}
f_i^\pm = \sum_{j \leq i} \left( E_{ji}^\pm \text{e}^{\lambda^\pm_j
  x/2} + F_{ji}^\pm \text{e}^{\kappa^\pm_j x/2} \right) + G_i^\pm ,
\\ \text{with}\quad \lambda_i^\pm = \frac{u_\pm}{D_i^\pm} \big(1 -
\sqrt{1 + 4D_i^\pm\Gamma_i^\pm/u^2_\pm}\big) , \nonumber
\\ \kappa_i^\pm = \frac{u_\pm}{D_i^\pm} \big(1 + \sqrt{1 + 4
  D_i^\pm\Gamma_i^\pm/u^2_\pm}\big) ,\nonumber
\label{eqn:solution}
\end{eqnarray}
where $G_i^\pm$ is the asymptotic value and $E^{+}_{ji}$ and
$F^{-}_{ji}$ are determined by the recursive relations:
\begin{align}
E_{ji}^{\pm} &= \frac{-4 \sum_{m \geq j} E_{mj}^{\pm} \Gamma_{j
    \rightarrow i}^{\pm}}{D_i^{\pm} \lambda_j^{\pm 2} - 2 u
  \lambda_j^{\pm} - 4 \Gamma_i^{\pm}} \, , \label{eqn:Eji}
\\ F_{ji}^{\pm} &= \frac{-4 \sum_{m \geq j} F_{mj}^{\pm} \Gamma_{j
    \rightarrow i}^{\pm}}{D_i^{\pm} \kappa_j^{\pm 2} - 2 u
  \kappa_j^{\pm} - 4 \Gamma_i^{\pm}} .
\label{eqn:Fji}
\end{align}
We require that the phase space distribution function converges to the
adopted primary composition $Y_i$ (at the injection energy $p_0$) far
upstream of the SNR shock:
\begin{equation}
f_i(x,p) \xrightarrow{x \rightarrow -\infty} Y_i \delta (p-p_0) \, ,
\quad \partial f_i/\partial x \, (x,p) \xrightarrow{x \rightarrow
  -\infty} 0 .
\end{equation}
We also require the solution to remain finite far downstream. As the
phase space density is continuous at the shock front, we connect the
solutions in both half planes to $f_i^0 = f_i(x=0,p)$ and find them to
be:
\begin{align}
f_i^{-} &= f_i^0 \text{e}^{\kappa^{-}_i x/2} + \sum_{j < i} F_{ji}^{-}
\left(\text{e}^{\kappa^{-}_j x/2} - \text{e}^{\kappa^{-}_i x/2}
\right) + Y_i \delta (p-p_0) \left( 1 - \text{e}^{\kappa^{-}_i x/2}
\right) \, ,
\label{eqn:solution2a} \\
f_i^{+} &= f_i^0 \text{e}^{\lambda^{+}_i x/2} + \sum_{j < i}
E_{ji}^{+} \left(\text{e}^{\lambda^{+}_j x/2} -
\text{e}^{\lambda^{+}_i x/2} \right) + G_i^{+} \left( 1 -
\text{e}^{\lambda^{+}_i x/2} \right) \, .
\label{eqn:solution2b}
\end{align}
Using Eqs.~\ref{eqn:cond1}-\ref{eqn:cond2}, we can linearly expand
$\lambda_i^{+}$ and $\kappa_i^{-}$ in Eq.~\ref{eqn:solution} and the
exponentials in Eqs.~\ref{eqn:solution2a}-\ref{eqn:solution2b},
\begin{equation}
\text{e}^{\lambda^{+}_i x/2} \simeq 1 - \frac{\Gamma^{+}_i}{u_{+}} x
\, , \quad \text{e}^{\kappa^{-}_i x/2} \simeq \big(1 +
\frac{\Gamma^{-}_i}{u_{-}} x \big) \text{e}^{u_{-}x/D_i} \,
\end{equation}
 to obtain:
\begin{equation}
f_i^{+} = f_i^0 + \frac{q_i^+(x=0) - \Gamma_i^{+} f_i^0 }{u_{+}} x \,
.
\label{eqn:fi+}
\end{equation}
where $q_i^\pm$ denotes the downstream/upstream source term:
$q^{\pm}_i = \sum_{j<i} f_j \Gamma^{\pm}_{j \rightarrow i}$.

Finally we integrate the transport equation over an infinitesimal
interval around the shock, assuming that $q_i^{+}/q_i^{-} =
\Gamma_i^{+} / \Gamma_i^{-} = n_{\text{gas}}^{+}/n_{\text{gas}}^{-} =
r$ and that $D_i^{+} \simeq D_i^{-}$:
\begin{align}
p \frac{\partial f_i}{\partial p} = - a f_i^0 - a (1+ r^2)
\frac{\Gamma_i^{-} D_i^{-}}{u_{-}^2} f_i^0 + a \left[ (1+ r^2)
  \frac{q_i^{-}(x=0) D_i^{-} }{u_{-}^2} + Y_i \delta(p-p_0) \right] \,
,
\label{eqn:DEfi0}
\end{align}
where $a = 3 r/(r - 1)$. This is readily solved by
\begin{align}
f_i^0(p) =& \int_0^p \frac{\mathrm{d}p'}{p'} \left( \frac{p'}{p}
\right)^{a} \text{e}^{-a (1 + r^2) (D_i^{-}(p) - D_i^{-}(p'))
  \Gamma_i^{-}/u_{-}^2} \nonumber \\ & \times a \left[ (1+ r^2)
  \frac{q_i^{-}(x=0) D_i^{-} (p')}{u_{-}^2} + Y_i \delta(p'-p_0)
  \right] .
\label{eqn:fi0} 
\end{align}
The Eqs.~\ref{eqn:fi+}-\ref{eqn:fi0} should be compared to
Eqs.~\ref{eqn:fpm}-\ref{eqn:fpm0} in Sec.~\ref{sec:AccnOfSecs} where
the loss terms $\Gamma_i f_i$ for ionisation was not taken into
account but cooling losses of the electrons and positron were
introduced in an {\it ad hoc} fashion by an exponential cut-off at
$E_\text{cut} \simeq 20 \, \text{TeV}$. This approach was justified
for electrons and positrons as cooling losses have been shown to
produce exactly such a functional behaviour with $E_\text{cut}$ in the
right range~\cite{Webb:1984}.  The exponential in our
Eq.~\ref{eqn:fi0} leads to a natural cut-off in both the primary and
secondary spectra above the energy predicted by Eq.~\ref{eqn:cond1}.
However, due to the approximations we have made, the
secondary-to-primary ratios cannot be predicted reliably for $4
\Gamma_i D_i/u^2 \gtrsim 0.1$, i.e. much beyond $\sim 1$~TeV.

Starting from the heaviest isotope, Eqs.~\ref{eqn:fi+} and
\ref{eqn:fi0} can be solved iteratively to obtain the injection
spectrum after integrating over the SNR volume,
\begin{equation}
N_i (E) = 4 \pi \int_0^{u_+ \tau_\text{SN}} \text{d}x\, p^2 f_i (p) \,
4 \pi \, x^2 .
\end{equation}

\subsection{Propagation of nuclei}

To account for the subsequent propagation of the nuclei through the
ISM we solve the transport equation in the leaky box model, see
Sec.~\ref{sec:LeakyBoxModel}, which reproduces the observed decrease
of secondary-to-primary ratios with energy in the range
$\sim1-100$~GeV by assuming an energy-dependent residence time. The
steady state cosmic ray densities $\mathcal{N}_i$ observed at Earth
are then given by recursion, starting from the heaviest isotope $i=1$,
\begin{equation}
\mathcal{N}_i = \frac { \sum_{j<i} \left( \Gamma^\text{spall}_{j
    \rightarrow i} + 1/ \varepsilon_{\text{k}} \tau_{j \rightarrow i}
  \right) \mathcal{N}_j + \mathcal{R}_\text{SN} N_i }{ 1 /
  \tau_{\text{esc}, i} + \Gamma_i } \, ,
\end{equation}
where $\mathcal{R}_\text{SN} \sim 0.03 \, \text{yr}^{-1}$ is the
Galactic supernova rate.

\section{Parameters}

We calculate the source densities $N_i$ and ambient densities
$\mathcal{N}_i$, taking into account all stable and metastable
isotopes from ${}^{64} \text{Ni}$ down to ${}^{46} \text{Cr}$/${}^{46}
\text{Ca}$ for the Ti/Fe ratio, and from ${}^{18} \text{O}$ down to
${}^{10} \text{Be}$ for the B/C ratio. Short lived isotopes that
$\beta^{\pm}$ decay immediately into (meta)stable elements are
accounted for in the cross-sections. The primary source abundances are
taken from Ref.~\cite{Engelmann:1990zz} and we have adopted an
injection energy of $1 \, \text{GeV}$ independent of the species. The
partial spallation cross-sections are from semi-analytical tabulations
and the total inelastic cross-sections is obtained from an empirical
formula~\cite{Silberberg:1973,Silberberg:1977,Silberberg:1990}. The
escape time is modelled according to the usual relation:
\begin{equation}
\tau_{\text{esc}, i} = (\rho \, c)^{-1} \, X_{\text{esc}, i} = (\rho
\, c)^{-1} \, X_{\text{esc}, i}^0 (E/Z_i)^{-\delta}
\end{equation}
where $X_{\text{esc}, i}$ is the column density traversed in the ISM
and $\rho = 0.02 \, \text{atom} \, \text{cm}^{-3}$ is the typical mass
density of hydrogen in the ISM. We have neglected spallation on helium
at this level of precision as its inclusion will have an effect $<10
\%$. The fit parameters are sensitive to the adopted partial
spallation cross-sections, for example $\delta \simeq 0.7$ for the
Ti/Fe ratio but $\sim 0.6$ for the B/C ratio.

The parameters are chosen as in Sec.~\ref{sec:SecondarySource}: $r=4$,
$u_{-} = 0.5 \times 10^8 \, \text{cm} \,\text{s}^{-1}$,
$n_\text{gas}^{-} = 2 \,\text{cm}^{-3}$ and $B = 1 \, \mu
\text{G}$. The diffusion coefficient in the SNR is
\begin{equation}
D_i (E) = 3.3 \times 10^{22} K_B \, B_{\mu}^{-1} E_{\text{GeV}}
Z_i^{-1} \, \text{cm}^2 \text{s}^{-1}
\label{eqn:D(p)Reminder}
\end{equation}
where the fudge factor $K_B$ is the ratio of the diffusion coefficient
to the Bohm value and is determined by fitting to the measured Ti/Fe
ratio.

\section{Results}

\begin{figure}[p]
\begin{center}
\includegraphics[width=0.7\textwidth]{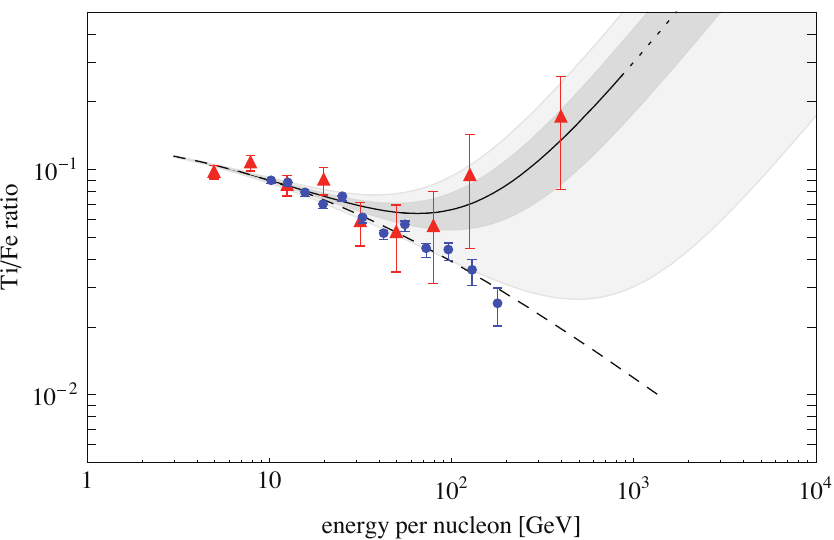}
\end{center}
\vspace{-0.5cm}
\caption[]{The Ti/Fe ratio in cosmic rays along with model predictions
  --- the `leaky box' model with production of secondaries during
  propagation only (dashed line), and including production and
  acceleration of secondaries in a nearby source (solid line - dotted
  beyond the validity of our calculation) for $K_B = 40$ as determined
  by fitting to the ATIC-2 data; the $1 \, \sigma$ and $2 \, \sigma$
  uncertainty bands are shown by the shaded dark grey and light grey
  areas, respectively. The data points are from ATIC-2
  (triangles)~\cite{Zatsepin:2009zr} and HEAO-3-C3 (circles)
  ~\cite{Vylet:1990yg}.}
\label{fig:Ti2Fe}
\end{figure}

\begin{figure}[p]
\begin{center}
\includegraphics[width=0.7 \textwidth]{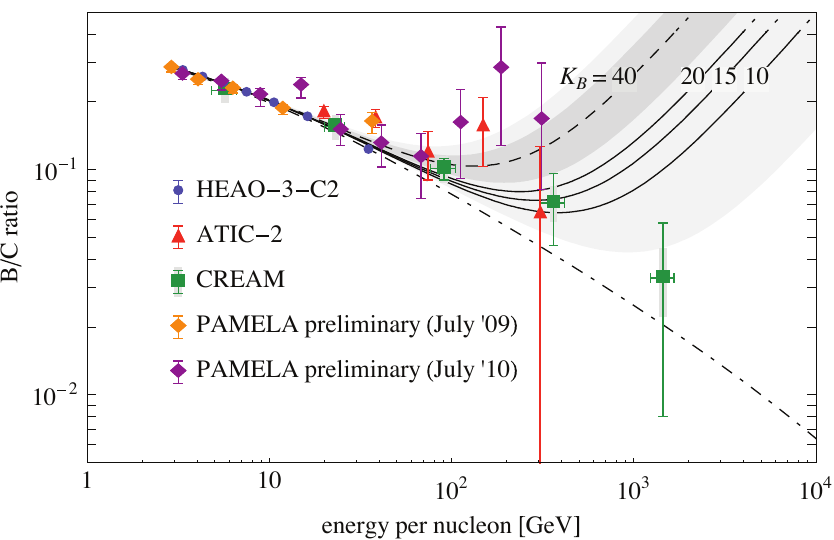}
\end{center}
\vspace{-0.5cm}
\caption[]{ The B/C ratio in cosmic rays along with model predictions
  --- the `leaky box' model with production of secondaries during
  propagation only (dot-dashed line), and including production and
  acceleration of secondaries in a nearby source (dashed line), $1 \,
  \sigma$ (dark grey) and $2 \, \sigma$ (light grey) uncertainty bands
  for $K_B = 40$ as determined from Ti/Fe; solid lines for $K_B = 10$,
  15 and 20 as suggested by the electron-positron fit (see
  Sec.~\ref{sec:Additional_Results}). The data points are from
  HEAO-3-C2 (circles)~\cite{Engelmann:1990zz}, ATIC-2
  (triangles)~\cite{Panov:2007fe}, CREAM (squares)~\cite{Ahn:2008my}
  and PAMELA (diamonds)~\cite{Picozza:2009,Sparvoli:2010}.}
\label{fig:myB2C}
\end{figure}

The calculated Ti/Fe ratio together with relevant experimental data is
shown in Fig.~\ref{fig:Ti2Fe}. The dashed line corresponds to the
leaky box model with production of secondaries during propagation only
and is a good fit to the (reanalysed) HEAO-3-C3
data~\cite{Vylet:1990yg}. The solid line includes production and
acceleration of secondaries inside the source regions which results in
an {\em increasing} ratio for energies above $\sim 50 \, \text{GeV}/n$
and reproduces well the ATIC-2 data \cite{Zatsepin:2009zr} taking $K_B
\simeq 40$. This is somewhat higher than the value of $K_B = 10
\mathellipsis 20$ determined from a fit of the total electron-positron
flux to the Fermi-LAT and HESS data (see
Sec.~\ref{sec:Additional_Results}). However, we note that the error
bars in the ATIC-2 data are rather large and therefore the uncertainty
in $K_B$ is also. In fact, the $1 \, \sigma$ and $2 \, \sigma$
confidence intervals are $K_B \in [27, 66]$ and $K_B \in [2.7, 90]$,
respectively. Therefore, the explanations for the rise in the positron
fraction and the excess in the total electron-positron flux is
consistent with the explanation for the rise in Ti/Fe. The value of
$K_B \simeq 20$ is also consistent with the antiproton-to-proton
ratio, see Sec.~\ref{sec:AntiprotonRatio}.

Fig.~\ref{fig:myB2C} shows the corresponding expectation for the B/C
ratio with the diffusion coefficient scaled proportional to rigidity
according to Eq.~\ref{eqn:D(p)Reminder}. The CREAM
data~\cite{Ahn:2008my} do show a downward trend as has been emphasized
recently~\cite{Simet:2009ne}, but the uncertainties are still large so
we await more precise measurements by PAMELA which has been directly
calibrated in a test beam~\cite{Campana:2008xj}.  Preliminary data
show indeed a rise above $\sim 100 \, \text{GeV}$ see
Fig.~\ref{fig:myB2C}.  Agreement with our prediction would confirm the
astrophysical origin of the positron excess as proposed in
Ref.~\cite{Blasi:2009hv} and Chapter~\ref{chp:additional} and thus
establish the existence of an accelerator of hadronic cosmic rays
within a few kpc.

\section{Conclusion}

We have presented a nice and interesting test of the proposed
acceleration of secondaries that could in principle explain the GCR
lepton excesses, by means of nuclear secondary-to-primary ratios. If
we saw a rise in such ratios, this would clearly point at the
importance of this effect, allow us to determine the normalisation and
perform cross checks with the positron analysis. Furthermore, we would
be able to extend our knowledge about the conditions prevailing in old
supernova remnants, for example the level of magnetic turbulence,
through this additional handle.

Unfortunately, at the moment, the experimental situation is not
clear. Although ATIC-2 has clearly observed a rise in Ti/Fe, the older
HEAO-3-C3 data are in strong disagreement. Therefore, the ultimate
test will be B/C for which the PAMELA experiment is continuously
taking data and the official result of their analysis is therefore
eagerly anticipated. Finally, the AMS-02 experiment~\cite{ams02} with
its superior charge identification, broad energy range and improved
statistics (due to the extension of the mission to 10 or even 18
years~\cite{Kounine:2010}) will clarify this issue once and for all.

%% file: 04haze/haze.tex
\chapter{Systematic effects in the extraction of the `WMAP haze'}
\label{chp:haze}

\section{Introduction}

Local cosmic ray measurements are not the only possible probe of the
galactic lepton population. Electrons and positrons of GeV and TeV
energies produce synchrotron radiation and gamma-rays by scattering
off the galactic magnetic fields and interstellar radiation fields
(ISRFs), respectively. Therefore every contribution in addition to the
standard electrons and positrons from supernova remnants (SNRs) must
also reflect in galactic diffuse radio and gamma-ray backgrounds.

We have already mentioned the possibility to constrain dark matter
(DM) models invented to explain the positron and electron-positron
excesses by gamma-ray and radio measurements. The bounds on the
annihilation cross section from diffuse measurements are at the moment
still at least an order of magnitude above the expectation for a
thermally produced weakly interacting massive particle (WIMP). These
studies are mostly restricted by modelling of the astrophysical
backgrounds, like diffuse emission. It is therefore necessary, to
improve our knowledge of these backgrounds to be able to subtract them
from the data and uncover possible, exotic contributions.

Some studies try to abbreviate this process by invoking proxies for
different physical processes that contribute to the
backgrounds. Following the assumed correlation, the data can be fitted
for and the background subtracted. One particular type of these
studies has found an excess in microwave sky maps -- the `WMAP haze'
(see Sec.~\ref{sec:Radio&Microwaves}).  A crucial ingredient of both
studies \cite{Finkbeiner:2003im,Bottino:2009uc} that identify a haze
is the extrapolation of the morphology of the synchrotron radiation
template from 408 MHz to the WMAP bands at 23 (K), 33 (Ka), 41 (Q), 61
(V) and 94 (W) GHz, i.e. over two orders of magnitude in frequency. In
fact the spatial distribution of the radiating CR electrons is likely
to differ significantly given their energy dependent diffusive
transport in the Galaxy. Instead of attempting such a bold
extrapolation, other studies, including the analysis by the WMAP
collaboration~\cite{Gold:2010fm}, employ the K-Ka difference map as a
tracer of synchrotron emission (despite some contamination by
free-free emission and an anomalous component which has been
interpreted (see, e.g.~\cite{deOliveiraCosta:2003az}) as spinning
dust~\cite{Draine:1998gq}). However although both maps are dominated
by synchrotron radiation, such a template could also contain any
unidentified radiation, such as a possible haze, and therefore cannot
exclude it.

CR transport in the Galaxy is dominated by diffusion through
interstellar magnetic fields with an {\em energy-dependent} diffusion
coefficient $D(E) = D_0 E^{\delta}$ where $\delta = 0.3 \mathellipsis
0.7$~\cite{Strong:2007nh}. Taking the energy loss rate $b (E) =
\mathrm{d}E/ \mathrm{d}t = b_0 E^2$ as is appropriate for synchrotron
and ICS, the diffusion length $\lambda$ is
%
\begin{equation*}
\lambda(E) \approx 5
\left(\frac{E}{\text{GeV}}\right)^{(\delta-1)/2}\,\text{kpc}\,,
\end{equation*}
for the standard values $D_0 = 10^{28}\,\text{cm}^{2}\text{s}^{-1}$
and $b_0 = 10^{-16}\,\text{s}^{-1}$~\cite{Strong:2007nh}. Therefore,
the distance that GeV energy electrons can diffuse is comparable to
the kpc scale on which the source distribution varies; moreover it
changes by a factor of 2.4 (1.5) for $\delta = 0.3$ (0.7) in the
energy range $\sim4-50$ GeV (corresponding to peak synchrotron
frequencies between 408 MHz and 50 GHz for a magnetic field of
$6\,\mu\text{G}$).  As a consequence the $\sim 50 \, \text{GeV}$
electrons will trace the source distribution much better than the
$\sim 4 \, \text{GeV}$ electrons which diffuse further away from the
sources and wash out their distribution.  The synchrotron map at 408
MHz {\em cannot} therefore be a good tracer of synchrotron radiation
at much higher, in particular WMAP, frequencies. Relying on such a
crude extrapolation of the morphology of synchrotron emission can thus
potentially introduce unphysical residuals. We estimate these by
simulating synchrotron sky maps at 408 MHz and the WMAP frequencies
and feeding these into the template subtraction process
\cite{Dobler:2007wv}. We show that this leads to residuals of the same
order as the claimed haze, which can in fact be matched for a
particular source distribution in the galactic disk. We conclude
therefore that the WMAP haze might be an artefact of inappropriate
template subtraction rather than evidence of dark matter annihilation.

\section{Template Subtraction}
\label{sec:TempSub}

The subtraction method is based on a multilinear regression of the CMB
subtracted WMAP data using foreground templates for free-free (f),
dust correlated (d) and synchrotron emission (s). Technically this can
be achieved by assembling the maps represented each by a vector of all
pixels, that is $\vec{f}$, $\vec{d}$ and $\vec{s}$, into one `template
matrix': $P = \left(\vec{f},\vec{d}, \vec{s}\right)$. (The template
for the haze, $\vec{h}$, is appended later.) The pseudo-inverse $P^+$
allows the determination of the coefficients $\vec{a} = P^+ \vec{w}$
that minimise the $\chi^2 = || \vec{w} - P\,\vec{a} ||^2 /\sigma^2$
for the different templates at the WMAP frequencies; $\sigma$ is the
mean measurement noise in each frequency band. For details see
Ref.~\cite{Dobler:2007wv}.

Since we are interested only in the effect of the electron diffusion
on the subtraction of the synchrotron foreground we do not use the
free-free and dust templates or radio sky maps that are strongly
affected by local structures such as Loop I
\cite{Berkhuijsen:1971}. Instead we simulate both the synchrotron sky
map at 408 MHz and the sky maps in the WMAP frequency range with the
{\tt GALPROP} code~\cite{Moskalenko:1997gh}. We adopt the same mask as
in Ref.~\cite{Dobler:2007wv} which excises pixels along the galactic
plane, around radio sources and in directions of excessive absorption.

To allow comparison with the results of Ref.~\cite{Dobler:2007wv} we
apply the same fitting procedure over the whole sky. In order to
determine the magnitude of the `haze' we append a template
\mbox{$\vec{h}=(1/\theta -1/\theta_0)$} to $P$ where $\theta =
\sqrt{\ell^2 + b^2}$ is in galactic coordinates and $\theta_0 =
45^\circ$. This corresponds to the ``FS8'' fit performed
in~\cite{Dobler:2007wv} and adding the haze back to the residual maps
gives the ``FS8 + haze'' maps. We determine the latitudinal profile of
the residual for $\ell = 0^\circ$ south of the galactic centre
direction. As our simulated maps do not contain any localised
structures, we do not need to divide the sky into several regions and
fit them independently, as was done with the ``RG8'' fit
\cite{Dobler:2007wv} . We have checked explicitly that doing so does
not change the profiles of the residual intensity or the spectral
indices.

We have checked that our procedure gives a residual `haze' in
agreement with Ref.~\cite{Dobler:2007wv} when we subtract the 408 MHz
survey sky map from the WMAP sky maps. Although with the CMB estimator
``CMB5'' we find a residual intensity of the same magnitude at 23 GHz,
its spectral index of about $-0.7$ is somewhat softer than in
Ref.~\cite{Dobler:2007wv}.

\section{Calculation}
\label{sec:diffusion_model}

The transport of CR electrons is governed by a diffusion-convection
equation (cf. Eq.~\ref{eqn:GCRtransport}),
\begin{align*}
\frac{\partial n}{\partial t} =& \vec{\nabla} \cdot \left( D_{xx}
\vec{\nabla} n - \vec{v} \,n \right) + \frac{\partial}{\partial p} p^2
D_{pp} \frac{\partial}{\partial p} \frac{1}{p^2} n -
\frac{\partial}{\partial p} \left( \dot{p} \, n - \frac{p}{3} \left(
\vec{\nabla} \cdot \vec{v} \right) n \right) + q \,,
\end{align*}
where $n \, \mathrm{d}p$ is the number density of electrons with $p
\in \left[p, p + \mathrm{d} p\right]$, $D_{xx} = D_{0 xx} (p/4 \,
\text{GeV})^\delta$ is the spatial diffusion coefficient, $\vec{v}$ is
the convection velocity, $D_{pp}$ is the momentum diffusion
coefficient and $q$ is the source power density. This equation is
numerically solved with the {\tt GALPROP} code {\tt v50.1p} in two
dimensions, that is assuming azimuthal symmetry around the galactic
centre and enforcing the boundary condition $n \equiv 0$ on a cylinder
of radius $R = 20\,\text{kpc}$ and half-height $z_\text{max}$ (see
below).

The source power density $q$ factorises into a source energy spectrum
$q_0 E^{-\alpha}$ and a spatial variation $\sigma(r)
\mathrm{e}^{-z/z_\text{scale}}$ with $z_\text{scale} =
0.2\,\text{kpc}$. For the radial part we consider two
possibilities. The distribution of SNRs is expected to be correlated
with that of pulsars which is inferred by Lorimer to be
\cite{Lorimer:2003qc}
\begin{equation}
\label{eqn:Lorimer}
\sigma_\text{Lorimer} (r) = 64.6
\left(\frac{r}{\text{kpc}}\right)^{2.35} \mathrm{e}^{-r/1.528
  \,\text{kpc}}\,.
\end{equation}
However, the determination of pulsar distances from their rotation
measures relies on knowledge of the thermal electron density
throughout the Galaxy and different distributions lead to different
functional forms for the inferred radial variation of the pulsar
density~\cite{Lorimer:2006qs}. Therefore we also consider an
exponential source distribution
\begin{equation}
\label{eqn:exponential}
\sigma_\text{exp}(r) = \sigma_0 \mathrm{e}^{-r/2\,\text{kpc}}\,,
\end{equation}
following Refs.\cite{Paczynski:1990} and~\cite{Sturner:1996}.

The normalisation $D_{0 xx}$, the scale height $z_\text{max}$ of the
CR halo and the spectral index $\delta$ of the diffusion coefficient
are usually determined from measurements of CR nuclei and nuclear
secondary to primary ratios (see
Sec.~\ref{sec:ObservationalResults}). The measurement of CR
`chronometers' like ${}^{10}\text{Be}/{}^9\text{Be}$ is still not
precise enough to break the degeneracy between $D_{0 xx}$ and
$z_\text{max}$ (see Sec.~\ref{sec:UnstableNuclearSecondaries}), so we
vary $z_\text{max}$ between $4\,\text{kpc}$ and $8\,\text{kpc}$ and
vary $D_{0 xx}$ only a little, checking that we have rough agreement
with the measured fluxes of nuclei and nuclear secondary-to-primary
ratios. On theoretical grounds~\cite{Ptuskin:2005ax} one expects a
spectral break in the diffusion coefficient at $\approx 1 \,
\text{GeV}$. We fix the break energy to 1 GeV and vary $\delta_1$ and
$\delta_2$ below and above the break (keeping $\delta_1\geq
\delta_2$), again trying to satisfy all local CR measurements.

The source electron spectrum is usually assumed to have a break around
4~GeV so we fix the electron source normalisation $q_0 \sigma_0$ and
the spectral indices $\alpha_1$ and $\alpha_2$ below and above the
break by fitting the propagated flux to the electron spectrum as
measured at Earth~\cite{Strong:2009xp,Abdo:2009zk}. We apply Solar
modulation in the spherical approximation~\cite{Gleeson:1968zz} with a
median potential of $\phi = 550\,\text{MV}$. Reacceleration and
convection play a role at energies below $10\,\text{GeV}$ and are
therefore important for the 408 MHz map. For the Alfv\`en velocity
$v_\mathrm{A}$ which determines the strength of reacceleration via
$D_{pp} \propto v_\mathrm{A}^2$ we consider the range
$0-50\,\text{km}\,\text{s}^{-1}$. {\tt GALPROP} assumes the convection
velocity to vary linearly with distance from the galactic plane and we
vary the slope ${\rm d} v_{\text{conv}}/{\rm d} z$ between 0 and
$20\,\text{km}\,\text{s}^{-1}\,\text{kpc}^{-1}$.

Since the random component of the galactic magnetic field is known to
dominate over the regular component~\cite{Beck:2008eb}, we neglect the
latter. For the radial dependence we adopt the usual exponential
fall-off where the radial scale $\rho$ and the (perpendicular
component of the) field strength $B_0$ at the galactic centre are
chosen to reproduce the 408 MHz sky map~\cite{Haslam:1982}.  Although
it was initially believed~\cite{Strong:1998fr} that an exponential
dependence on $z$ could give a satisfactory fit to the 408 MHz
latitude profile, the galactic field model was later
refined~\cite{Orlando:2009xg} by considering different,
non-exponential behaviours which in fact give better fits.  We
therefore apply the method described in Ref.~\cite{Philipps:1981} of
determining the emissivity dependence on $r$ for galactic longitude
$\ell = \pm 180^\circ$ (towards the galactic anti-centre). With an
estimate for the electron density this translates into a
$z$-dependence of the form $a + b \exp{[(-|z|/\xi)^\kappa]}$ and this
is iterated to convergence where we find $a/b = 0.27$, $\xi = 0.51$
and $\kappa = 0.68$.

\section{Results}
\label{sec:Results}

\subsection{Lorimer source distribution}

The parameters of the diffusion model, the magnetic field and the
electron source spectrum have been adjusted as described above and the
values are shown for the Lorimer source distribution,
Eq.~\ref{eqn:Lorimer}, in Table~\ref{tbl:parameters}.

\begin{table}[!bt]
\centering
\begin{threeparttable}[b] 
\caption{Parameters of source and diffusion models.}
\label{tbl:parameters}
\begin{tabular*}{0.65\columnwidth}{@{\extracolsep{\fill}} c c c}
\hline\hline & Lorimer & exponential \\ \hline Source &
\multirow{2}{*}{{\em c.f.} Eq.~\ref{eqn:Lorimer}} &
\multirow{2}{*}{{\em c.f.} Eq.~\ref{eqn:exponential}} \\ [-1ex]
distribution & & \\ $\alpha_1, \alpha_2$ & 1.2, 2.2 & 1.2, 2.2
\\ $D_{0 xx}$ & $5.75 \times 10^{28}\,\text{cm}^2\,\text{s}^{-1}$ &
$5.75 \times 10^{28}\,\text{cm}^2\,\text{s}^{-1}$ \\ $z_\text{max}$ &
4\,\text{kpc} & 8\,\text{kpc} \\ $\delta_1, \delta_2$ & 0.34, 0.34 &
0.1, 0.4 \\ $v_{\text{A}}$ & $50 \, \text{km}\,\text{s}^{-1}$ & $36 \,
\text{km}\,\text{s}^{-1}$ \\ ${\rm d} v_{\text{conv}}/{\rm d} z$ & $10
\, \text{km}\,\text{s}^{-1} \, \text{kpc}^{-1}$ & $15 \,
\text{km}\,\text{s}^{-1} \, \text{kpc}^{-1}$ \\ $B_0$ & $6.3 \, \mu
\text{G}$ & $6.8 \, \mu \text{G}$ \\ $\rho$ & $5 \, \text{kpc}$ & $50
\, \text{kpc}$ \\ \hline\hline
\end{tabular*}
\end{threeparttable}
\end{table}

\begin{figure}[t]
\includegraphics[scale=0.56]{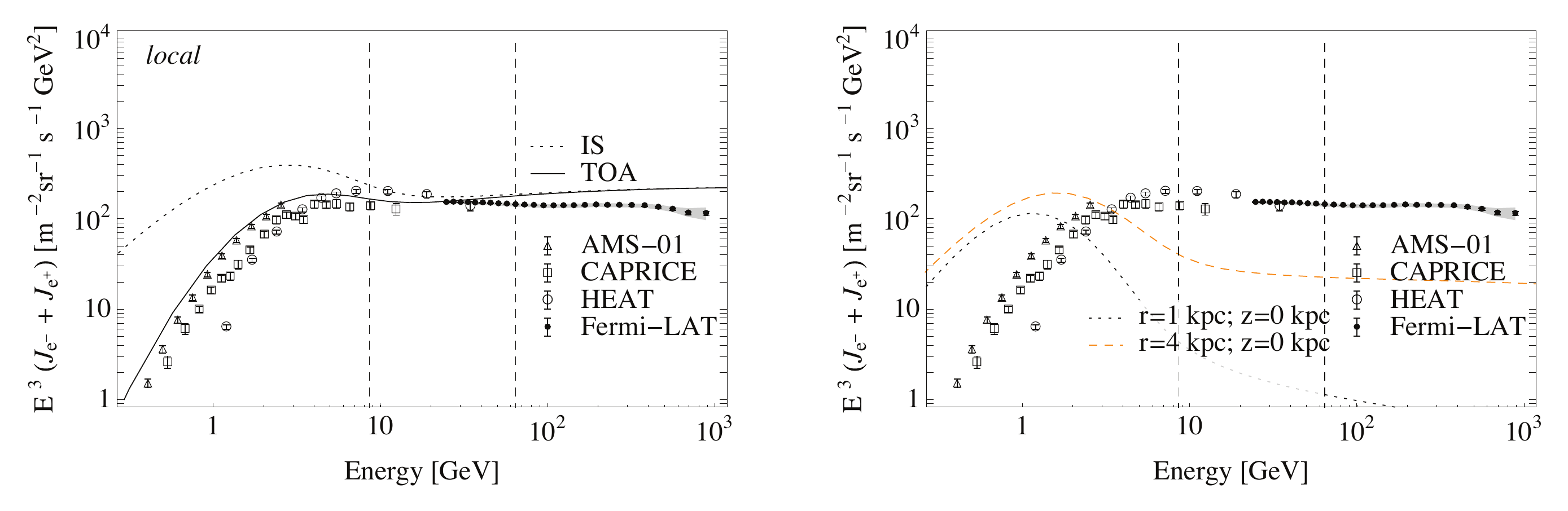}
\caption{{\bf Left:} The electron (plus positron) flux measured
  locally by AMS-01, CAPRICE, HEAT~\cite{Strong:2009xp} and Fermi-LAT
  ~\cite{Abdo:2009zk}, compared with the expectation for the Lorimer
  source distribution, Eq.~\ref{eqn:Lorimer}; the dotted line is the
  calculated interstellar flux while the solid line is its Solar
  modulated value (with $\phi = 550$~MV). The dashed vertical lines
  show the energy corresponding to peak synchrotron frequencies of 408
  MHz and 23 GHz for the local magnetic field.  {\bf Right:} The
  calculated electron (plus positron) flux at the positions $\{ (r, z)
  \} = \{ (1, 0), (4, 0) \}$ (in kpc).}
\label{fig:pElPlusPos_50p_901019}
\end{figure}

\begin{figure}[!ht]
\includegraphics[width=\textwidth]{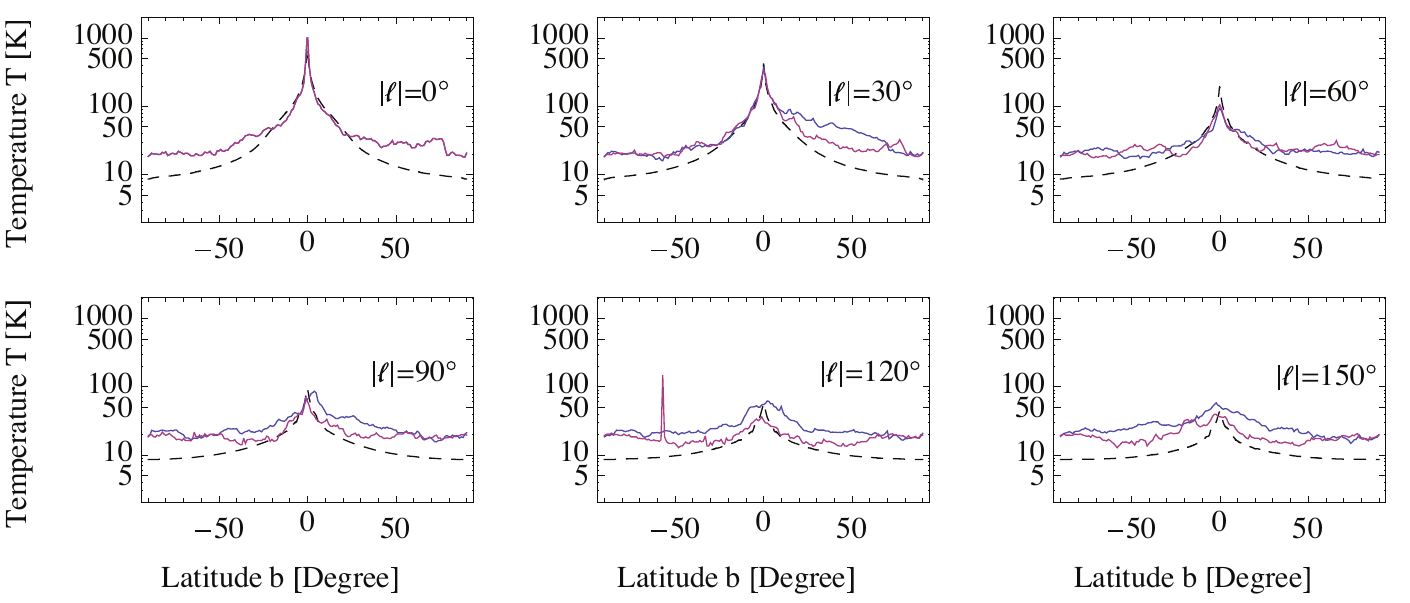}
\caption{The calculated latitudinal profile of galactic synchrotron
  radiation at 408 MHz (black dashed line) for galactic longitudes
  $|\ell|~=~0^\circ, 30^\circ, 60^\circ, 90^\circ, 120^\circ$ and
  $150^\circ$. The red (blue) solid line is the observed profile
  ~\cite{Haslam:1982} for positive (negative) $\ell$.}
\label{fig:pLats_50p_901019}
\end{figure}

The electron flux measured locally and at the positions $\{ (r, z) \}
= \{ (1, 0), (4, 0), (0, 4) \}$ (in kpc) are shown in
Fig.~\ref{fig:pElPlusPos_50p_901019}. We note that close to the
galactic centre the electron flux responsible for synchrotron
radiation at 408 MHz is not only much softer but also suppressed by
over an order of magnitude with respect to its locally measured
value. Fig.~\ref{fig:pLats_50p_901019} shows the latitudinal profiles
of the synchrotron radiation at 408 MHz; in general, the fit is good
for $b \lesssim 50^\circ$ but underestimates the emission at larger
latitudes. It has been shown~\cite{Philipps:1981} that this can
potentially be overcome by increasing the scale height of the
synchrotron emissivity at larger galactic radii. The remaining
discrepancies between the simulated and measured profiles are probably
due to the assumption of rotational symmetry. This leads to an
underestimation of the synchrotron radiation along tangents of the
spiral arms and an overestimation between them. For example, the
Carina arm is tangent at $75^\circ$ and the Sagittarius arm at
$-40^\circ$, so both $\ell = +60^\circ$ and $\ell = -60^\circ$ are
between spiral arms and thus slightly overestimated, in particular in
the galactic plane. It is also clear that point sources (that have not
been subtracted from the 408 MHz data) are not accounted for in our
calculation (e.g., Fornax at $\ell \simeq 120^\circ$, $b \simeq
-57^\circ$).

\begin{figure}[t]
\centering \includegraphics[scale=1]{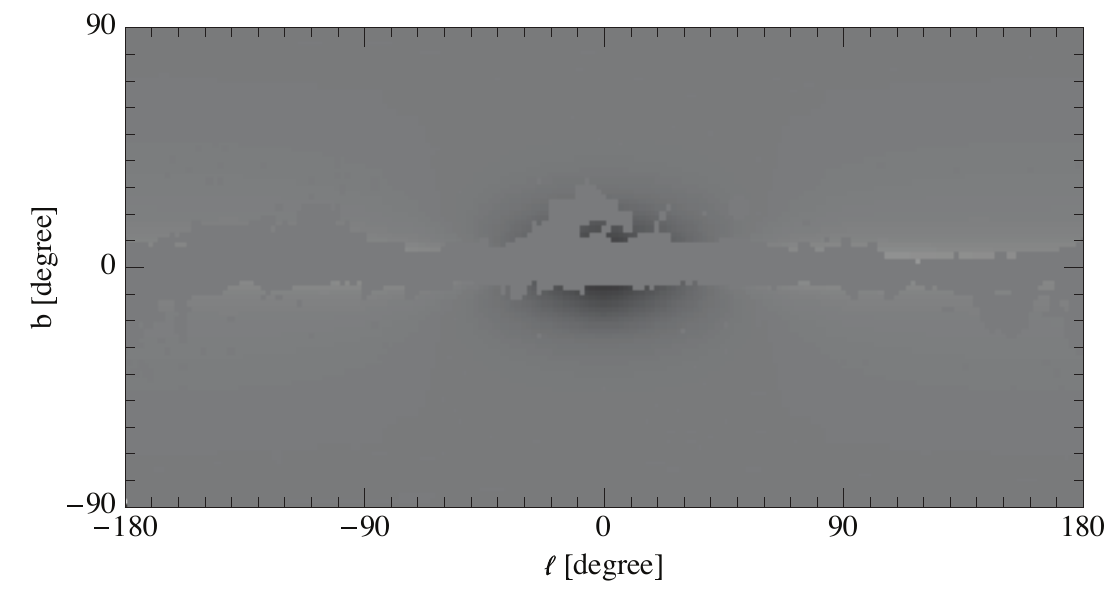}
\caption{Residual sky map in galactic coordinates for the Lorimer
  source distribution, Eq.~\ref{eqn:Lorimer}. The grey scale is linear
  from $-4$ to $+4 \, \text{kJy} \, \text{sr}^{-1}$ (corresponding to
  $-0.25$ to $+0.25 \, \text{mK}$ at 22.8 GHz).}
\label{fig:residual_skymap_Lorimer}
\end{figure}

\begin{figure}[ht]
\centering \includegraphics[width=0.45\textwidth]{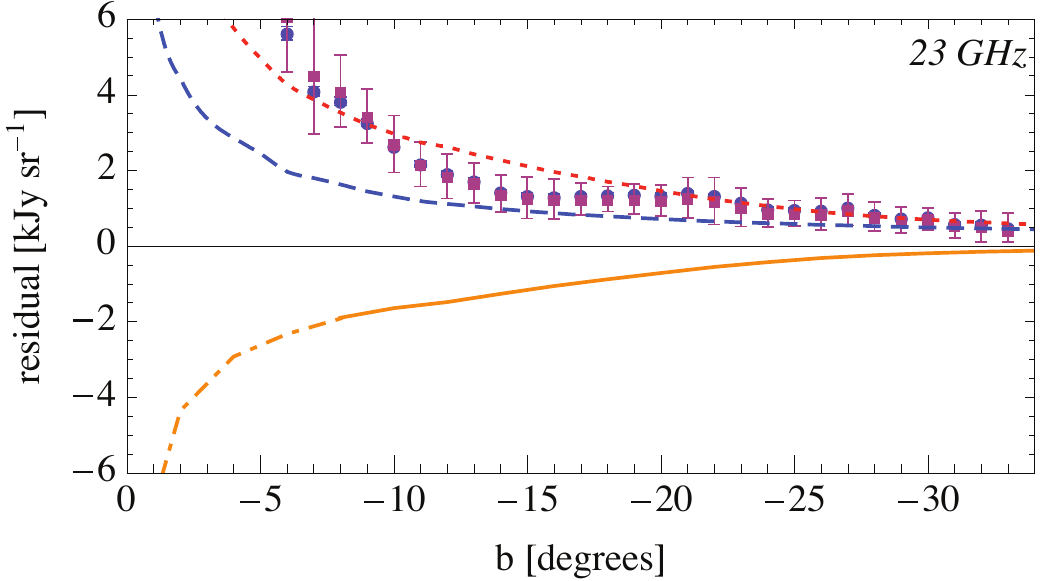}
\includegraphics[width=0.45\textwidth]{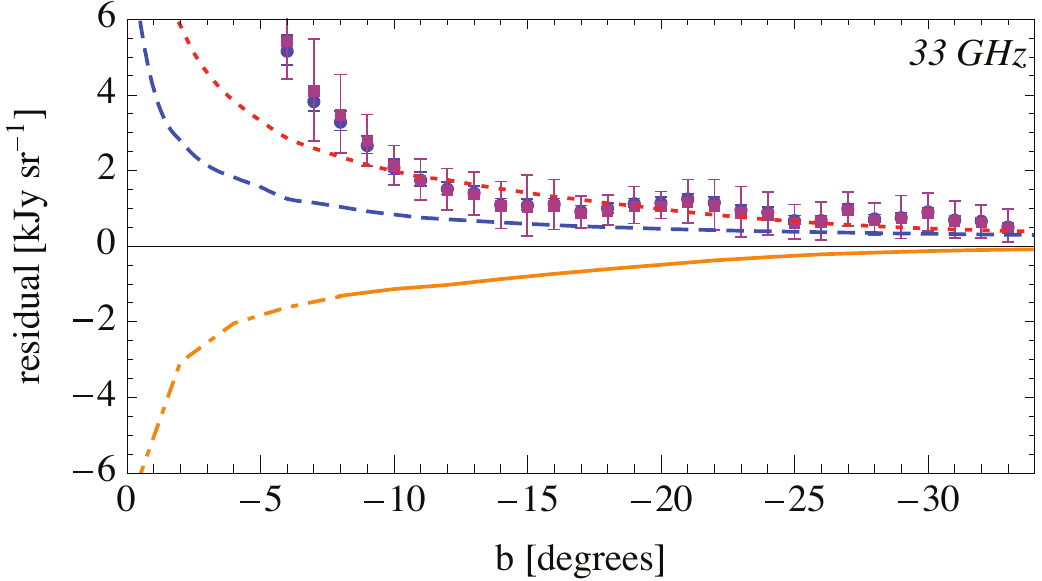}
\caption{Latitudinal profile of the K band residual outside (solid
  curve) and under (dot-dashed curve) the mask at 23 GHz (left panel)
  and 33 GHz (right panel). The square (circle) data points are the
  `haze' as extracted in Ref.~\cite{Dobler:2007wv}
  (\cite{Hooper:2007kb}). The dotted line shows the extrapolated
  emission at 23 (33) GHz from scaling the simulated 408 MHz emission
  and the dashed line shows the actual simulated 23 (33) GHz
  emission.}
\label{fig:residual_profile_Lorimer}
\end{figure}

The sky map of the residual $r(\ell,b)$
(Fig.~\ref{fig:residual_skymap_Lorimer}) shows a deficit for $|\ell|
\leq 40^\circ$ and $|b| \leq 20^\circ$. Further away from the galactic
centre direction there is a slight excess. The residual specific
intensity (Fig.~\ref{fig:residual_profile_Lorimer}) is of opposite
sign but its absolute value is of the same order of magnitude as the
`haze' at 23 and 33 GHz.

\subsection{Exponential source distribution}

\begin{figure}[tb]
\includegraphics[width=\textwidth]{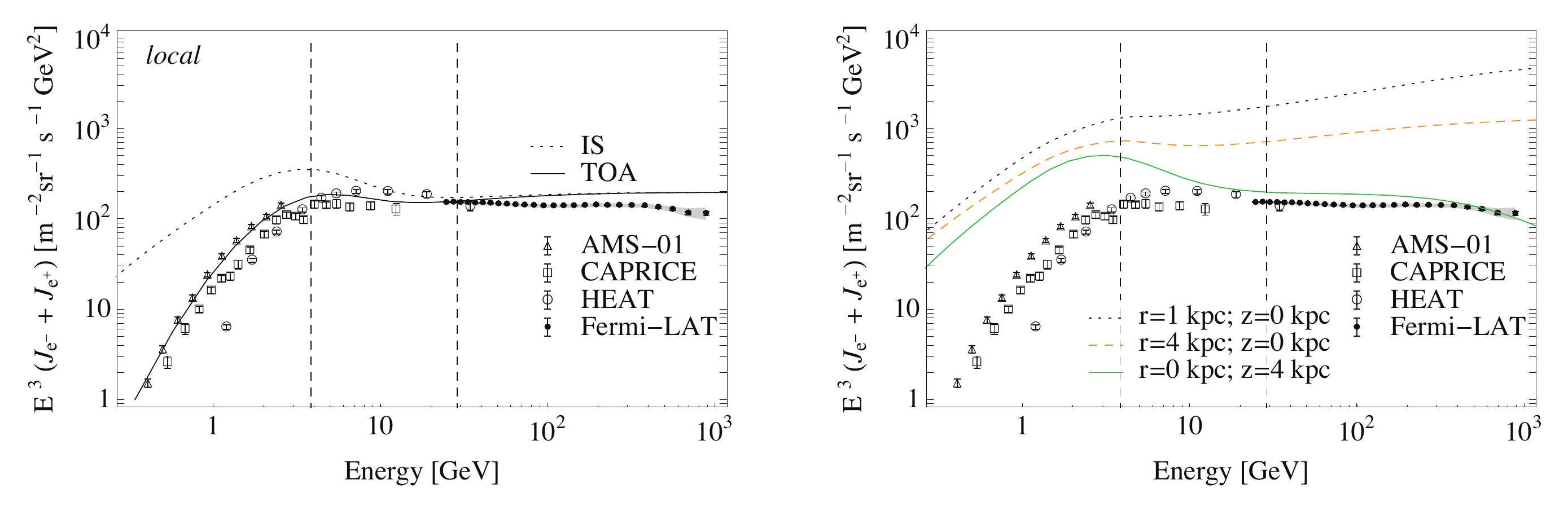}
\caption{Same as in Fig.~\ref{fig:pElPlusPos_50p_901019}, but for the
  exponential source distribution, Eq.~\ref{eqn:exponential}.}
\label{fig:pElPlusPos_50p_910025b}
\end{figure}

\begin{figure}[tb]
\includegraphics[width=\textwidth]{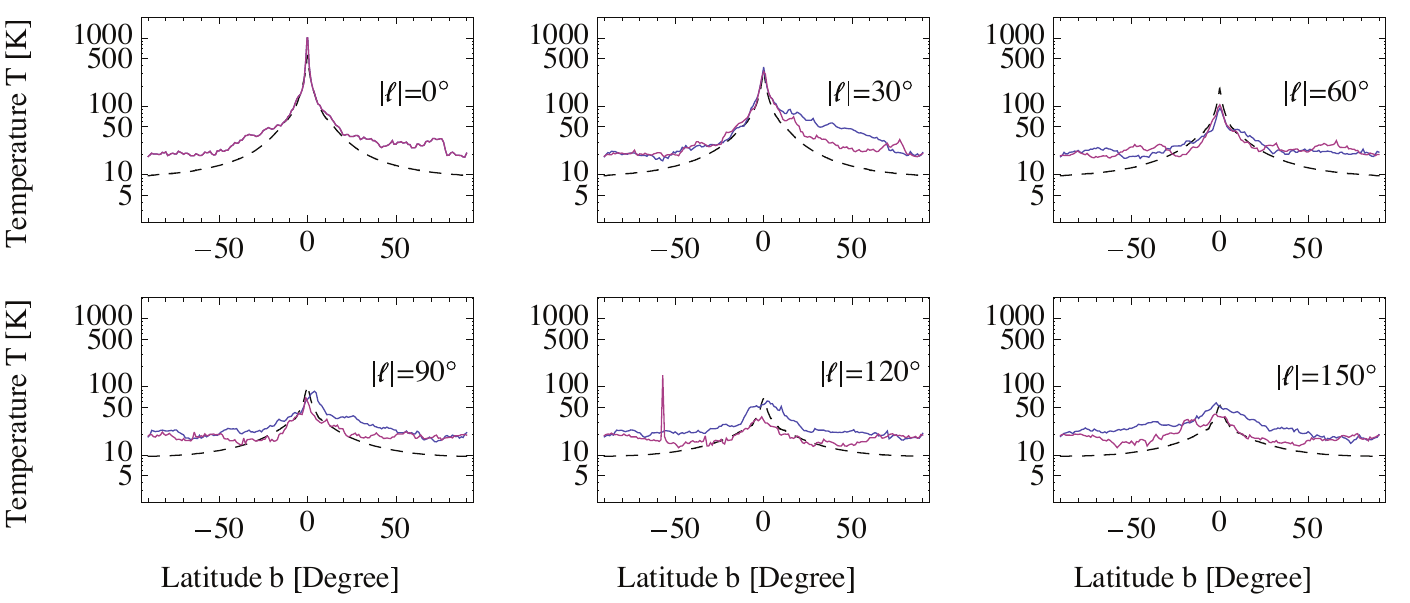}
\caption{Same as in Fig.~\ref{fig:pLats_50p_901019}, but for the
  exponential source distribution, Eq.~\ref{eqn:exponential}.}
\label{fig:pLats_50p_910025b}
\end{figure}

For the exponential source distribution, Eq.~\ref{eqn:exponential},
the electron fluxes are shown in
Fig.~\ref{fig:pElPlusPos_50p_910025b}. Close to the galactic centre,
it is larger by about an order of magnitude than measured locally and
slightly harder. The latitudinal profiles of the synchrotron radiation
at 408 MHz are shown in Fig.~\ref{fig:pLats_50p_910025b}.

\begin{figure}[t]
\centering \includegraphics[scale=1]{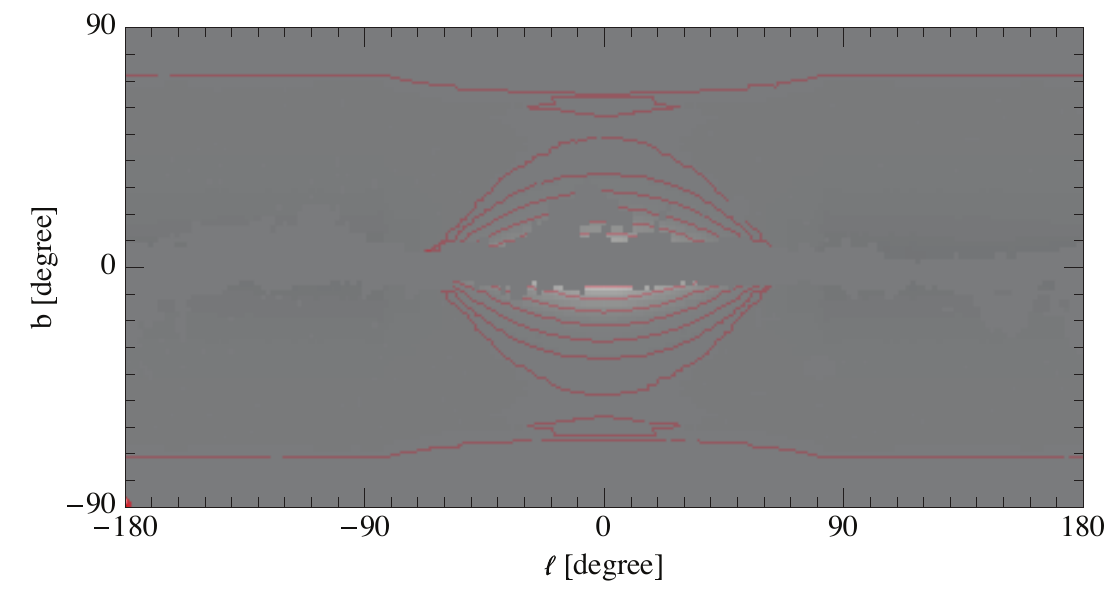}
\caption{Residual sky map in galactic coordinates for the exponential
  source distribution, Eq.~\ref{eqn:exponential}. The grey scale is
  linear from $-4$ to $+4 \, \text{kJy} \, \text{sr}^{-1}$
  (corresponding to $-0.25$ to $+0.25 \, \text{mK}$ at 22.8 GHz). The
  contour lines are logarithmically spaced in intensity between $0.02$
  and $2 \, \text{kJy} \, \text{sr}^{-1}$.}
\label{fig:residual_skymap_exponential}
\end{figure}

The residual sky map contains a roughly spherical excess around the
centre of the map, although somewhat more extended in longitude than
in latitude (see Fig.~\ref{fig:residual_skymap_exponential}).

\begin{figure}[ht]
\centering \includegraphics[width=0.45\textwidth]{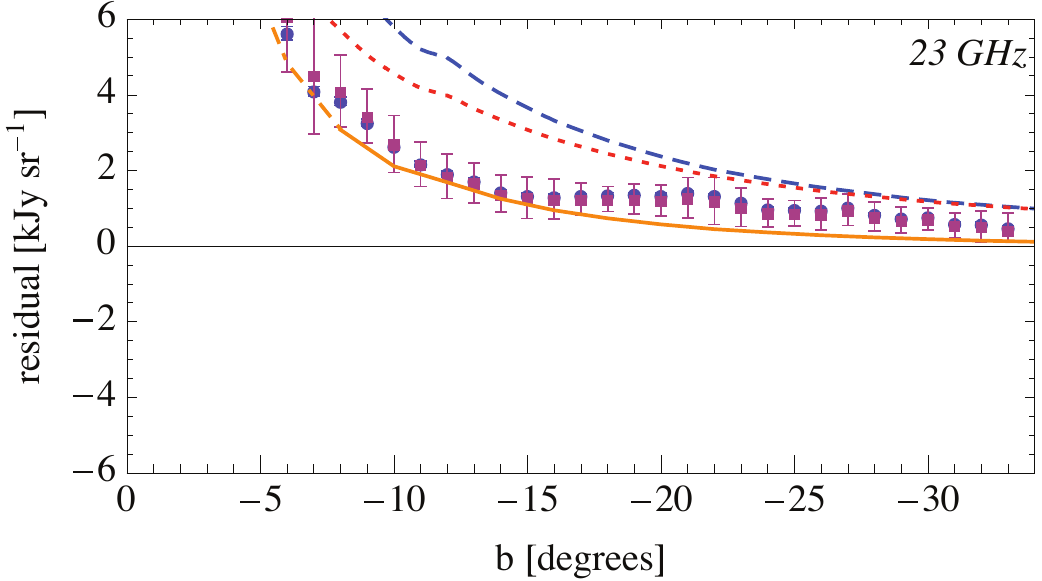}
\includegraphics[width=0.45\textwidth]{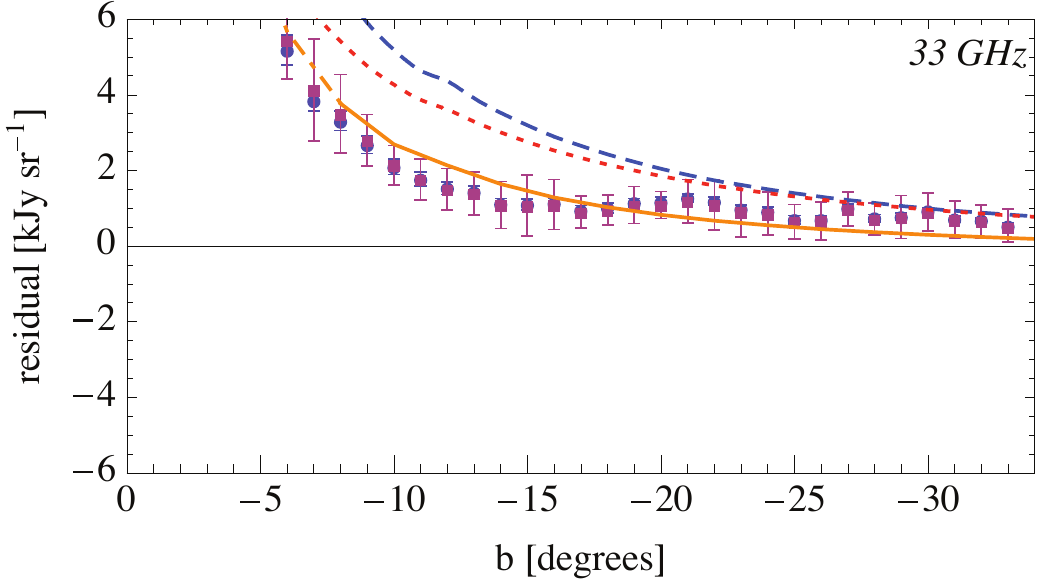}
\caption{Same as in Fig.~\ref{fig:residual_profile_Lorimer}, but for
  the exponential source distribution, Eq.~\ref{eqn:exponential}. We
  have added an offset to the calculated residual of $+11.8 \, \vec{h}
  \, \text{kJy} \, \text{sr}^{-1}$ ($+23.7 \, \vec{h} \, \text{kJy} \,
  \text{sr}^{-1}$) in the 23 GHz (33 GHz) band reflecting the
  systematic uncertainty from chance correlations between the `haze'
  template and the CMB.}
\label{fig:residual_profile_exponential}
\end{figure}

We note that the systematic uncertainty of the residual intensity (as
determined from real sky maps) induced by chance correlations between
the `haze' template and the CMB has been estimated in
Ref.~\cite{Dobler:2007wv} and can be read off their Fig. 8 as $\pm
11.8 \, \vec{h} \, \text{kJy} \, \text{sr}^{-1}$ ($\pm 23.7 \, \vec{h}
\, \text{kJy} \, \text{sr}^{-1}$) in the 23 GHz (33 GHz) band. We
therefore allow for an offset of our calculated residual relative to
the `haze' template in this range when fitting the residuals from real
sky maps.  The residual intensity
(Fig.~\ref{fig:residual_profile_exponential}) {\em matches} the
claimed WMAP haze.

\begin{figure}[p]
\centering
\vspace{0cm} \includegraphics[scale=1]{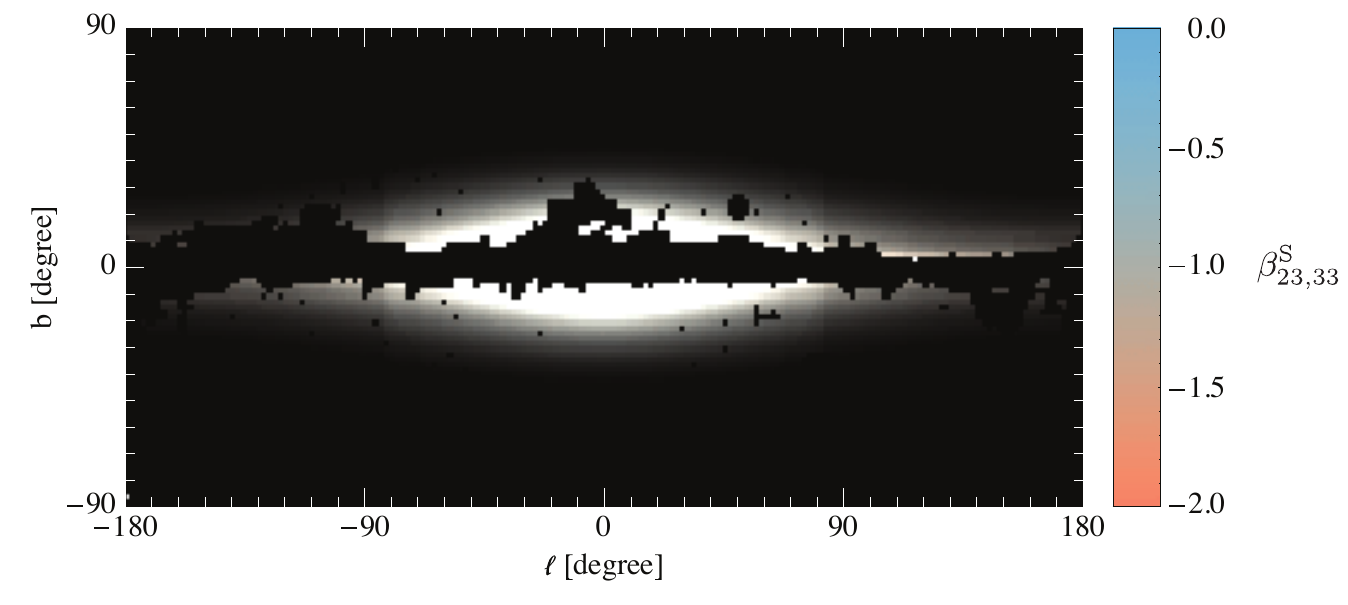} \\
\vspace{0cm} \includegraphics[scale=1]{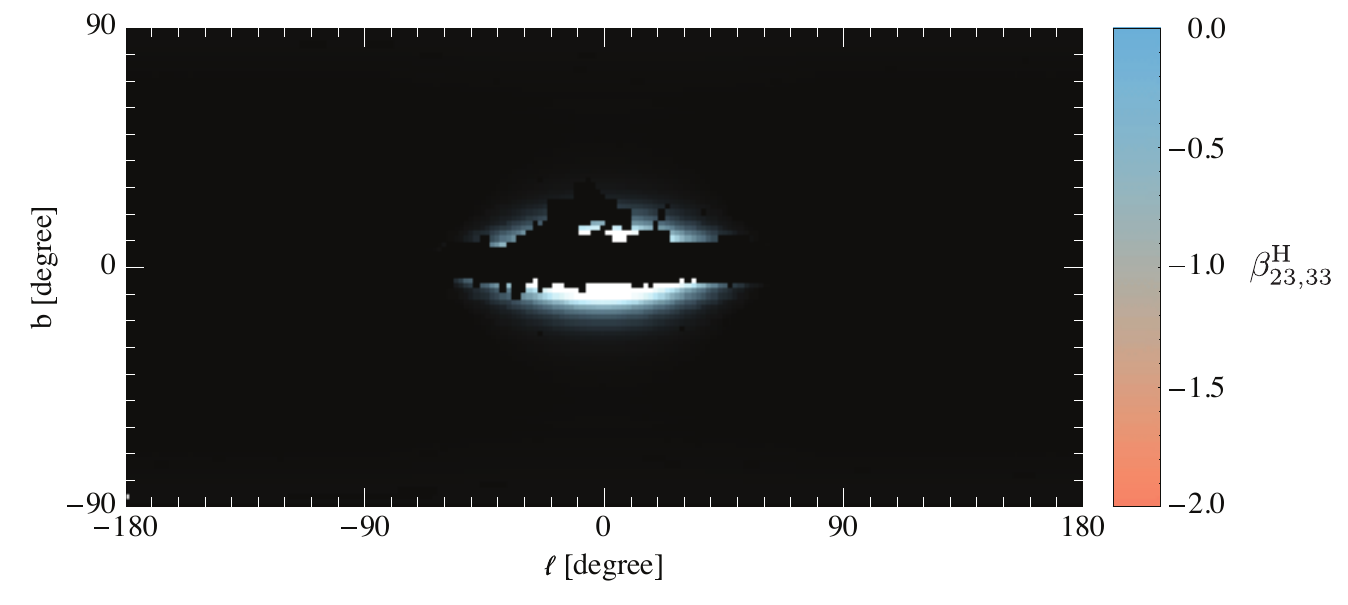} \\[-0.6cm]
\caption[]{Colour maps of spectral indices between 23 and 33 GHz
  defined in Eq.~\ref{eqn:DefSpecInd} scaled by the 23 GHz intensity
  for synchrotron + residual (top panel) and residual only (bottom
  panel).}
\label{fig:SpecIndMaps}
\end{figure}

\begin{figure}[p]
\centering \includegraphics[scale=0.8]{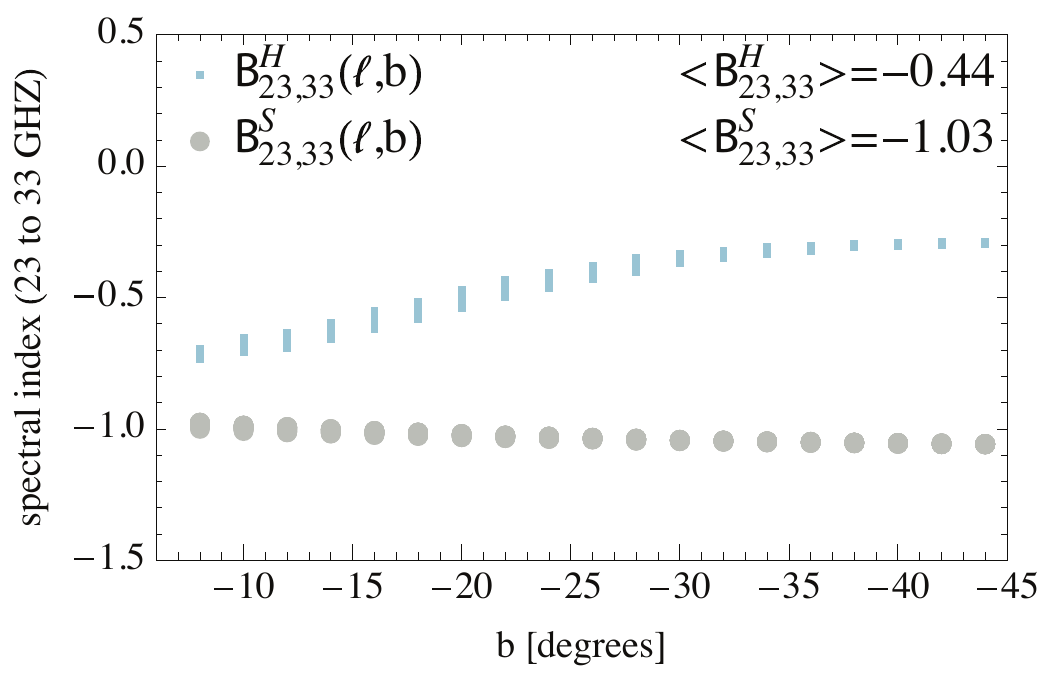}
\caption[]{Spectral indices of the unmasked pixels in the region south
  of the galactic centre ($b \in [-45 {}^\circ, 0{}^\circ], \, \ell
  \in [-25{}^\circ, 25{}^\circ]$) as a function of latitude for
  residual + synchrotron (large beige circles) and residual alone
  (small blue squares). The average spectral indices, $\beta_\text{S}$
  and $\beta_\text{H}$, are shown in the upper right corner.}
\label{fig:SpecIndLat}
\end{figure}

To compare our results to those of Ref.~\cite{Dobler:2007wv}, we also
determine the average spectral index (for details see Appendix
\ref{sec:SpecIndicesSkyMaps}) in a region south of the galactic
centre, $b \in [-45 {}^\circ, 0{}^\circ], \, \ell \in [-25{}^\circ,
  25{}^\circ]$. The colour maps of spectral indices scaled by
intensity are shown in Fig.~\ref{fig:SpecIndMaps}, both for the
synchrotron + residual and for the residual alone. Not only is the
synchrotron emission much more disk-like than the residual, but the
spectral index of the residual is also considerably harder than the
synchrotron spectral index. This is to be compared with Fig. 7 of
Ref.~\cite{Dobler:2007wv} which exhibits the same qualitative
behaviour.

Furthermore, we show the spectral index for the unmasked pixels in the
region south of the galactic centre (as defined above) as a function
of latitude in Fig.~\ref{fig:SpecIndLat}, again both for the
synchrotron + residual and for the residual alone. With average
indices of $\langle \beta_{23,33}^\text{H} \rangle = -0.44$ for the
residual and of $\langle \beta_{23,33}^\text{S} \rangle = -1.03$ for
the residual + synchrotron in this region, we find that the residual
index is harder than the synchrotron index by 0.6, which is in
excellent agreement with the findings of
Ref.~\cite{Dobler:2007wv}. The values of our different model
parameters are shown in Table~\ref{tbl:parameters}.

\section{Discussion}

\begin{figure}[b]
\includegraphics[width=0.45\textwidth]{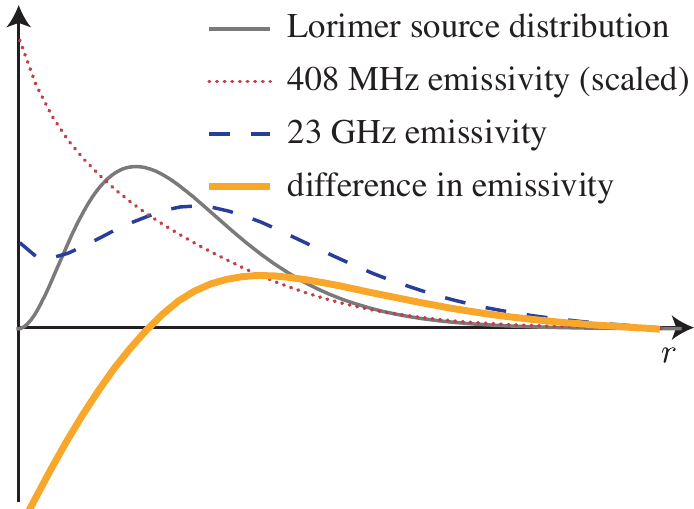}
\includegraphics[width=0.45\textwidth]{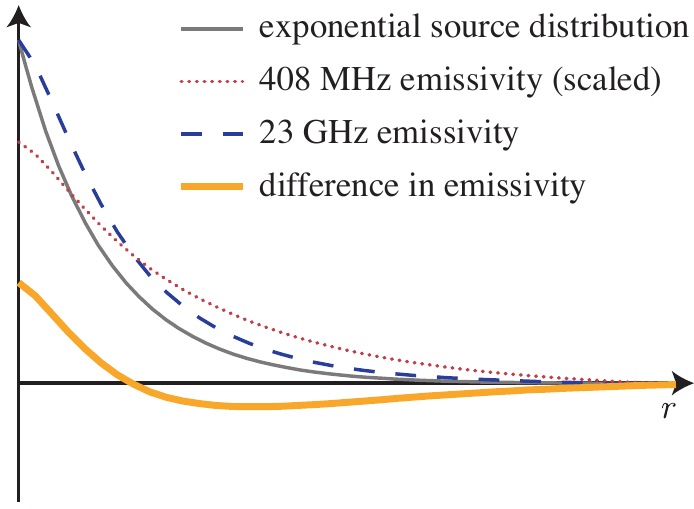}
\caption[]{The (scaled) synchrotron emissivity at 408~MHz and 23~GHz,
  and their difference, for the Lorimer (left panel) and the
  exponential (right panel) source distribution.}
\label{fig:difference}
\end{figure}

To qualitatively understand these results, consider the longitudinal
profile of the synchrotron intensity $I(\ell, b)$; for simplicity let
us constrain ourselves to the galactic plane, i.e. $b \equiv 0$. The
intensity in any direction $\ell$ is given by the integral of the
synchrotron emissivity over the line of sight and this samples the
radial distribution of the relativistic electron density in the range
$r \in [d \sin\ell, \,R]$, where $d$ is the distance of the Sun from
the galactic centre. Since the fitting procedure minimises the square
of the difference in the maps, the sign and size of the residual is
determined not by the absolute difference but by the difference in the
{\em radial slope} of the emissivity $\varepsilon(r)$ at 408~MHz and
the WMAP frequencies. The difference in the slopes reflects the energy
dependence of the electron diffusion --- higher energy electrons lose
their energy more quickly, hence their emissivity traces the source
distribution more closely than does the emissivity of low energy
electrons.

For the pulsar source distribution, the low energy electrons peak at
the galactic centre whereas the high-energy electrons peak further
away (see left panel of Fig.~\ref{fig:difference}). This leads to a
deficit for small radii (translating to small longitudes) and a slight
excess further away from the galactic centre (see also
Fig.~\ref{fig:residual_skymap_Lorimer}). For the exponential source
distribution the radial distribution of synchrotron emissivity is
steeper at higher energies. The template subtraction therefore yields
a residual with an excess around the centre direction and a deficit
further away along the galactic plane (see right panel of
Fig.~\ref{fig:difference}).

We note that the size and morphology of the residual is thus sensitive
not only to the source distribution but also to the parameters of the
diffusion model. For instance, decreasing the Alfv\`en speed below the
value given above reduces the importance of reacceleration, and
therefore effectively limits the number of GeV electrons around the
galactic centre where otherwise energy losses dominate.

\section{A Related Study}

A related study~\cite{McQuinn:2010ju}, published at the same time as
the present work~\cite{Mertsch:2010ga}, also tries to determine
whether the `WMAP haze' could have a purely astrophysical origin and
how to distinguish between this and a DM explanation. As there are
some important differences, we will briefly comment on this work.

At first sight, the astrophysical explanation invoked in
Ref.~\cite{McQuinn:2010ju}, a large number of supernova remnants at
the galactic centre, might look not too different from our model with
the exponential source distribution. However, while we also agree on
the importance of diffusion-loss steepened electron spectra for
producing the haze there is a major difference between our approaches
--- while the authors of Ref.~\cite{McQuinn:2010ju} consider the haze
to be {\em physical}, we argue that it might in fact be an artefact of
the foreground subtraction. Our models are also more constrained
insofar as we reproduce the observed radio emission at 408 MHz and
match the direct measurements of the electron spectrum at the Solar
position. Furthermore, we allow for spatial dependence of the
$\vec{B}$ field, and convection and reacceleration of cosmic ray
electrons, which are all essential in order explain all these datasets
simultaneously.

\section{Conclusion}

We have investigated systematic effects in WMAP foreground subtraction
stemming from the na\"ive extrapolation of the 408 MHz map. To this
end we have considered two illustrative cosmic ray diffusion models
assuming different source distributions, the first one based on a
pulsar survey, and the second one exponential in galactocentric
radius. Both models are able to reproduce the synchrotron radiation at
408 MHz, the locally measured electron flux and are furthermore
consistent with nuclear cosmic ray fluxes and secondary-to-primary
ratios. When our `foreground' 408 MHz map is subtracted from the 23
GHz map, we find a residual whose size and morphology depends on the
source and diffusion model adopted. Thus the energy-dependent
diffusion of relativistic electrons makes the 408~MHz sky map a {\em
  bad} tracer of synchrotron radiation at microwave frequencies, as
had been suspected earlier~\cite{Bennett:2003ca}. Such a template
subtraction produces a residual, which for certain values of the
source and propagation model can reproduce the `WMAP haze' in
intensity and spectrum.

The residual obtained from the exponential source distribution does
not perfectly reproduce the morphology found in
Ref.~\cite{Dobler:2007wv} (although it is {\em not} disk-like but
rather clustered around the galactic centre). However, a quantitative
assessment of the discrepancy is not straightforward, mainly because
Ref.~\cite{Dobler:2007wv} does not provide any objective measure,
e.g. the ellipticity of equal intensity contours. On the other hand,
even the numerical {\tt GALPROP} model we employed for our analysis is
very likely too simple to fully capture the complexity of synchrotron
emission in the Galaxy. For instances, not only the source density but
also the galactic magnetic field is supposed to be correlated with the
galactic spiral arms, which will break the symmetry in $r$ (and hence
in $\ell$) and can therefore considerably modify the
morphology. Furthermore, much of the `diffuse' synchrotron emission
from the disk may originate in the shells of old supernova remnants
which have grown very large in their radiative phase
\cite{Sarkar:1980}. Exactly the same argument concerning the
energy-dependent diffusion length that we applied to the cosmic ray
source distribution can be applied to such localised structures
too. Therefore the 408 MHz survey sky map is not expected to trace the
emission from the latter at higher frequencies either. One can easily
imagine that such localised structures (of which Loop I may be a
nearby example) can at least in part modify the morphology of the
residual and bring the simulated map into agreement with the one
determined from the subtraction of real templates.

%% file: 05epilogue/epilogue.tex
\clearpage
\phantomsection
\addcontentsline{toc}{chapter}{Epilogue}
\chapter*{Epilogue}
\label{chp:epilogue}

Nearly two years after the excesses in the positron fraction and the
total electron-positron flux were reported by the PAMELA and Fermi
satellites respectively, the dark matter explanations proposed for
these `anomalies' seem to be almost ruled out. One the one hand, the
necessarily powerful injection of additional electrons and positrons
from annihilating/decaying dark matter in an extended halo is in
conflict with radio/microwave and $\gamma$-ray observations. One the
other hand, a broad range of astrophysical effects have been
identified which can account for these excesses and would naturally
dominate over plausible dark matter signals. Furthermore, signatures
that have been believed to be typical of dark matter have in fact been
found to be rather generic: features in the total electron-positron
flux at tens and hundreds of GeV can easily be induced by the
discreteness of astrophysical sources. Even sharp shoulders in the
energy spectrum can be due to a nearby source with a hard spectrum,
like a pulsar. Conversely, it appears that an arbitrary spectral
signal, even if it has nothing to do with dark matter, can be fitted
in a dark matter scenario by harnessing the multitude of free or
uncertain model parameters.

Prospects for dark matter indirect detection in charged lepton cosmic
ray channels looks therefore rather bleak at the moment. In the long
run with the advent of advanced satellite detectors such as AMS-02, we
might hope for a better understanding of the astrophysical backgrounds
down to perhaps even the per cent level. This would allow us to
subtract them and uncover possible dark matter contributions, which in
WIMP scenarios would naturally be expected to contribute at this
level. Other high energy channels might, however, provide better
prospects. Photons, for example, do not only sample a much larger
volume than locally measured charged leptons which, as we have seen,
originate within a few kiloparsecs of the Solar system. Even more
importantly, because of their rectilinear propagation, it is possible
to focus on specific targets that have little or no astrophysical
backgrounds, e.g. the Galactic halo at high latitude or prominent
substructures such as dwarf spheroidal galaxies. The Fermi satellite
as well as the forthcoming \v{C}erenkov Telescope Array will provide
important data in this connection.

Of course, the attention given to the lepton excess because of the
claimed connection with particle dark matter has played an important
sociological role in helping the astroparticle community to appreciate
the complexities of galactic cosmic ray physics. The first hints of an
anomalous positron fraction were reported over 30 years ago, but only
a few prescient works have taken this signal seriously and tried to
explain it, e.g. by the contribution from nearby pulsars. We hope very
much therefore that the momentum which cosmic ray physics has gained
of late will foster improvements in their astrophsyical modelling and
help us to be open-minded and alert to other possible anomalies that
might show up in the data. An example could be the recent observation
by the CREAM detector of different spectral indices for protons and
helium nuclei.  Possible interpretations range from different source
populations for protons and heavier nuclei, to secondary production
and acceleration of helium in supernova remnants, in much the same way
as has been suggested in this thesis for secondary electrons and
positrons.

Although dark matter thus seems to be ruled out as an explanation for
the lepton excesses this puzzle is far from being solved. Not only is
it unclear which sources are responsible for the additional positrons,
one might also wonder whether these excesses are local
(i.e. particular to the environment of the Solar system) or global
(i.e. similarly present in other regions of the Galaxy). In this
context, we would like to stress the two-fold role of the nuclear
component of galactic cosmic rays: Firstly, if a \emph{rising}
secondary-to-primary ratio (e.g. B/C) is observed, this will clearly
show that the acceleration of secondaries is the likely cause of the
rise in the positron fraction. Secondly, as the local fluxes of stable
nuclei obtain contributions from larger distances in the Galaxy than
do charged leptons, such a rise would imply that the secondaries are
also produced and accelerated in cosmic ray sources elsewhere in the
Galaxy.

It has also been suggested that anisotropies in the arrival directions
of charged leptons in cosmic rays can help identifying the origin of
the additional positrons. We believe however that the uncertainties
here are too large and that anisotropies are therefore unlikely to
settle the issue. As mentioned above, models that explain the
increasing positron fraction and excessive total electron-positron
flux by a (few) nearby pulsar(s), usually find anisotropies of at most
a few per mil and rely on assumptions concerning the homogeneity of
energy losses and diffusion. It is therefore conceivable that small
variations of the diffusion model can drastically alter the
predictions. A systematic orientation of the local magnetic fields,
for example, can shift the predominant arrival directions in such a
way that no correlation with known objects can be made; or it can lead
to a spurious anisotropy even for an isotropic source distribution.

The only possible way to investigate the high energy electrons and
positrons \emph{elsewhere} in the Galaxy is to look for their
secondary radiation, i.e. radio/microwaves synchrotron radiation on
Galactic magnetic fields and $\gamma$-rays from inverse-Compton
scattering on interstellar radiation fields. As we have seen, a prime
application of this is again the study of dark matter annihilation or
decays into leptons. Of course, this raises the important question of
how to improve the modelling of diffuse galactic backgrounds which are
of general interest also for other fields of physics, e.g.  studies of
the polarisation of the cosmic microwave background (CMB). At the
moment, large efforts are being made by the Fermi collaboration in
explaining the astrophysical contribution to the $\gamma$-ray
sky. Their approach accounts not only for the ICS contribution using
the {\tt GALPROP} model but also uses a very flexible fitting of ISM
densities in galactocentric rings to the $\pi^0$ decay generated
$\gamma$-rays. First results show that the residuals thus obtained are
very small but the obvious question is whether this is because of too
many model degrees of freedom. Theoretical models might not have fewer
parameters but can at least provide some insight into their physical
significance. At the moment, however, such models are not able to
incorporate, let alone predict, local structure such as the radio
`loops' clearly visible in sky maps over a wide range of
frequencies. Furthermore, a high degree of degeneracy is to be
expected, for example between the GCR source distribution and the
structure of the magnetic fields or ISRFs. Attempts at solving the
inverse problem, i.e. determining the three-dimensional distribution
of emissivity would help in extracting the relativistic
electron-positron density throughout the Galaxy as a function of the
assumed radiation backgrounds and would bring us closer to identifying
individual sources of GCRs.

On a more general level and irrespective of whether the explanations
we have suggested for the apparent signals will be confirmed by future
data, the moral is that we need both better understanding of the
astrophysical sources and of the propagation of galactic cosmic rays.

Many of these efforts will probably be data-driven. For example,
Fermi-LAT is presently closing a crucial energy gap in GeV
$\gamma$-rays that IACTs could not address. This will allow looking
for a specific bump in the $\gamma$-ray spectra of SNRs that would
implicate pion production and hence hadronic acceleration
processes. In general, many arbitrary extrapolations from TeV energies
as well as the underlying physics models will be tested. In addition,
the improved statistics with respect to, e.g. the EGRET satellite will
allow sampling of the spectra of even comparatively faint and distant
sources. Furthermore, the superior spatial resolution will allow
and/or necessitate going beyond zero- or one-dimensional source models
and considering the internal structure of the sources. It might turn
out that some common assumptions about the sources are in fact
prejudices and cannot be sustained any longer.

With respect to improving the cosmic ray propagation models, what is
probably needed is a completely new approach. Many important steps
towards realistic models have been taken in the last couple of
years. For example, the inclusion of Klein-Nishina corrections to the
interactions of electrons and positrons with radiation fields has been
recognised to lead to spectral effects at the highest energies. Two
other improvements aim at a more realistic modelling of the ISM and
its effect on cosmic propagation: Firstly, the structure of the
large-scale ordered magnetic field is being given due importance for
synchrotron radiation studies. Secondly, the spatial dependence of
energy loss rate and diffusion coefficients is beginning to be
accounted for in numerical propagation codes. However, what these
models lack is a clear physical intuition for \emph{how} these
quantities should vary across the Galaxy. Here, the hitherto followed
approach of calibrating diffusion models by measurements of local
cosmic ray nuclei necessarily reaches its limits: Stable nuclear
cosmic rays average over a large volume of the cosmic ray halo
(typically several kiloparsecs at GeV energies) and can therefore not
resolve structures at smaller scales. At the highest energies,
however, local sources become more important and therefore the large
bulk of the galactic volume remains inaccessible. Generally speaking,
it is difficult to imagine how an enlarged set of parameters that
necessarily accompanies every more complex propagation model could be
constrained by the same number of observations.

One possible resort could be to devise new sets of observables which
are particularly sensitive to some of the parameters. We have already
encountered one such example in our analysis of the systematic effects
of template subtraction. The degeneracy encountered here was that
between the source distribution and the magnetic field
configuration. If one can neglect exotic contributions, a new
observable like the relative differences of maps at different
frequencies could be used to pin down one particular combination of
source distribution and magnetic fields.

The more radical approach would be to search for a new propagation
model altogether. This would need to provide a self-consistent
explanation for many different mechanisms that today's heuristic
models invoke in an {\it ad hoc} fashion. The combination and
description of cosmic rays, magnetic fields and the interstellar
medium in one framework will require deeper theoretical efforts as
well as immense computational power. One can hardly overestimate the
difficulties such a program would face, nevertheless it would be a
fitting challenge to undertake on the eve of the centenary of the
discovery of cosmic radiation by Victor Hess in 1912.

%% file: 06backmatter/backmatter.tex
\appendix

\chapter{Moments of Functions of Random Variables}
\label{app:Moments}

Suppose we have two random variables, $x$ and $y$ with the probability
densities $f_x(x)$ and $f_y(y)$. A function $\phi(x,y)$ of $x$ and $y$
is also a random variable,
\begin{equation}
z = \phi(x,y) \, , \quad \frac{\mathrm{d} z}{\mathrm{d} x}
= \phi'(x,y) \, .
\end{equation}
Let's further assume that $\phi$ has an inverse function with respect
to $x$,
\begin{equation}
\phi^{-1}(y,z) = x \,, \quad \frac{\mathrm{d} x}{\mathrm{d} z} = \frac{1}{\phi'(\phi(y,z),y)} \, .
\end{equation}

The $m$-th moment of $z$ is then
\begin{align}
\langle z^m \rangle &= \int_{-\infty}^{\infty} \mathrm{d} z \, z^m f_z(z) = \int_{-\infty}^{\infty} \mathrm{d} z \, z^m \frac{\mathrm{d} F}{\mathrm{d} z} = \int_{-\infty}^{\infty} \mathrm{d} z \, z^m \frac{\mathrm{d}}{\mathrm{d} z} \left[ \mathop{\int \int}_{D_z} \mathrm{d} y  \, \mathrm{d} x \, f(x,y) \right] \\
&= \int_{-\infty}^{\infty} \mathrm{d} z \,
z^m \frac{\mathrm{d}}{\mathrm{d}
z} \left[ \int_{-\infty}^{\infty} \mathrm{d}
y \int_{-\infty}^{\phi^{-1}(y,z)} \mathrm{d} x \, f(x,y) \right] \\
&= \int_{-\infty}^{\infty} \mathrm{d}
y \int_{-\infty}^{\infty} \mathrm{d} z \,
z^m \underbrace{\frac{\mathrm{d}}{\mathrm{d}
z} \left[ \int_{-\infty}^{\phi^{-1}(y,z)} \mathrm{d} x \,
f(x,y)\right]}_{
\begin{array}{c} 
	= \frac{ \mathrm{d} \phi^{-1}}{\mathrm{d} z} \frac{\mathrm{d}}
	{ \mathrm{d} \phi^{-1}} \int_{-\infty}^{\phi^{-1}(y,z)} \mathrm{d}
	x f(x,y) \\ = \frac{1}{\phi'} f(\phi^{-1},y)
\end{array}} \\
&= \int_{-\infty}^{\infty} \mathrm{d}
y \int_{-\infty}^{\infty} \mathrm{d} x \frac{\mathrm{d} z}{\mathrm{d}
x} \phi^m(x,y) \frac{1}{\phi'} f(x,y) \\
&= \int_{-\infty}^{\infty} \mathrm{d}
y \int_{-\infty}^{\infty} \mathrm{d} x \, \phi^m (x,y) f(x,y) \, .
\end{align}

\chapter{Determination of Spectral Indices From Synchrotron Sky Maps}
\label{sec:SpecIndicesSkyMaps}

In general, a spectral index $\beta(\vec{x})$ between two different
frequencies, $\nu_1$ and $\nu_2$, can be defined for each given pixel
$\vec{x}$ by assuming a power law behaviour of the specific intensity,
$I(\nu, \vec{x})$,
\begin{equation}
\label{eqn:DefSpecInd}
\frac{I(\nu_2, \vec{x})}{I(\nu_1, \vec{x})} = \left( \frac{\nu_2}{\nu_1} \right)^{\beta(\vec{x})} \, .
\end{equation} 
However, it turns out that the template method applied to the WMAP
data and the 408~MHz sky map leads to a residual with {\em negative}
intensities for some pixels (see e.g., Fig. 6 of
Ref. \cite{Dobler:2007wv}), partly due to over-subtraction and partly
because the sky maps are mean-subtracted. We also find negative
intensities for some pixels when applying the template subtraction to
our mock microwave data and radio template. This does not necessarily
imply that the residual is not physical but that a global offset
$\Delta I (\nu)$ exists between the residual intensity, $I'$, as
determined from the template subtraction and the intensity of the {\it
actual,} possibly physical residual, $I$:
\begin{equation}
\Delta I (\nu) \equiv I (\nu, \vec{x}) - I' (\nu, \vec{x}) \, .
\end{equation} 
This makes the determination of the spectral index non-trivial.

At first sight, the analysis presented in Ref. \cite{Dobler:2007wv}
seems to avoid this difficulty by determining the average spectral
index in the region south of the galactic centre from the average
ratio $r'$ of the intensities at two different frequencies $\nu_1$ and
$\nu_2$, e.g. $\nu_1 = 23 \, \text{GHz}$ and $\nu_2 = 33 \,
\text{GHz}$. This ratio can be determined from a scatter plot of the
pairs of residual intensities $\{I'_{\nu_1}, I'_{\nu_2}\}$ (as
determined from the template subtraction), to which a straight line,
$I'_{\nu_2}(I'_{\nu_1}) = r' I'_{\nu_1} + \hat{I}'_{\nu_2}$, is
fitted, allowing for the ordinate offset $\hat{I}'_{\nu_2}$ because of
the unknown global offset $\Delta I (\nu)$. The average spectral index
$\langle \beta'_{\nu_1,\nu_2} \rangle$ defined by this procedure is
then simply $\log{(r')}/\log{(\nu_2/\nu_1)}$. Alternatively, if the
spectral index is determined from a scatter plot of the {\em actual}
residual intensities $\{I_{\nu_1}, I_{\nu_2}\}$, then there is no
ordinate offset, so the straight line is $I_{\nu_2}(I_{\nu_1}) = r
I_{\nu_1}$ and the {\em actual} average spectral index $\langle
\beta_{\nu_1,\nu_2} \rangle = \log{(r)}/\log{(\nu_2/\nu_1)}$.

In general, these two descriptions cannot be expected to give a
similar spectral index. Even assuming that with an appropriate `haze'
template $\vec{h}$ the amount of over-subtraction is much smaller than
the offset due to the use of mean-subtracted maps, the answer is in
general different. In this case, the offset is simply the mean over
the $n$ pixels, $\Delta I (\nu) = \langle I (\nu, \vec{x})
\rangle$. The coordinate system $\{I'_{\nu_1}, I'_{\nu_2}\}$ is
therefore centred at the centre of gravity of the data $\{I(\nu_1,
\vec{x}), I(\nu_2, \vec{x})\}$, and the ordinate offset
$\hat{I}'_{\nu_2}$ is zero. As usual, the slope of the linear
regression $I'_{\nu_2}(I'_{\nu_1}) = r' I'_{\nu_1}$ is
\begin{equation}
  r' = \frac{\sum_i I'_{\nu_1}(\vec{x}_i) I'_{\nu_2}(\vec{x}_i) -
n \langle I'_{\nu_1}(\vec{x}_i) \rangle
\langle I'_{\nu_2}(\vec{x}_i) \rangle}{\sum_i 
I'^2_{\nu_1}(\vec{x}_i) - n \langle
I'_{\nu_1}(\vec{x}_i) \rangle^2} \, .
\end{equation}
Unless the covariance of $I'_{\nu_1}$ and $I'_{\nu_2}$ is much larger
than the product of their mean values, which is for example the case
if the spectral index is constant in the region of interest, this is
in general different from the slope $r$ of the straight line
$I_{\nu_2}(I_{\nu_1}) = r I_{\nu_1}$,
\begin{equation}
r = \frac{\sum_i I'_{\nu_1}(\vec{x}_i) I'_{\nu_2}(\vec{x}_i)}{\sum_i
I'^2_{\nu_1}(\vec{x}_i)} \, .
\end{equation}
However, since we cannot determine the offset $\Delta I(\nu)$ from
data, we need to {\em define} an offset $\Delta I (\nu)$. We choose it
to be:
\begin{equation}
\Delta I(\nu) = \min_{\vec{x}} \left[ I'(\nu, \vec{x}) \right] \, ,
\end{equation} 
such that the intensity is always positive, allowing us to define the
spectral index in each pixel. (The exact value chosen for $\Delta
I(\nu)$ is actually $(1 + 10^{-3}) \min \left[ I'(\nu, \vec{x})
\right]$ to prevent the spectral index from diverging in the pixel
where $I(23 \, \text{GHz})$ attains its minimum.)

\clearpage
\phantomsection
\addcontentsline{toc}{chapter}{Bibliography}
\begin{singlespace}
\bibliography{06backmatter/bibliography}{}
\bibliographystyle{utphys}
\end{singlespace}